\def\asec{$^{\prime\prime}$}
\def\Lya{Ly$\alpha$~}

\def\IRACcol{[$3.6\umu{\rm m}] - [4.5\umu{\rm m}$]~}
\def\chone{[$3.6\umu{\rm m}$]~}
\def\chtwo{[$4.5\umu{\rm m}$]~}
\documentclass[useAMS,usenatbib]{mn2e}
\usepackage{upgreek}
\usepackage{graphicx}
\usepackage{url}
\usepackage{amsmath}
\usepackage{amssymb}
\usepackage{subfig}
\usepackage{caption}
\usepackage{verbatim}
\usepackage{color, soul}
\sethlcolor{yellow} %% Then do \hl{ }

\title[The bright end of the galaxy luminosity function at $z \simeq 7$]{The bright end of the galaxy luminosity function at ${\bf z \simeq 7}$: before the onset of mass quenching?}  
\author[R. A .A. Bowler et al.]{R. A. A. Bowler$^{1}$\thanks{E-mail:
raab@roe.ac.uk}, J. S. Dunlop$^{1}$, R. J. McLure$^{1}$, A. B. Rogers$^{1}$, H. J. McCracken$^2$, \and
B. Milvang-Jensen$^{3}$, H. Furusawa$^{4}$, J. P. U. Fynbo$^{3}$, Y. Taniguchi$^{5}$, J. Afonso$^{6,7}$, \and M. N. Bremer$^{8}$, O. Le F\`{e}vre$^{9}$\\ 
$^{1}$SUPA\thanks{Scottish Universities Physics Alliance}, Institute for Astronomy, University of Edinburgh, Royal Observatory, Edinburgh, EH9 3HJ \\
$^{2}$Institut d'Astrophysique de Paris, UMR 7095 CNRS, Universit\'{e} Pierre et Marie Curie, 98 bis Boulevard Arago, 75014 Paris, France \\
$^{3}$Dark Cosmology Centre, Niels Bohr Institute, University of Copenhagen, Juliane Maries Vej 30, 2100 Copenhagen, Denmark \\
$^{4}$Astronomical Data Center, National Astronomical Observatory of Japan, Mitaka, Tokyo 181-8588, Japan \\
$^{5}$Research Institute for Space and Cosmic Evolution, Ehime University, 2-5 Bunkyo-cho, Matsuyama 790-8577, Japan \\
$^{6}$Centro de Astronomia e Astrof\'{\i}sica da Universidade de Lisboa, Observat\'{o}rio Astron\'{o}mico de Lisboa, Tapada da Ajuda, 1349-018, Portugal \\
$^{7}$Faculdade de Cincias da Universidade de Lisboa, Ed. C8. Campo Grande, 1749-016, Lisbon, Portugal \\
$^{8}$H H Wills Physics Laboratory, Tyndall Avenue, Bristol BS8 1TL \\
$^{9}$Laboratoire d'Astrophysique de Marseille, CNRS and Aix-Marseille Universit\'{e}, 38 rue Fr\'{e}d\'{e}ric Joliot-Curie, 13388 Marselle Cedex 13, France \\
}

\begin{document}
%\vspace{-1in}
\date{}

\pagerange{\pageref{firstpage}--\pageref{lastpage}} \pubyear{2013}

\maketitle
\label{firstpage}

\begin{abstract}
We present the results of a new search for bright star-forming galaxies at redshift $z\simeq 7$ within the UltraVISTA DR2 and UKIDSS UDS DR10 data, which together provide 1.65\,deg$^2$ of near-infrared imaging with overlapping optical and \emph{Spitzer} data.
Using a full photometric-redshift analysis to identify high-redshift galaxies and reject contaminants, we have selected a sample of 34 luminous ($-22.7<M_{UV}<-21.2$) galaxies with $6.5<z<7.5$. 
Crucially, the deeper imaging provided by UltraVISTA DR2 confirms all of the robust objects previously uncovered by~\citet{Bowler2012}, validating our selection technique. 
Our new expanded galaxy sample includes the most massive galaxies known at $z\simeq 7$, with $M_{\star}\simeq 10^{10}{\rm M_{\odot}}$, and the majority are resolved, consistent with larger sizes ($r_{1/2}\simeq 1-1.5$\,kpc) than displayed by less massive galaxies.
From our final robust sample, we determine the form of the bright end of the rest-frame UV galaxy luminosity function (LF) at $z\simeq 7$, providing strong evidence that it does not decline as steeply as predicted by the Schechter-function fit to fainter data.
We exclude the possibility that this is due to either gravitational lensing, or significant contamination of our galaxy sample by active galactic nuclei (AGN). 
Rather, our results favour a double power-law form for the galaxy LF at high redshift, or, more interestingly, a LF which simply follows the form of the dark-matter halo mass-function at bright magnitudes. 
This suggests that the physical mechanism which inhibits star-formation activity in massive galaxies (i.e. AGN feedback or some other form of `mass quenching') has yet to impact on the observable galaxy LF at $z\simeq 7$, a conclusion supported by the estimated masses of our brightest galaxies which have only just reached a mass comparable to the critical `quenching mass' of $M_{\star}\simeq 10^{10.2}\,{\rm M_{\odot}}$ derived from studies of the mass function of star-forming galaxies at lower redshift.

\end{abstract}

\begin{keywords}galaxies: evolution - galaxies: formation - galaxies: high-redshift.
\end{keywords}

\section{Introduction}

The study of galaxies at high redshift is crucial for understanding the early and subsequent stages of galaxy evolution in the Universe. 
Within the last decade, the number of galaxies known at $z > 6$ has increased to samples of hundreds, led by observations taken with the Wide Field Camera 3 on the \emph{Hubble Space Telescope} (WFC3/\emph{HST}).
The key feature of WFC3 that makes it so successful at detecting high-redshift galaxies, is the unrivalled sensitivity in the near-infrared, which allows the detection of Lyman-break galaxies (LBGs) at $z > 6.5$ by the redshifted rest-frame UV light from these star-forming galaxies.
The accurate selection of LBGs at high redshifts relies on a measurement of the strong spectral break at the wavelength of the Lyman-$\alpha$ line (1216\AA), produced by absorption from the integrated neutral Hydrogen along the line of sight.
The resulting spectral energy distribution (SED) can then be identified in multiwavelength imaging as an ``optical-dropout'' galaxy with either colour-colour selection or a SED fitting analysis.

The deepest near-infrared image ever taken in the \emph{Hubble} Ultra Deep Field (HUDF), now reaches depths of ${\rm m}_{\rm AB} \simeq 30$ over 4.5 arcmin$^2$ (UDF12 observing programme;~\citealp{Koekemoer2012}).
The imaging confirms that the low-luminosity, early galaxies uncovered in the HUDF are compact (half-light radius, $r_{1/2} < 0.5\,$kpc;~\citealp{Ono2013}) and have similar colours to local star-forming galaxies (rest-frame UV slope $\beta_{} \simeq -2$, where $F_{\lambda} \propto \lambda^{\beta}$;~\citealp{Dunlop2013}). 
The detection of an increasing number of galaxies at $z = 6-8$, including the first sample of galaxies at $ z > 9$~\citep{Ellis2013, Oesch2013}, within the UDF12 and other surveys~\citep{Zheng2012, Coe2013} has allowed the determination of the rest-frame UV luminosity function at high redshift.
The LF characterises the number density of galaxies per comoving volume element as a function of luminosity, and hence is an important measurement for charting the evolution of galaxies (e.g.~\citealp{Bouwens2011a}). 
A commonly used parameterisation of the LF, which well describes the number densities of galaxies at low redshift, is the Schechter function, where $\phi(L) = \phi^*(L/L^*)^\alpha e^{-L/L^*}$.
The Schechter function form exhibits a power-law slope to faint luminosities described by the index $\alpha$, and an exponential cut-off at luminosities exceeding the characteristic luminosity $L^*$.
The extremely faint galaxies ($M_{\rm UV} > -17$) detected within the ultra-deep imaging of the UDF12 programme, have constrained the faint-end slope of the $z \simeq 7$ LF to be very steep with $\alpha = -1.9$ at $z = 7$~\citep{Mclure2013a, Schenker2013}.
The slope of the faint-end of the LF is key for ascertaining the role LBGs play in reionizing the Universe, as only by extrapolating the number densities of galaxies beyond the faintest galaxy detected even in the HUDF, can the ionising photon budget be met by early galaxies~\citep{Robertson2013, Salvaterra2011}.

The form of the LF, and the characteristic break luminosity, cannot be accurately constrained using galaxy samples from the HUDF alone however; samples of intrinsically-rarer bright galaxies are required.
There are several key programmes from~\emph{HST} that have detected significantly brighter galaxies at $z \simeq 7$ around the apparent break luminosity; the Cosmic Assembly Near-IR Deep Extragalactic Legacy Survey (CANDELS,~\citealp{Grogin2011, Koekemoer2011}) is extremely powerful at detecting galaxies analogously to those in the HUDF, the Cluster Lensing and Supernova Survey with Hubble (CLASH,~\citealp{Postman2012,Zheng2012, Coe2013}) uses foreground clusters to detect gravitationally lensed galaxies, and the Brightest of the Reionizing Galaxies (BoRG,~\citealp{Trenti2011, Bradley2012b}) programme has allowed the detection of bright galaxies specifically at $z = 8$.
Despite the clear success of these \emph{HST} programmes at selecting galaxies at $z > 6.5$, the brightest $z \simeq 7$ galaxies detected in the CANDELS imaging to date are only slightly brighter than the characteristic luminosity ($L \sim 3L^*$,~\citealp{Mclure2013a}).
To detect the very brightest galaxies, degree-scale surveys are needed, and these can only be completed efficiently from the ground with wide field-of-view near-infrared cameras.

Theoretically, the form and redshift dependence of the LF contains important information on the key physical processes that govern early galaxy formation and subsequent evolution.
To match the observed luminosity- and mass-functions at $z = 0$, simulations of the build-up of galaxies require that, at some stage, the growth of galaxies in the most massive dark-matter haloes be suppressed by some mechanism, such as feedback from a central AGN. 
In effect, the challenge is to match the steep exponential decline of the galaxy stellar mass (and luminosity) functions at high mass/luminosity, as parameterised through the Schechter function described above. 
Interestingly, recent studies of the stellar mass function of {\it star-forming galaxies} have now shown that the characteristic mass above which this steep decline sets in appears to be essentially independent of redshift out to $z \simeq 3$ (e.g.~\citealp{Ilbert2013}). 
This has led some authors to infer the presence of a characteristic ``quenching mass'', i.e. a stellar mass above which a galaxy is likely to have its star-formation activity strongly suppressed by some physical mechanism. 
As discussed in ~\citet{Peng2010}, current data indicate that the quenching mass above which galaxies rapidly cease forming stars and leave the ``main-sequence'' of star-forming galaxies is $M_{\star} \sim 10^{10.2}\,{\rm M}_{\sun}$. 
If this mass quenching, whatever its physical origin, really does set in at a physical threshold which is independent of redshift, then we might reasonably expect the form of the galaxy UV LF to start to diverge from a simple Schechter function at very early epochs (when very few galaxies will have grown to the relevant stellar mass). 
The ability to test such key ideas, and potentially better constrain the (still unclear) physical origin of  ``mass quenching'', provides additional strong motivation for determining the form of the bright end of the galaxy LF at the highest redshifts.

Further interest in the detailed properties of $z \simeq 7$ galaxies has been generated by the follow-up~\emph{HST} and Atacama Large Millimeter/Submillimeter Array imaging of the spectroscopically confirmed Lyman-$\alpha$ emitter (LAE) at $z = 6.595$~\citep{Ouchi2013}, nicknamed `Himiko', that was first discovered by~\citet{Ouchi2009a}.
When observed at ground-based resolution this galaxy appears as a single bright extended source with ${\rm m}_{\rm AB} \simeq 25$ (at $\lambda \simeq 1\umu$m), however ~\emph{HST}/WFC3 imaging reveals the galaxy to be an apparent triple merger system, where each component has roughly the characteristic luminosity $L^*$.
Although an extreme galaxy, Himiko illustrates the potential for detailed study of rare and bright $z \simeq 7$ galaxies that can only be efficiently detected in ground-based narrow- or broad-band survey fields.

Here we extend the work presented in~\citet{Bowler2012} using the second data release (DR2) of UltraVISTA, which provides deeper imaging in the $Y$, $J$, $H$ and $K_s$-filters, in strips covering approximately half of the full 1.5 deg$^2$ of the UltraVISTA DR1 data~\citep{McCracken2012}.
The improved photometry over 70\% of the field searched in~\citet{Bowler2012} allows us to check the previous candidates and hence validate our selection methodology.
We also incorporate the UKIDSS Ultra Deep Survey (UDS), which has a comparable depth and area of $J$, $H$ and $K$-band imaging to the UltraVISTA DR2.
By combining the deep near-infrared survey data with the optical and mid-infrared data currently available in the UltraVISTA and UDS fields, we have assembled the widest area (1.65 deg$^2$) of imaging available for the robust selection of $z \simeq 7$ Lyman-break galaxies.

We begin with a summary of the multiwavelength data utilised here from the UltraVISTA and UDS fields in Section~\ref{sect:data}, followed by the details of our candidate selection in Section~\ref{sect:candsel}.  
The resulting sample of galaxies is presented in Section~\ref{sect:cand}, with physical properties derived from our SED fitting analysis in Section~\ref{sect:properties}.
In Section~\ref{sizes} we investigate the sizes of the galaxies in our sample, including an analysis of \emph{HST} imaging of four galaxies in our sample that lie within the region of the UltraVISTA field imaged as part of the CANDELS programme.  
We calculate the binned luminosity function for our sample in Section~\ref{sect:lf}, where we also carefully consider the potential effect of gravitational lensing.
We discuss our results and compare them to previous work at $z = 5$ and $z = 6$ in Section~\ref{sect:discussion}, which also includes a prediction of the level of contamination of our sample by high-redshift quasars.
The astrophysical implications of our results are considered further in Section~\ref{sect:implications}, and our conclusions are summarised in Section~\ref{sect:conc}.
All magnitudes quoted are in the AB system~\citep{Oke1974,Oke1983} and we assume a cosmology with $H_{0}$~=~70~kms$^{-1}$Mpc$^{-1}$, $\Omega_{\rm m}$~=~0.30 and $\Omega_{\Lambda}~=~0.70$ throughout.

%% COSMOS field overlapping image
\begin{figure}                                                                                                                             
\includegraphics[width = 0.5\textwidth,  trim = 0cm 1.5cm 0.5cm 2cm, clip = true]{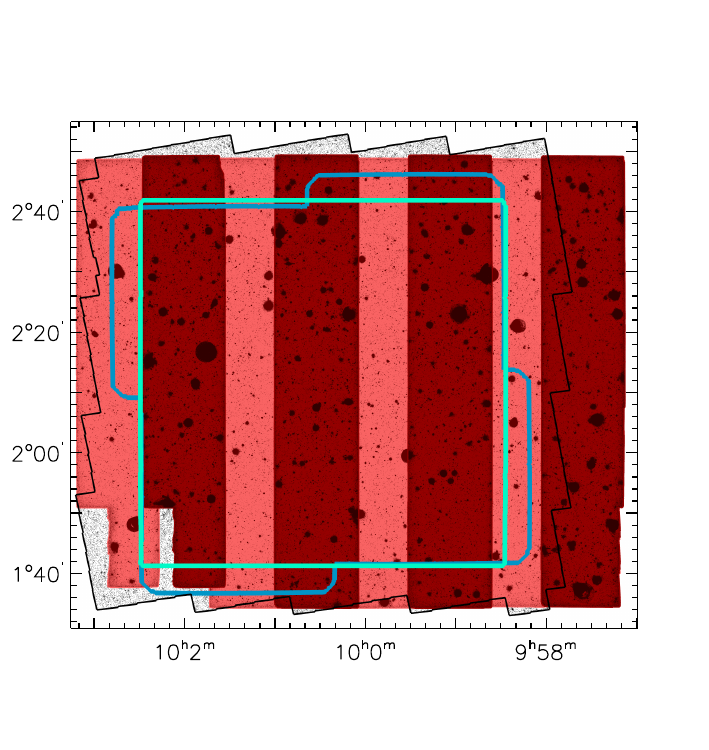}
\caption{The footprint of the UltraVISTA/COSMOS field showing the key multiwavelength data used here.  
The large red rectangle shows the year-one 1.5\,deg$^2$ of near-infrared imaging from UltraVISTA DR1, with the deeper strips comprising the second data release shown in dark red.
The Subaru $z'$-band mosaic, formed from four individual Suprime-Cam pointings, is indicated by the blue outline, and the 2\,deg$^2$ of \emph{HST}/ACS $I_{814}$-band imaging from the COSMOS survey is shown as the large jagged outline.
The overlap with the DR2 strips in dark red and the central green square, which is the $1$\,deg$^2$ area of a single pointing of CFHT/MegaCam, defines the $\simeq\!0.7$\,deg$^2$ area searched in this study. }
\label{fig:UVISTAcoverage}
\end{figure}

\section{Data}\label{sect:data}

The two multiwavelength survey fields analysed in this paper contain a wealth of observations from X-ray to radio wavelengths.
In the following section we describe the specific datasets utilised here for the selection of $ z > 6.5$ galaxies, most importantly the near-infrared data from the UltraVISTA DR2 and UDS DR10.
The coverage maps of the different wavelength data are shown in Figs.~\ref{fig:UVISTAcoverage} and~\ref{fig:UDScoverage}, and a summary of the available broad- and narrow-band filters utilised to image each field can be found in Table~\ref{table:depths}.
The final area of overlapping multiwavelength imaging over the two fields, excluding large stellar diffraction haloes that were masked, comprised 1.65 deg$^2$, with 0.62 deg$^2$ from the UltraVISTA `ultra-deep' survey, 0.29 deg$^2$ from the UltraVISTA `deep' component and 0.74 deg$^2$ in the UDS field.

\subsection{The COSMOS/UltraVISTA field}

\subsubsection{UltraVISTA near-infrared imaging}

The analysis presented in this paper relies on the first and second data releases of the ongoing UltraVISTA survey\footnote{\url{http://www.eso.org/sci/observing/phase3/data_releases/}}, which consists of $Y, J, H$ and $K_s$ imaging with the Visible and Infrared Camera (VIRCAM) on the VISTA telescope within the Cosmological Evolution Survey (COSMOS) field.
The first data release, described in detail by~\citet{McCracken2012}, provided near-infrared imaging over the maximum area of the programme covering $1.5\,$deg$^2$.
DR2 provides deeper data in strips that cover $\sim$70\% of the the full field as shown in Fig.~\ref{fig:UVISTAcoverage}.
Integration times for the DR2 $Y$, $J$, $H$ and $K_s$ bands range from 29-82 hours per pixel, compared with 11-14 hours per pixel from DR1.
Throughout this paper we refer to the DR2 imaging within the strips as the `ultra-deep' part of the survey, and the DR1 imaging over the full field as the `deep' part.

\subsubsection{Auxiliary optical and mid-infrared imaging}

The auxiliary data used in this paper is described in full by~\citet{Bowler2012}, but here we briefly describe the key datasets that are shown in Fig.~\ref{fig:UVISTAcoverage}.
The UltraVISTA survey lies within the  multiwavelength imaging taken as part of the COSMOS survey~\citep{Scoville2007}, which covers a total of 2 deg$^2$ on the sky.
Specifically we use optical imaging from the CFHTLS T0006 data release, which defines the maximal area of our search centred on RA $10^{\rm h} 00^{\rm m} 28^{\rm s}.00$, Dec. $+2^{\circ}12'30''$, with additional deep $z'$-band data from Subaru Suprime-Cam.
Mid-infrared imaging over the COSMOS field by \emph{Spitzer}/IRAC exists from two programmes; the \emph{Spitzer} Extended Deep Survey (SEDS;~\citealp{Ashby2013}) and the \emph{Spitzer} Large Area Survey with Hyper-Suprime-Cam (SPLASH, PI: Capak).
The SPLASH data consist of 438 individual exposures in the 3.6$\umu$m and 4.5$\umu$m bands, available as calibrated Level-2 files on the \emph{Spitzer} Legacy Archive.
We created a mosaic of the SPLASH images by first background subtracting the frames using {\sc SExtractor} with a large background mesh size, before combining the frames using the software package {\sc Swarp}.
The SEDS data was also background subtracted and incorporated into the SPLASH mosaic using {\sc Swarp}.
The photometric and astrometric accuracy was confirmed by comparing to the shallower \emph{Spitzer}/IRAC imaging across the field from the S-COSMOS survey~\citep{Sanders2007}.
Finally the field is also covered to single-orbit depth in the $I_{814}$-band by the \emph{HST} Advanced Camera for Surveys (ACS;~\citealp{Koekemoer2007, Scoville2007a, Massey2010}).

\subsubsection{Data processing and consistency}\label{dataUVISTA}

All images were resampled to the pixel grid of the CFHTLS data (0.186-arcsec/pixel) using the {\sc iraf} package {\sc sregister}, once the astrometric solution had been matched to that of the UltraVISTA $Y$-band image using the {\sc iraf} package {\sc ccmap}.

Zeropoints of the full set of multiwavelength imaging were checked by inspecting the colours of flat-spectrum objects defined by the colour bridging the central band.  
For example, a sample of objects was extracted with aperture corrected flat $z' - J$ colours (e.g. $|z' - J| < 0.05$), and the $z' - Y$ colours of these objects were examined with the expectation that they should also be close to zero.
The optical bands were found to have zeropoint offsets of $ < 0.05 \,$mag, however when comparing the DR1 UltraVISTA data utilised in~\citet{Bowler2012} to the optical imaging we found the $Y$-band magnitudes to be brighter than expected by $ 0.06 \,$mag.
The re-reduction of the full 1.5\,deg$^2$ field encompassing the strips for the second data release, which included improved $Y$-band calibrations, entirely compensates for this observed colour difference in the DR1.
Hence, when comparing the depths of the full-field of imaging in Table~\ref{table:depths} and the magnitudes of the 10 objects selected in~\citet{Bowler2012} in Section~\ref{Bowler2012}, note that there are small changes in the $Y$-band photometry as a result of this zeropoint change.
The zeropoints of the individual Subaru $z'$-band tiles were also adjusted slightly ($\Delta {\rm m} \simeq 0.04$) to be consistent with the single pointing of $z$-band imaging from CFHT/MegaCam.

\subsection{The UKIDSS Ultra Deep Survey field}

%% coverage map of the UDS field
\begin{figure}                                                                                                                              
\includegraphics[width = 0.5\textwidth,  trim = 0cm 0cm 0.5cm 0cm, clip = true]{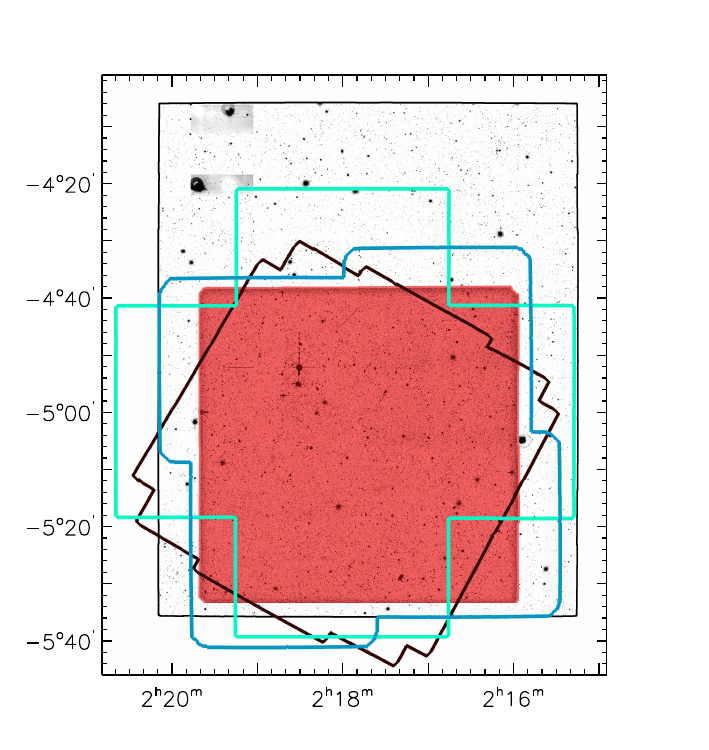}
\caption{
The footprint of the UDS field, showing the UKIRT near-infrared data as the red square sitting within the large rectangle of $Y$-band imaging from the VISTA VIDEO survey.
Data from Subaru Suprime-Cam are shown as the green and blue outlines, where the blue outline defines a $z'$-band mosaic of four pointings as in Fig.~\ref{fig:UVISTAcoverage}.
The green cross-shaped outline shows the $B, V, R, i$ and $z'$-imaging from the original Subaru programme~\citep{Furusawa2008}, where the mosaic is formed from five separate pointings.
Finally, the footprint of \emph{Spitzer} data from SpUDS is shown as the black diamond.
The total area of overlapping Subaru optical and UKIRT near-infrared imaging is $\simeq$ 0.8 deg$^2$.
}
\label{fig:UDScoverage}
\end{figure}

\subsubsection{UKIRT and VISTA near-infrared imaging}

In this study, we use the 10th data release (DR10) of near-infrared imaging in the UDS field, from the UKIRT Infrared Deep Sky Survey (UKIDSS; ~\citealp{Lawrence2007}).
The UKIDSS UDS  consists of deep imaging ($m_{\rm AB} \sim 25$, $5\sigma$, see Table~\ref{table:depths}) in the $J, H$ and $K$-bands over a total area of 0.8 deg$^2$ centred on RA 02$^{\rm h}$17$^{\rm m}$48$^{\rm s}$, Dec. $-05^\circ$05$^\prime$57\asec (J2000).
Data release 10 was made public in January 2013 and is available on the WFCAM science archive\footnote{\url{http://surveys.roe.ac.uk/wsa/}}.

The UDS field lies within the \emph{XMM-Newton} large-scale structure field, where there exists $Y$-band imaging from the VISTA Deep Extragalactic Observations survey (VIDEO;~\citealp{Jarvis2013}) as can be seen in Fig.~\ref{fig:UDScoverage}.
The VIDEO survey is a public survey with VISTA that aims to cover 12\,deg$^2$ in the $Z, Y, J, H$ and $K_s$-bands over three separate fields.
Imaging in the $Y$-band is key for the robust selection of $z \simeq 7$ galaxies in the UltraVISTA and UDS datasets, as the filter straddles the position of the spectral break and hence can separate genuine high-redshift galaxies from dwarf stars and low-redshift galaxy contaminants that can have identical, red, $z'- J$ colours.
Furthermore, the presence of cross-talk artefacts (described further in Section~\ref{udsselection}) in the UKIRT $J$, $H$ and $K$ imaging, makes the presence of a detection in a bluer band independent from the UKIRT data an essential condition for confirming the reality of the high-redshift candidates.
When compared to the UltraVISTA $Y$-band imaging however, the current release of the VISTA VIDEO data is substantially shallower by 1 mag (see Table~\ref{table:depths}), which reduces the capabilities of the data for selecting $z \sim 7 $ sources and makes the selection function for the UDS field different to the UltraVISTA field.

\subsubsection{Auxiliary optical and mid-infrared imaging}

Optical imaging over the field was provided by the Subaru Suprime-Cam as part of the Subaru/\emph{XMM-Newton} Deep Survey (SXDS;~\citealp{Furusawa2008}).
We also obtained additional $z'$-band data in four Suprime-Cam pointings, each with $8-15$ hrs of integration time (Furusawa et al. in preparation).
The astrometry of the individual tiles was matched to that of the UKIRT $J$-band image using the {\sc iraf} package {\sc ccmap}.
The tiles were background subtracted using {\sc SExtractor}~\citep{Bertin1996} and combined into a mosaic with the software {\sc swarp}~\citep{Bertin2002}, where overlapping regions were combined with the {\sc weighted} keyword, using weight maps produced by {\sc SExtractor}.
Finally, a science image on the pixel scale of the binned UKIRT imaging (0.2684-arcsec/pixel) was created from the mosaic (on the native Subaru pixel size of 0.202-arcsec/pixel) using the {\sc iraf} package {\sc sregister}.

In addition to the broad-band filters presented above, we also included data taken with the $N\!B921$ filter on the Subaru Suprime-Cam in our analysis~\citep{Sobral2012}.
The $N\!B921$ filter is positioned to the red side of the Subaru Suprime-Cam $z'$-band filter and hence allows better constraints on the photometric redshift for objects at $6.5~<~z~<~7.0$.
The individual Suprime-Cam tiles were combined into a mosaic covering the full UDS field using the same method as for the $z'$-band mosaic described above.

The UDS field is covered by \emph{Spitzer}/IRAC data from the \emph{Spitzer} UKIDSS Ultra Deep Survey (SpUDS) programme (PI Dunlop), which has comparable depth to S-COSMOS (see Table~\ref{table:depths}) with an integration time per pointing of 1200 seconds.
The central $\sim600\,$arcmin$^2$ of the UDS field is covered by deeper \emph{Spitzer} imaging from SEDS, and we used a mosaic that included both the SpUDS and SEDS data for our analysis\footnote{\url{http://www.cfa.harvard.edu/SEDS/data.html}}, taking into account the varying depths across the field using local depths.

\subsubsection{Data processing and consistency}

All images were mapped onto the astrometric solution and pixel grid of the binned UKIRT $J$-band imaging, with a pixel scale of 0.2684 arcsec/pixel.
Consistency within $ < 0.05$ mag was found between the zeropoints, with the exception being the VISTA VIDEO $Y$-band image which we found to be $\sim0.1$ mag brighter than predicted from the sample of objects with flat $z'-J$ colours.
We also observed an identical offset when comparing magnitudes with the Very Large Telescope/HAWK-I $Y$-band imaging in the field (Fontana et al. in preparation).
Hence we shifted the zeropoint of the VIDEO imaging to produce fainter magnitudes by 0.1 mag, with the expectation that future reduction of the data will largely correct this offset with improved calibration, as was the case with the UltraVISTA $Y$-band data.

\subsection{Image depths}\label{imagedepths}

%% Table of depths
\begin{table*}
\caption{The median 5$\sigma$ limiting magnitudes for the relevant optical and near-infrared data used in this study, obtained from the median of local depths calculated from apertures placed in blank regions of each image (see Section~\ref{imagedepths} for more details).  
All ground-based magnitudes were calculated within the 1.8-arcsec diameter circular aperture used for photometry here.
For the \emph{HST}/ACS $I_{814} $ depth, we used a 0.6-arcsec diameter aperture and the IRAC $3.6\,\mu$m and $4.5\,\mu$m values were calculated in a 2.8-arcsec diameter aperture.
We present the depths of the near-infrared data from UltraVISTA separately for the `ultra-deep' and `deep' parts of the survey. 
Note that the depths for the `deep' part shown here are $\sim 0.4$ mags deeper than the results from the original data release presented in~\citet{Bowler2012} and~\citet{McCracken2012}.
This is a result of improvements in the photometric calibration of the VISTA data, the smaller apertures used (1.8-arcsec diameter as opposed to 2-arcsec results presented previously) and an improved global depth derived from the local depths.}

\begin{tabular}{ l | c c l l | c l }
     \hline
     
     \multicolumn{4}{l}{COSMOS/UltraVISTA} & \multicolumn{3}{l}{UKIDSS UDS}\\
    
     \hline
 Filter  & m$_{5\sigma, \rm AB}$  & m$_{5\sigma, \rm AB}$ &Source & Filter  & m$_{5\sigma, \rm AB}$ & Source \\
  
  & deep & ultra-deep & & & &  \\
  \hline
  $u^*$ & 27.2 &  & CFHT/MegaCam & $B$ & 27.9 & Subaru/Suprime-Cam\\
  $g$ & 27.3  & & CFHT/MegaCam & $V$ & 27.6 & Subaru/Suprime-Cam\\
  $r$ &  27.0 &&  CFHT/MegaCam & $R$ & 27.3 & Subaru/Suprime-Cam\\
  $i$ &  26.7 & & CFHT/MegaCam & $i$ & 27.2  & Subaru/Suprime-Cam\\
  $I_{814}$ &  26.7 & & \emph{HST}/ACS &  &  & \\
  
  $z$ &  25.5 & & CFHT/MegaCam & &   &  \\
  $z'$ &  26.7 & & Subaru/Suprime-Cam & $z'$ & 26.5  & Subaru/Suprime-Cam\\
   &  & &  & $N\!B921$ & 26.1 & Subaru/Suprime-Cam\\
  $Y$ &  25.1 & 25.8 & UltraVISTA & $Y$ & 24.8 & VISTA VIDEO\\
  $J$ & 24.9 & 25.3 & UltraVISTA & $J$ & 25.7  & UKIRT/WFCAM\\
  $Y+J $& 25.3 & 25.9 & UltraVISTA &  &  &  \\
  $H$ & 24.6 & 24.9 & UltraVISTA & $H$ & 25.1  & UKIRT/WFCAM\\
  $K_{s}$ &  24.1 & 25.0  & UltraVISTA & $K$ & 25.3 & UKIRT/WFCAM\\
  $J_{125}$ & & 26.8 & \emph{HST}/WFC3 & $J_{125}$ & 26.9 & \emph{HST}/WFC3 \\
$H_{160}$ & & 27.0 & \emph{HST}/WFC3 & $H_{160}$ &  27.0 & \emph{HST}/WFC3 \\
  $3.6\umu$m & 25.3 & & \emph{Spitzer}/SPLASH & $3.6\umu$m & 24.4, 25.2   & \emph{Spitzer}/SpUDS, SEDS \\
  $4.5\umu$m & 25.1 & & \emph{Spitzer}/SPLASH & $4.5\umu$m & 24.2, 25.0  & \emph{Spitzer}/SpUDS, SEDS\\

\hline
\end{tabular}
\label{table:depths}
\end{table*}

The careful determination of imaging depths across all bands is crucial to obtain accurate errors for use in the SED fitting analysis, and also in making appropriate preliminary magnitude cuts in the selection process.
However, defining global limiting depths for data over degree-scale fields in the optical and near-infrared becomes problematic, as the combined pointings may have different integration times and seeing.
Therefore, we obtained local depths over each image from the clipped median absolute deviation of the 200 closest apertures to each point.
Apertures were placed randomly on the blank regions of the images that had been background subtracted using {\sc SExtractor}, where blank regions were defined using the segmentation map of each image.
The medians of the local depths for the imaging utilised here are presented in Table~\ref{table:depths}, within the 1.8-arcsec diameter circular aperture used for the  photometry in this paper.

\subsection{Determination of the enclosed flux}\label{enclosed_flux}

We expect the high-redshift galaxies detected here to be close to unresolved in the available ground-based imaging, which typically has seeing of 0.8 arcsec (see Section~\ref{sizes}).
However, the variations in seeing throughout the multiwavelength data, along with potential variations across each individual mosaic, result in a different fraction of the enclosed flux in a given fixed circular aperture, which must be corrected for to ensure accurate colours and magnitudes.
To extract a point spread function (PSF) for each image, we selected stars using the BzK-diagram (uzK-diagram for the UltraVISTA dataset) as defined in~\citet{Daddi2004}.
For each star we extracted a postage-stamp from a background-subtracted image, before centring using the centroid coordinates from {\sc SExtractor}. 
In a range of magnitude bins, these stars were then further background-subtracted and normalised, and a median was taken to form a high signal-to-noise PSF.
The curve of growth was then measured on the stack for each magnitude bin (typically from $ {\rm m}_{\rm AB} = 18 - 25$, $\Delta {\rm m} = 1.0$), and the curves were visually inspected to exclude bins where the stars were saturated and to identify any possible trend with magnitude.
Enclosed flux values in a 1.8-arcsec diameter aperture were typically 80\% across the optical to near-infrared data, with the exception of the $Y$ and $J$-band imaging from VISTA VIRCAM which has extended wings (as noted in~\citealp{McCracken2012}) and hence encloses only $\sim$70\%.
For the IRAC imaging, the enclosed flux values quoted in the \emph{Spitzer}/IRAC handbook were used.

\section{Candidate Selection}\label{sect:candsel}

\subsection{Initial detection and photometry}\label{sect:initialdet}

The primary catalogues for the UltraVISTA and UDS fields were created using SExtractor v2.8.6~\citep{Bertin1996}, run in the dual-image mode to create multiwavelength catalogues. 
The UltraVISTA catalogue was selected as in~\citet{Bowler2012} in a $Y+J$ inverse-variance weighted stack, with additional objects included from $Y$ and $J$-selected catalogues to ensure we are sensitive to very blue and red spectra up to $z = 7.5$.
In the UDS field, where the $J$-band imaging is substantially deeper than the $Y$-band, we combined $Y$ and $J$-band selected catalogues to form a master catalogue rather than creating a $Y+J$ stacked image. 
Magnitudes were measured in circular apertures of diameter 1.8-arcsec to deliver high signal-to-noise whilst ensuring that the measurements are robust against any astrometric differences between bands (typically $< 0.1$\asec).
We also simultaneously created catalogues with photometry measured in 1.2-arcsec diameter circular apertures, which were used for SED fitting of stellar templates with the aim of increasing the signal-to-noise for true point sources.

The \emph{Spitzer}/IRAC measurements were made in 2.8-arcsec diameter circular apertures, using images that had been background subtracted using a large filter size by {\sc SExtractor}.
The majority of our high-redshift candidates presented in the next section are isolated and hence the large aperture photometry is sufficiently accurate.
However, when the photometry is confused, we have excluded the 3.6$\umu$m and 4.5$\umu$m bands from the SED fitting process.
We have flagged confused and hence unreliable IRAC magnitudes (which translate into unreliable stellar mass estimates) in Tables~\ref{table:allmags},~\ref{table:allprops} and~\ref{table:properties}.

\subsection{UltraVISTA DR2 selection}

The initial catalogue for the UltraVISTA dataset consisted of 278916 objects within the $\simeq 0.7$ deg$^2$ `ultra-deep' part of UltraVISTA imaging that overlaps with the multiwavelength auxiliary data shown in Fig.~\ref{fig:UVISTAcoverage}.
We then required that an object be detected at greater than $5\sigma$ significance in either the $Y$ or $J$-bands, where the $5\sigma$ limit here was taken as the median local depth from the deepest of the three strips ($Y$ = 25.8, $J = 25.4$ mag, 5$\sigma$, 1.8-arcsec diameter circular aperture).
Using our local depth estimates to compensate for the large diffraction halos around stars in the CFHT/Mega-Cam imaging, we then required the candidate to be undetected in the $u^{*}gri$-bands at the $2\sigma$-level.
These optical non-detection requirements will inevitably result in a minority of real $z \simeq 7$ candidates being rejected at this stage (e.g. a true optical dropout galaxy has a probability of $\sim 0.04$ of being excluded here, assuming Gaussian distributed random noise).
The number of objects lost at this stage is small however, and such a cut is standard practice to reduce the number of candidates for further analysis.
The incompleteness of our sample as a result of these significance cuts, was carefully simulated using injection/recovery simulations, and folded into our LF analysis (see Section~\ref{compsim}).
The result of the described cuts was a sample of 1188 galaxy candidates, which was further reduced to 589 candidates with the removal of artefacts in the UltraVISTA imaging close to the area of missing data (seen in Fig.~\ref{fig:UVISTAcoverage}) and in the haloes/diffraction spikes produced by bright stars.

\subsection{UKIDSS/UDS selection}\label{udsselection}

The raw catalogue from the UDS dataset consisted of 248191 objects over the full area of $J, H$ and $K$-band imaging shown in Fig.~\ref{fig:UDScoverage}.
We then applied the initial criteria that the candidates must be brighter than the $5\sigma$-level in the $J$-band and undetected at $2\sigma$-significance in the $i$-band, leaving 24797 sources.
The available $Y$-band imaging in the UDS field is of insufficient depth to be competitive with the UltraVISTA $Y$-band imaging for the selection of galaxies, however it is essential to remove potential `cross-talk' artefacts that occur only within the UKIRT imaging and hence can closely mimic a $z$-dropout LBG at $z \ge 7 $.
Cross-talk appears at constant pixel separation from all objects in the UKIRT/WFCAM imaging, which is 128 pixels on the native 0.4-arcsec/pixel scale (51.2-arcsec), and can appear many multiples of this distance away from the source object.
Although the brightest occurrences are easily distinguishable from real objects by their `doughnut' appearance, for fainter objects or artefacts a large distance away from the source, it can be very difficult if not impossible to distinguish cross-talk from a high-redshift galaxy.
Hence, we apply the condition that candidates must be brighter than the $2\sigma$-level in the VISTA VIDEO $Y$-band, and clearly visible in either the $Y$-image or the $z'$-band imaging, to ensure a detection in data obtained from independent telescopes.
By further insisting that the objects must lie within the region of overlapping Subaru optical and UKIRT near-infrared data (see Fig.~\ref{fig:UDScoverage}), and are further than 100 pixels from the edge of the UKIRT imaging, where the signal-to-noise drops significantly, we remove the majority of objects leaving only 202.

\subsection{Visual inspection}

In the final step before photometric redshift fitting, the candidates were visually inspected in the $z'$, $Y$ and $J$-images to remove obvious artefacts such as diffraction spikes, remaining cross-talk in the case of the UDS, and sources within the bright haloes around stars in the VISTA imaging.
The $i$-band images were also inspected, and objects with any detection here (that may have escaped the $2\sigma$-level cut applied above) were removed as indicative of galaxies at $z < 6.5$.
The final samples containing the remaining high-redshift candidates for SED fitting consisted of 400 galaxies in the UltraVISTA data and 36 in the UDS field (where the lower number of objects remaining in the UDS is due to the additional $Y$-band detection requirement).
All near-infrared photometry was corrected to a constant enclosed flux level of 84\% (the enclosed flux in a 1.8-arcsec diameter aperture for the UltraVISTA CFTHLS $z$-band imaging), using the enclosed flux values derived for each image as described in Section~\ref{enclosed_flux}.
The 1$\sigma$-errors on the photometry were calculated from the nearest 200 blank apertures to each object, using the method described in Section~\ref{imagedepths}.

\begin{table*}
\caption{The photometry for the sources in our final UltraVISTA and UDS samples is presented in the upper and lower sections of the table respectively.
The magnitudes presented here were based on 1.8-arcsec diameter circular apertures except for the \emph{Spitzer}/IRAC channels where 2.8-arcsec diameter circular apertures were used.
The photometry has been corrected to the 84\% enclosed flux level using the appropriate point-source correction.
Where the flux is below 2$\sigma$ significance, as defined by the local depth derived for each object, we replaced the magnitude with the limiting 2$\sigma$ magnitude.
The errors displayed were derived from the median absolute deviation of the fluxes from the closest 200 blank sky apertures.
The presence of a dagger symbol in the right-hand column indicates that the \chone and \chtwo photometry is confused, and the number corresponds to the~\citet{Bowler2012} galaxies with the order consistent with table 2 in~\citet{Bowler2012}.
The `Himiko' galaxy from~\citet{Ouchi2013} referenced in the text is marked with an `H'.}

 \begin{tabular}{l l l c c c c c c c c l}

\hline

ID & RA & DEC & $z'$ & $N\!B921$ & $Y$ & $J$ & $H$ & $K$ &$3.6\umu$m & $4.5 \umu$m &  B12\\

  \hline

136380 & 09:59:15.89 & +02:07:32.0 & $ >  27.9$ &  -  & $  25.5_{-  0.1}^{+  0.2}$ & $  25.3_{-  0.2}^{+  0.3}$ & $  25.1_{-  0.2}^{+  0.3}
$ & $  25.6_{-  0.3}^{+  0.4}$ & $ >  25.5$ & $ >  25.3$ &     \\
28495 & 10:00:28.13 & +01:47:54.4 & $ >  28.1$ &  -  & $  25.4_{-  0.2}^{+  0.2}$ & $  25.1_{-  0.2}^{+  0.2}$ & $  25.0_{-  0.2}^{+  0.3}$
 & $  25.2_{-  0.2}^{+  0.3}$ & $  24.8_{-  0.3}^{+  0.3}$ & $  24.2_{-  0.2}^{+  0.2}$ &     \\
 268511 & 10:00:02.35 & +02:35:52.4 & $ >  27.6$ & & $  25.0_{-  0.2}^{+  0.3}$ & $  25.0_{-  0.3}^{+  0.5}$ & $  25.4_{-  0.4}^{+  0.6}$& $ >  25.0$ & $ >  25.6$ & $ >  25.2$ & 6\\
268037 & 09:59:20.69 & +02:31:12.4 & $ >  27.6$ &  -  & $  25.1_{-  0.2}^{+  0.2}$ & $  25.5_{-  0.3}^{+  0.4}$ & $ >  25.6$ & $  24.9_{-  0.2}^{+  0.2}$ & $ >  25.5$ & $ >  25.2$ &     \\
65666 & 10:01:40.69 & +01:54:52.5 & $ >  27.6$ &  -  & $  24.9_{-  0.1}^{+  0.2}$ & $  24.7_{-  0.2}^{+  0.2}$ & $  24.6_{-  0.2}^{+  0.3}$
 & $  24.8_{-  0.2}^{+  0.3}$ & $ >  25.1$ & $  24.6_{-  0.2}^{+  0.3}$ & 4   \\
211127 & 10:00:23.77 & +02:20:37.0 & $ >  27.6$ &  -  & $  25.1_{-  0.2}^{+  0.2}$ & $  25.3_{-  0.2}^{+  0.3}$ & $ >  25.8$ & $  25.3_{-  
0.3}^{+  0.4}$ & $  24.6_{-  0.2}^{+  0.2}$ & $  23.9_{-  0.2}^{+  0.2}$ &  \textdagger  \\
137559 & 10:02:02.55 & +02:07:42.0 & $ >  27.5$ &  -  & $  25.4_{-  0.1}^{+  0.2}$ & $  25.6_{-  0.3}^{+  0.4}$ & $  25.5_{-  0.4}^{+  0.6}
$ & $  26.1_{-  0.4}^{+  0.7}$ & $ >  25.3$ & $ >  25.1$ &     \\
282894 & 10:00:30.49 & +02:33:46.3 & $ >  27.7$ &  -  & $  25.5_{-  0.2}^{+  0.3}$ & $  25.8_{-  0.4}^{+  0.6}$ & $ >  25.9$ & $  25.5_{-  
0.4}^{+  0.6}$ & $  25.0_{-  0.3}^{+  0.4}$ & $  24.7_{-  0.2}^{+  0.3}$ &     \\
238225 & 10:01:52.31 & +02:25:42.3 & $ >  27.6$ &  -  & $  25.0_{-  0.2}^{+  0.2}$ & $  25.0_{-  0.2}^{+  0.2}$ & $  25.0_{-  0.2}^{+  0.3}
$ & $  24.9_{-  0.2}^{+  0.3}$ & $  24.7_{-  0.2}^{+  0.3}$ & $ >  25.2$ & 3   \\
305036 & 10:00:46.79 & +02:35:52.9 & $ >  27.7$ &  -  & $  25.3_{-  0.1}^{+  0.1}$ & $  25.2_{-  0.2}^{+  0.2}$ & $  24.9_{-  0.3}^{+  0.3}
$ & $  25.1_{-  0.3}^{+  0.4}$ & $  23.9_{-  0.2}^{+  0.2}$ & $  23.9_{-  0.2}^{+  0.2}$ &  \textdagger  \\
35327 & 10:01:46.18 & +01:49:07.7 & $  27.5_{-  0.4}^{+  0.5}$ &  -  & $  25.3_{-  0.2}^{+  0.2}$ & $  25.8_{-  0.3}^{+  0.5}$ & $  25.7_{-
  0.4}^{+  0.7}$ & $ >  25.5$ & $ >  25.3$ & $ >  25.3$ &     \\
304416 & 10:00:43.37 & +02:37:51.6 & $  26.6_{-  0.2}^{+  0.3}$ &  -  & $  24.3_{-  0.1}^{+  0.1}$ & $  24.2_{-  0.1}^{+  0.1}$ & $  24.1_{
-  0.1}^{+  0.1}$ & $  24.2_{-  0.1}^{+  0.1}$ & $  23.4_{-  0.2}^{+  0.2}$ & $  23.4_{-  0.2}^{+  0.2}$ & 1\textdagger  \\
185070 & 10:00:30.19 & +02:15:59.8 & $  27.4_{-  0.3}^{+  0.4}$ &  -  & $  25.4_{-  0.1}^{+  0.2}$ & $  25.3_{-  0.2}^{+  0.2}$ & $  25.5_{
-  0.4}^{+  0.6}$ & $  25.1_{-  0.2}^{+  0.2}$ & $  23.7_{-  0.2}^{+  0.2}$ & $  24.7_{-  0.2}^{+  0.3}$ &  \textdagger  \\
169850 & 10:02:06.48 & +02:13:24.2 & $  26.1_{-  0.1}^{+  0.1}$ &  -  & $  24.5_{-  0.1}^{+  0.1}$ & $  24.5_{-  0.1}^{+  0.1}$ & $  24.6_{
-  0.2}^{+  0.2}$ & $  24.6_{-  0.2}^{+  0.2}$ & $ >  25.1$ & $ >  25.1$ & 2   \\
304384 & 10:01:36.86 & +02:37:49.2 & $  26.4_{-  0.2}^{+  0.3}$ &  -  & $  25.0_{-  0.3}^{+  0.4}$ & $  24.6_{-  0.2}^{+  0.2}$ & $  24.9_{
-  0.3}^{+  0.4}$ & $  24.8_{-  0.3}^{+  0.4}$ & $  24.6_{-  0.2}^{+  0.3}$ & $  24.7_{-  0.3}^{+  0.3}$ & 5\textdagger  \\
279127 & 10:01:58.50 & +02:33:08.5 & $  26.1_{-  0.1}^{+  0.1}$ &  -  & $  24.8_{-  0.1}^{+  0.1}$ & $  24.6_{-  0.1}^{+  0.2}$ & $  25.4_{
-  0.3}^{+  0.5}$ & $  25.5_{-  0.3}^{+  0.5}$ & $  25.3_{-  0.4}^{+  0.6}$ & $  25.0_{-  0.3}^{+  0.4}$ &     \\
170216 & 10:02:03.82 & +02:13:25.1 & $  26.8_{-  0.3}^{+  0.3}$ &  -  & $  25.5_{-  0.2}^{+  0.3}$ & $  26.0_{-  0.4}^{+  0.6}$ & $ >  25.6
$ & $  25.4_{-  0.3}^{+  0.4}$ & $ >  25.1$ & $ >  25.0$ &     \\
104600 & 10:00:42.13 & +02:01:57.1 & $  26.3_{-  0.1}^{+  0.2}$ &  -  & $  25.0_{-  0.1}^{+  0.2}$ & $  25.0_{-  0.2}^{+  0.2}$ & $  24.7_{
-  0.2}^{+  0.2}$ & $  25.4_{-  0.3}^{+  0.4}$ & $  23.5_{-  0.2}^{+  0.2}$ & $  23.6_{-  0.2}^{+  0.2}$ &  \textdagger  \\
268576 & 10:00:23.39 & +02:31:14.8 & $  26.8_{-  0.2}^{+  0.3}$ &  -  & $  25.5_{-  0.2}^{+  0.3}$ & $  25.6_{-  0.3}^{+  0.3}$ & $ >  25.8
$ & $  25.9_{-  0.4}^{+  0.6}$ & $  24.4_{-  0.2}^{+  0.2}$ & $ >  25.5$ &     \\
2103 & 10:01:43.13 & +01:42:55.0 & $  26.1_{-  0.1}^{+  0.2}$ &  -  & $  25.1_{-  0.2}^{+  0.2}$ & $  25.3_{-  0.3}^{+  0.4}$ & $ >  25.3$ 
& $  24.9_{-  0.3}^{+  0.4}$ & $  24.8_{-  0.3}^{+  0.4}$ & $ >  25.4$ &     \\
179680 & 09:58:39.76 & +02:15:03.3 & $  26.0_{-  0.2}^{+  0.2}$ &  -  & $  25.0_{-  0.2}^{+  0.2}$ & $  24.9_{-  0.2}^{+  0.2}$ & $  24.8_{
-  0.2}^{+  0.3}$ & $  24.6_{-  0.2}^{+  0.3}$ & $  24.0_{-  0.2}^{+  0.2}$ & $  24.5_{-  0.2}^{+  0.2}$ &  \textdagger  \\
18463 & 09:58:49.36 & +01:46:02.1 & $  26.3_{-  0.1}^{+  0.2}$ &  -  & $  25.4_{-  0.1}^{+  0.1}$ & $ >  26.0$ & $ >  25.8$ & $  25.9_{-  0
.4}^{+  0.7}$ & $  24.8_{-  0.2}^{+  0.3}$ & $  24.7_{-  0.2}^{+  0.3}$ &  \textdagger  \\
122368 & 10:01:53.46 & +02:04:59.9 & $  26.5_{-  0.2}^{+  0.3}$ &  -  & $  25.5_{-  0.2}^{+  0.2}$ & $ >  26.2$ & $ >  25.9$ & $ >  25.8$ &
 $  24.9_{-  0.3}^{+  0.5}$ & $ >  25.0$ &     \\
583226 & 10:00:46.89 & +01:58:46.9 & $  26.4_{-  0.2}^{+  0.2}$ &  -  & $  25.5_{-  0.2}^{+  0.2}$ & $  25.7_{-  0.4}^{+  0.6}$ & $  25.2_{
-  0.3}^{+  0.5}$ & $  25.6_{-  0.4}^{+  0.7}$ & $  24.5_{-  0.2}^{+  0.2}$ & $ >  25.4$ &     \\
82871 & 10:01:43.04 & +01:58:01.1 & $  25.9_{-  0.1}^{+  0.1}$ &  -  & $  25.1_{-  0.1}^{+  0.1}$ & $  25.6_{-  0.2}^{+  0.3}$ & $  25.4_{-
  0.3}^{+  0.3}$ & $  25.2_{-  0.3}^{+  0.4}$ & $  25.0_{-  0.3}^{+  0.4}$ & $ >  25.2$ &     \\
68240 & 09:59:16.85 & +01:55:22.1 & $  25.4_{-  0.1}^{+  0.1}$ &  -  & $  24.7_{-  0.1}^{+  0.1}$ & $  24.6_{-  0.1}^{+  0.2}$ & $  24.9_{-
  0.2}^{+  0.3}$ & $  24.7_{-  0.2}^{+  0.2}$ & $  23.7_{-  0.2}^{+  0.2}$ & $  24.0_{-  0.2}^{+  0.2}$ &     \\
271028 & 10:00:45.17 & +02:31:40.2 & $  25.9_{-  0.1}^{+  0.1}$ &  -  & $  25.5_{-  0.1}^{+  0.2}$ & $  25.2_{-  0.3}^{+  0.3}$ & $  25.3_{
-  0.4}^{+  0.6}$ & $  25.4_{-  0.2}^{+  0.3}$ & $  24.3_{-  0.2}^{+  0.2}$ & $  24.7_{-  0.2}^{+  0.3}$ &  \textdagger  \\
30425 & 10:00:58.01 & +01:48:15.3 & $  25.2_{-  0.1}^{+  0.1}$ &  -  & $  24.8_{-  0.1}^{+  0.2}$ & $  24.6_{-  0.1}^{+  0.1}$ & $  25.0_{-
  0.3}^{+  0.4}$ & $  25.1_{-  0.3}^{+  0.4}$ & $  23.5_{-  0.2}^{+  0.2}$ & $  24.9_{-  0.4}^{+  0.6}$ & 9   \\
234429 & 09:58:36.65 & +02:24:56.4 & $  25.7_{-  0.2}^{+  0.3}$ &  -  & $  25.3_{-  0.3}^{+  0.3}$ & $ >  25.6$ & $ >  25.4$ & $ >  25.2$ &
 $ >  25.2$ & $ >  25.3$ &     \\
328993 & 10:01:35.33 & +02:38:46.3 & $  25.6_{-  0.1}^{+  0.1}$ &  -  & $  25.4_{-  0.3}^{+  0.5}$ & $  25.5_{-  0.3}^{+  0.5}$ & $ >  25.4
$ & $ >  25.4$ & $  24.2_{-  0.2}^{+  0.2}$ & $  24.1_{-  0.2}^{+  0.2}$ &     \\
  
  \hline
35314 & 02:19:09.49 & -05:23:20.6 & $  26.7_{-  0.2}^{+  0.2}$ & $ >  26.8$ & $  25.2_{-  0.3}^{+  0.5}$ & $  25.1_{-  0.1}^{+  0.2}$ & $  25.3_{-  0.3}^{+  0.3}$ & $  25.5_{-  0.2}^{+  0.2}$ & $  25.3_{-  0.4}^{+  0.5}$ & $ >  24.7$ & \\
118717 & 02:18:11.50 & -05:00:59.4 & $  26.8_{-  0.4}^{+  0.6}$ & $  26.5_{-  0.3}^{+  0.4}$ & $  25.0_{-  0.3}^{+  0.5}$ & $  25.3_{-  0.2}^{+  0.2}$ & $  25.3_{-  0.3}^{+  0.3}$ & $  25.0_{-  0.2}^{+  0.2}$ & $  23.8_{-  0.1}^{+  0.2}$ & $  23.6_{-  0.2}^{+  0.2}$ & \textdagger \\
88759 & 02:17:57.58 & -05:08:44.8 & $  25.8_{-  0.1}^{+  0.1}$ & $  24.0_{-  0.1}^{+  0.1}$ & $  25.5_{-  0.4}^{+  0.7}$ & $  25.1_{-  0.1}^{+  0.1}$ & $  25.5_{-  0.3}^{+  0.4}$ & $  24.9_{-  0.1}^{+  0.2}$ & $  23.9_{-  0.2}^{+  0.2}$ & $  24.7_{-  0.4}^{+  0.7}$ & H\\
87995 & 02:18:50.86 & -05:08:57.8 & $  26.3_{-  0.1}^{+  0.2}$ & $  25.6_{-  0.1}^{+  0.1}$ & $  25.0_{-  0.3}^{+  0.5}$ & $  25.3_{-  0.1}^{+  0.2}$ & $  25.3_{-  0.2}^{+  0.3}$ & $  25.3_{-  0.2}^{+  0.3}$ & $  24.9_{-  0.3}^{+  0.5}$ & $  24.8_{-  0.4}^{+  0.6}$ & \textdagger\\
   
     \hline  	
     	\end{tabular}
 \label{table:allmags}
\end{table*}

\subsection{Photometric redshift analysis}\label{photozs}

The final step in selecting our sample of $z \sim 7$ galaxies involves fitting the available multiwavelength data points using a photometric redshift fitting routine.
By incorporating optical, near- and mid-infrared photometry, we can select good high-redshift galaxy candidates and identify possible low-redshift galaxy interlopers (where the Balmer or 4000\AA ~break is confused with the Lyman-break) or galactic dwarf stars whose spectrum peaks in the near-infrared.

We fitted~\citet{Bruzual2003} models assuming a~\citet{Chabrier2003} initial-mass function, using the Le Phare photometric redshift code~\citep{Arnouts1999,Ilbert2006}\footnote{\url{http://www.cfht.hawaii.edu/~arnouts/lephare.html}}.
At each redshift, stellar populations were constrained to be older than 10 Myr and younger than the age of the Universe.
We fitted models with an exponentially-decreasing star-formation history with a characteristic timescale, $50{\rm\,Myr} \le \tau \le 10\,{\rm Gyr}$, for two metallicities ($Z = 1/5\, {\rm Z}_{\sun}$ and ${\rm Z}_{\sun}$).
Note that a constant star-formation and burst model can be closely reproduced by the longest and shortest age $\tau$ models respectively.
Internal dust reddening was calculated using the~\citet{Calzetti2000} attenuation law, and parameterised by values of the rest-frame $V$-band attenuation in the range $0.0 \le A_V \le 4.0$.
Absorption by the intergalactic medium was applied using the prescription of~\citet{Madau1995}.

The presence of a Lyman-$\alpha$ emission line within the spectrum can significantly alter the photometric redshift derived (up to $\Delta z \sim 0.5$) when fitting to broad-band photometry, and hence can cause genuine galaxies at $z \ge 6.5$ to be excluded from our sample.
In addition to the models described above, we also separately fitted templates where Lyman-$\alpha$ emission was added to the full template set, with rest-frame equivalent width in the range $0.0 \le EW_0 \le 240\,$\AA.
The continuum level was estimated from the mean value of the continuum in the wavelength range $\lambda = 1250-1300$ \AA~before the reddening was applied.

Contamination by cool galactic stars can be a significant problem when using ground-based data for high-redshift studies~\citep{Dunlop2013b}, particularly when there is insufficient wavelength sampling of the SED around the predicted Lyman-break.
To ascertain how well our galaxy candidates could be described by stellar templates, we fitted the reference stellar spectra from the SpeX library\footnote{\url{http://pono.ucsd.edu/~adam/browndwarfs/spexprism/}} with spectral types from M4 to T8.
The dwarf spectra do not extend to the wavelengths of the \emph{Spitzer}/IRAC bands  and so these were excluded during the fitting processes, although they can be taken into account in the selection via the mid-IR colours (see~\citealt{Bowler2012}).

Using the redshift-$\chi^2$ distributions from our photometric redshift fitting procedure, we required an acceptable fit above z = 6 (determined by $\chi^2 < 10$ and $11.3$ for the UDS and UltraVISTA respectively, which corresponds to $2\sigma$ significance given the degrees of freedom in the fitting).
In an effort to remove low-redshift contaminants from the sample we also excluded all objects that had a low-redshift solution ($z < 4.5$) within $\Delta \chi^2 = 4$ of the high-redshift ($z > 6$) solution.
In the final stage of selection we performed careful visual checks of a stack of the optical bands blueward of the $z'$-band to ensure there was no residual optical flux that would imply a lower redshift solution.
These further steps resulted in samples of 53 and 18 remaining candidates in the UltraVISTA and UDS datasets respectively.

Armed with a reduced sample of candidates that were acceptable as $z > 6$ objects, we performed further SED fitting including Ly$\alpha$-emission and stellar templates.
We also included photometry measured in the $N\!B921$ filter for candidates within the UDS field, as the narrow-band sits half-way through the $z'$-band filter and, with or without allowing for \Lya emission in the SED, allows a more precise determination of the photometric redshift.
At this stage we also fitted to the photometry including the IRAC bands at $3.6\umu$m and $4.5\umu$m, to exclude dusty low-redshift solutions which have SEDs that rise rapidly to longer wavelengths, and also to obtain a more accurate estimate of the galaxy masses.
We compared the fits with and without the inclusion of the IRAC data to ensure objects were not excluded due to nebular emission lines being present in the spectra (see Section~\ref{nebularemission} for further discussion), although in practice the IRAC imaging is sufficiently shallow compared to the near-infrared data that the unusual IRAC colours indicative of nebular emission, affect the $\chi^2$ values derived from stellar only SED models very little (see Fig.~\ref{figure:seds}).
The results of careful consideration of the SED fits, along with final visual identification and removal of subtle near-infrared artefacts, resulted in the sample of 34 galaxies presented in the next section.

The strength of our selection methodology is demonstrated by the confirmation, using the deeper near-infrared imaging analysed in this work, of 9 of the 10 candidates we originally selected from the first data release of the UltraVISTA dataset in~\citet{Bowler2012} as high-redshift galaxies (see Section~\ref{Bowler2012}).
Furthermore, in the UDS dataset we have independently selected the brightest $z = 6.6$ LAE in the field, which was initially found using narrow-band imaging (where it is $> 1 \,$ mag brighter than in the continuum bands that we use for selection) by~\citet{Ouchi2008}.

\section{Candidate galaxies}\label{sect:cand}

The observed photometry for the final sample of 34 galaxies in the UltraVISTA and UDS fields is presented in Table~\ref{table:allmags} and the photometric redshifts and best-fitting parameters, such as the rest-frame equivalent width for \Lya and the dust attenuation, are presented in Table~\ref{table:allprops}.
In Figure~\ref{figure:seds} we present postage-stamp images of each candidate and the best-fitting galaxy and star SEDs.
All the following tables, SED fits and postage-stamps show the candidates ordered by best-fitting photometric redshift without \Lya emission included in the fitting, where candidates have been split by field.
With the goal of including all potential $z > 6.5$ galaxies, in the following tables we have included galaxy candidates that have best fitting photometric redshifts in the range $6.5 < z < 7.5$ only with the inclusion of \Lya emission in the SED,  in addition to the galaxies that do not require \Lya to be robustly at $z > 6.5$.
The presence of the spectroscopically confirmed galaxy `Himiko' at z = 6.595 in the sample motivates this inclusive approach, as without the  inclusion of \Lya in the SED, the galaxy would have been excluded from the sample on the basis of a best fitting photometric redshift of $z = 6.38_{-0.05}^{+0.03}$.

\begin{table*}

\caption{The best-fitting photometric redshift and model parameters for the DR2 UltraVISTA and UDS samples are presented in the upper and lower sections of the table respectively.
Where the IRAC photometry is considered confused with a nearby object we exclude the \chone and \chtwo bands from the fitting procedure, and present the resulting $\chi^2$ value labelled with a dagger to illustrate that there are different degrees of freedom, and hence acceptable $\chi^2$ values, for these objects.
We order the galaxies by the best-fitting photometric redshift in column 2.
Columns 7-10 of the table show the redshift, $\chi^2$-value, rest-frame equivalent width and $A_V$ value when we introduce the possibility of Ly$\alpha$-emission in the fits.
The candidate labelled `H' is Himiko, which has an extremely bright $N\!B921$ flux that can only be well-fitted with models including \Lya emission, hence the unacceptable $\chi^2$ value for continuum-only fitting seen in column three.
The best-fitting stellar template, where we fit spectral types M4$-$T8, is given in column 11 with the $\chi^2$ in column 12.
Where a FWHM value has been measured, corresponding to an object selected by {\sc SExtractor} in that band, it is displayed in the three columns on the right-hand side.
In the final column, we flag the candidates from~\citet{Bowler2012} that have been re-selected here, with a number corresponding to the order of the candidates in~\citet{Bowler2012}.
}

 \begin{tabular}{l c r c c c r r c c c r r r r l }
\hline

& \multicolumn{4}{l}{No \Lya} & \multicolumn{5}{l}{With \Lya}  &  \multicolumn{2}{l}{Star} & \multicolumn{3}{c}{FWHM} & \\
\hline
    
    \multicolumn{1}{c}{ID} & \multicolumn{1}{c}{$z $} & \multicolumn{1}{c}{$\chi^2$} & $A_V$ & ${\rm Z} $& $z$ & $\chi^2$ & ${\rm EW}_0$ & $A_V$ & ${\rm Z} $& Stellar & $\chi^2$ & $z'$ & $ Y$ & $J$ & B12 \\ 
     & &  & $/{\rm mag}$ & $ /{\rm Z}_{\sun}$& & & \multicolumn{1}{c}{/\AA} & $/{\rm mag}$ & $/{\rm Z}_{\sun}$& Type & &  \multicolumn{3}{c}{/arcsec}  &  \\ 

 \hline

   136380 & $ 7.21_{-0.21}^{+0.10}$ & $ 1.7\phantom{\dagger} $&  0.0 & 1.0 & $ 7.24 $ &  1.6 &  10 & 0.0 & 1.0 & T3 &  17.2 &  -  &   1.0 &   1.3 &  \\
       28495 & $ 7.19_{-0.14}^{+0.10}$ & $ 3.5\phantom{\dagger} $&  0.1 & 1.0 & $ 7.69 $ &  1.7 & 180 & 0.0 & 1.0 & T1 &  23.2 &  -  &   2.1 &   2.2 &  \\
           268511 & $ 7.12_{-0.11}^{+0.14}$ &  $2.6\phantom{\dagger}$ &  0.0 & 0.2 & $ 7.25 $ &  2.2 &  80 & 0.0 &   0.2 & T8 &   9.2 &   1.7 &   1.4 &   0.5  & 6\\
      268037 & $ 7.07_{-0.12}^{+0.14}$ & $ 9.5\phantom{\dagger} $&  0.2 & 1.0 & $ 7.55 $ &  8.6 & 200 & 0.3 & 1.0 & T8 &  22.5 &  -  &   2.0 &   0.7 &  \\
       65666 & $ 7.04_{-0.11}^{+0.16}$ & $ 5.5\phantom{\dagger} $&  0.4 & 1.0 & $ 7.04 $ &  5.5 &   0 & 0.4 & 1.0 & T4 &  25.0 &   1.7 &   1.1 &   1.6 & 4\\
      211127 & $ 7.03_{-0.11}^{+0.12}$ & $ 3.2^\dagger$ &  0.0 & 0.2 & $ 7.20 $ &  2.4 & 160 & 0.0 & 0.2 & T8 &  29.0 &  -  &   1.5 &   1.1 &   \\
      137559 & $ 7.03_{-0.16}^{+0.14}$ & $ 1.9\phantom{\dagger} $&  0.0 & 0.2 & $ 7.15 $ &  1.8 &  60 & 0.0 & 0.2 & T8 &   9.8 &  -  &   1.7 &  -  &  \\
      282894 & $ 7.01_{-0.15}^{+0.14}$ & $ 8.5\phantom{\dagger} $&  0.0 & 0.2 & $ 7.31 $ &  6.6 & 240 & 0.4 & 0.2 & T8 &  12.5 &  -  &   1.6 &   0.7 &  \\
      238225 & $ 6.98_{-0.12}^{+0.12}$ & $ 3.1\phantom{\dagger} $&  0.5 & 1.0 & $ 7.01 $ &  3.0 &  10 & 0.5 & 1.0 & T3 &  23.2 &   1.0 &   1.0 &   1.6 & 3\\
      305036 & $ 6.95_{-0.22}^{+0.23}$ & $ 0.7^\dagger$ &  0.0 & 1.0 & $ 7.04 $ &  0.7 &  40 & 0.0 & 1.0 & T3 &  11.6 &  -  &  -  &  -  &   \\
       35327 & $ 6.88_{-0.13}^{+0.10}$ & $ 1.5\phantom{\dagger} $&  0.0 & 0.2 & $ 7.05 $ &  0.8 & 100 & 0.0 & 0.2 & T8 &  19.0 &  -  &   1.5 &  -  &  \\
      304416 & $ 6.85_{-0.10}^{+0.09}$ & $ 2.4^\dagger$ &  0.0 & 1.0 & $ 6.85 $ &  2.4 &   0 & 0.0 & 1.0 & T3 &  27.2 &  -  &   1.7 &   1.9 & 1 \\
      185070 & $ 6.77_{-0.19}^{+0.14}$ & $ 1.1^\dagger$ &  0.1 & 0.2 & $ 7.01 $ &  1.1 &  90 & 0.2 & 0.2 & T2 &  32.0 &  -  &   1.3 &   1.3 &   \\
      169850 & $ 6.70_{-0.06}^{+0.05}$ & $ 4.6\phantom{\dagger} $&  0.2 & 1.0 & $ 6.86 $ &  4.4 &  50 & 0.2 & 1.0 & M6 &  28.0 &   1.4 &   1.7 &   2.2 & 2\\
      304384 & $ 6.64_{-0.22}^{+0.12}$ & $ 1.9^\dagger$ &  0.8 & 0.2 & $ 6.65 $ &  1.9 &   0 & 0.8 & 0.2 & T3 &   6.3 &   1.3 &   1.1 &   1.7 & 5 \\
      279127 & $ 6.59_{-0.06}^{+0.05}$ & $ 8.8\phantom{\dagger} $&  0.0 & 0.2 & $ 6.59 $ &  8.8 &   0 & 0.0 & 0.2 & M6 &  25.9 &   2.6 &   2.8 &   1.9 &  \\
      170216 & $ 6.55_{-0.17}^{+0.14}$ & $ 2.5\phantom{\dagger} $&  0.2 & 1.0 & $ 6.95 $ &  1.6 & 200 & 0.5 & 1.0 & M6 &  10.6 &   1.5 &   1.2 &  -  &  \\
      104600 & $ 6.54_{-0.08}^{+0.07}$ & $10.1^\dagger$ &  0.4 & 1.0 & $ 6.54 $ & 10.1 &   0 & 0.4 & 1.0 & M6 &  16.7 &   1.9 &   1.5 &   1.7 &   \\
      268576 & $ 6.51_{-0.12}^{+0.14}$ & $ 9.2\phantom{\dagger} $&  0.0 & 0.2 & $ 6.93 $ &  8.6 & 180 & 0.0 & 0.2 & M6 &  13.4 &   1.4 &   1.4 &   1.2 &  \\
        2103 & $ 6.41_{-0.12}^{+0.11}$ & $ 7.0\phantom{\dagger} $&  0.0 & 1.0 & $ 6.90 $ &  6.3 & 240 & 0.6 & 1.0 & M7 &  17.3 &   1.1 &   2.8 &   1.3 &  \\
      179680 & $ 6.40_{-0.26}^{+0.16}$ & $ 3.7^\dagger$ &  0.4 & 0.2 & $ 6.93 $ &  3.0 & 240 & 1.0 & 0.2 & M6 &  20.3 &   1.6 &   1.0 &   0.9 &   \\
       18463 & $ 6.38_{-0.10}^{+0.07}$ & $ 7.3^\dagger$ &  0.0 & 0.2 & $ 6.71 $ &  6.1 & 170 & 0.0 & 0.2 & M7 &  37.4 &   1.3 &   1.6 &  -  &   \\
      122368 & $ 6.36_{-0.16}^{+0.14}$ & $ 6.4\phantom{\dagger} $&  0.0 & 0.2 & $ 6.94 $ &  5.4 & 240 & 0.0 & 0.2 & M7 &  19.5 &   1.8 &   1.7 &  -  &  \\
      583226 & $ 6.33_{-0.16}^{+0.15}$ & $ 5.0\phantom{\dagger} $&  0.0 & 1.0 & $ 6.66 $ &  4.7 & 140 & 0.0 & 1.0 & M7 &  10.7 &   1.5 &   1.7 &   1.9 &  \\
       82871 & $ 6.31_{-0.08}^{+0.09}$ & $ 6.1\phantom{\dagger} $&  0.0 & 1.0 & $ 6.81 $ &  4.0 & 240 & 0.1 & 1.0 & M7 &  29.3 &   1.0 &   1.1 &   1.0 &  \\
       68240 & $ 6.29_{-0.10}^{+0.08}$ & $ 3.7\phantom{\dagger} $&  0.0 & 0.2 & $ 6.71 $ &  3.1 & 200 & 0.0 & 0.2 & M7 &  20.6 &   1.8 &   2.0 &   2.3 &  \\
      271028 & $ 6.21_{-0.19}^{+0.10}$ & $ 0.5^\dagger$ &  0.7 & 1.0 & $ 6.50 $ &  0.3 & 110 & 0.8 & 0.2 & M7 &   8.2 &   1.3 &   1.8 &   1.6 &   \\
       30425 & $ 6.20_{-0.08}^{+0.10}$ & $15.3\phantom{\dagger} $&  0.0 & 1.0 & $ 6.59 $ & 13.6 & 150 & 0.0 & 0.2 & M7 &  31.9 &   1.3 &   2.1 &   1.3 & 9\\
      234429 & $ 6.11_{-0.21}^{+0.22}$ & $ 4.3\phantom{\dagger} $&  0.1 & 0.2 & $ 6.64 $ &  3.1 & 240 & 0.0 & 1.0 & M7 &  16.1 &   1.1 &   0.0 &  -  &  \\
      328993 & $ 6.01_{-0.21}^{+0.19}$ & $ 0.9\phantom{\dagger} $&  0.5 & 0.2 & $ 6.58 $ &  0.1 & 240 & 0.1 & 1.0 & M7 &  21.7 &   0.9 &  -  &   0.7 &  \\
   
       \hline
       
         35314 & $ 6.69_{-0.07}^{+0.10}$ &  $6.9\phantom{\dagger}$ &  0.2 & 0.2 & $ 6.90 $ &  6.6 &  80 & 0.3 & 0.2 & M6 &  18.7 &   2.0 &  -  &   1.1 &  \\
      118717 & $ 6.51_{-0.08}^{+0.05}$ & $ 1.0^\dagger $ &  0.0 & 0.2 & $ 6.61 $ &  1.0 &  30 & 0.0 & 0.2 & M9 &   6.8 &   0.9 &  -  &   1.4 &  \\
       88759 & $ 6.38_{-0.05}^{+0.03}$ & $55.8\phantom{\dagger}$ &  0.0 & 1.0 & $ 6.52 $ & 12.7 & 150 & 0.1 & 1.0 & M8 & 109.8 &   2.2 &  -  &   2.1 &  H \\
       87995 & $ 6.48_{-0.23}^{+0.03}$ & $ 1.4^\dagger $ &  0.0 & 1.0 & $ 6.59 $ &  1.0 &  60 & 0.0 & 0.2 & M8 &  29.5 &   0.8 &  -  &   1.3 &  \\

       \hline
     \end{tabular}
     \label{table:allprops}
 \end{table*}

\subsection{UltraVISTA DR2}

From the $0.62$ deg$^2$ of the `ultra-deep' UltraVISTA DR2 data, we found 29 candidate $6.5 < z < 7.5$ galaxies.
These candidates are listed in the upper part of Tables~\ref{table:allmags} and~\ref{table:allprops}, and include seven of the eight candidates presented in~\citet{Bowler2012} that lie within the regions of the image covered by the ultra-deep data.
The table also includes a single object from~\citet{Bowler2012} that lies outside the new ultra-deep UltraVISTA imaging we analyse here, but is confirmed as a high-redshift galaxy with the improved photometry now available (see further discussion in Section~\ref{Bowler2012}).
Of the 29 candidates within the ultra-deep strips, 11 have best-fitting photometric redshifts at $ z > 6.5$ only when Ly$\alpha$-emission is included in the templates, although in some cases the candidate still has a reasonable probability of being at $z > 6.5$ with continuum-only fitting (as illustrated by the error bars presented and the $\chi^2$ vs. redshift insets in the SED figures).
The majority of the candidates are detected in the \emph{Spitzer}/IRAC bands; in the cases where the photometry was contaminated by nearby lower-redshift objects we have excluded these bands from the fitting procedure and flagged the object in Tables~\ref{table:allmags}, ~\ref{table:allprops} and~\ref{table:properties}.
When excluding probable galactic stars from the sample based on the stellar template fitting, the size information obtained from the full-width at half-maximum (FWHM) measurement described in Section~\ref{sizes} was taken into account.
Candidates excluded as stars were retained for size comparisons with the final sample as shown in Fig.~\ref{fwhm}.

\subsection{UDS}

Within the 0.74 deg$^2$ of overlapping optical and near-infrared imaging in the UDS field, we found four candidates for galaxies at $z > 6.5$ as listed in the lower section of Tables~\ref{table:allmags} and~\ref{table:allprops}.
Of the four candidate $z > 6.5$ galaxies, two have best-fitting models at $z > 6.5$ only when the fitting allows \Lya emission, one of which is the previously identified and spectroscopically confirmed galaxy `Himiko'~\citep{Ouchi2013} at z = 6.595.  
For Himiko, we find a best-fitting photometric redshift of $z = 6.52$ when including the narrow-band $N\!B921$ photometry and allowing for \Lya emission in the fits, and a value of $z = 6.38_{-0.05}^{+0.03}$ with continuum-only fits.
The slightly lower redshift compared to that found spectroscopically for Himiko ($z = 6.52$ vs. $z = 6.595$), is a result of the exact redshift being very sensitive to the calibration between the $z'$ and the $N\!B921$, as the strength of any line emission must be able to reproduce the $z' - N\!B921$ colour observed.
The effect is compounded by the slightly longer-wavelength peak of the \Lya line compared with the central wavelength of the narrow-band filter, which is optimised to select LAEs with $z = 6.56 \pm 0.05$.
We find a best-fitting $EW_0 = 160 \pm 60$\AA~ for Himiko, which although large compared to that found spectroscopically ($\sim 80$\AA), is consistent with the original estimate from the broad- and narrow-band photometry made by~\citet{Ouchi2009} of $EW_0 = 100^{+302}_{-43}$\AA.

\subsection{\citet{Bowler2012} candidates revisited}\label{Bowler2012}

In~\citet{Bowler2012} we presented ten candidate $6.5~<~z~<~7.5$ galaxies in the first year `deep' component of the UltraVISTA survey, within the full 1 deg$^2$ of overlapping data shown in Fig.~\ref{fig:UVISTAcoverage}.
Using the new reductions of the full field that contain the DR2 imaging, we obtained improved photometry and photometric redshifts for all ten galaxies, which we include in the Appendix.
Although only eight of the original ten candidates lie within the ultra-deep strips, the imaging available for the remaining two objects includes more exposures in the DR2 release than those utilised in~\citet{Bowler2012}, and improvements in the calibration and pipeline has resulted in deeper data by up to 0.35 mag in the case of the $H$-band.
The depth improvements in the $Y$- and $J$-bands over the full area are moderate however ($< 0.1$ mag) and hence we do not repeat the selection process over the full 1 deg$^2$ here.
The improved photometry shown in Table~\ref{table:photometry} was taken from the $Y+J$ selected catalogue created as part of the selection process described in Section~\ref{sect:initialdet}.
When comparing the photometry between DR1 and DR2, the $Y$-band magnitudes are fainter by $\gtrsim$ 0.1 mag, a consequence of the zeropoint shifts in the data reduction process in part, with the remaining magnitude offset a likely result of the use of smaller apertures here (1.8-arcsec diameter vs. 2.0-arcsec) on resolved galaxies (see Section~\ref{sizes}) .

The four candidates in the top `robust' category are all confirmed as high-redshift galaxies by the new imaging, which forms effectively an independent check of the photometry given the difference in integration times between DR1 and DR2.
Comparing the best-fitting parameters, all four have lower best-fitting redshifts by $\Delta z \leq 0.1$.
Inspection of the magnitudes and SED plots between the DR1 and DR2 data indicates that the shift in best-fitting redshift is a consequence of the drop in $Y$ and $J$-band magnitudes closer in-line with the $z'$-band photometry, requiring the Lyman-break to additionally fill the $z'$-band to reproduce the now bluer $z'-Y$ colour.
The stellar fits for these candidates get worse for three of the four `robust' candidates, with the $\chi^2$ value for object ID 218462 only marginally decreasing from $\chi^2 = 23.9$ to $\chi^2 = 23.2$.

Of the four out of six further candidates from~\citet{Bowler2012} that lie in the deep strips, two are present in our final sample (277880 and 28400) with consistent photometry and resulting photometric redshift results as presented.
The weakest candidate from~\citet{Bowler2012}, with ID 2233, was not present in the new $Y+J$ selected catalogue.
Visual inspection shows 2233 to be faint in the near-infrared imaging, but bright and elongated in the $z'$-band.
There is a tentative detection in the smoothed $i$-band image suggesting, when coupled with the photometric redshift analysis, that this object is likely at a slightly lower redshift of $6.0 < z <  6.5$.
The remaining candidate (271105), which had the lowest $\chi^2$ value for the stellar fits in~\citet{Bowler2012}, is best fitted as a star, with the characteristic `hook' in the SED clearly seen in Fig.~\ref{fig:B12sedfits}.

Only two of the original ten candidates (268511 and 95661) do not lie within the deep strips, however we remeasured the photometry using the improved data reductions of the full field.
Candidate 95661 was included in the~\citet{Bowler2012} sample as a result of the large errors on the photometric redshift of $z = 6.13^{+0.38}_{-0.27}$, but is now excluded from our new sample as a $z < 6.5$ galaxy with best-fitting photometric redshift of $z = 6.25^{+0.10}_{-0.13}$.
Finally, object 268511 remains a good $z > 6.5$ LBG candidate, although as it is effectively only detected in the $Y$-band there is still the possibility of it being a transient.
Hence,  we include candidate 268511 in the final sample as our only candidate from the region of the full 1 deg$^2$ of overlapping multiwavelength imaging not covered by the ultra-deep DR2 imaging.

In conclusion, of the ten candidate $z > 6.5$ galaxies presented previously in~\citet{Bowler2012}, all four of the `robust' candidates are reconfirmed here.
Two of the three `insecure/contaminant' galaxies are still present in the sample, with the third candidate being excluded as a star based on the improved photometry.
Only one of the three `insecure' candidates is retained in our final $z \simeq 7$ sample, with the two excluded galaxies now confirmed to lie at a slightly lower redshift  in the range $6.0 < z < 6.5$.
We present the improved photometry and derived SED properties for all ten candidates in the Appendix, along with the SED fits and postage-stamp images of the three candidates from~\citet{Bowler2012} that are not included in the final sample in this paper.

\section{Galaxy Properties}\label{sect:properties}

The extensive multiwavelength photometry available within the UltraVISTA and UDS fields allows an investigation of the physical properties of the galaxies, which further exploits the SED fitting analysis utilised in the selection of our sample.
Here we present the derived physical properties including stellar masses, star-formation rates (SFRs) and specific star-formation rates (sSFRs).
The section ends with an investigation into the rest-frame optical colours of the galaxies including the potential effects of nebular line emission.

\begin{table*}
\caption{The physical properties of the galaxies in our $z \simeq 7$ sample derived from the best-fitting SED models.
The stellar mass, SFR and sSFR were determined from the best-fitting $\tau$-model with the errors derived from the points with $\Delta \chi^2 = 1$ from the minimum $\chi^2$ value, after minimising over all other variables.
A sSFR of $100$ Gyr$^{-1}$ corresponds to the maximum attainable in the set of models we use, which always corresponds to an uncertain mass estimate.
We also present the SFR$_{\rm UV}$ derived using the~\citet*{Madau1998} prescription, which converts the observed rest-frame UV flux into a SFR directly.
The absolute rest-frame UV magnitudes of our galaxies were measured on the best-fitting model (corrected to total magnitudes using a point-source correction) in the rest-frame, when observed through a top-hat filter of width $\Delta \lambda = 100$\AA~centred on 1500\AA.
Note that the absolute magnitudes presented here have not been corrected for gravitational lensing magnification, as discussed in Section~\ref{lensing}.
In the final two columns we present the rest-frame UV slope index $\beta_{}$, calculated by fitting a power law to the measured $YJHK$ or $JHK$ photometry.
A missing value for $\beta_{}$ indicates that the object has large uncertainties in one or more bands and hence a secure value could not be obtained.
}
\label{table:properties}
\begin{tabular}{l r l l c c c c  l}

\hline
ID & log($M_{\star}/{\rm M}_{\sun}$) & SFR & sSFR & SFR$_{\rm UV}$ & $M_{1500}$ & $\beta_{YJHK}$ & $\beta_{JHK}$ & B12 \\
& & /${\rm M}_{\sun}{\rm yr}^{-1}$  & /${\rm Gyr^{-1}}$ & /${\rm M}_{\sun}{\rm yr}^{-1}$ &  /mag & & &  \\

\hline
  136380 & $ 8.4^{+ 0.6}_{- 0.1}$ & $\phantom{0}27^{+22}_{-16}$ & $100^{+0}_{-88}$ & $  19$ & $ -22.0$ &  $-2.4 \pm  0.4 $ &  $-2.4 \pm  0.6 $ &   \\
    28495 & $10.1^{+ 0.1}_{- 0.7}$ & $\phantom{0}23^{+190}_{-15}$ & $\phantom{00}2^{+68}_{-1}$ & $  21$ & $ -22.1$ &  $-2.0 \pm  0.4 $ &  $-2.0 \pm  0.5 $ &   \\
   268511 & $ 8.5^{+ 0.5}_{- 0.1}$ & $\phantom{0}31^{+9}_{-19}$ & $100^{+0}_{-87}$ & $  20$ & $ -22.1$ &  $-3.5 \pm  0.9 $ &  $-3.9 \pm  1.2 $ & 6 \\
   268037 & $ 8.6^{+ 0.5}_{- 0.2}$ & $\phantom{0}43^{+29}_{-31}$ & $100^{+0}_{-92}$ & $  18$ & $ -21.9$ &  $-1.8 \pm  0.4 $ &  $-0.8 \pm  0.6 $ &   \\
    65666 & $ 9.2^{+ 0.4}_{- 0.3}$ & $\phantom{0}89^{+87}_{-72}$ & $\phantom{0}54^{+46}_{-49}$ & $  27$ & $ -22.4$ &  $-2.0 \pm  0.3 $ &  $-2.1 \pm  0.5 $ & 4 \\
   211127 & $ 8.8^{+ 1.3}_{- 0.4}$ & $\phantom{0}17^{+62}_{-11}$ & $\phantom{0}28^{+72}_{-27}$ & $  18$ & $ -21.9$ &  $-2.7 \pm  0.5 $ &  $-2.0 \pm  0.8 $ &  \textdagger\\
   137559 & $ 8.4^{+ 0.9}_{- 0.1}$ & $\phantom{0}20^{+28}_{-14}$ & $\phantom{0}91^{+9}_{-86}$ & $  14$ & $ -21.7$ &  $-2.8 \pm  0.6 $ &  $-2.8 \pm  0.4 $ &   \\
   282894 & $ 9.9^{+ 0.2}_{- 0.3}$ & $\phantom{00}8^{+25}_{-5}$ & $\phantom{00}1^{+5}_{-0.5}$ & $  12$ & $ -21.5$ &  $-2.8 \pm  1.0 $ &  --  &   \\
   238225 & $ 9.0^{+ 0.6}_{- 0.2}$ & $110^{+48}_{-98}$ & $100^{+0}_{-98}$ & $  25$ & $ -22.3$ &  $-1.9 \pm  0.4 $ &  $-1.8 \pm  0.5 $ & 3 \\
   305036 & $10.4^{+ 0.6}_{- 1.6}$ & $\phantom{0}10^{+459}_{-8}$ & $0.4^{+100}_{-0.3}$ & $  20$ & $ -22.1$ &  $-1.6 \pm  0.3 $ &  $-1.8 \pm  0.6 $ &  \textdagger\\
    35327 & $ 8.3^{+ 0.8}_{- 0.1}$ & $\phantom{0}21^{+21}_{-14}$ & $100^{+0}_{-95}$ & $  15$ & $ -21.7$ &  $-3.5 \pm  1.1 $ &  $-2.9 \pm  1.1 $ &   \\
   304416 & $10.5^{+ 0.5}_{- 1.0}$ & $\phantom{0}26^{+560}_{-16}$ & $0.9^{+99}_{-0.7}$ & $  47$ & $ -23.0$ &  $-1.9 \pm  0.2 $ &  $-1.9 \pm  0.3 $ & 1\textdagger\\
   185070 & $10.5^{+ 0.2}_{- 1.6}$ & $\phantom{0}10^{+270}_{-9}$ & $0.3^{+100}_{-0.2}$ & $  17$ & $ -21.9$ &  $-1.8 \pm  0.3 $ &  $-1.7 \pm  0.5 $ &  \textdagger\\
   169850 & $ 8.9^{+ 0.2}_{- 0.1}$ & $\phantom{0}78^{+16}_{-38}$ & $100^{+0}_{-69}$ & $  35$ & $ -22.7$ &  $-2.2 \pm  0.2 $ &  $-2.2 \pm  0.3 $ & 2 \\
   304384 & $ 9.4^{+ 1.4}_{- 0.4}$ & $250^{+120}_{-250}$ & $100^{+0}_{-100}$ & $  26$ & $ -22.3$ &  $-1.9 \pm  0.4 $ &  $-2.3 \pm  0.6 $ & 5\textdagger\\
   279127 & $ 9.2^{+ 0.2}_{- 0.6}$ & $\phantom{0}21^{+54}_{-9}$ & $\phantom{0}12^{+88}_{-8}$ & $  25$ & $ -22.3$ &  $-2.8 \pm  0.4 $ &  $-3.8 \pm  0.8 $ &   \\
   170216 & $ 9.0^{+ 0.7}_{- 0.8}$ & $\phantom{0}15^{+86}_{-12}$ & $\phantom{0}17^{+83}_{-15}$ & $  11$ & $ -21.4$ &  $-2.0 \pm  0.6 $ &  $-0.8 \pm  1.2 $ &   \\
   104600 & $ 8.9^{+ 1.3}_{- 0.2}$ & $\phantom{0}74^{+84}_{-67}$ & $100^{+0}_{-100}$ & $  20$ & $ -22.1$ &  $-2.2 \pm  0.3 $ &  $-2.4 \pm  0.5 $ &  \textdagger\\
   268576 & $ 9.6^{+ 0.3}_{- 0.2}$ & $\phantom{00}4^{+15}_{-1}$ & $\phantom{00}1^{+6}_{-0.6}$ & $  12$ & $ -21.5$ &  $-2.5 \pm  0.6 $ &  $-2.7 \pm  1.0 $ &   \\
     2103 & $ 9.4^{+ 0.3}_{- 0.7}$ & $\phantom{0}10^{+130}_{-4}$ & $\phantom{00}4^{+96}_{-2}$ & $  16$ & $ -21.8$ &  $-2.0 \pm  0.6 $ &  $-1.2 \pm  0.9 $ &   \\
   179680 & $10.5^{+ 0.6}_{- 1.5}$ & $\phantom{0}22^{+610}_{-21}$ & $0.6^{+99}_{-0.5}$ & $  20$ & $ -22.1$ &  $-1.5 \pm  0.4 $ &  $-1.5 \pm  0.6 $ &  \textdagger\\
    18463 & $ 8.2^{+ 1.2}_{- 0.1}$ & $\phantom{0}17^{+11}_{-12}$ & $100^{+0}_{-98}$ & $  12$ & $ -21.5$ &  --  &  --  &  \textdagger\\
   122368 & $ 9.5^{+ 0.2}_{- 0.7}$ & $\phantom{00}7^{+7}_{-5}$ & $\phantom{00}2^{+9}_{-1}$ & $   9$ & $ -21.2$ &  $-4.1 \pm  1.2 $ &  $-2.5 \pm  2.0 $ &   \\
   583226 & $ 9.8^{+ 0.2}_{- 0.7}$ & $\phantom{00}7^{+160}_{-5}$ & $\phantom{00}1^{+99}_{-0.7}$ & $  12$ & $ -21.5$ &  $-2.0 \pm  0.5 $ &  $-1.8 \pm  0.8 $ &   \\
    82871 & $ 9.3^{+ 0.3}_{- 0.7}$ & $\phantom{0}11^{+76}_{-5}$ & $\phantom{00}5^{+95}_{-3}$ & $  15$ & $ -21.7$ &  $-2.3 \pm  0.4 $ &  $-1.3 \pm  0.7 $ &   \\
    68240 & $10.0^{+ 0.3}_{- 0.1}$ & $\phantom{00}8^{+43}_{-3}$ & $0.8^{+4}_{-0.3}$ & $  25$ & $ -22.3$ &  $-2.2 \pm  0.3 $ &  $-2.3 \pm  0.5 $ &   \\
   271028 & $ 9.0^{+ 1.4}_{- 0.3}$ & $\phantom{0}97^{+140}_{-94}$ & $100^{+0}_{-100}$ & $  12$ & $ -21.5$ &  $-1.9 \pm  0.4 $ &  $-2.2 \pm  0.6 $ &  \textdagger\\
    30425 & $ 9.9^{+ 0.2}_{- 0.6}$ & $\phantom{0}12^{+229}_{-4}$ & $\phantom{00}2^{+98}_{-0.7}$ & $  22$ & $ -22.2$ &  $-2.2 \pm  0.4 $ &  $-3.0 \pm  0.7 $ & 9 \\
   234429 & $ 8.9^{+ 0.6}_{- 0.8}$ & $\phantom{00}9^{+62}_{-6}$ & $\phantom{0}12^{+88}_{-10}$ & $  10$ & $ -21.3$ &  $-3.4 \pm  1.1 $ &  --  &   \\
   328993 & $10.3^{+ 0.1}_{- 0.4}$ & $\phantom{0}28^{+76}_{-25}$ & $\phantom{00}2^{+10}_{-1}$ & $  11$ & $ -21.4$ &  $-2.0 \pm  0.7 $ &  $-1.9 \pm  1.1 $ &   \\
   \hline
    35314 & $ 8.8^{+ 0.7}_{- 0.4}$ & $\phantom{0}35^{+47}_{-27}$ & $\phantom{0}62^{+38}_{-59}$ & $  20$ & $ -22.0$ &  $-2.5 \pm  0.4 $ &  $-2.6 \pm  0.4 $ &   \\
   118717 & $10.2^{+ 0.6}_{- 0.9}$ & $\phantom{00}3^{+320}_{-1}$ & $0.2^{+100}_{-0.1}$ & $  16$ & $ -21.8$ &  $-1.8 \pm  0.4 $ &  $-1.6 \pm  0.4 $ &  \textdagger\\
    88759 & $10.3^{+-0.0}_{- 0.3}$ & $\phantom{0}17^{+2}_{-8}$ & $0.8^{+0.2}_{-0.7}$ & $  21$ & $ -22.1$ &  $-1.8 \pm  0.3 $ &  $-1.9 \pm  0.3 $ &   \\
    87995 & $10.0^{+ 0.5}_{- 1.4}$ & $\phantom{0}13^{+210}_{-10}$ & $\phantom{00}1^{+99}_{-1}$ & $  16$ & $ -21.8$ &  $-2.0 \pm  0.5 $ &  $-2.0 \pm  0.5 $ &  \textdagger\\

 \hline

\end{tabular}
\end{table*}

\subsection{Stellar populations}
\subsubsection{Stellar masses, SFRs and sSFRs}
We present the stellar mass, SFR and sSFR for each object in the full sample in Table~\ref{table:properties}, derived from the range of $\tau$-models we fitted to the photometry.
Around half (19/34) of the galaxies in our sample have masses close to that found for fainter galaxies with $ M_{\star} \sim 10^{9} {\rm M}_{\sun}$~\citep{McLure2011}.
However, as would be expected for our more luminous sample, the remaining 15 galaxies have significantly higher masses in the range  $9.5 < {\rm log}(M_{\star}/{\rm M}_{\sun}) < 10.5$.
We note however, that masses below ${\rm log}(M_{\star}/{\rm M}_{\sun}) = 9.5$ become increasingly uncertain, as they often arise from non-detections in the~\emph{Spitzer}/IRAC bands that correspond to the rest-frame optical part of the galaxy SED (and hence better trace the stellar mass of the galaxy).
The majority of the galaxies have SFRs in the range $2~<~{\rm SFR}/{\rm M}_{\sun}{\rm yr}^{-1} < 40$ from the best-fitting $\tau$-model, which is consistent with the SFR determined directly from the UV luminosity via the~\citet{Madau1998} formalism.
For the 15 galaxies that have masses at $9.5 < {\rm log}(M_{\star}/{\rm M}_{\sun}) < 10.5$, we find a mean sSFR of $1.0 \pm 0.1$ Gyr$^{-1}$ where the error quoted is the standard error on the mean.
The errors on any individual measurement of the sSFR are large as a result of degeneracies in the SED fitting process, where in general a very young model with significant attenuation cannot be distinguished from an older model with little dust.
Previous studies with less massive galaxies ($M_{\star} \sim 10^{9} {\rm M}_{\sun}$) found higher values of sSFR than we find in our sample.
For example, \citet{SmitR.2013} put a lower limit of sSFR $ > 4 $ Gyr$^{-1}$ from a small sample of $z = 6.8$ lensed LBGs, which is consistent with the value of sSFR $\simeq 10$ Gyr$^{-1}$ found by~\citet{Stark2013}.
We do not include models with nebular emission lines in our SED fitting analysis, which can result in underestimated masses by a factor of $\simeq $\,2.4--4.4~\citep{Stark2013}, where the larger value assumes a continued evolution in the EW distribution of the contaminating lines beyond $z = 5$.
Although correcting the masses of our sample by the factors derived by~\citet{Stark2013} would bring our results into line with previous results, the inclusion of nebular emission lines into the SED models not only affects the resulting stellar mass estimate, but can result in younger ages and lower dust attenuation which subsequently affects the SFR estimate~\citep{DeBarros2012}.
~\citet{CurtisLake2013} found that sSFRs at $ z \simeq 6$ were increased by a factor of at most two when nebular emission lines were included, and therefore it is likely that our sample shows a genuinely lower sSFR than lower-mass galaxies at $z \simeq 7$, with a sSFR $\lesssim 2$ Gyr$^{-1}$.
The lower value of sSFR we find for our sample of $M_{\star} \sim 10^{10} {\rm M}_{\sun}$ galaxies is consistent with a galaxy formation model where star formation is most efficient in smaller galaxies (e.g.~\citealp{Kauffmann2003}).

From the subset of galaxies in our sample that have lower and hence more uncertain masses, there are several objects that have very low masses coupled with high best-fitting SFRs $\simeq 30 - 100 \,{\rm M}_{\sun}{\rm yr}^{-1}$ and hence extreme sSFR $\simeq 100 \,{\rm Gyr}^{-1}$.
Closer inspection of the best-fitting SEDs for these candidates show that they tend to have very blue SEDs and non-detections in the \emph{Spitzer}/IRAC bands, which results in the SED fitting procedure fitting the youngest possible age (10 Myr) in an attempt to reproduce the steep spectral slope.
While it is plausible that the extreme sSFRs observed could be genuine, resulting from a brief burst of star-formation in the galaxy, insufficient depth in the rest-frame optical region of the spectrum is more likely the cause of the large sSFRs.
The near-infrared data available in the UltraVISTA and UDS fields is substantially deeper than the \emph{Spitzer}/IRAC imaging, and the photometric errors on many of these galaxies are sufficiently large to allow a SED fit with a more moderate SFR and mass within the errors, as illustrated by the large uncertainties shown in Table~\ref{table:properties}.

With improved photometry for the~\citet{Bowler2012} sample of galaxies, we find that the derived physical properties of the objects are more extreme in the analysis presented here.
The high SFR and sSFR for several of the~\citet{Bowler2012} objects are also a result of extremely young best-fitting ages as was the case for fainter galaxies described above, where the close-to-flat near-infrared photometry can typically be best-fitted using a model with a young age, high SFR and a large dust-attenuation.
The coupling between high dust-attenuation and a low age in the best-fitting SEDs of these galaxies, is likely a consequence of degeneracies in the fitting process, where the currently available photometry could also be acceptably fit with a model including zero dust attenuation and a larger age.

By requiring zero dust attenuation using the~\citet*{Madau1998} prescription we find more moderate values of the SFR, which then produce slightly lower sSFRs $\sim 10-60 \,{\rm Gyr}^{-1}$.
In the case of the top candidate in~\citet{Bowler2012} which now has a more moderate SFR, further tension in the fitting was introduced in~\citet{Bowler2012} by including confused IRAC photometry, which is excluded from the SED fitting procedure here.

\subsubsection{Rest-frame UV slope}
The rest-frame UV slope $\beta_{}$ was calculated for the galaxies in our sample by fitting a power law to the $YJHK$ or $JHK$ photometry following the method of~\citet{Rogers2013}.
The error on the derived slope is reduced by including the $Y$-band.
However, above $z = 6.8$ the $Y$-band is an unreliable continuum measure, due to the Lyman-break moving through the filter and potential contamination by Ly$\alpha$-emission, and hence we quote the $JHK$ value for comparison.
The median values of $\beta_{} = -2.1$ and $\beta_{} = -2.0$ including the $YJHK$ and $JHK$ bands respectively, are consistent with the value found for fainter galaxies at $z \sim 7$ using the same fitting method ($\beta_{} = -2.1 \pm 0.2$; ~\citealp{Rogers2013, Dunlop2013}) and using the colours directly ($\beta_{} = - 2.33 \pm 0.33$;~\citealp{Bouwens2013}).
However, there is evidence that a colour-magnitude relation extends to $z = 7$~\citep{Bouwens2013}, and in which case we would expect the relatively luminous galaxies presented here to have redder rest-frame UV slopes on average.
The relation derived by~\citet{Bouwens2013} would predict an average $\beta_{} = -1.6$ at $M_{\rm UV} \simeq -22$ and $\beta_{} = -1.4$ at $M_{\rm UV} \simeq -23$.
The $\beta_{}$ values for the faintest galaxies presented here at $M_{1500} > -22$ have large errors and exhibit a wide scatter, however the two brightest galaxies (304416 and 169850) that were originally selected in~\citet{Bowler2012} are detected at $\sim 10\sigma$ in the near-infrared bands and have absolute magnitudes of  $M_{1500} \sim -22.7$ (correcting for gravitational lensing as discussed in Section~\ref{lensing}).
Both candidates show bluer $\beta_{}$ values than predicted by the relation from~\citet{Bouwens2013}, with $\beta_{YJHK} = -1.9 \pm 0.2$ and $\beta_{YJHK} = -2.2 \pm 0.2$ respectively.
However, the most recent results at $z=5$ suggest that the bluest galaxies have a similar $\beta_{}$ at all luminosities, with an increased scatter to redder values with increasing luminosity, possibly as a results of greater dust attenuation or age spread~\citep{Rogers2013a}.
The bluer values of $\beta_{}$ we find for the two brightest members of our sample are coupled with best-fitting models that show low values of dust extinction ($A_V = 0-0.2$), and therefore these galaxies could be extreme examples of the generally redder population of galaxies at $M_{1500} \simeq -22.5$.
Our results contrast with~\citet{Willott2013}, who found a redder rest-frame UV slope of $\beta_{} = -1.44 \pm 0.1$ in a stack of bright $M_{1350} \sim -22$ galaxies at $z =6$, which was attributed to dust-reddening.
Further imaging within the ultra-deep UltraVISTA survey regions over the next few years will allow more accurate $\beta_{}$ measurements for a larger sample of objects, and hence a meaningful constraint on the very bright-end of the colour-magnitude relation at $z \simeq 7$ .

\subsection{Nebular emission}\label{nebularemission}

\begin{figure}
\includegraphics[width = 0.5\textwidth, trim = 0.5cm 0.5cm -0.5cm 0.5cm ]{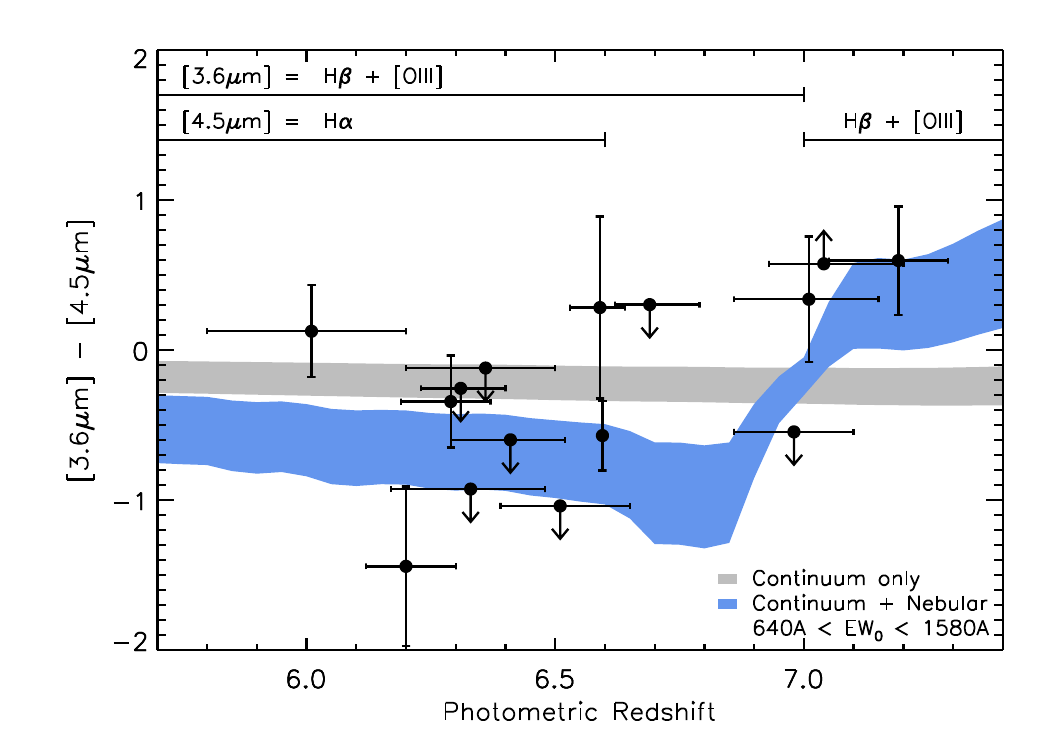}
\caption{The \IRACcol colours plotted against photometric redshift for the UltraVISTA DR2 and UDS candidates.
The grey and blue shaded regions represent the predicted \IRACcol colours as a function of redshift as described in the text, for~\citet{Bruzual2003} models with or without strong emission lines added respectively.
Candidates with confused IRAC photometry, defined as when flux from a nearby object enters the 2.8-arcsec diameter circular aperture of the candidate, are not plotted here.
If an object was not detected at greater than $2\sigma$ significance in either band then it has been excluded from the plot.
Limit arrows represent objects that were undetected at less than $2\sigma$ significance in the appropriate band, and here the magnitude in the undetected band has been replaced by the locally-determined $2\sigma$ limiting depth for that object.
The dominant emission lines present in each IRAC filter at that particular redshift are shown by the marks at the top of the plot.
The data point at $z = 6.6$ with extremely small errors on the redshift is the spectroscopically-confirmed LAE Himiko.
}
\label{fig:nebularemission}
\end{figure}

\begin{figure*}

\subfloat{\includegraphics[width = 0.5\textwidth]{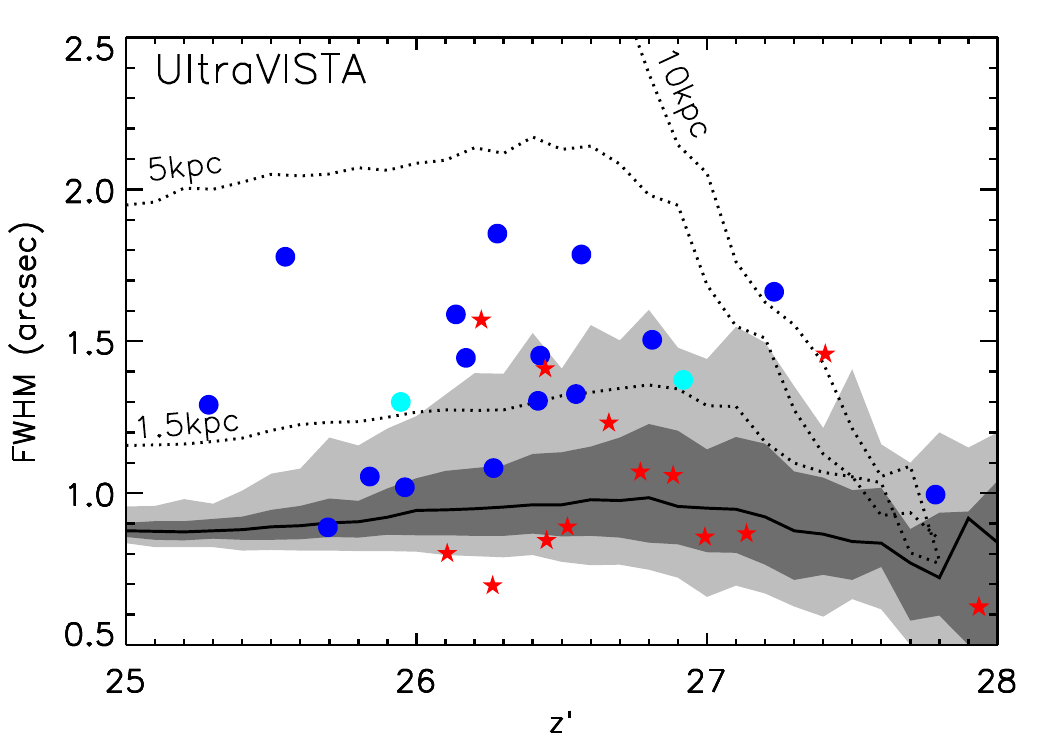}} 
\subfloat{\includegraphics[width = 0.5\textwidth]{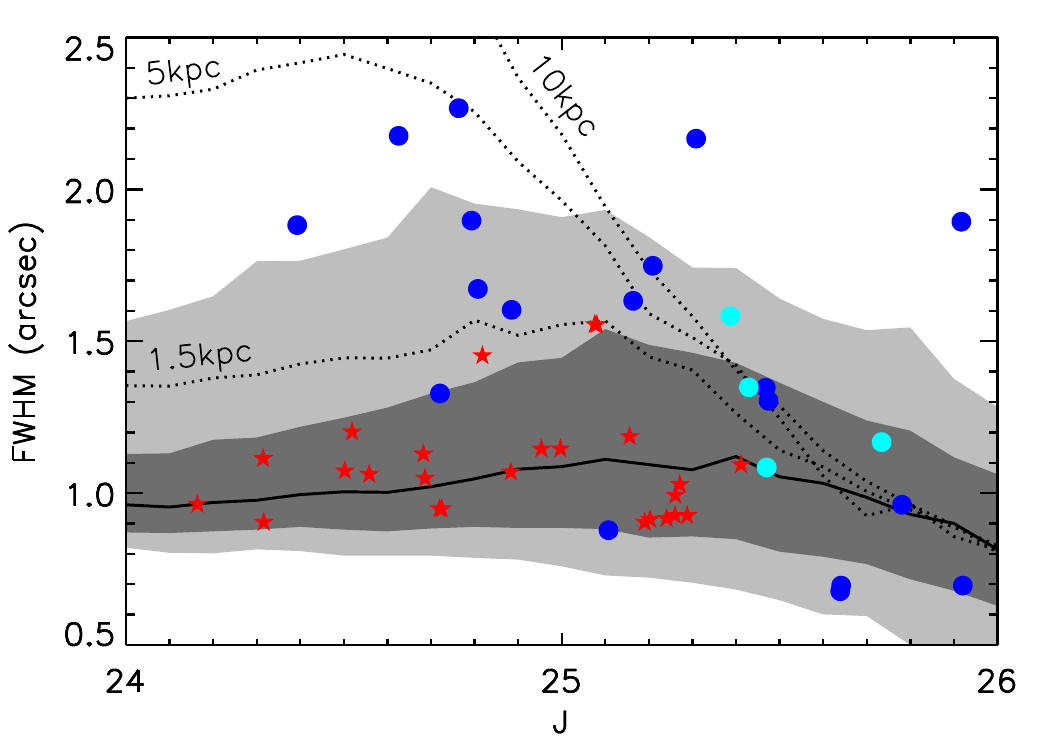}}\\
\vspace{-0.5cm}
\subfloat{\includegraphics[width = 0.5\textwidth]{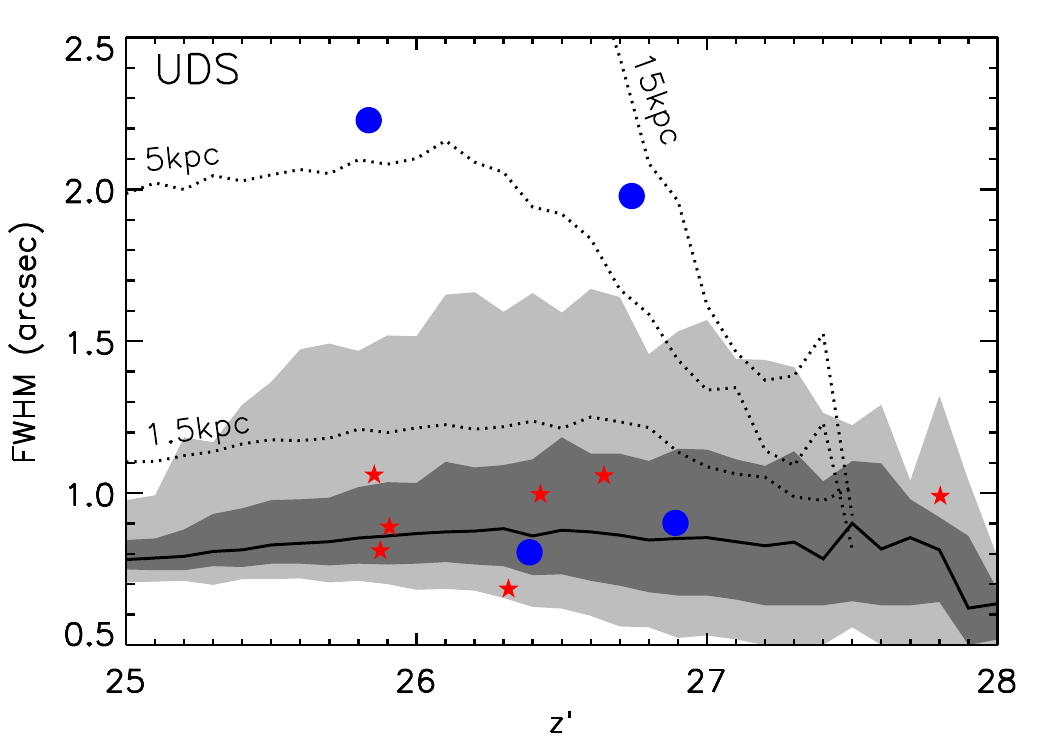}}
\subfloat{\includegraphics[width = 0.5\textwidth]{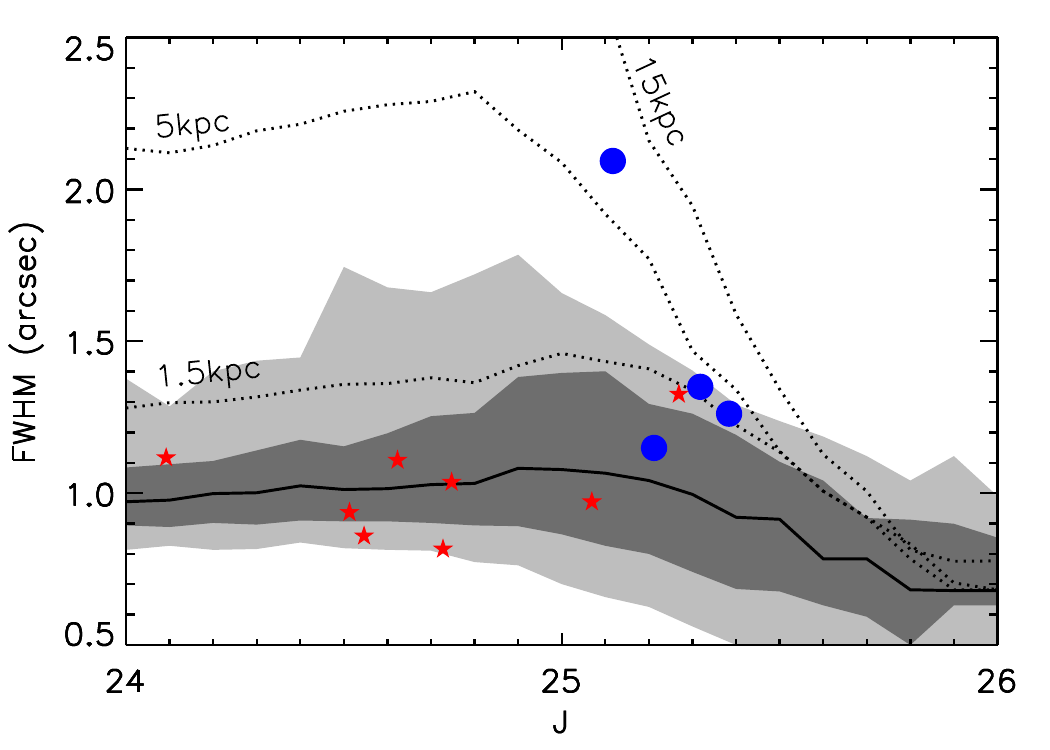}} \\

\caption{The measured FWHM as a function of magnitude for the UltraVISTA and UDS galaxy samples in the top and bottom row respectively.
Each column shows the results for the $z'$ and $J$-bands, with the galaxy candidates presented as the blue circles.
The four galaxies that lie within the COSMOS CANDELS imaging are shown as cyan circles.
The magnitudes presented for the real and artificial sources are raw 1.8-arcsec diameter circular aperture values.
The red stars show the FWHM of faint galactic dwarf stars, that were originally selected as high-redshift galaxy candidates but later excluded as stars based on the SED fitting analysis.
The black curve shows the median value of recovered FWHM as it varies with the magnitude of injected point sources, and the dark and light grey bands show the 68\% and 95\% confidence intervals respectively.
The dotted lines show the median value of recovered FWHM for Sersic profiles with $r_{1/2} =$ 1.5, 5.0 and 10.0 or 15.0 kpc, that were convolved with the PSF and injected into the images.}
\label{fwhm}
\end{figure*}

There is growing evidence for the presence of nebular emission lines in the spectra of high-redshift galaxies~\citep{Stark2013,SmitR.2013}.
Aside from the physical repercussions for the star-formation environment at high redshift, the contamination of the rest-frame optical photometry by nebular emission lines can increase the stellar mass estimate and hence artificially suppress the derived sSFR~\citep{Stark2013}. 
In contrast to the majority of LBGs known, our candidates are sufficiently bright to be detected in the \emph{Spitzer}/IRAC \chone and \chtwo-filters, and several galaxies within our sample show an unusual \IRACcol colour that cannot be reproduced by continuum fits for high or low-redshift galaxies.
The \IRACcol colours of dwarf stars can reproduce such colours (see Figure 3 of~\citealp{Bowler2012}), and the stellar locus in \IRACcol colour would roughly follow the predicted nebular emission curve.
The dependence occurs because cooler T-dwarfs have extremely red spectra (\IRACcol $ \sim 0.0 \mbox{--} 2.0$) and hence become confused with higher redshift candidates, whereas the M and L dwarfs tend to have colours of \IRACcol $ \sim -0.5$ and best-fit galaxy redshifts of $z_{\rm phot} < 6.7$.
However, we expect the degree of stellar contamination in our sample to be very low, as a result of our SED fitting analysis in combination with FWHM measurements (see Section~\ref{fwhm}) that can aid the exclusion of point sources for the candidates with the best (but still significantly worse that the high-redshift models) stellar fits.

For comparison with our observations, we predict the \IRACcol colours for continuum only and continuum + nebular emission by adding bright nebular emission lines to stellar continuum models.
The underlying continuum was taken from a~\citet{Bruzual2003} model with a constant SFH, Chabrier IMF,  ages of either 10 or 100 Myr and metallicities of either $1/5\, {\rm Z}_{\sun}$ or $1/50 \,{\rm Z}_{\sun}$.
The model parameters were chosen to produce a range of plausible \IRACcol colours, while including the bluest values that could be exhibited by a realistic star-forming galaxy at high redshift in the absence of nebular emission lines.
The expected colours for a given rest-frame equivalent width of H$\beta_{}$~and [OIII] combined were estimated assuming the H$\beta_{}$~to [OIII]$\lambda$4959, 5007 and H$\beta_{}$~to [OII]$\lambda$3727 ratios for a metallicity of $1/5 \,{\rm Z}_{\sun}$ calculated by~\citet{Anders2003}, and assuming the H$\alpha$ to H$\beta_{}$~ratio of 2.87 from~\citet{Osterbrock2006}.
The combined $EW_0$ of H$\beta_{}$~to [OIII] $\lambda$4959, 5007 was chosen to be a minimum of 637 \AA~at $z = 6.8$, to correspond to the lower limit for the $EW_0$ of the mean $z = 7$ galaxy derived by~\citet{SmitR.2013}.
An upper value for the $EW_0$ was taken to be 1582\AA, the value derived from the four bluest galaxies in the~\citet{SmitR.2013} sample.
The $EW_0$ was allowed to evolve with redshift according to $EW ({\rm H}\beta_{} + {\rm [OIII]})  \propto (1+z)^{1.8}$ which has been derived from lower-redshift results~\citep{Fumagalli2012}.
Although we did not calculate the relative line strengths using a full recombination analysis, our results closely mimic those presented in~\citet{Wilkins2013} where the full analysis was undertaken.

Our observed \IRACcol colours plotted against the best-fitting photometric redshift (without Ly$\alpha$ emission included), along with the predicted colours from our models are shown in Fig.~\ref{fig:nebularemission}.
We excluded candidates that had confused IRAC photometry, and those that were undetected at less than $2\sigma$ significance in both IRAC filters, resulting in 15 galaxies remaining.
Around half of the galaxies with isolated IRAC detections are consistent with continuum-only models, however the remaining galaxies show deviations from the approximately flat colour and follow the predictions for continuum + nebular emission line models.
~\citet{DeBarros2012} found that 65\% of $z \sim 3-6$ galaxies showed signs of strong nebular emission lines, which would agree with our results that a fraction of our sample show IRAC colours consistent with no nebular contamination, assuming that the galaxies that are detected in the IRAC bands are representative of the population as a whole.
Our results are also consistent with~\citet{SmitR.2013}, who found evidence for strong nebular emission in the majority of seven lensed galaxies around $z_{\rm phot} \simeq 6.7$.
Future deconfusion analysis of the `ultra-deep' imaging regions and spectroscopic confirmation of the candidates would  allow tighter constraints on the prevalence and strength of nebular emission lines in the SEDs of bright $z \sim 7$ galaxies.

The presence of strong nebular emission lines in the SEDs of our galaxies could affect the best-fitting SED `stellar-only' model that we determine and therefore the $M_{1500}$ we derive.
Although we do not explicitly fit with nebular emission lines in our SED fitting analysis, we do consider the SED fits both with and without the \emph{Spitzer}/IRAC bands that could be contaminated by the nebular emission lines (see Figure~\ref{fig:nebularemission}).
Crucially, the rest-frame UV slope from which we measure the $M_{1500}$ of each galaxy, is predominantly constrained by the four near-infrared filters ($Y$, $J$, $H$ and $K$) in both the UltraVISTA and UDS fields, and the data in the \chone and \chtwo filters is at least 0.5 mag shallower and therefore contributes significantly less weight to the resulting best-fitting model.
As a simple check for how much nebular emission lines could affect our measured $M_{1500}$, we calculated the $M_{1500}$ from fits that included and excluded the \chone and \chtwo data, and found no systematic difference between the derived $M_{1500}$ values.

\section{Galaxy Sizes}\label{sizes}

Lyman-break galaxies are known to be smaller at high-redshift, with the median half-light radius for faint (0.1-1$L^*$) galaxies being $ r_{1/2} < 1$ kpc at $z = 7$~\citep{Oesch2010b,Huang2013}.
Confirmation of the existence of a clear size-magnitude relation (as observed at lower redshift) becomes challenging at $z >6$, because of the restricted luminosity baseline over which sizes can be accurately measured.
For example, the few tens of galaxies at $ z > 7$ detected and measured in the HUDF by~\citet{Oesch2010b} and more recently in the UDF12 dataset by~\citet{Ono2013} typically have sizes at the limit of the resolution of~\emph{HST} ($ < 0.1$ arcsec, which  corresponds to a physical size of $ d < 0.5$ kpc at $z = 7$).
By incorporating the wider-area CANDELS data, ~\citet{Grazian2012a} were able to determine a size-magnitude relation at $z = 7$ extending to $ M_{\rm UV} \simeq -21$.
An extrapolation of the size-magnitude relations of~\citet{Oesch2010b} and~\citet{Grazian2012a} would predict $1.0 \lesssim r_{1/2} \lesssim 1.5$kpc for the bright galaxies detected here ($-22 \gtrsim M_{\rm UV} \gtrsim -23$ at $z = 7$).
Converting the half-light radius into a full-width at half-maximum (FWHM) depends on the functional form of the surface-brightness profile; here we assume a Sersic profile with Sersic index $n = 1.5$ following~\citet{Oesch2010b}.

\subsection{FWHM measurements}

We measured the FWHM of the galaxies in the final UltraVISTA and UDS samples shown in Table~\ref{table:allprops} using {\sc SExtractor}, where the value is calculated from a Gaussian fitted to the core photometry.
Note that fitting a Gaussian to an extended Sersic profile causes an overestimate of the FWHM, and can be sensitive to the deblending procedure in {\sc SExtractor}, which can cause nearby objects to be grouped and the resulting FWHM measurement to be larger than for the central object alone.
The FWHM of the UltraVISTA and UDS galaxies in the $z'$ and $J$-bands are shown in Fig.~\ref{fwhm}, along with the measured FWHM of simulated stars and galaxy profiles within each field.
The $Y$-band imaging has poorer seeing than the $J$-band, and is of insufficient depth in the UDS image for FWHM measurements, and hence is not used.
To simulate the range of sizes of recovered point spread functions (PSFs) and typical galaxy profiles, we injected and recovered simulated source profiles into the images.
The PSF derived using the method described in Section~\ref{enclosed_flux}, and Sersic galaxy profiles with a given $r_{1/2}$, were scaled to an input 1.8-arcsec circular aperture magnitude and injected into blank regions of the $z'$ and $J$-band images.
Simulated galaxy profiles were created with intrinsic $r_{1/2} = 1.5, 5.0, 10.0 \,{\rm and} \,15.0$ kpc, and were then convolved with the PSF (modelled here as a Moffat function obtained by fitting to the stacked PSF).
The injected images were then analysed using {\sc SExtractor}, and the FWHM and aperture magnitude measurement was extracted for each recovered source,  thus providing a consistent size measurement for comparison with our sample.
In general, the recovered FWHM increases towards fainter magnitudes as a result of noise in the measurement, with the 68\% and 95\% confidence levels showing asymmetry as a result of the increased probability of detection of a source that has been boosted in magnitude and size.
As one would expect, the FWHM measurements become increasingly unreliable to fainter magnitudes, where the median FWHM turns over and starts to decrease as a result of the bias in selection; only sources that sit on a noise spike at such faint magnitudes will be detected and included.

Of the 30 galaxies in our final UltraVISTA sample,  approximately two-thirds are consistent with having $r_{1/2}~\geq~1.5~{\rm kpc}$ assuming a Sersic profile with $n = 1.5$.
The $z'$-band imaging shows a clearer separation of the measured FWHM of galaxies in our sample and that from PSFs, due to the increased depth and better seeing available compared to the $J$-band imaging.
Several of our brightest galaxies, including the brightest candidates from~\citet{Bowler2012}, are clearly resolved in the ground-based imaging, showing FWHM values that would suggest intrinsic sizes up to $r_{1/2} \sim 5 $ kpc.
Although the majority of the UltraVISTA galaxies are also consistent with being unresolved in the ground-based imaging, as shown by the 95\% percentile displayed in Fig.~\ref{fwhm}, the distribution of FWHM values away from the locus of recovered genuine points sources suggests otherwise.
Stellar contamination of the sample is strongly ruled out by the SED fitting, and given that the current known size-magnitude relation would predict a FWHM that is only just resolvable from the ground, it is entirely plausible that we would find unresolved galaxies with the available seeing.
This point is reinforced in Section~\ref{sect:candels}, where we analyse \emph{HST} imaging of four galaxies in the UltraVISTA sample that are consistent with being a point source in the ground-based data (highlighted in cyan on Fig.~\ref{fwhm}), and find strong evidence that they have extended low surface brightness features.

From the final UDS sample of four galaxies, two appear resolved in the $z'$-band imaging (see Fig.~\ref{fwhm}), including the known extended z = 6.595 galaxy Himiko, which is the only UDS candidate to appear clearly resolved in the $J$-band imaging.
Himiko has an extent of $\sim 9$ kpc in the continuum measured from higher resolution~\emph{HST} imaging, consistent with the FWHM measured here, and a diffuse \Lya halo greater than 17 kpc across.
There are no sources in the UltraVISTA sample that have FWHM suggesting that they exceed 5 kpc in size, supporting the conclusion that Himiko is an extremely rare triple merger system and not a typical ${\rm m}_{\rm AB} \sim 25$ galaxy at $z = 7$.

Our results agree with \citet{Willott2013}, who found that around half of their bright $z \sim 25$ LBGs at $z =6$ were resolved in the CFHTLS ground-based imaging which had a seeing of 0.85-arcsec.
They found a median galaxy FWHM $ = 1.1$-arcsec, which corresponded to an intrinsic FWHM = $0.35$-arcsec or a $r_{1/2} = 2$ kpc.
 \citet{Willott2013} noted that, from high-resolution imaging of several members of their sample with \emph{HST}, many of the galaxies appeared clumpy and extended with features up to 6 kpc from the centre.
 Similarly, the two brightest $z_{850}$-drops presented in~\citet{Ono2013} are formed of multiple components.
 These results suggest that mergers or interactions may be prevalent in bright galaxies at high redshift, although it is possible that these low surface brightness features are also present in fainter galaxies beyond the current detection limits.
 
\subsection{\emph{HST} imaging from CANDELS}\label{sect:candels}

\begin{figure*}

\includegraphics[width = \textwidth]{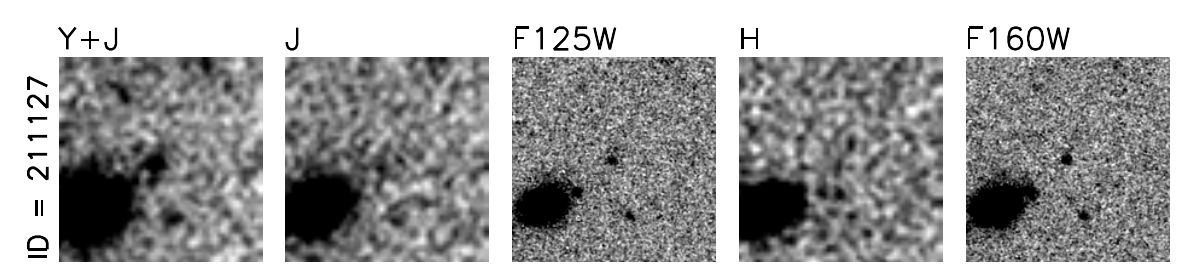}\\
\includegraphics[width = \textwidth, trim = 0 0 0 1.0cm, clip = true]{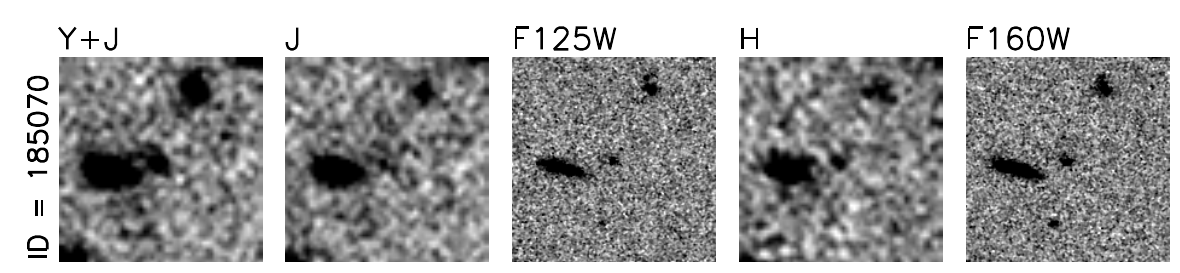}\\
\includegraphics[width = \textwidth, trim = 0 0 0 1.0cm, clip = true]{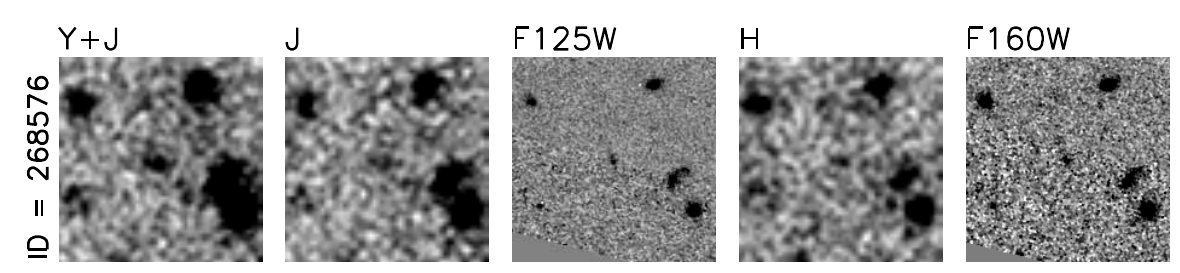}\\
\includegraphics[width = \textwidth, trim = 0 0 0 1.0cm, clip = true]{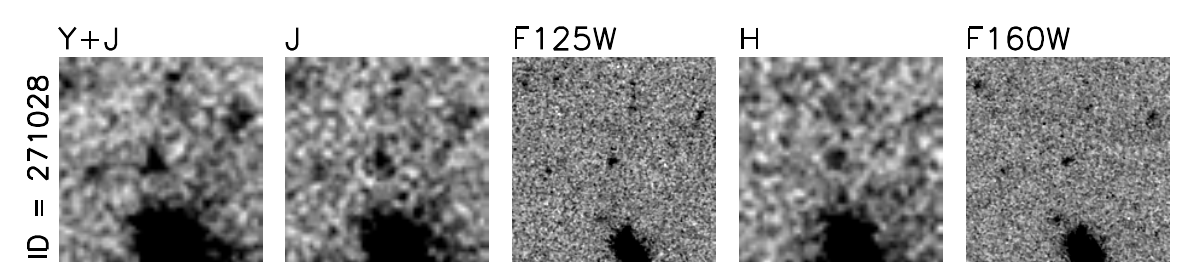}

\caption{Ground-based and \emph{HST} postage-stamp near-infrared images of the four galaxy candidates that lie with the CANDELS COSMOS imaging.
For each object the UltraVISTA $Y+J$ detection image is shown on the left, followed by the UltraVISTA $J$-band, WFC3 $J_{125}$, UltraVISTA $H$ and WFC3 $H_{160}$ stamps.
Each stamp is 10 x 10 arcsec$^2$, with North to the top and East to the left.
Three of the objects presented are fairly close to a low-redshift extended galaxy, however with the improved resolution provided by \emph{HST} the objects are  confirmed as clearly separate objects.
Note that object 268576 lies near the edge of the WFC3 imaging, and hence the errors on the photometry and size measurements are correspondingly higher.}
\label{fig:candels}
\end{figure*}

Of the 30 candidates within the 1 deg$^2$ of UltraVISTA imaging, four lie within the 200 arcmin$^2$ of the CANDELS COSMOS field~\citep{Grogin2011} which lies close to the centre of the CFHTLS pointing that defines our maximal survey area, and hence in the left-central strip of the `ultra-deep' UltraVISTA imaging.
The CANDELS COSMOS field consists of a mosaic of $4 \times 11$ pointings of \emph{HST}/WFC3 providing F125W ($J_{125}$) and F160W ($H_{160}$) imaging, and ACS imaging taken with the F606W ($V_{606}$)and F814W ($I_{814}$) filters in the parallel observations~\citep{Koekemoer2011}.
In Fig.~\ref{fig:candels} we present postage-stamps of the four objects in the $J_{125}$ and $H_{160}$ images, with the corresponding VISTA $J$ and $H$ images for comparison.
The measured photometry and FWHM values from {\sc SExtractor} are shown in Table~\ref{table:candels}.

Of the four candidates we find in the UDS, the LAE Himiko lies within the 145 arcmin$^2$ of CANDELS UDS imaging.
The detailed analysis of Himiko from the \emph{HST} imaging was presented by~\citet{Ouchi2013}, who found the galaxy to consist of three components in an apparent triple-merger system.
A triple-merger system is very rare at any redshift, and the UltraVISTA galaxies that lie within the CANDELS COSMOS imaging do not show any evidence of multiple components.
The four galaxies in the CANDELS imaging however are fainter than Himiko, with absolute magnitudes in the range $ -21.9 \leq M_{1500} \leq -21.5 $ compared to $M_{1500} = -22.1$, and the brightest candidates presented here are comparably extended from the FWHM measurements in the ground-based imaging.
Our sample includes galaxies brighter or comparable to the luminosity of Himiko, and hence follow-up with \emph{HST} could determine how many of the most luminous $z = 7$ galaxies are in merger systems and would constrain the size-magnitude relation at the very bright end.

In the UltraVISTA imaging the four candidates, highlighted in Fig.~\ref{fwhm}, show FWHM that are consistent with being unresolved at ground-based resolution, with the exception of 271028 which appears marginally resolved in the $z'$-band imaging.
The typical resolution of the \emph{HST}/WFC3 $J_{125}$ and $H_{160}$ imaging in the CANDELS fields is $0.20 \pm 0.01$-arcsec~\citep{Koekemoer2011}, and hence of the four candidates, three immediately appear resolved in the WFC3 data on the basis of the FWHM measurements.
The brightest object in the WFC3 imaging, 211127, is close to being unresolved with a FWHM $ = 0.22$ that only slightly exceeds the value expected for a point source.
The remaining galaxies have larger FWHM in the range 0.3--0.5 arcsec, which is supported by the observed elongation of several galaxies in the images shown in Fig.~\ref{fig:candels}.
Note that object 268576 lies near the edge of the image and hence the size and photometry measurements are subject to large uncertainties.

\subsubsection{Photometry}

\begin{table*}
\caption{The measured photometry and errors for the four $z \simeq 7$ galaxies in our sample that lie within the area of \emph{HST} imaging provided by the CANDELS COSMOS survey.
The measurements from the UltraVISTA $J$ and $H$ data were made in a 1.8-arcsec diameter circular aperture and corrected to the 80\% enclosed flux level.
The $J_{125}$ and $H_{160}$ photometry was measured in a 0.6-arcsec diameter circular aperture, which corresponds to 80\% enclosed flux for a point source.
The object ID 211127 appears barely resolved in the WFC3 imaging based on a measured FWHM close to the typical resolution of 0.20-arcsec, but as shown in Fig.~\ref{fig:surfacebrightness}, all four galaxies are clearly resolved in the~\emph{HST} imaging.}
\label{table:candels}
\begin{tabular}{l c c c c c c c}
\hline
ID & $J$ & $J_{125}$ & $H$ & $H_{160}$ & FWHM $J$  &  FWHM $J_{125}$  & $M_{1500}$\\
&  &  &  &  & /arcsec & /arcsec & \\
\hline

  211127 & $ 25.4^{+  0.3}_{ -0.2}$ & $ 25.6^{+  0.1}_{ -0.1}$ & $ >  25.8$ & $ 25.6^{+  0.1}_{ -0.1}$ &   1.1 &  0.21& $-21.9$ \\
  185070 & $ 25.3^{+  0.2}_{ -0.2}$ & $ 25.7^{+  0.1}_{ -0.1}$ & $  25.6_{ -0.4}^{+  0.6 }$ & $ 25.6^{+  0.1}_{ -0.1}$ &   1.3 &  0.31 & $-21.9$ \\
  268576 & $ 25.6^{+  0.3}_{ -0.3}$ & $ 25.7^{+  0.1}_{ -0.1}$ & $ >  25.9$ & $ 26.0^{+  0.1}_{ -0.1}$ &   1.2 &  0.48& $-21.5$ \\
  271028 & $ 25.3^{+  0.3}_{ -0.3}$ & $ 25.8^{+  0.1}_{ -0.1}$ & $  25.3_{ -0.4}^{+  0.6 }$ & $ 26.0^{+  0.1}_{ -0.1}$ &   1.6 &  0.37 & $-21.5$ \\

\hline
\end{tabular}
\end{table*}

When comparing the measured aperture magnitudes shown in Table~\ref{table:candels}, we find an offset between the photometry measured in the UltraVISTA $J$ and \emph{HST}/WFC3 $J_{125}$ imaging which is inconsistent with the galaxies being point sources.
Excluding object 268576 near the edge of the imaging, we find a mean offset of 0.4 mag between the raw 0.6-arcsec diameter aperture photometry measured on the $J_{125}$ image (80\% enclosed flux for a point source) and the 1.8-arsec diameter aperture photometry measured on the $J$-band image ($\sim70$\% enclosed flux) when corrected to match the enclosed flux level of the WFC3 imaging.
Of course the most extended galaxies in the higher resolution \emph{HST} imaging will similarly appear extended in the ground-based imaging, as we have shown the majority of the galaxies in our sample appear resolved in the $z'$ and $J$-band data.
However, because the circular apertures we used for photometry in the space- and ground-based data are designed to enclose around 70-80\% of the flux of a point source, in the case of a resolved galaxy, more of the flux will spread beyond the aperture in the higher resolution \emph{HST} imaging compared to the ground-based imaging where the seeing dominates over the intrinsic FWHM.
We created model galaxy profiles to predict the magnitude offset we would expect between the VISTA and WFC3 imaging for an extended profile as compared to a point source.
A magnitude offset of 0.4 mag could be reproduced if the simulated galaxy with an exponential profile (Sersic index $n = 1$) had a half-light radius of $r_{1/2} = 1$ kpc.
Such a profile would result in 0.05 mag being lost from the ground-based aperture (increased to 0.1 mag for a profile with $r_{1/2} = 1.5$ kpc).
The ground-based imaging provides a closer measure of the true total flux of the galaxy than the results of using small apertures on higher resolution \emph{HST} imaging, when the galaxy is assumed to be unresolved in both cases.

Finally, we measured the photometry for our sample using larger apertures, to empirically determine what the effect of assuming our galaxies are point sources in the ground-based imaging has on the total magnitudes derived. 
We used 3-arcsec diameter circular apertures (corrected to total assuming a point source) and two further magnitudes from {\sc SExtractor}; the MAG\_AUTO which returns an estimate of the total magnitude using Kron apertures and MAG\_ISOCOR which uses isophotal apertures corrected to total using a Gaussian approximation to the galaxy profile.
For the brightest objects, where Kron and isophotal magnitudes correspond to a sufficiently accurate total magnitude, we found an offset of $\sim 0.1$ mag compared with the total magnitude initially measured in a 1.8-arcsec aperture and then corrected to total, which is in agreement with that found for a simulated Sersic profile with $r_{1/2} = 1.5 $ kpc as described above.
Hence, from our ground-based imaging measurements, we expect at most a 0.1 mag brightening of our candidates to correct for missing flux beyond the apertures we use.
In the case of the $J_{125}$ and $H_{160}$ imaging however, the offset is significantly larger and can result in an underestimation of the total magnitude of the galaxy by $\sim 0.4$ mag for photometry measured in a 0.6-arcsec diameter circular aperture.

\subsubsection{Surface-brightness profiles}

\begin{figure}

\includegraphics[width = 0.5\textwidth]{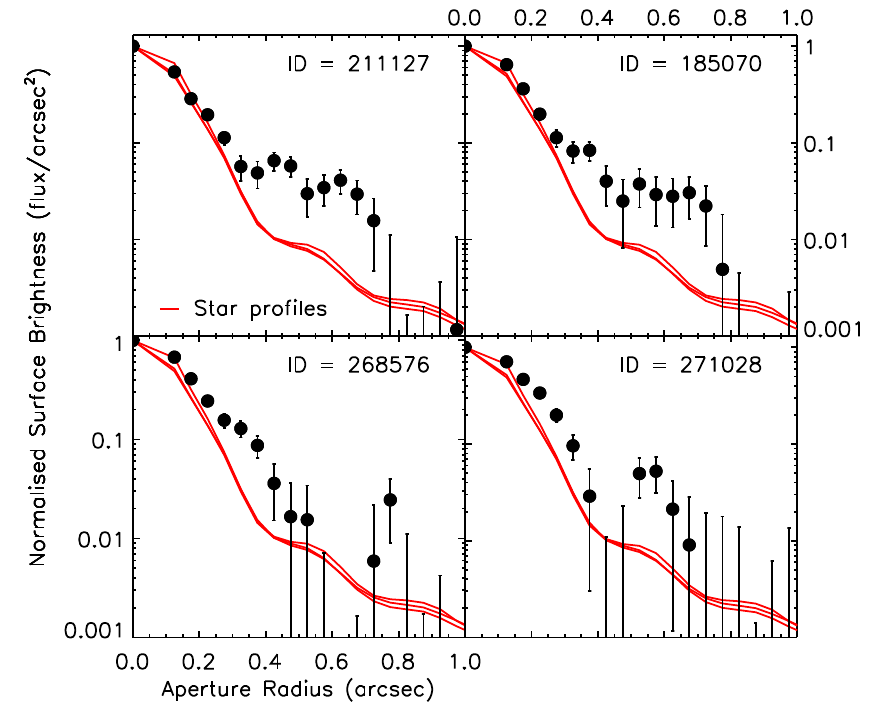}

\caption{The $J_{125}$ surface-brightness profiles of the four galaxies in the UltraVISTA sample that lie within the CANDELS COSMOS \emph{HST} imaging.
Stellar profiles are shown for comparison (red lines), and all the galaxies and stars have been normalised to a peak flux of 1.0 in a 0.2-arcsec diameter aperture.
All four galaxies show extended profiles.
}
\label{fig:surfacebrightness}
\end{figure}

The surface-brightness profiles for the four galaxies in our sample that lie within the CANDELS COSMOS imaging are shown in Fig.~\ref{fig:surfacebrightness}.
Each galaxy was normalised within the central circular aperture of radius 0.1 arcsec, and the normalised surface brightness was calculated in annuli of width 0.05 arcsec.
We extracted several unsaturated stars from the imaging and computed the surface brightness for comparison with the galaxy candidates.
The profiles show that the first three galaxies show compact cores, surrounded by extended emission that is observable to a maximum radius of 0.8-arcsec.
For object 211127, which has a FWHM consistent with being a point source in the $J_{125}$ image, the presence of extended emission further rules out a galactic dwarf star (which is strongly disfavoured from the SED fitting analysis).
The compact cores observed for galaxies 211127 and 185070 are consistent with being unresolved in the \emph{HST} imaging and could suggest a bulge component or active nucleus (further discussed in Section~\ref{qlf}). 
The largest galaxy in our sample as measured by both the UltraVISTA $J$-band and $J_{125}$ FWHM values, 271028, has a core that is significantly larger than the stellar point source surface brightness profile.

\section{The Luminosity Function}\label{sect:lf}

The result of the analysis presented above is a sample of 34 $z \simeq 7$ galaxies with $M_{1500} \leq - 21.2$,  selected from the 1.65 deg$^2$ of imaging within the UltraVISTA and UDS fields.
In this section we use our sample of galaxies to calculate the very bright end ($M_{1500} \simeq -22.5$) of the luminosity function at $z = 7$.
Before we proceed to calculate the binned luminosity function points however, we must calculate how complete our selection process is and hence ascertain the true number density of sources our sample implies.
At the bright end of the LF the use of realistic simulations becomes even more important, because the steeply-declining function can cause flux boosting of a significant number of galaxies into the sample from below the flux density limit of the data.

\subsection{Completeness simulations}\label{compsim}

The sample of galaxies presented here will suffer incompleteness due to a combination of blending with foreground/background objects and photometric scattering which can shift galaxies over/below the detection threshold of the survey, as well as potentially changing the colours in such a way as to cause rejection as a contaminant.
Hence, to calculate the LF accurately, the numbers of galaxies in each magnitude bin must be corrected for these effects.

We calculated the completeness of our two survey fields by injecting sources, assumed to be unresolved in ground-based imaging (see Section~\ref{sizes}), that mimic high-redshift LBGs, and recovering them using the same method used for the selection of galaxies in this paper.
We first populated a grid of absolute magnitude vs. redshift space with steps of $\Delta M = 0.1$ and $\Delta z = 0.05$, assuming a linearly evolving Schechter function with redshift as described in~\citet{McLure2009} and~\citet{Mclure2013a}.
To determine the effect of a less steep functional form on the completeness corrections, we also ran simulations where a double power law was used to populate the input $M_{\rm UV}-z$ plane (see Section~\ref{binnedlf}).
When populating the input $M_{\rm UV} - z$ grid, the LF was integrated down to one magnitude fainter than the median $5\sigma$-limit of each field (e.g. an apparent magnitude limit of $Y = 27$ for the UltraVISTA DR2 strips), to account for scattering into the sample from below the magnitude limit of the survey.
For each galaxy in the grid we randomly assigned a rest-frame UV slope $\beta_{}$, drawing the value from a Gaussian distribution centred on $\beta_{} = -2.0$ with a standard deviation of $\sigma = 0.2$.
We then calculated the observed total magnitude in each band for a galaxy model with that $\beta_{}$, $M_{\rm UV}$, and $z$ combination, taking models from the~\citet{Bruzual2003} library.
To inject the galaxy into each band, the PSF (as determined using the method described in Section~\ref{enclosed_flux}) was scaled to the apparent magnitude in each band and added into the images at a random position.
For each survey field, we performed the simulation on 4 to 5 subsections to provide a representative result given the different depths of the individual mosaic panels and strips in the UltraVISTA and UDS imaging, while keeping computing time reasonable.
We then ran {\sc SExtractor} on the full optical and near-infrared images with the artificial sources injected, and selected objects from the resulting catalogues that passed the $Y$, $J$, or $Y+J$ cuts imposed on the genuine galaxy catalogues described in Section~\ref{sect:candsel}.
A non-detection in the $i$-band was required, using local depths for the UltraVISTA ultra-deep data and global depths for candidates in the UDS to match our selection criterion used for the selection of real sources.
Finally, SED fitting was performed and candidates were retained if the $\chi^2$ value was acceptable, and no low-redshift solution had a $\chi^2$ within $\Delta \chi^2 < 4$ from the minimum. 
In total we added 750000 artificial sources to each field through multiple runs of the simulation, with roughly $10000 - 50000$ candidates being retrieved depending on the functional form assumed for the input LF.
The resulting completeness values were $\simeq 70-80$\% for the two brighter LF bins we use, where the loss of input objects is dominated by the blending with foreground sources.
For the faintest objects in our sample the completeness falls as low as $\simeq 50$\%, as a result of objects at the $5\sigma$ limit of the survey being scattered below the detection limit, coupled with up-scattered objects (that for the brighter magnitude bins helps to balance the loss of objects scattered to fainter magnitudes), having poorer SED fits and hence being more likely excluded as a high-redshift object.

\subsection{The binned luminosity function}\label{binnedlf}

\begin{figure}
\includegraphics[width = 0.5\textwidth, trim = 0.7cm 0cm 0cm 0cm, clip = true]{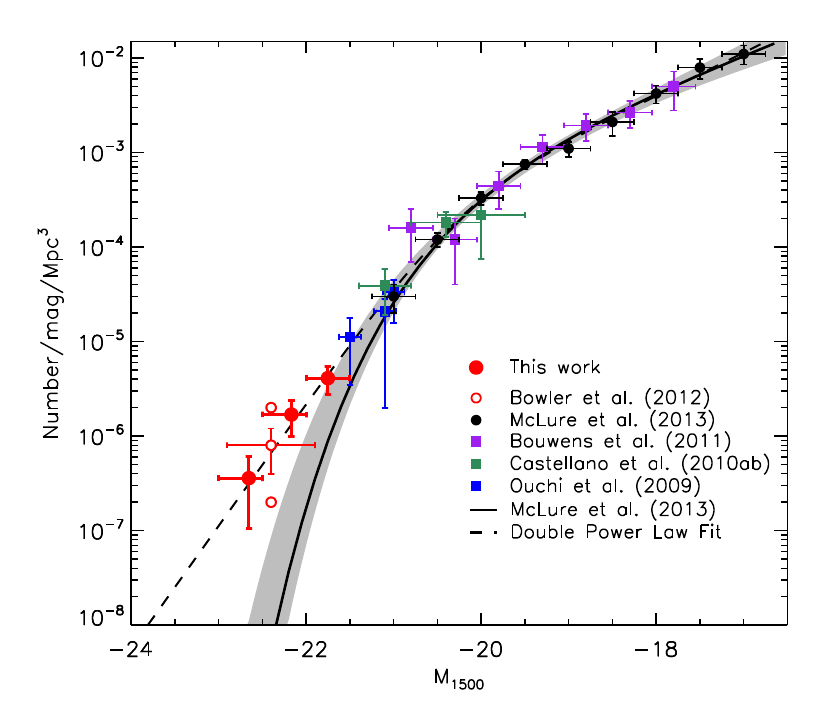}

\caption{ The $z = 7$ UV ($\sim 1500$\AA) luminosity function.
The results from our sample of galaxies from the UltraVISTA DR2 and UDS fields are shown as the red filled circles. 
The previous estimate calculated in~\citet{Bowler2012} is shown as the open red circle, where the upper and lower circle represent the prediction if ten or one of the candidates were confirmed to be at $ z > 6.5$.
The best-fitting Schechter function at $z = 7$ from~\citet{Mclure2013a} is plotted as the black line, and the best-fitting double power law is shown as the dashed line.
By varying the Schechter function parameters ($M^{\star}, \phi^{\star}$ and $\alpha$), a one-sigma confidence limit on the best-fitting LF can be obtained, and is shown as the grey shaded region.
Data points determined by~\citet{Mclure2013a} and~\citet{Bouwens2011a} are shown extending to $M_{\rm UV} = -17$.
The remaining data points were obtained from wider-area ground-based surveys by~\citet{Ouchi2009} and~\citet{Castellano2010a, Castellano2010b}.
}
\label{fig:lf}
\end{figure}

The luminosity function in a given absolute magnitude bin,  $\Phi(M)$, derived from our data over the full redshift range of our survey ($6.5 < z < 7.5$) was estimated using the classic $1/V_{\rm max}$ method of~\citet{Schmidt1968}: 
\begin{equation}\label{equationLF}
\Phi(M) = \sum_{i = 1}^{N} \frac{C(M_i, z_i) }{V_{{\rm max}, i}}
\end{equation}
where the sum is over the $N$ objects in the sample within the magnitude bin.
In the simplest form, the luminosity function is found by summing $1/V_{\rm max}$ for all $N$ galaxies in the bin, where $V_{\rm max}$ is the maximum volume the galaxy could occupy and still be included in the sample.
The $V_{\rm max}$ was calculated by redshifting the best-fitting SED for each galaxy in the sample until it was fainter than the required  $Y$ and/or $J$-limit for detection in each field (with $z_{\rm max} \leq 7.5$), and summing the resulting volumes.
The incompleteness is taken into account via the correction factor $C(M_i, z_i)$, which depends on the absolute magnitude and redshift of the galaxy.

When calculating the binned LF from our sample, we exclude the 0.29 deg$^2$ of shallower data within the UltraVISTA field not covered by the ultra-deep strips, as the volume is small compared with that from the UltraVISTA ultra-deep and UDS fields combined.
The inclusion of the deep UltraVISTA field in the LF calculation leaves the fainter bins unchanged, as none of these objects could have been detected in the shallower data, and reduces the brightest bin by 18\%.
To be conservative in our LF estimation, galaxies were only included if the best-fitting photometric redshift without \Lya emission was in the range  $6.5 < z < 7.5$.
The absolute UV magnitude was calculated from the best-fitting SED to each galaxy (observed in the rest frame using a top-hat filter of width $\Delta \lambda = 100$\AA~ centred on 1500\AA), without any correction for dust attenuation.
We chose three 0.5 magnitude wide bins centred on $M_{1500}= -22.75$, $-22.25$ and $ -21.75$ to span the range of absolute magnitudes within our sample.
The brighter bin at $M_{1500}= -22.75$ is occupied by two galaxies, the top two candidates presented in~\citet{Bowler2012}, and the fainter bins contain 6 and 9 galaxies respectively.
We note that if all the galaxies listed in Table~\ref{table:allprops} were included (i.e. also those which require Ly$\alpha$-emission in the SED to lie at $z > 6.5$) the faintest bins would rise by $\simeq 0.1\,$ dex.

To calculate the correction required for each galaxy due to the incompleteness of the survey, we compared the results when using an evolving Schechter function and several double power law functions.
The exponential cut-off at the bright end of the Schechter function results in a significant number of galaxies being scattered into the sample from below the limit of the survey.
The number of these scattered galaxies dominates over the intrinsic number of galaxies in the bin, resulting in a completeness value that exceeds one and a reduction in the final number density derived.
However, even correcting the number density down as a result of scattering, we find an excess of galaxies above the Schechter function prediction, implying that the density of galaxies we find cannot be accounted for by such a function.
One way to reconcile our results with a Schechter functional form, given that we find an excess of galaxies at $M_{1500} \lesssim -22.0$, would be to assume that the characteristic magnitude, $M^*$, is brighter than the current determination of the $z = 7$ parameters~\citep{Mclure2013a, Schenker2013}.
However, the result of shifting $M^*$ to a brighter value would bring the fit into conflict with the data points around the knee of the function, in particular the points from~\citet{Ouchi2009} and the brightest bin from~\citet{Mclure2013a}.
Hence we proceed to calculate the completeness corrections assuming a shallower decline at the bright end of the LF as implied by our data, using a double power law (DPL).
Our DPL function, which is the parameterisation commonly used to fit the quasar luminosity function (see Section~\ref{qlf}), has the following functional form:
\begin{equation}
   \phi (M) = \frac{\phi^*}{10^{0.4 (\alpha + 1)(M - M^*)} + 10^{0.4(\beta_{} + 1)(M-M^*)}}.
 \end{equation}
where $\phi^*$, $M^*$ and $\alpha$ are the normalisation, the characteristic magnitude and the faint-end slope in common with the Schechter function, and the bright-end slope is described by the power $\beta_{}$.
We carried out completeness simulations for DPL functions for bright-end slopes in the range $-4.6 \leq \beta_{} \leq -4.0$.
The results of the simulations are relatively insensitive to the steepness of the DPL function, with the results changing by less than 4\% in the faintest bins and the brightest bin remaining unchanged.
The data suggest a more moderate value of $\beta_{} = -4.3$, which we use for our final results presented in Fig.~\ref{fig:lf} and Table~\ref{table:lf}.

Finally, if the underlying LF was a Schechter function, the absolute magnitude distribution should be dominated by galaxies at the limiting depth of the survey, however we find a more uniform distribution in the range $ -22.5 < M_{\rm UV} < -21.5$.
We calculate the probability of obtaining the absolute magnitude distribution we find by drawing a sample at random from the output of our simulations, and determining the number of times the distribution has the same number or more galaxies in the brighter 0.5 mag section of the $M_{1500}= -22.0$ bin as compared to the fainter section.
For a Schechter function we find this distribution in only 3\% of cases, whereas for the DPL we find this distribution in 18\% of the samples drawn. 
The detection of two extremely bright $M_{1500}\simeq -22.7$ galaxies further supports our claim of a shallower functional form for the LF.

\begin{table*}
\caption{The binned LF points at $z = 7$ derived from our combined UltraVISTA and UDS analysis.
We use three 0.5 mag wide bins centred on $M_{1500}= -21.75, -22.25, -22.75$ and plot the point at the position of the mean completeness corrected absolute magnitude in that bin, which is displayed in the second column.
The errors derived on the number density are the Poisson errors on the number counts.
}
\label{table:lf}
\begin{tabular}{c c c c c}

\hline
$M_{1500}$ range & $M_{1500}$ weighted & $\phi$ & $\phi_{\rm upper}$ &   $\phi_{\rm lower}$ \\
 /mag & /mag & /mag$^{-1}$Mpc$^{-3}$ & /mag$^{-1}$Mpc$^{-3}$ & /mag$^{-1}$Mpc$^{-3}$\\
\hline

$ - 21.5 < M < -22.0 $ & $    -21.75$ & $4.10 \times 10^{-6}$ & $5.47 \times 10^{-6}$ & $2.73 \times 10^{-6} $ \\
$ - 22.0 < M < -22.5 $ & $    -22.17$ & $1.69 \times 10^{-6}$ & $2.38 \times 10^{-6}$ & $1.00 \times 10^{-6} $ \\
$ - 22.5 < M < -23.0 $ & $    -22.66$ & $3.59 \times 10^{-7}$ & $6.13 \times 10^{-7}$ & $1.05 \times 10^{-7} $\\

\hline

\end{tabular}
\end{table*}

\subsection{Gravitational lensing}\label{lensing}

Given that the galaxies in our sample are the brightest known to date at $z = 7$, we must consider the possibility that they are gravitationally lensed.
There are two scenarios to consider, strong lensing by a lower-redshift galaxy directly along the line-of-sight of our candidate and moderate lensing by galaxies close to the line-of-sight.
The first case is ruled out in our sample by the deep optical non-detections of our galaxies.
The second case, of galaxies close to the line-of-sight causing moderate magnifications has been discussed before at $ z = 5-6$ by~\citet{McLure2006}, who found a moderate magnification of $< 0.1$ mag for several candidates in their sample, and~\citet{Willott2013} who concluded that the brightest candidate in their sample was lensed by $\sim 0.25$ mag.
Following the method detailed in~\citet{McLure2006}, we calculated the magnification due to gravitational lensing from galaxies within 10-arcsec of each high-redshift galaxy.
The velocity dispersion, $\sigma_V$, of each potential lensing galaxy was estimated using the Faber-Jackson relation for early-type galaxies from~\citet{Bernardi2003}, which gives a correlation between the absolute magnitude in the rest-frame $i$-band and $\sigma_V$ at $z = 0$.
The magnification can then be calculated by assuming the gravitational potential is an isothermal sphere.
The absolute magnitude in the $i$-band was estimated from the nearest band to the redshifted $i$-band from a $K$-band selected {\sc SExtractor} catalogue, where the total magnitude was estimated using MAG\_AUTO.
We assumed the redshift of the lensing galaxy from the COSMOS Photometric Redshift Catalogue~\citep{Ilbert2008} for the UltraVISTA sample, and the best-fitting photometric redshift for the UDS sample, where we found the $z_{\rm phot}$ using our photometric redshift fitting routine described in Section~\ref{photozs} using the $K$-band selected catalogue.

All of the 34 candidates within our sample have at least one low-redshift galaxy within 10 arcsec on the sky, and several galaxies have up to seven nearby low-redshift galaxies.
The largest magnification comes from the closest galaxies as one would expect, with galaxies around 2.5-arcsec separation typically providing a magnification of 0.1 mag.
Our selection procedure naturally removed genuine high-redshift galaxies in the wings of low-redshift galaxies, with potentially high-magnification factors,  as the foreground galaxy will contaminate the photometry of the candidate resulting in rejection.
If the magnification factors from all of the galaxies within 10 arcsec of the high-redshift galaxy are combined, we find that galaxies in our sample can appear up to 0.3 mag brighter as a result of gravitational lensing.
For example, the brightest galaxy in our sample at a measured $M_{1500}= -23.0$, has an estimated lensing boost of 0.3 mag as a result of multiple objects close to the line-of-sight.
To present the most conservative estimate of the bright-end of the $z = 7$ LF, we demagnify all of the galaxies in our sample according to our estimate of the magnification, and use the corrected magnitudes in our derivation of the LF points.
Furthermore, we present the LF points with a weighted central magnitude for each bin shown in Table~\ref{table:lf}, calculated from the mean  completeness corrected absolute magnitude after demagnification.

\section{Discussion}\label{sect:discussion}

The results of our LF analysis are shown in Fig.~\ref{fig:lf}.
Our results at bright magnitudes $ M_{1500}< -21.5$ are clearly in tension with the best-fitting Schechter function fit to fainter galaxies, and therefore we also fit a DPL to our derived points and the fainter bins from~\citet{Mclure2013a}, with the parameterisation presented in Table~\ref{table:dplparams}.
Early work presented in~\citet{Bowler2012} indicated that there may be an excess of galaxies at the bright end of the $ z = 7$ LF, which has been reproduced here with a confirmation of the four brightest galaxies and the majority of the fainter candidates (see the discussion in Section~\ref{Bowler2012} for more details).
In particular, our two brightest bins contain our most secure candidates (with the two galaxies at $ M_{1500}\sim -22.5$ now detected at $> 10\sigma$ significance) and therefore provide a strong challenge to the Schechter LF fitted to the fainter datapoints.
The tension between the ground- and space-based observations could be compounded by the effect of missing flux from the use of small apertures and a point-source correction in the \emph{HST} images.
As discussed in Section~\ref{sizes}, we find an offset of $\sim$0.4 mag between the total magnitudes (and therefore absolute magnitude) derived from the UltraVISTA imaging and \emph{HST} CANDELS imaging, when the galaxies are treated as point sources.
The size-magnitude relation would suggest that this effect is not a problem for the faintest and therefore smallest galaxies that population the LF, however around the knee of the function at $M_{1500}\simeq -21$, the absolute magnitudes of the galaxies could be underestimated.
Other ground-based analyses support a shallower decline at the bright-end of the LF, with the brightest bin of~\citet{Ouchi2009} higher (but still consistent with) the best-fitting Schechter function.

Our derivation of the bright-end of the LF at $z = 7$ is not in conflict with the current sample of galaxies detected in the CANDELS survey, which in total provides $\sim 800$ arcmin$^2$ of imaging.
Using the DPL fit to our data points at $z = 7$, we would predict  $\sim 4$ galaxies at $M_{1500}< -21.5$ in the CANDELS wide survey.
\citet{Finkelstein2013} has suggested that there may be an excess of $ z \simeq 7 $ high-SFR galaxies in the CANDELS Great Observatories Origins Deep Survey North (GOODS-N), where they found an unusually bright $z = 7.51$ LBG with $ H_{160}= 25.6$ ($M_{\rm UV} = -21.2$).
\citet{Ono2012} found one galaxy at $M_{\rm UV} = -21.8$ that lies within the CANDELS GOODS-N wide field, that was spectroscopically confirmed to be at $z = 7.21$.
The LAE Himiko lies within the CANDELS UDS field ($M_{1500} = -22.1$) and of the four new galaxies we find in the CANDELS COSMOS field, two have photometric redshift at $z > 6.5$ and  $M_{1500} < -21.5$.
The above examples shows that there exist secure detections of $M_{1500} < -21.5$ LBGs at $z \simeq 7$ within the CANDELS fields, with a number in agreement with our predictions, and also illustrates the potential for follow-up observations of the brightest high-redshift galaxies as all the previously known $M_{1500}< -21.5$ examples have been spectroscopically confirmed.

We note here that when finished, the VISTA VIDEO survey~\citep{Jarvis2013} should provide the necessary $Z$ and $Y$-band depth to search for extremely bright $z=7$ LBGs over a total area of 12 deg$^2$, and therefore determine whether our DPL extrapolation holds to brighter magnitudes.

\subsection{Comparison with ${\bf z = 5}$ and ${\bf z = 6}$ results}

\begin{figure*}
\subfloat{\includegraphics[width = 0.5\textwidth, trim = 0.7cm 0cm 0cm 0cm, clip = true]{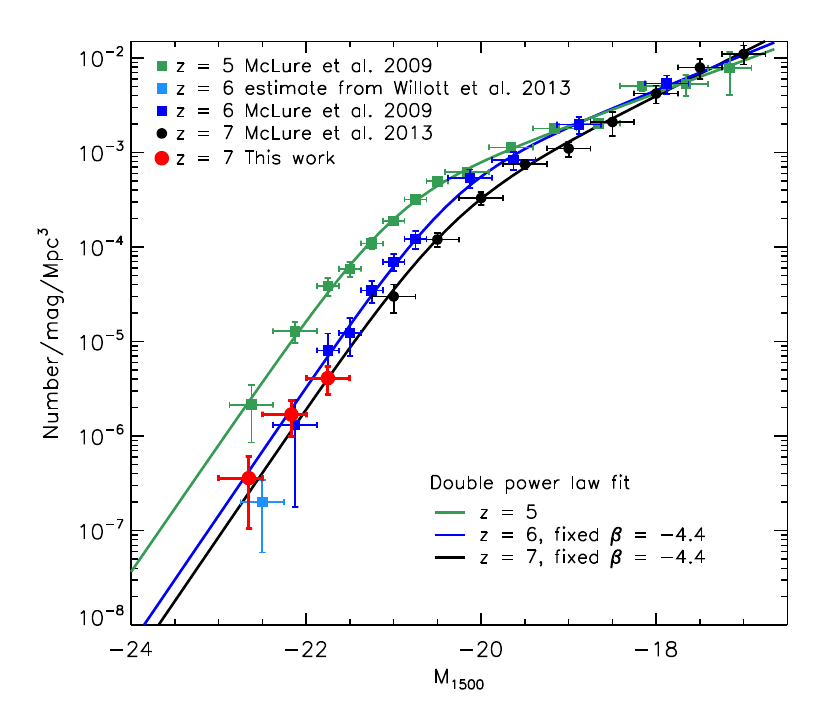}}
\subfloat{\includegraphics[width = 0.5\textwidth, trim = 0.7cm 0cm 0cm 0cm, clip = true]{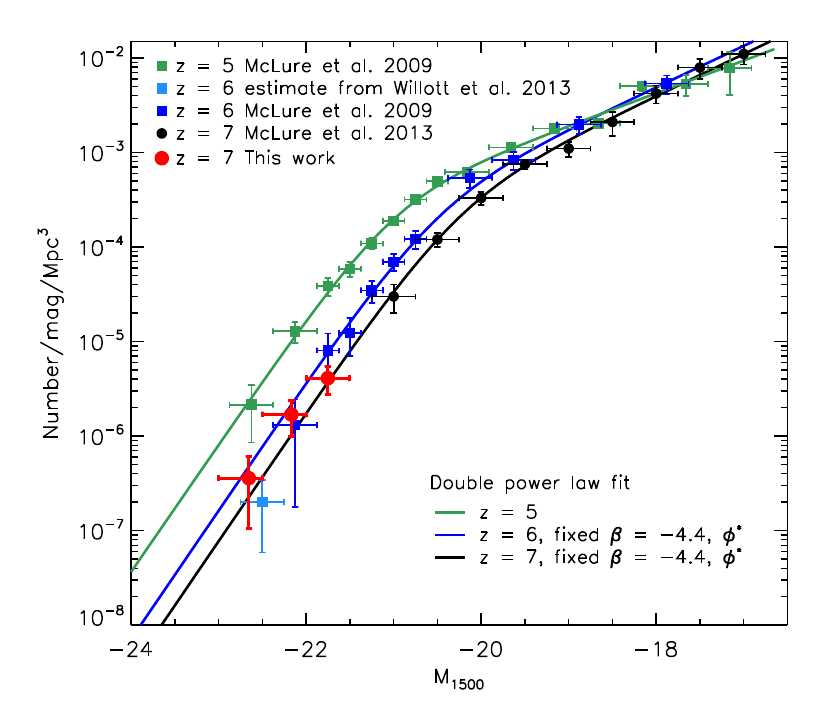}}
\caption{ The $z = 5$, $z = 6$ and $z = 7$ UV ($\sim 1500$\AA) luminosity function points with the best-fitting double power law fits.
The filled red circles show our results for bright galaxies in UltraVISTA and the UDS at $z = 7$, and the black points show the results at $z = 7$ from \citet{Mclure2013a}.
At $z = 5$ and $z = 6$ we plot the results of~\citet{McLure2009} that were obtained from a similar photometric redshift fitting approach to the work presented here.
We estimate the number density of $M_{1500} = -22.5$ galaxies from the work of~\citet{Willott2013} and plot this point in light blue.
The left-hand panel shows DPL fits where the bright-end slope $\beta_{}$ has been fixed to the best-fitting value at $z = 5$, and the right-hand panel shows the results when both $\beta_{}$ and $\phi^*$ are fixed to the value at $z = 5$.
The best-fitting parameters in each case are presented in Table~\ref{table:dplparams}.}
\label{fig:lfdpl}
\end{figure*}

\begin{table}
\caption{The best-fitting double power law parameters for the fits shown in Fig.~\ref{fig:lfdpl} at $z = 5, 6$ and $7$.
The upper part of the table shows the results at $z = 7$ when all the DPL parameters are allowed to vary (shown in Fig.~\ref{fig:lf}).
The central section shows the results when fixing the value of $\beta_{} = -4.4$ to the best-fitting value at $z = 5$, and the lower section shows the results with $\phi^*$ also fixed.}
\label{table:dplparams}
\begin{tabular}{c c c c c}
\hline
$z$ &$ \phi^*$ & $M^*$ & $\alpha$ & $\beta_{}$ \\
& /mag$^{-1}$Mpc$^{-3}$ & /mag & & \\
\hline
$7.0$ &$3.6 \times 10^{-4}$ & $-20.3$ & $-2.1$ & $-4.2$ \\
\hline
$ 7.0 $ & $3.1 \times 10^{-4}$ & $-20.4$ & $-2.2$ & $-4.4$ \\
$ 6.0 $ & $6.1\times 10^{-4}$ & $-20.3$ & $-1.9$ & $-4.4$ \\
$ 5.0 $ & $3.9 \times 10^{-4}$ & $-21.0$ & $-1.9$ & $-4.4$ \\
\hline
$ 7.0 $ & $3.9 \times 10^{-4}$ & $-20.3$ & $-2.1$ & $-4.4$ \\
$ 6.0 $ & $3.9\times 10^{-4}$ & $-20.5$ & $-2.1$ & $-4.4$ \\
$ 5.0 $ & $3.9\times 10^{-4}$ & $-21.0$ & $-1.9$ & $-4.4$ \\
\hline
\end{tabular}
\end{table}

We present the best constraints available on the $z = 5$ and $z = 6$ LF in Fig.~\ref{fig:lfdpl}.
In particular, the number density at the bright end has been determined by~\citet{McLure2009} who exploited the DR8 release of near-infrared imaging within the  UDS field to select a sample of LBGs at $z =5$ and $z=6$.
For comparison with our results at $z = 7 $, we fit DPLs to the binned LF points at $z = 5,6$ and $7$ from~\citet{McLure2009, Mclure2013a} and this work using a simple $\chi^2$-minimising routine and present the results in Table~\ref{table:dplparams} and Fig.~\ref{fig:lfdpl}.
At $ z = 5 $ we allow all the DPL parameters to be varied in the fitting procedure ($\phi^*$, $M^*$, $\alpha$ and $\beta_{}$), and for $z = 6$ and $z = 7$ we fix the value of the bright-end slope to the best-fitting value at $z = 5$ ($\beta_{} = -4.4$).
Fixing $\beta_{}$ ensures that we are not overly-fitting to the bright points that are affected by large Poisson errors and cosmic variance.
We find that the reduced $\chi^2$ values for the double power law fits are improved compared to a Schechter function, providing additional motivation for the choice of functional form.
By alternately fixing $\phi^*$ and $M^*$ we find that $M^*$-evolution is preferred in the fitting procedure, supporting previous conclusions in favour of pure `luminosity evolution' at $ z = 5-7$, and we display the fits with $\phi^*$ fixed in Fig.~\ref{fig:lfdpl}.
The results of~\citet{McLure2009} around $M_{1500}\simeq -22$ are not in conflict with our high derived number density, and would support very little evolution in the bright-end of the LF between $z = 6$ and $z = 7$.

Our results however, appear in conflict with the brightest bin at $M_{1350} = -22.5$ from~\citet{Willott2013},  who used a step-wise maximum-likelihood analysis to conclude that there was an exponential cut-off at the bright-end of the $z = 6$ LF.
From a sample of 40 $i$-dropout LBGs within a total area of 4 deg$^2$ from the four separate CFHT Legacy Survey fields, ~\citet{Willott2013} calculated a number density of $2.66^{+5.12}_{-1.75} \times 10^{-8}$ /mag/Mpc$^3$  at $M_{1350}=-22.5$, which, if unchanged at $z = 7$, would predict an order of magnitude fewer objects than we find in the UltraVISTA and UDS fields.
On closer inspection of the sample obtained from the CFHTLS field however, there exist at least two candidates at $ M_{1350} < -22.25$ that should be represented in the LF derivation, which would appear to contradict the space density calculated by~\citet{Willott2013}.
~\citet{Willott2013} note that they are unable to actually measure a volume density as low as $\sim 10^{-8}$ /mag/Mpc$^3$ from their dataset, and the anomalously low position of their brightest data-point could be an artefact of their chosen method of fitting the LF.
Simply assuming the two brightest galaxies (WHM5 and WHM29) that occupy the $M_{1350} = -22.5$ bin could have been selected in the full survey volume of $10^7$ Mpc$^3$, we estimate the measured value of the number density from the CFHTLS analysis and present this point in Fig.~\ref{fig:lfdpl}. 
The brightest bin from this work and~\citet{Willott2013} both contain a small number of objects and hence are sensitive to Poisson errors and cosmic variance, however within the errors, our determination of the bright-end of the $z = 7$ LF is not in conflict with that found at $z = 6$ by~\citet{Willott2013}, assuming our simple binned estimate for the bright bin at $M_{1350} = -22.5$.
Finally, the best-fitting DPL we find at $z = 6$ would predict 4 galaxies in the brightest bin of the 4 deg$^2$ of data utilised by~\citet{Willott2013}, and hence our results are consistent within Poisson errors.

\subsection{Cosmic variance}

The derived LF can vary depending on the dimensions of the survey field observed,  as a result of the underlying large-scale density fluctuations within the Universe.
We find more galaxies within the UltraVISTA ultra-deep data than in the UDS field, which covers a similar area of sky.
The interpretation is complicated, however, by the lower selection efficiency of the UDS dataset compared to UltraVISTA, as a result of the substantially shallower $Y$-band imaging in the field.
Moreover, because we require that a candidate galaxy be detected at $> 2\sigma$ significance in the $Y$-band, our sample from the UDS field is significantly less complete for galaxies at $z > 6.8$ where the break starts to cut into the $Y$-band filter.
We note that although all of the UltraVISTA candidates would be detected at the $2\sigma$-level in the UDS $Y$-band, only the two brightest candidates would be detected at the $5\sigma$ level and hence likely included in the sample.

Cosmic variance affects the number counts of more massive and hence rarer galaxies more severely, however for small number counts the Poisson error can be more significant.
We explore the sources of error on our observed number counts using the Cosmic Variance Calculator v1.02\footnote{\url{http://casa.colorado.edu/~trenti/CosmicVariance.html}}~\citep{Trenti2008}, assuming a Sheth-Tormen halo mass function, $\sigma_{8}$ = 0.9 and a halo filling factor of 1.0.
From the LF points derived from our sample, we estimated the true number of galaxies present in the UltraVISTA and UDS survey volumes for each 0.5 mag bin.
Then, by inputting the completeness and survey dimensions, taking the UltraVISTA field to be a single rectangle with $0.65 \times 1.0$ deg$^2$ for simplicity, we can retrieve the error on the number counts and the relative contribution of Poisson noise and cosmic variance. 
For the brightest bin centred at $M_{1500}= -22.75$, we find the predicted number counts in each field to be $N = 1 \pm 1$ as one would expect from simple Poisson errors.
The cosmic variance for such a small number of objects is dwarfed by the Poisson uncertainty, however it still contributes $\sim 10$ \% of the total error.
Similarly, the central bin we calculate at $ M = -22.25$ has a predicted number of $4 \pm 3$ galaxies in each field where 30\% of the error is a result of cosmic variance.
This prediction is consistent with the 5 galaxies we find in this magnitude range in the UltraVISTA ultra-deep imaging and 1 galaxy in the UDS.
The faintest bin we calculate is very incomplete for the UDS field and so we do not compare the number counts between the two fields here.
Therefore, we conclude that for our sample, the errors due to Poisson noise dominate over the cosmic variance, which contributes at most 30\% of the total error on the number counts.

\subsection{Contribution of faint ${\bf z = 7}$ quasars}\label{qlf}
%% QLF figure
\begin{figure}
\includegraphics[width = 0.5\textwidth, trim = 0.63cm 0cm 0cm 0cm, clip = true]{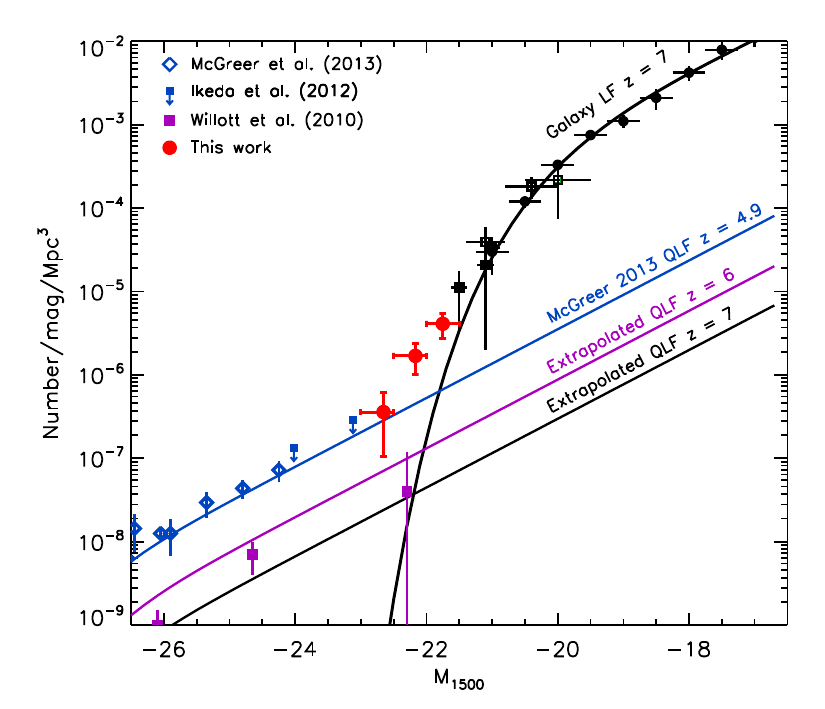}
\caption{The $z = 7$ galaxy LF from~\citet{Mclure2013a} with the $z = 7$ data points from Fig.~\ref{fig:lf} shown in black and the data points determined from our analysis shown in red.
The constraints on the $z = 5$ quasar LF from~\citet{McGreer2013} are shown as the blue diamonds, with additional upper limits at fainter magnitudes from~\citet{Ikeda2012}, where we have shifted the data point at $M_{1500}= -23.0$ by 0.1 mag for clarity.
Overplotted in blue is the best fitting double power law at $z = 5$ presented in~\citet{McGreer2013}.
 The purple squares show the constraints on the QLF at $z = 6$ from~\citet{Willott2009}.
 The QLF at $z = 6$ and $z = 7$ is estimated from evolving the $z = 5$ QLF shown using a LEDE evolution model for log($\phi^*$) taken from~\citet{McGreer2013}, with fixed $\alpha = -2.03$ and ${\rm M}^*= -27.0$.
}
\label{fig:qlf}
\end{figure}

Around the peak of the quasar number density at  $z = 3$, there is evidence that the very brightest end of the galaxy LF is contaminated by quasars~\citep{BianFuyan2013}.
There are large uncertainties in the faint-end of the $z > 3$  quasar LF, and the faint-end of the $z = 7$ QLF is completely unknown due to a lack of datasets with the required depth over an adequate area on the sky for detection (see~\citealp{Willott2009}).
In Fig.~\ref{fig:qlf}, we compare the galaxy LF at $z =7$ to the known QLF at $z = 5$ and $z =6$ and the extrapolation of the $z = 7$ QLF beyond the absolute magnitude of the faintest quasar found at $z = 7 $ with $M_{1700} = -25.5$~\citep{Venemans2013}.
The bright-end of the QLF can be estimated from the four $z > 6.5$ quasars known, which includes one from the UKIDSS-LAS found by~\citet{Mortlock2011} and three from the VISTA VIKING survey~\citep{Venemans2013}.
At $z = 6$, around ten quasars are known~\citep{Willott2009}, however again the constraints on the faint-end slope are weak, illustrated by the large error bars on the faintest bin shown in Fig.~\ref{fig:qlf} (with ${\rm M}_{1500} \sim -22$), which contains only one quasar.

To ascertain the level of contamination of our sample by quasars, we compare the number densities of bright LBGs at $z = 7$ to the predicted QLF at $z = 7$ by evolving the $z = 5$ QLF determined by~\citet{McGreer2013} using the evolution model presented in their paper.
A double power law form is typically used when fitting the QLF, as described in Section~\ref{binnedlf}.
When fitting a DPL to the data points, both~\citet{McGreer2013} and~\citet{Willott2009} fix the value of the faint-end slope, $\alpha$, as a response to the large uncertainties in the faint-end determination.
Lower redshift $z < 3$ results tend to favour $\alpha = -1.5$~\citep{Croom2009}, with a tenuous steepening observed with $\alpha \simeq -1.7$ to higher redshifts~\citep{Masters2012}.
\citet{McGreer2013}  found a further steepening of the faint-end slope, with the best-fitting DPL with a fixed $\alpha = -1.8$ still under-predicting the number of faint quasars found.
If the bright-end slope is instead fixed to $\beta = -4.0$, \citet{McGreer2013} found a best fitting model with a steeper $\alpha = -2.03^{+0.15}_{-0.14}$.
Therefore, in Fig.~\ref{fig:qlf} we plot the best fitting function with $\alpha = -2.03$ from~\citet{McGreer2013}, to provide an upper limit on the number of faint $z = 5$ quasars. 

 The $z = 7$ LF parameters are predicted  following the Luminosity Evolution Density Evolution (LEDE) model described in~\citet{McGreer2013}, using a fixed $\alpha = -2.03$ and $M_* = -27.0$.
 The model predicts that log($\phi^*$) evolves linearly with redshift with gradient $k = \delta {\rm log}(\phi^*) / \delta z = -0.47$, derived from fitting to the measured parameters from $z = 2.2 - 4.9$~\citep{Fan2001}.
The strength of the evolution when extended to higher redshift is supported by \citet{Venemans2013} at $z = 7$ who found $k = -0.49^{+0.28}_{-0.74}$.
We display the data at $z = 6$ from~\citet{Willott2009}, overlayed with an extrapolated model using the same method as for the $z=7$ model, again to show an upper limit on the expected number density of quasars here.
As can be seen in Fig.~\ref{fig:qlf}, the LEDE model is consistent with the $z = 6$ LF from~\citet{Willott2013} at least given the uncertainties at the faint end, although see the detailed discussions in~\citet{McGreer2013}.
To estimate the number of quasars that could contaminate our sample, and assuming that the quasar SEDs are indistinguishable from the LBGs using the selection here, we integrate the QLF within the three LF points we calculated.
We predict $0.3$ quasars in the fainter bin at $M_{1500}= -21.75$, and $0.2$ and $0.1$ in the brighter bins at $M_{1500}= -22.25$ and $M_{1500}= -22.75$ respectively.
Hence, from the analysis of the LF and QLF we conclude that contamination of our sample by quasars is minimal, a result obviously consistent with our finding that the vast majority of our $z \simeq 7$ objects are spatially resolved (see Section~\ref{sizes}).
However, given the large uncertainties in the faint-end slope of the QLF, the possibility of some low-level contamination is not completely ruled out.
For example \citet{Willott2009} calculated an upper limit of two quasars per deg$^2$ in the UltraVISTA survey, when assuming the most extreme LF parameters from the range of acceptable fits to the QLF at $z = 6$.
There is evidence that quasars may contaminate bright $z = 6$ LBG samples on the order of $\sim 10$\%, for example the sample of ten LBGs presented in by~\citet{McLure2009} and spectroscopically confirmed by~\citep{CurtisLake2012}, included one Type I quasar identified by the broadened \Lya line~\citep{Willott2009}.

Note that at high redshift, the observed QLFs only account for Type I unobscured quasars with both broad-line and narrow-line components, and there is evidence from X-ray surveys that unobscured quasars only account for 25\% of the total number at $z = 4$~\citep{Masters2012}.
If the ratio of Type I to Type II quasars persists to high-redshift, the results here are likely a lower limit on the number densities of quasars at $ z = 7$.

\subsubsection{Radio and X-ray signatures of high-redshift quasars}

The rest-frame UV colours of LBGs and quasars at high-redshift are impossible to distinguish with the current photometric accuracy~\citep{BianFuyan2013}, and identical colour-colour cut selection criterion are often used for the selection of galaxies and quasars at $z > 5$~\citep{Willott2009}.
Here, for completeness, we consider whether radio or X-ray emission could be detected from a quasar selected as a galaxy and included in our sample, with the data available in the COSMOS and UDS fields.
The COSMOS field is imaged by the \emph{Chandra}-COSMOS survey~\citep{Elvis2009}, which has a limiting depth of $1.9 \times 10^{-16} $ ergs/s/cm$^2$/Hz in the 0.5-2.0 keV channel and $7.3 \times 10^{-16} $ ergs/s/cm$^2$/Hz in the 2-10 keV channel.
In the radio, the COSMOS field is covered by the Very Large Array (VLA)-COSMOS survey~\citep{Schinnerer2010}, with a sensitivity of 12$\umu{\rm Jy}$ per beam.
The UDS field has X-ray imaging from the Subaru/XMM-Newton deep survey~\citet{Ueda2008}, to depths of $6 \times 10^{-16} $ ergs/s/cm$^2$/Hz in the 0.5-2.0 keV channel and $5 \times 10^{-15} $ ergs/s/cm$^2$/Hz in the 2-10 keV channel.

We find no radio or X-ray counterparts for any of the galaxies within our sample, when comparing to the publicly available catalogues within each field derived from the datasets described above.
We also perform a stack of the objects in the VLA-COSMOS imaging and again find no detection to a limit of $\sim$ 12$\,\umu \,{\rm Jy}$ per beam.
The average quasar SEDs from~\citet{Shang2011} suggests that if one of our candidates was a typical radio-loud quasar, it would just be detectable in the X-ray and radio imaging (see Figure 6 of~\citealp{Ouchi2009a}).
Furthermore, a bright radio-loud quasar such as J1429+5447 at $z = 6.21$~\citep{Willott2009, Frey2011}, would be detected at high-significance in the available radio data.
Therefore, although a non-detection in the radio and X-ray for the objects in our sample rules out the possibility that the majority of the objects have strong active nuclei, our sample could still contain $\simeq 1$ radio-quiet quasar.
For comparison, the $z = 7.1$ quasar discovered by~\citet{Mortlock2011}, which is substantially brighter than our LBGs with $M_{\rm UV} = -26.6$, has been detected in the X-ray by~\citet{Page2013} and~\citet{Moretti2014} with fluxes of $5.7 \pm 1.2  \times 10^{-16} $ ergs/s/cm$^2$ and $9.3_{-1.1}^{+1.6}  \times 10^{-16} $ ergs/s/cm$^2$ respectively in the 0.5-2.0 keV channel, but not in the radio with a $3\sigma$ upper limit of 23.1 $\umu{\rm Jy}$ per beam~\citep{Momjian2013}.

\section{Astrophysical Implications}\label{sect:implications}

As discussed in the introduction to this paper, the Schechter functional form, with its steep exponential decline at high luminosity/mass, provides a good description of the galaxy LF and mass function (MF) observed at low-redshift~\citep{MonteroDorta2009}.
Moreover, recent work extending the study of the galaxy luminosity and stellar mass function out to higher redshift, indicate that a Schechter function (or double Schechter function) still provides a good description of the data out to at least $z \simeq 3$~\citep{Ilbert2013, Muzzin2013}. 
However, the results we have presented here suggest that this may not be the case at $z \simeq 7$.

It is thus worth briefly considering whether our derived $z = 7$ galaxy LF is physically reasonable, and what the inability of a Schechter function to reproduce the bright end of the LF might mean. 
As already discussed in Section~\ref{sect:lf}, and shown in Fig.~\ref{fig:lf} one way to describe the apparent lack of a steep exponential decline at the bright end is to parameterise the LF as a double power-law fit, which well describes the full range of available data at $z \simeq 7$.
However, the physical meaning of such a double power law is unclear, and moreover it is important to check that the number density of bright $z \simeq 7$ galaxies inferred from our study is not physically unreasonable given the expected number density of appropriate dark matter halos expected to exist at these early times.

\begin{figure}
\includegraphics[width = 0.5\textwidth, trim = 0.7cm 0cm 0cm 0cm, clip = true]{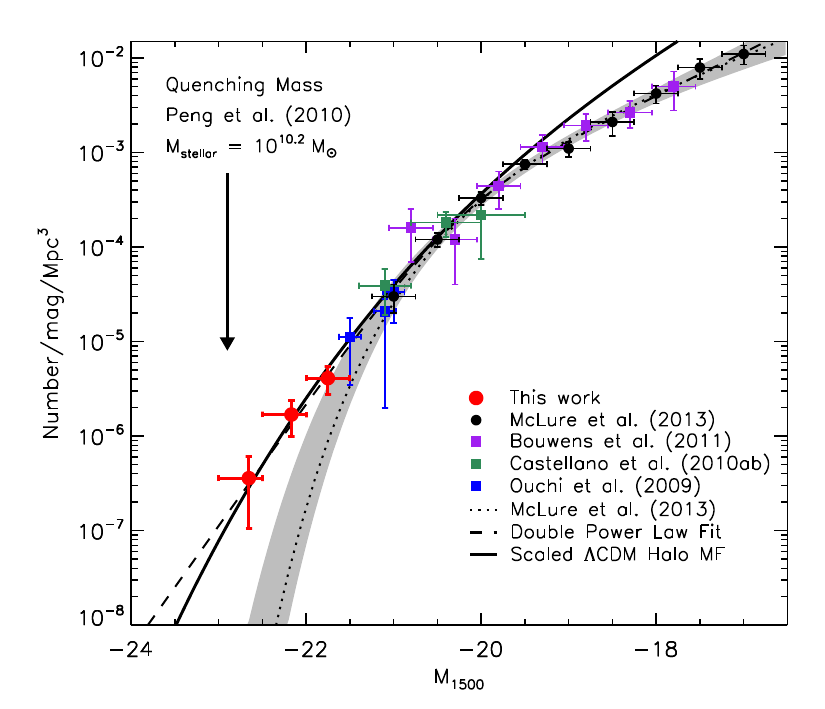}

\caption{ The $z = 7$ UV ($\sim 1500$\AA) luminosity function showing a scaled $\Lambda$CDM halo mass function as described in the text (solid black line).
The results from our sample of galaxies from the UltraVISTA DR2 and UDS fields are shown as the red filled circles. 
Data points from other studies are as described in the caption for Fig.~\ref{fig:lf}
The best-fitting Schechter function at $z = 7$ from~\citet{Mclure2013a} is plotted as the dotted black line, and the best-fitting double power law to our data points and those from~\citet{Mclure2013a} is shown as the dashed line.
The one-sigma confidence limit on the Schechter function parameters ($M^{\star}, \phi^{\star}$ and $\alpha$) is shown as the grey shaded region.
We highlight the quenching mass derived by~\citet{Peng2010} with an arrow, after converting to a UV luminosity using the typical mass-to-light ratio displayed by our sample.
}
\label{fig:hmf}
\end{figure}

We therefore conclude by showing, in Fig.~\ref{fig:hmf}, a comparison of the $z \simeq 7$
$\Lambda$CDM dark-matter halo mass function, scaled via a constant mass-to-light ratio, with our new observational determination of the $z \simeq 7$ galaxy LF. We produced the halo mass function using the code provided by~\citet{Reed2007} using our chosen cosmology and $\sigma_{8} = 0.9$, but the basic results are not strongly influenced by the precise choice of code or parameters within current uncertainties.
We then simply scaled the halo mass function into a UV luminosity function using a constant mass-to-light ratio, set by assuming that a galaxy with $M_{1500} = -22.4$ has a stellar mass of $M_* \simeq 10^{10}{\rm M_{\sun}}$ (as supported by our data) and a dark matter halo mass to stellar mass ratio of 30 (e.g.~\citealp{Behroozi2013}).
As can be seen from Fig.$\,10$, the result is a predicted LF which, without any additional shifting or fitting, does an excellent job of reproducing our new $z \simeq 7$ LF from the previously inferred break luminosity at $M_{1500} \simeq -20$ out to our brightest luminosity bin. 
Interestingly, over this range, it is evidently indistinguishable from our double power-law fit, confirming that it provides an excellent representation of the data. It is significantly shallower than the exponential decline shown by the pre-existing Schechter function fit, and only starts to deviate from the bright-end power law at very bright magnitudes (thus suggesting that extrapolation of the double power law brightward of $M_{1500} \simeq -23$ will over-predict the number of extremely bright galaxies to be found in future wider area surveys).

This interesting result has a number of potentially important implications.
First, it confirms that the number density of bright galaxies revealed in this study is not unreasonable. 
Neither is the inferred bright-end slope, as this essentially parallels the decline in the number density of appropriate-mass dark-matter halos. Second, it suggests that while the process (e.g. supervovae feedback) which limits star-formation in faint galaxies appears to be in operation at these early times (as evidenced from the difference between the slope of the halo mass function and the UV LF at faint magnitudes), the mechanism that limits high-mass galaxy growth may have yet to impact on the form of the LF at $z \simeq 7$, at least over the luminosity/mass range probed here. 
Perhaps this is because AGN have yet to grow to the masses and hence luminosities required to eject gas available for future star formation, and certainly there is little evidence for AGN within out galaxy sample (see Section~\ref{sect:discussion}). 
However, without over speculating we can at least say that, whatever the physical mechanism which ultimately limits the masses of star-forming galaxies, our results are certainly consistent with the redshift invariant `mass quenching' argument proposed by~\citet{Peng2010}. Since the estimated masses of our brightest galaxies have only just reached a mass comparable to the proposed critical `quenching mass' of $M_* = 10 ^{10.2} {\rm M_{\odot}}$, it is perhaps
to be expected that the quenching of star-formation activity in galaxies which causes them to leave the `main sequence' will only be revealed at lower redshifts and/or higher stellar masses than probed by the sample presented here.

Finally, we note that the above discussion implicitly attempts to relate our observational determination of the UV LF to the underlying mass function. The extent to which the form of the LF actually mirrors the form of the underlying stellar mass function of course remains unclear, and may only really be resolved at these extreme redshifts with the deep, high-resolution near-mid-infrared data anticipated from the James Webb Space Telescope.

\section{Conclusions}\label{sect:conc}

We present the results of a new search for bright star-forming galaxies at $z > 6.5$ utilising the very latest data within the UltraVISTA and UKIDSS Ultra Deep Survey (UDS) fields.
The 1.65\,deg$^2$ of available overlapping optical, near- and mid-infrared data was analysed using a photometric redshift fitting method, which enables the identification of high-redshift galaxies and the rejection of contaminants such as low-redshift galaxies and galactic dwarf stars.
In total we present 34 galaxies, 29 from within the ultra-deep UltraVISTA imaging, one from the deep UltraVISTA region, and four from the UDS field.
With the improved photometry available, we reselect seven of the ten candidates presented in~\citet{Bowler2012} as $6.5 <z < 7.5$ galaxies.
Of the remaining original candidates,  two are confirmed as galaxies at a slightly lower redshift of $ 6.0 < z < 6.5$ and the final candidate is now best fitted as a T-dwarf.

From the best-fitting model to the photometry of each galaxy we calculated the stellar mass, SFR and sSFR.
Our sample contains some of the most massive galaxies at $z = 7$, with $M_{\star} \simeq 10^{10} {\rm M}_{\sun}$, which show a low sSFR compared to lower mass galaxies at $z = 7$, with an upper limit of sSFR $\lesssim 2$ Gyr$^{-1}$.
We find no evidence for a redder rest-frame UV slope $\beta_{}$ for our sample (median $\beta_{JHK} = -2.0$), as would be expected by an extrapolation of the colour-magnitude relation at lower redshift to $z = 7$.
We measure the sizes of the galaxies in our sample and find that although the majority are consistent with being uresolved in the ground-based imaging, a significant number have a larger FWHM suggesting $r_{1/2} \gtrsim 1.5 $ kpc.
For four galaxies that lie within \emph{HST} imaging from the CANDELS COSMOS survey, we find an offset ($\sim 0.4$ mag) between the total magnitudes when the object is assumed to be a point source in both the ground and space-based imaging.
Inspection of surface-brightness profiles shows that the galaxies have an extended profile, which can lead to an underestimate of the galaxies total and therefore absolute magnitude when small apertures are used with the assumption of a point-source profile in \emph{HST} imaging.

From our final sample we determine the form of the bright end of the rest-frame UV galaxy luminosity function (LF) at $z \simeq 7$.
We use a $1/V_{\rm max}$ estimator to determine the binned LF points at $M_{1500} = -22.75, -22.25$ and $ -21.75$, folding in the completeness of our selection methodology using injection and recovery simulations.
In our determination of the LF, we take into account that some of our galaxies are gravitationally lensed by low-redshift galaxies along the line-of-sight, with a typical brightening of $0.1-0.3$ mag.
We find that the bright end of the $z = 7$ LF does not decline as steeply as predicted by the Schechter function fitted to fainter data, and can be better described by a double power law.
The possibility of significant contamination of our sample by high-redshift quasars can be excluded,  with a predicted number of $< 1$ in the UltraVISTA and UDS imaging, calculated from an extrapolation of the $z = 5$ QLF to $z = 7$.
From the observed UV LF at $z = 5$ and $z =6$, we show that a DPL fit can provide a good description of the data and that the bright end of the LF at $z = 6$ and $z = 7$ shows little evolution.
Our results at the bright end of the LF mimic the prediction from the scaled dark matter halo MF, suggesting that the physical mechanism which inhibits star-formation activity in massive galaxies (for example AGN feedback) has yet to become efficient at $z \simeq 7$.
The interpretation of our results agrees with the phenomenological model of~\citet{Peng2010}, which would suggest that the most massive $ z \simeq 7 $ galaxies in our sample have only just reached the critical ``quenching mass" of $M_{\star} = 10 ^{10.2}\,{\rm M_{\odot}}$, above which star-formation activity is strongly suppressed.

\section*{Acknowledgements}
RAAB and JSD acknowledge the support of the European Research Council via
the award of an Advanced Grant. 
JSD acknowledges the support of the Royal Society via a Wolfson Research Merit Award.  
JSD acknowledges the contribution of the EC FP7 SPACE project ASTRODEEP (Ref.No: 312725).
RJM acknowledges the support of the Leverhulme Trust via the award of a Philip Leverhulme research prize.
RJM acknowledges the support of the European Research Council via
the award of a Consolidator Grant (PI McLure).
ABR acknowledges the support of the UK Science \& Technology Facilities Council.
HM acknowledges support of the PNCG. 
BMJ and JPUF acknowledge support from the ERC-StG grant EGGS-278202.  
The Dark Cosmology Centre is funded by the Danish National Research Foundation.
JA gratefully acknowledges support from the Science and Technology Foundation (FCT, Portugal) through the research grant PTDC/FIS-AST/2194/2012 and PEst-OE/FIS/UI2751/2011.
OLF acknowledges support fromthe ERC grant ERC-2010-AdG-268107-EARLY.

This work is based on data products from observations made with ESO Telescopes at the La Silla Paranal Observatories under ESO programme ID 179.A-2005 and on data products produced by TERAPIX and the Cambridge Astronomy survey Unit on behalf of the UltraVISTA consortium. This study was based in part on observations obtained with MegaPrime/MegaCam, 
a joint project of CFHT and CEA/DAPNIA, at the Canada-France-Hawaii Telescope (CFHT) which is operated by the National Research Council (NRC) of Canada, the Institut National des Science de l'Univers of the Centre National de la Recherche Scientifique (CNRS) of France, and the University of Hawaii. This work is based in part on data products produced at TERAPIX and the Canadian Astronomy Data Centre as part of the Canada-France-Hawaii Telescope Legacy Survey, a collaborative project of NRC and CNRS. This work is based in part on observations made with the NASA/ESA {\it Hubble Space Telescope}, which is operated by the Association 
of Universities for Research in Astronomy, Inc, under NASA contract NAS5-26555.
This work is based in part on observations made with the {\it Spitzer Space Telescope}, which is operated by the Jet Propulsion Laboratory, 
California Institute of Technology under NASA contract 1407. We thank the staff of the Subaru telescope and Yuko Ideue and Yutaka Ihara for their assistance with the $z'$-band imaging utilised here.

\label{lastpage}

\bibliographystyle{mn2e}
 \bibliography{library_abbrv.bib} 

%% Now include SED plots
\appendix
\section{Images and SED fits}

In this appendix we present postage-stamp images and the best-fitting galaxy and star SED fits for the galaxies in our sample.
The 30 galaxies from the UltraVISTA field are shown in Fig.~\ref{figure:seds}, followed by the four UDS galaxies in Fig.~\ref{figure:udsseds}.

\begin{figure*}

\includegraphics[width = 0.265\textwidth, trim = 0cm -0.5cm 0cm 0.0cm, clip = true]{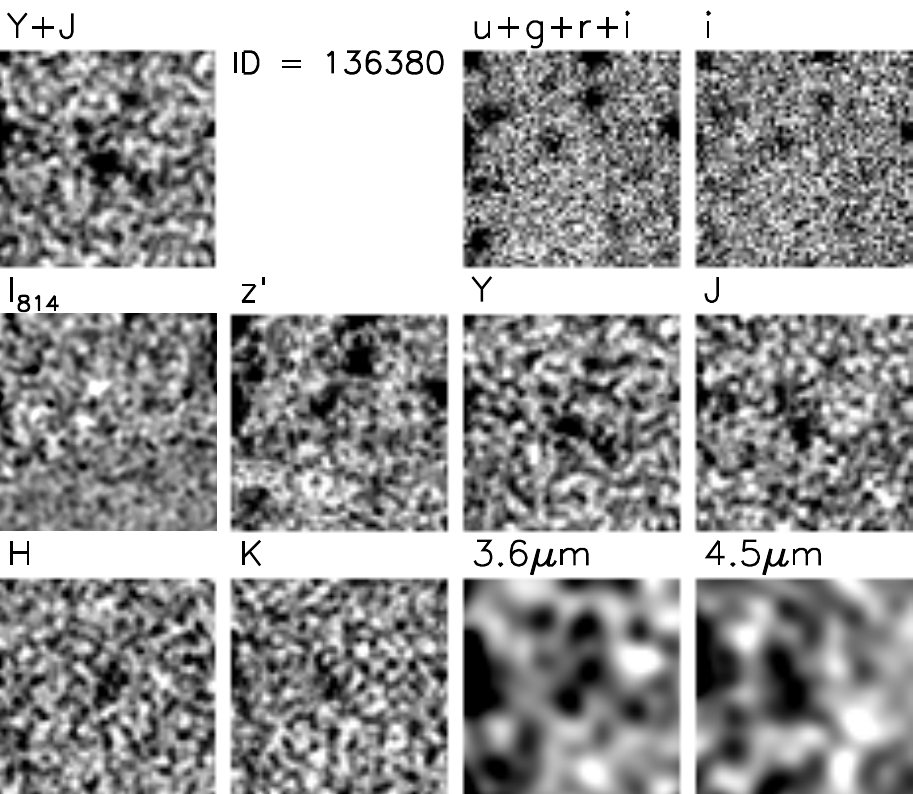} \hspace{0.3cm}
\includegraphics[width = 0.31\textwidth]{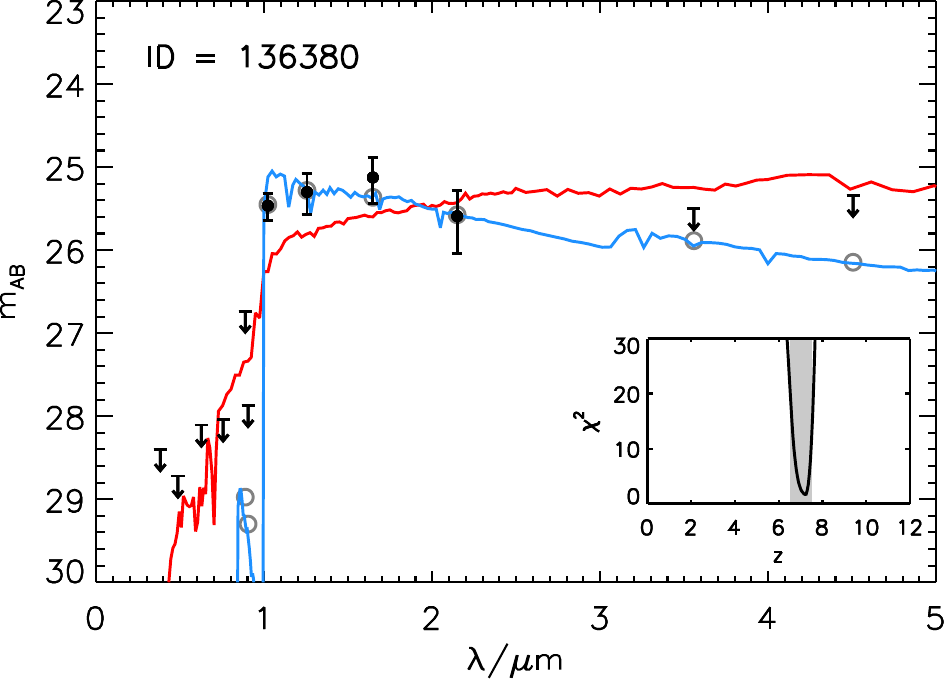} \hspace{0.3cm}
\includegraphics[width = 0.31\textwidth]{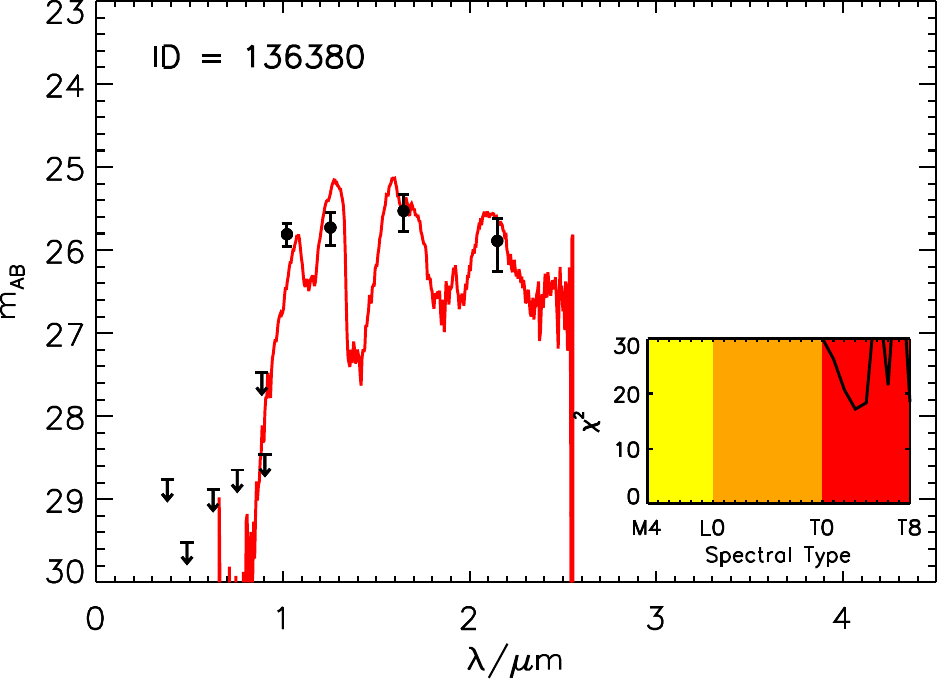} \hspace{0.3cm}\\ 
\vspace{0.2mm}
\includegraphics[width = 0.265\textwidth, trim = 0cm -0.5cm 0cm 0.0cm, clip = true]{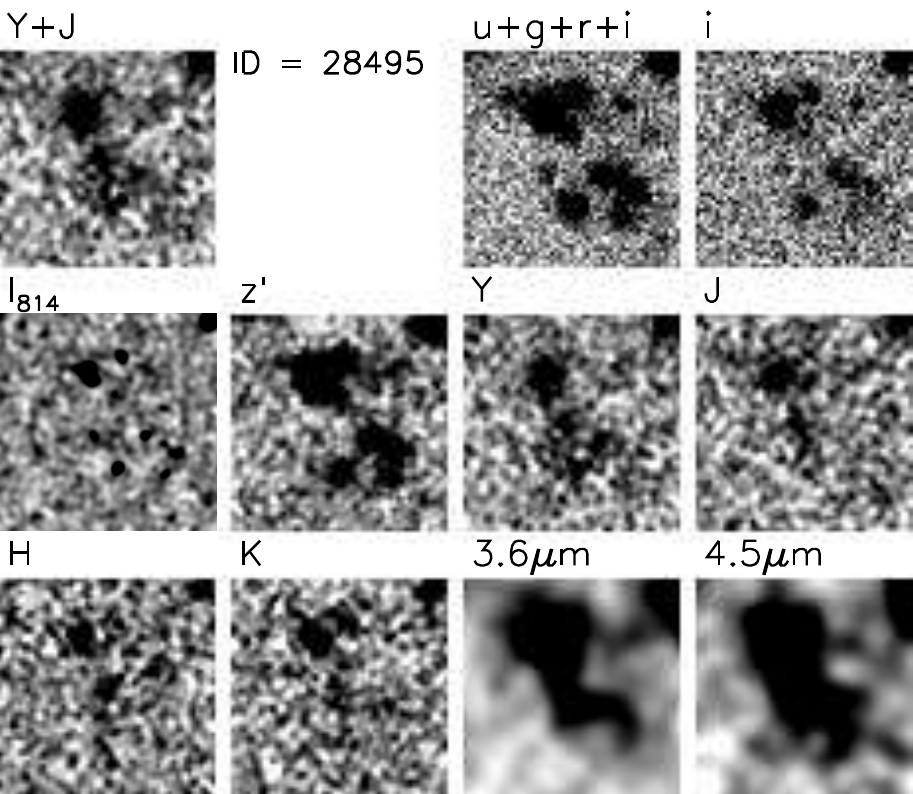} \hspace{0.3cm}
\includegraphics[width = 0.31\textwidth]{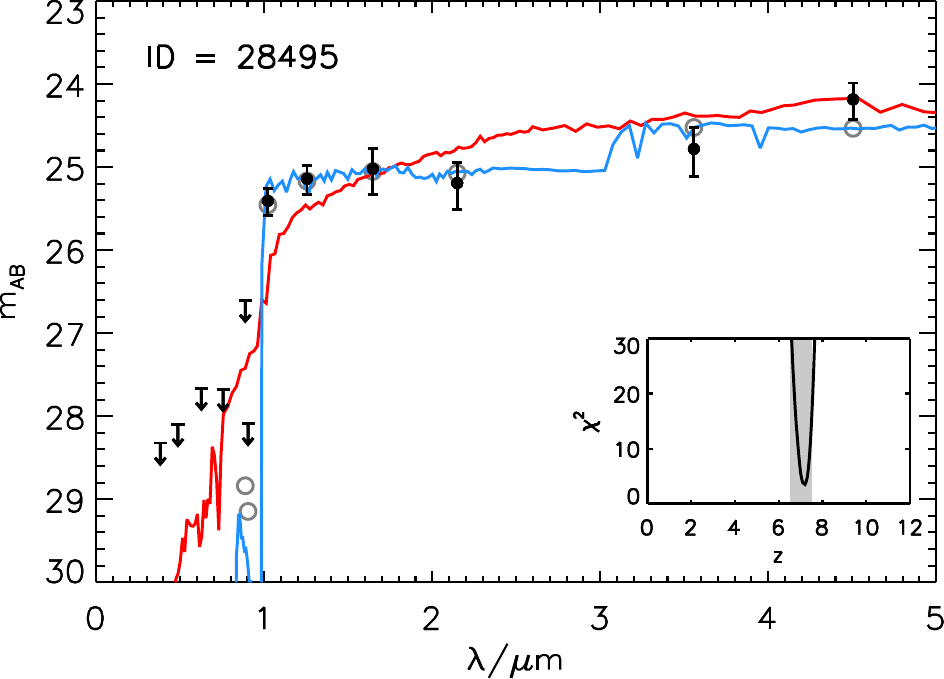} \hspace{0.3cm}
\includegraphics[width = 0.31\textwidth]{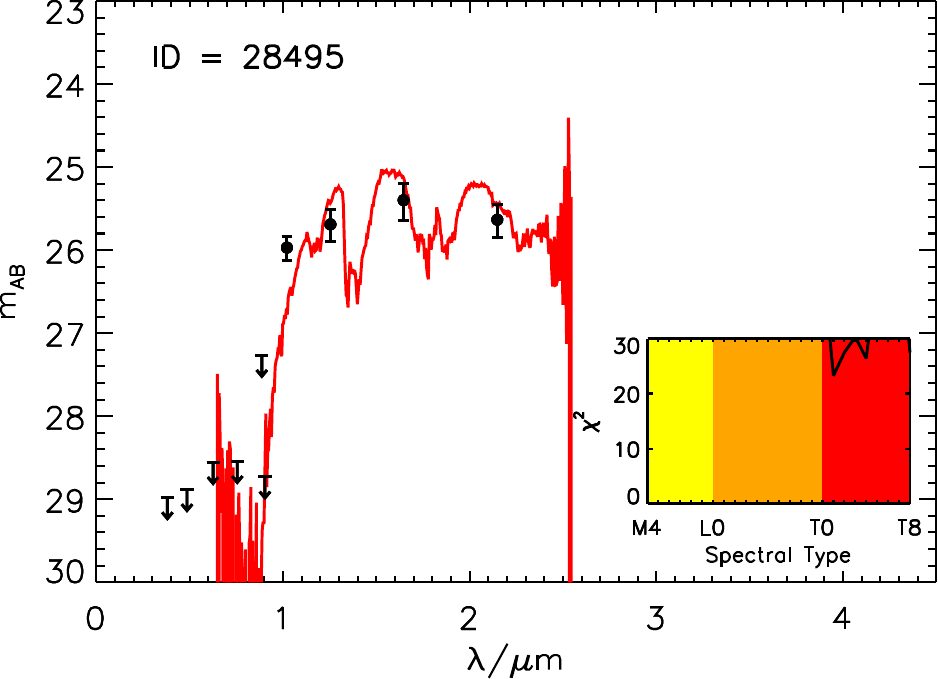} \hspace{0.3cm}\\ 
\vspace{0.2mm}

\includegraphics[width = 0.265\textwidth, trim = 0cm -0.5cm 0cm 0.0cm, clip = true]{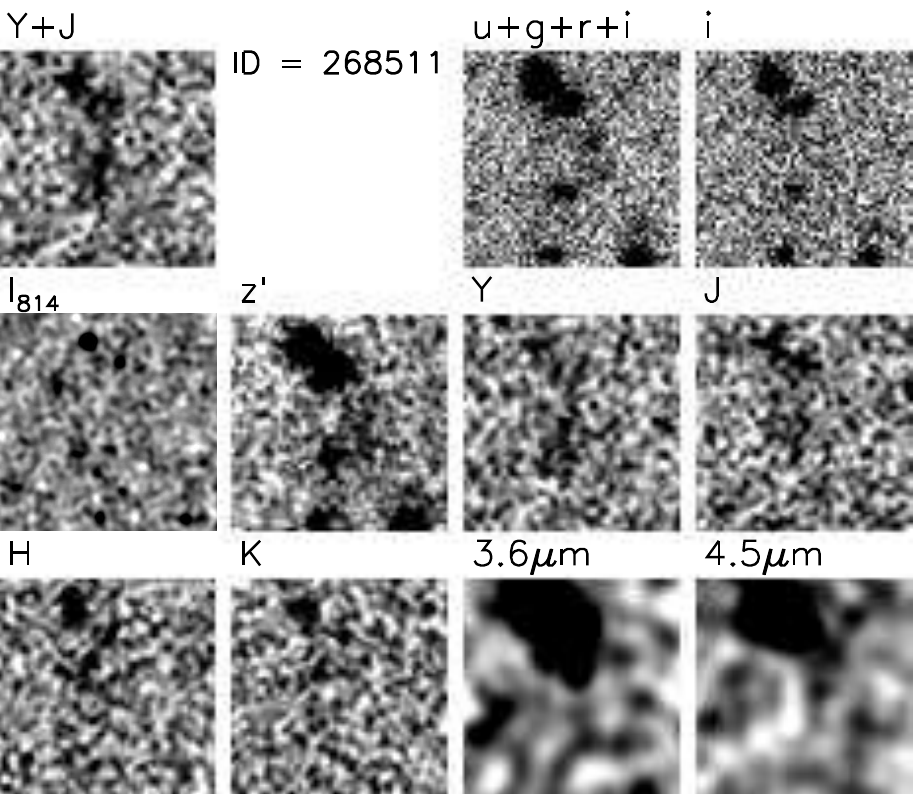} \hspace{0.3cm}
\subfloat{\includegraphics[width=0.31\textwidth, trim = 0.6cm 0cm 1.25cm 1.2cm, clip = true]{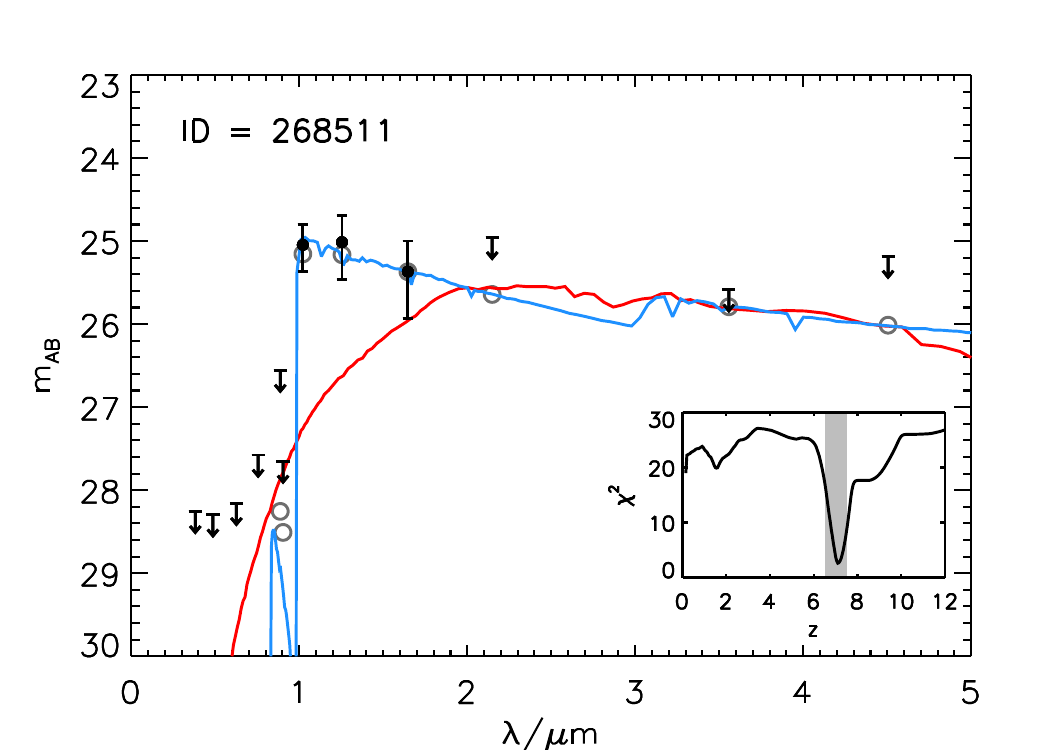}}\hspace{0.3cm}
\subfloat{\includegraphics[width=0.31\textwidth, trim = 0.6cm 0cm 1.25cm 1.2cm, clip = true]{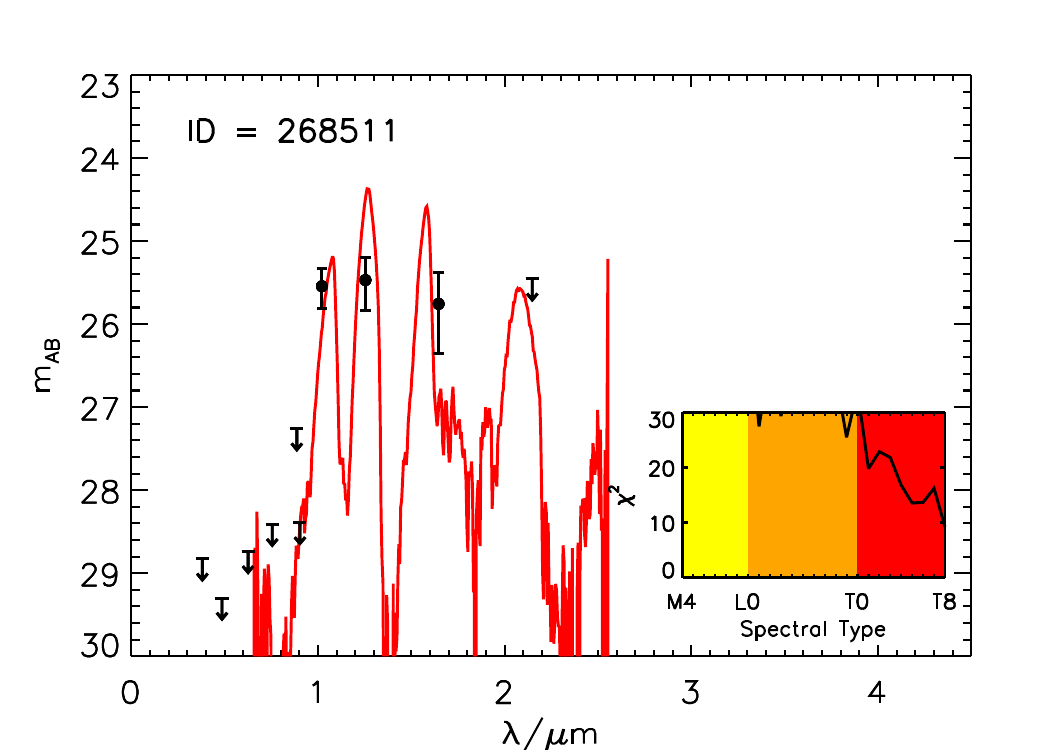}} \\

\includegraphics[width = 0.265\textwidth, trim = 0cm -0.5cm 0cm 0.0cm, clip = true]{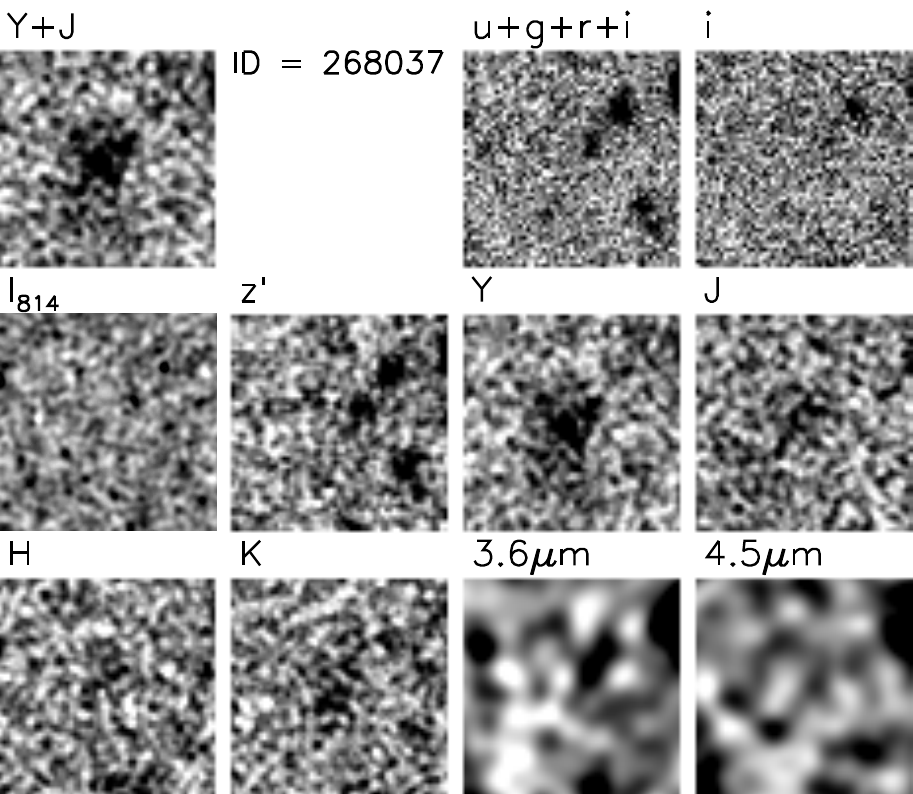} \hspace{0.3cm}
\includegraphics[width = 0.31\textwidth]{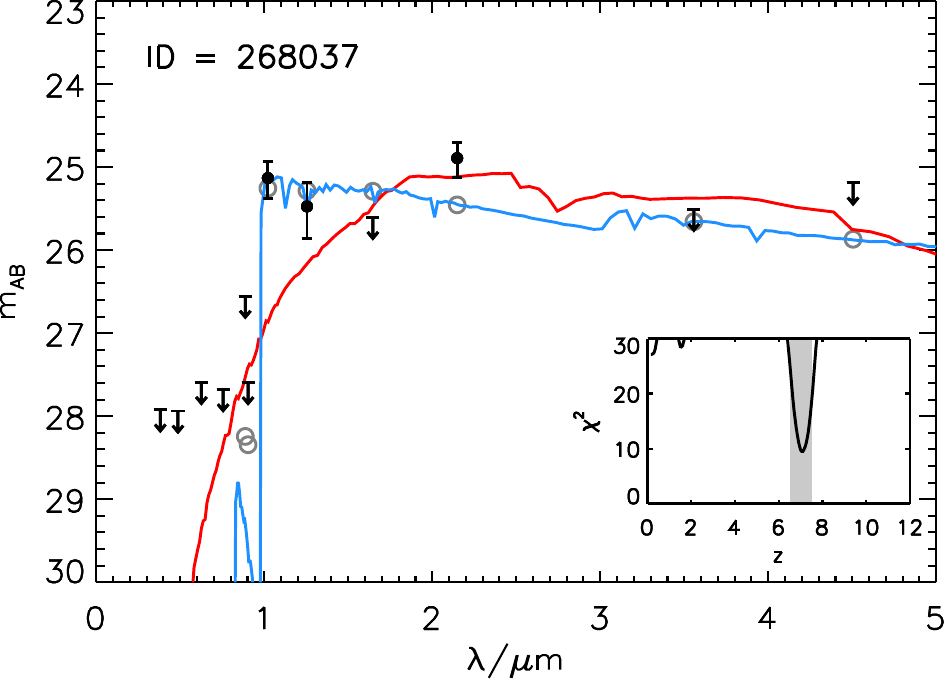} \hspace{0.3cm}
\includegraphics[width = 0.31\textwidth]{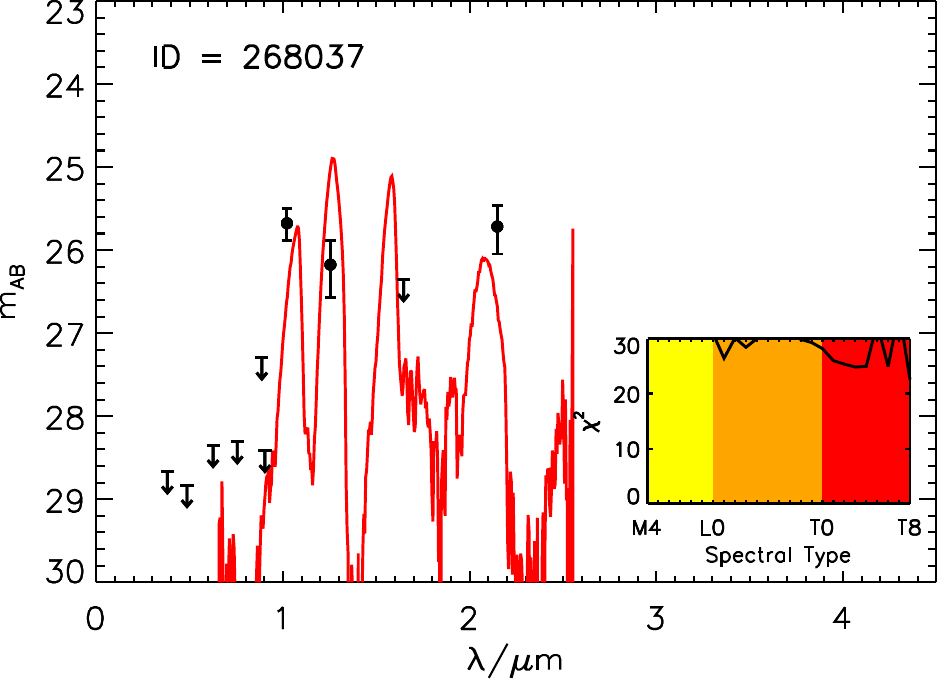} \hspace{0.3cm}\\ 

\caption{Postage-stamp images and galaxy and star SED fits to the 30 galaxies from the UltaVISTA field.
The stamps shown to the left are $10 \times 10$-arcsec, with the grey scale determined from saturating all pixels that exceed $3\sigma$ from the background.
The measured photometry and errors are shown as the black points, with the grey circles showing the predicted photometry from the best-fitting high-redshift model. 
In the central plots, the best fitting low-redshift ($ z < 4.5$) and high-redshift solutions are shown as the red and blue lines respectively.
On the right, the best-fitting stellar templates are presented, where here the photometry was measured in smaller 1.2-arcsec diameter apertures.
The insets on each plot show the chi-squared distribution as a function of redshift or stellar type, with the grey band on the redshift-$\chi^2$ plot showing the range of redshifts covered by our sample ($6.5 <z <7.5$).}\label{figure:seds}
\end{figure*} 

\begin{figure*}

\includegraphics[width = 0.265\textwidth, trim = 0cm -0.5cm 0cm 0.0cm, clip = true]{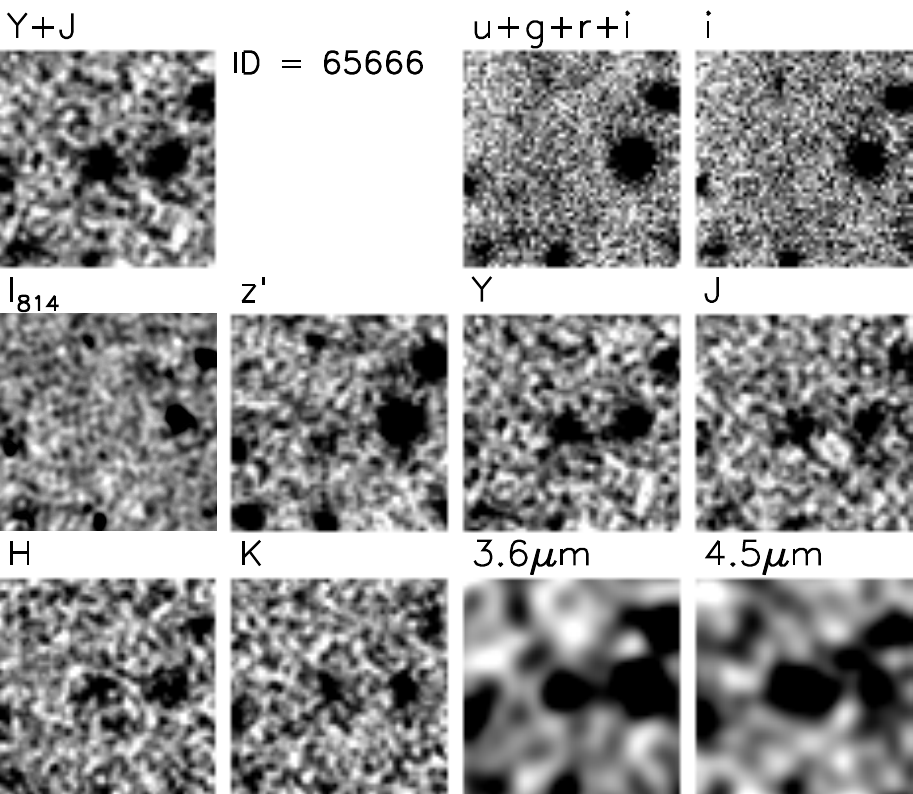} \hspace{0.3cm}
\includegraphics[width = 0.31\textwidth]{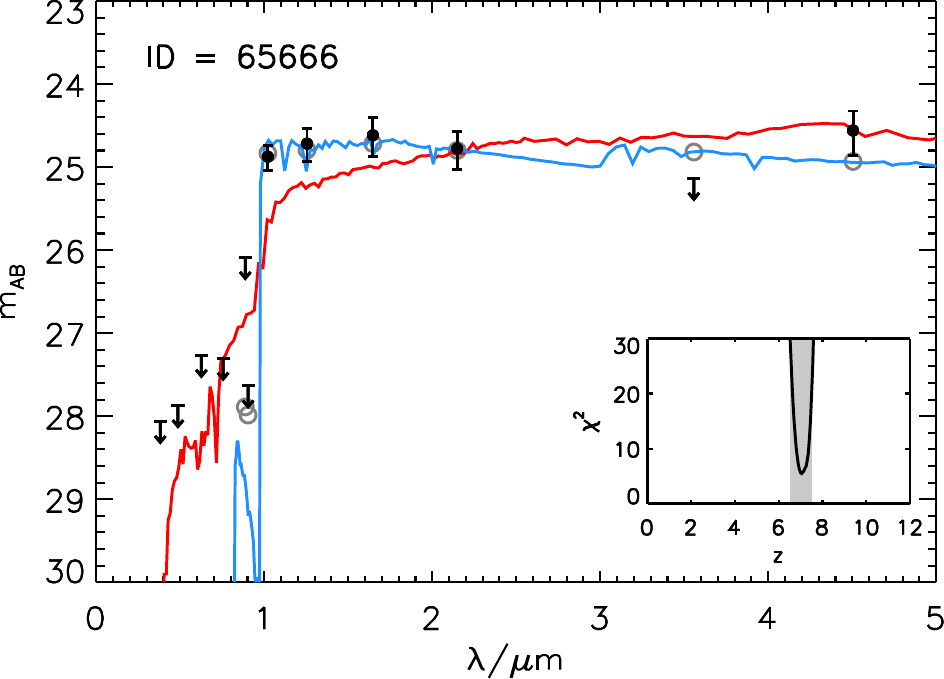} \hspace{0.3cm}
\includegraphics[width = 0.31\textwidth]{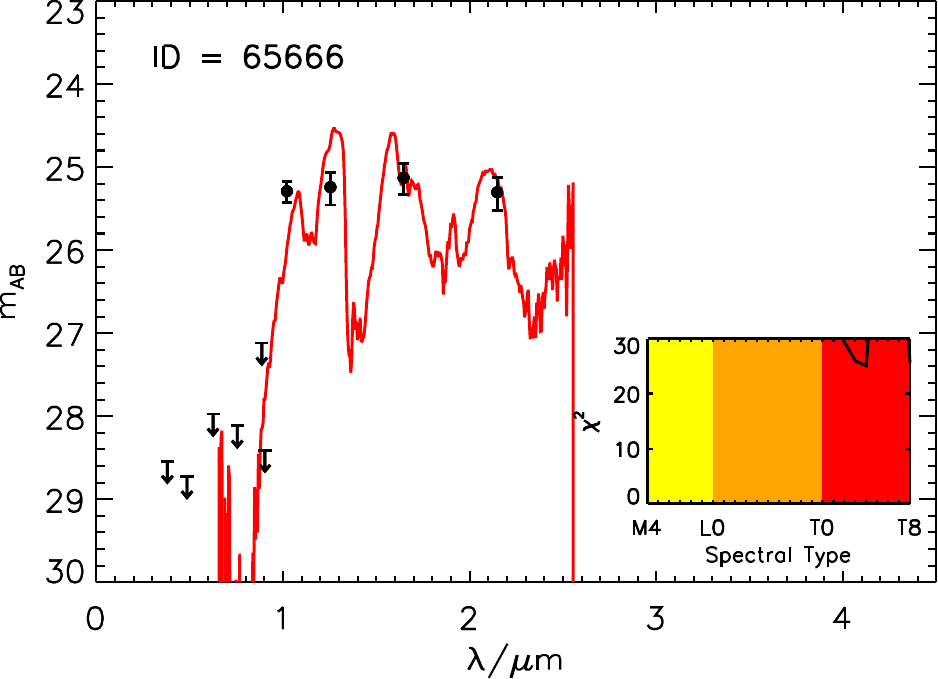} \hspace{0.3cm}\\ 

\includegraphics[width = 0.265\textwidth, trim = 0cm -0.5cm 0cm 0.0cm, clip = true]{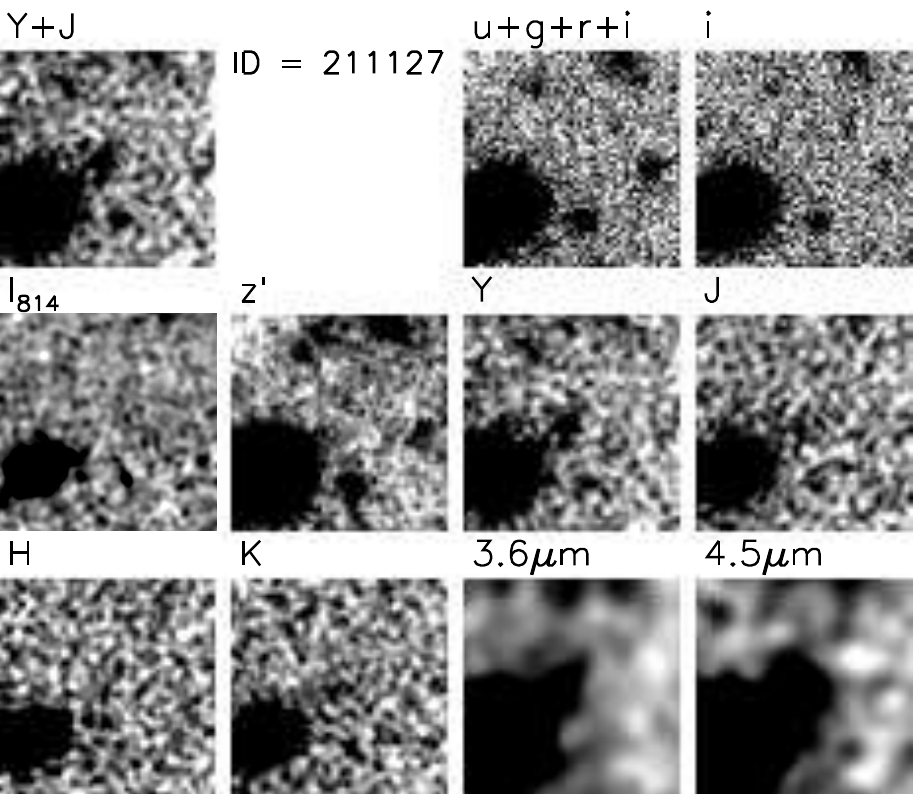} \hspace{0.3cm}
\includegraphics[width = 0.31\textwidth]{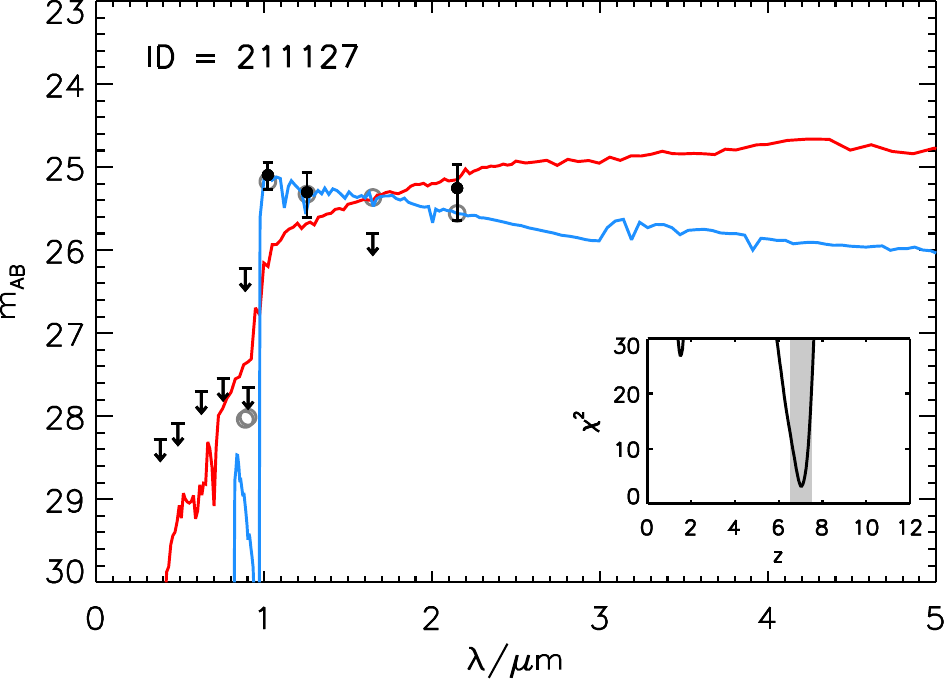} \hspace{0.3cm}
\includegraphics[width = 0.31\textwidth]{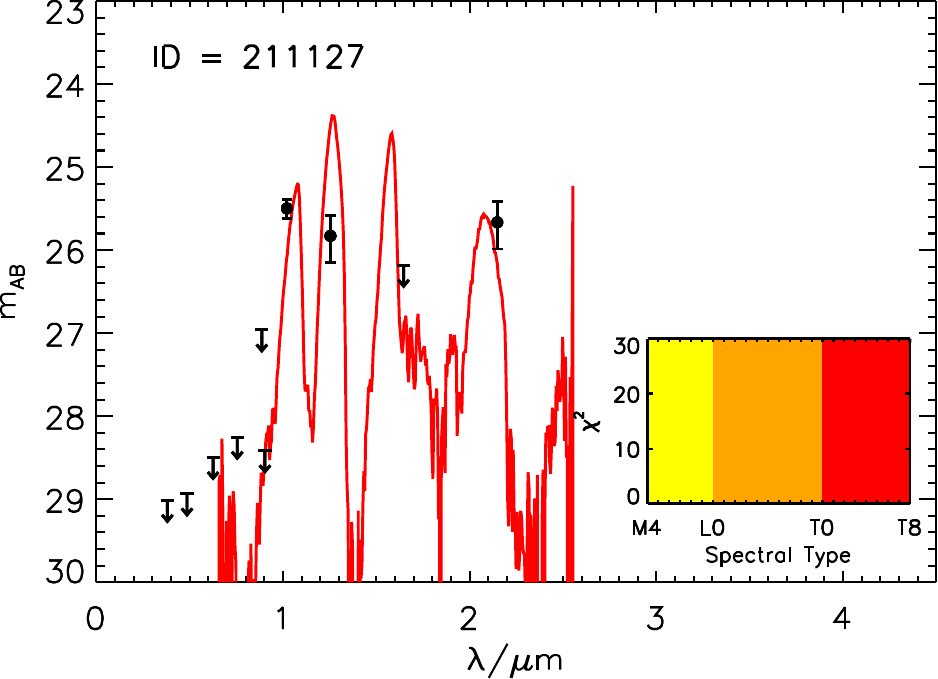} \hspace{0.3cm}\\ 
\vspace{0.2mm}
\includegraphics[width = 0.265\textwidth, trim = 0cm -0.5cm 0cm 0.0cm, clip = true]{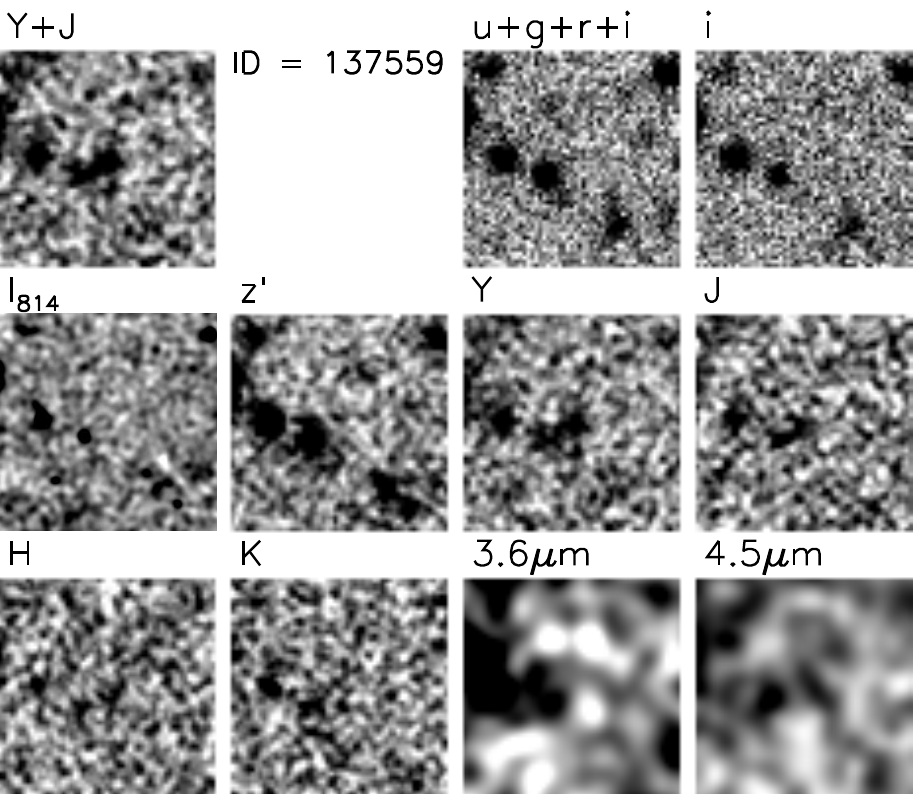} \hspace{0.3cm}
\includegraphics[width = 0.31\textwidth]{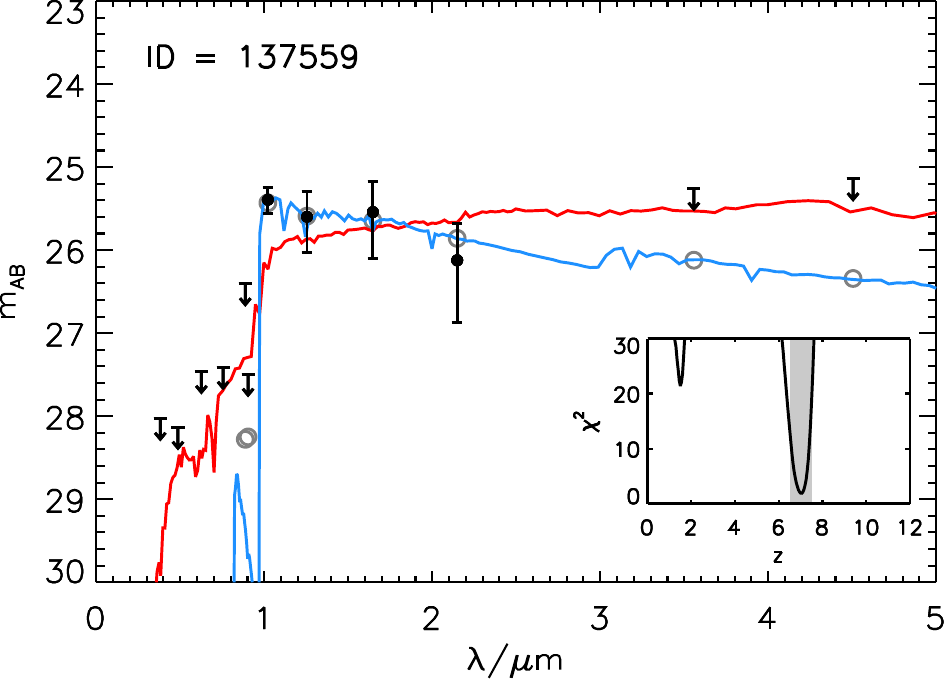} \hspace{0.3cm}
\includegraphics[width = 0.31\textwidth]{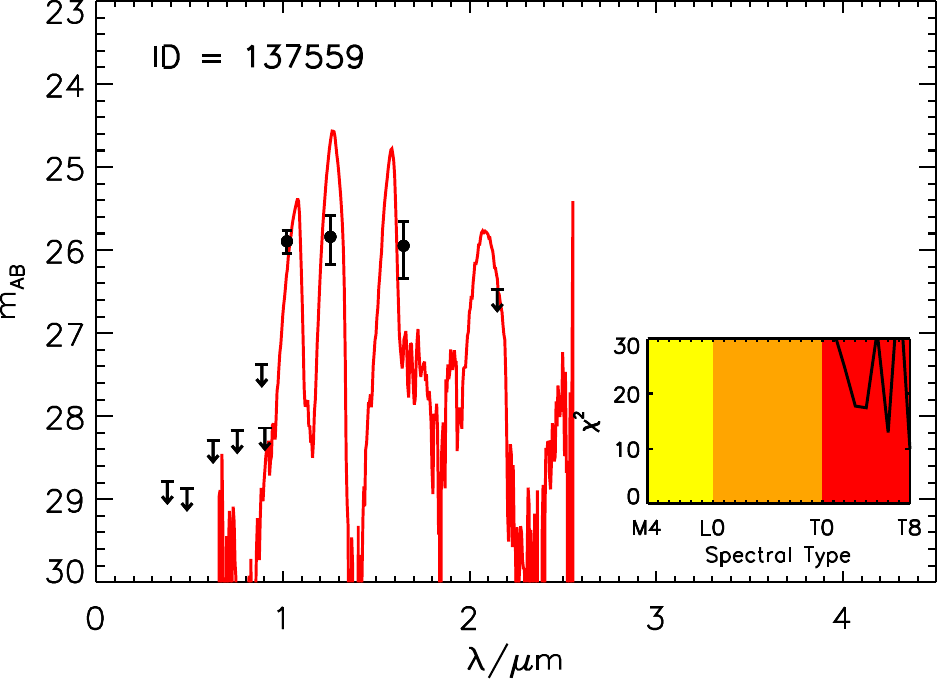} \hspace{0.3cm}\\ 
\vspace{0.2mm}
\includegraphics[width = 0.265\textwidth, trim = 0cm -0.5cm 0cm 0.0cm, clip = true]{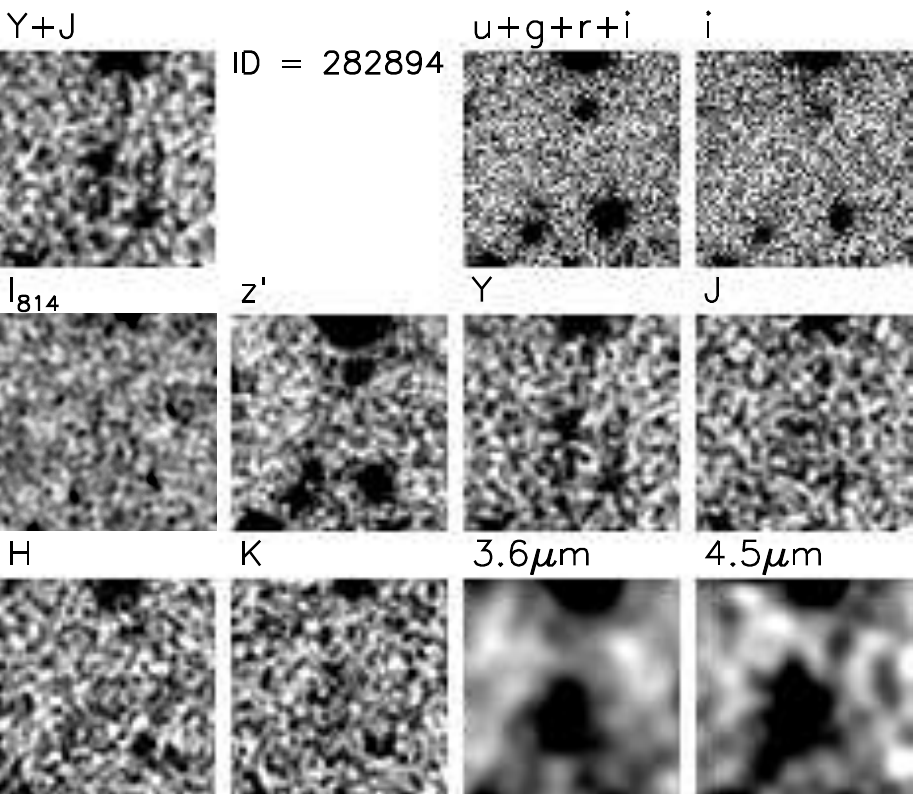} \hspace{0.3cm}
\includegraphics[width = 0.31\textwidth]{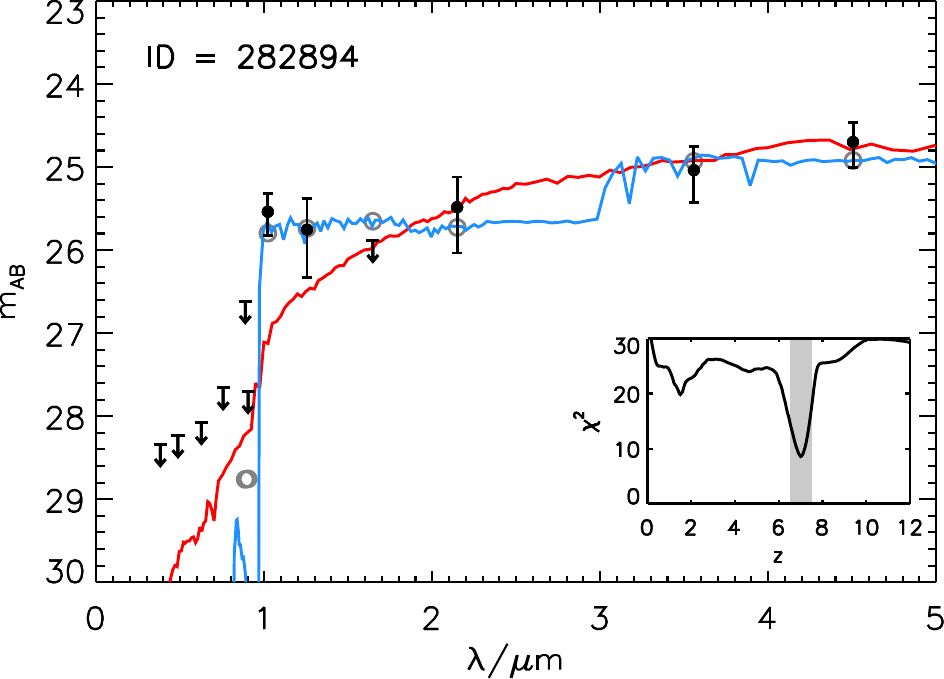} \hspace{0.3cm}
\includegraphics[width = 0.31\textwidth]{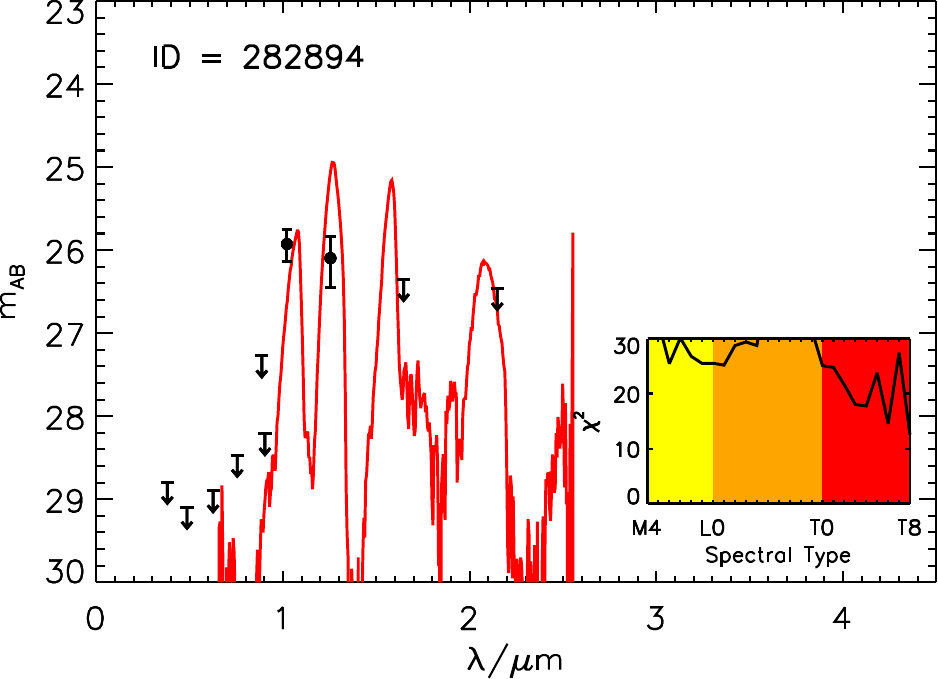} \hspace{0.3cm}\\ 
\vspace{0.2mm}
\includegraphics[width = 0.265\textwidth, trim = 0cm -0.5cm 0cm 0.0cm, clip = true]{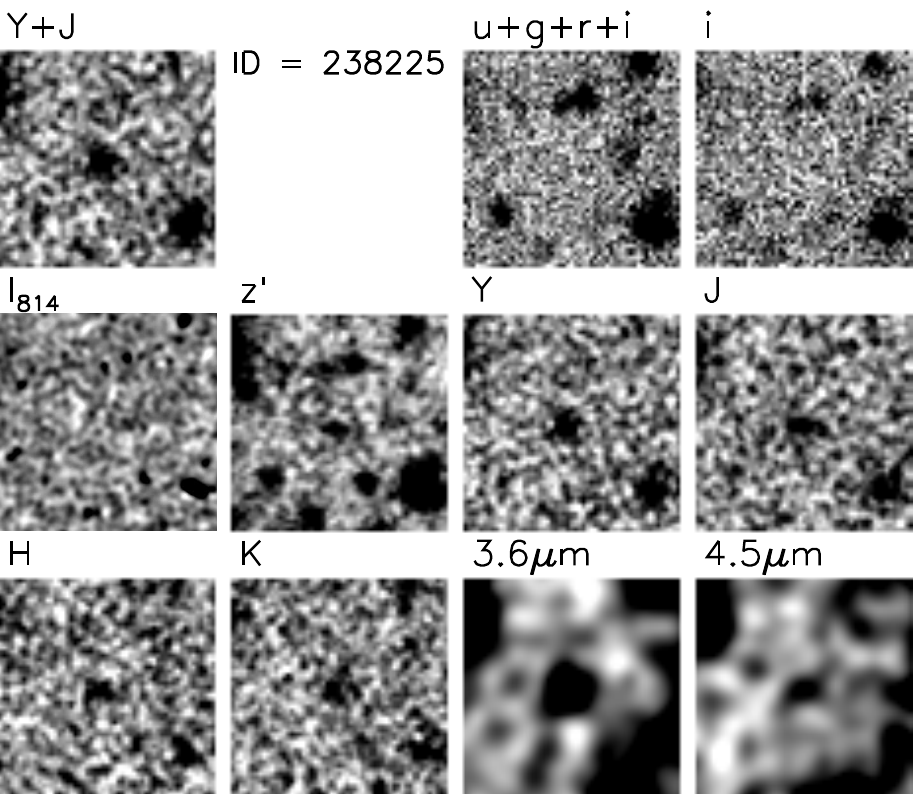} \hspace{0.3cm}
\includegraphics[width = 0.31\textwidth]{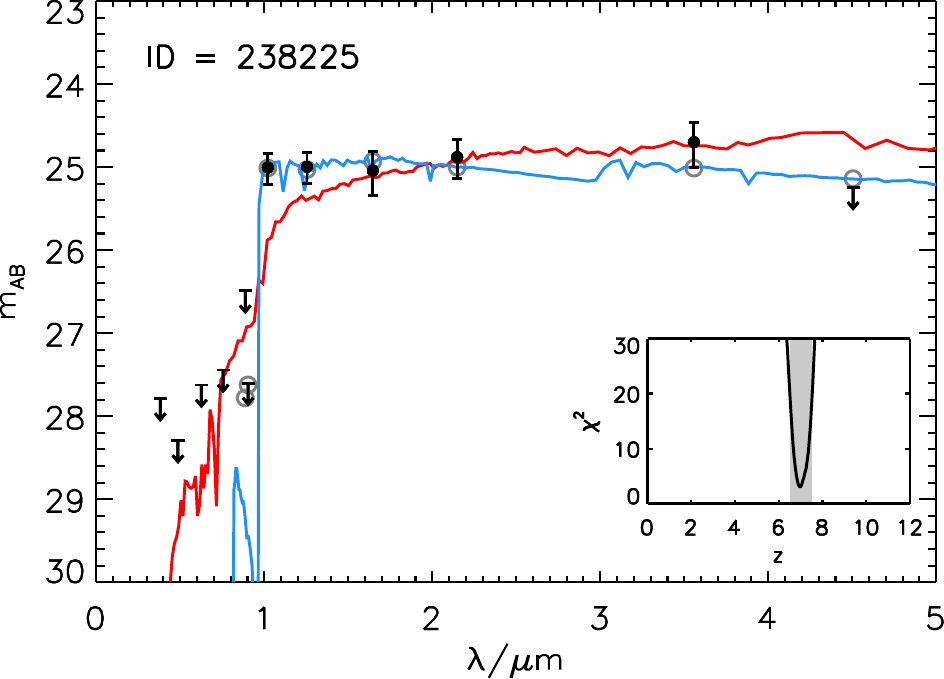} \hspace{0.3cm}
\includegraphics[width = 0.31\textwidth]{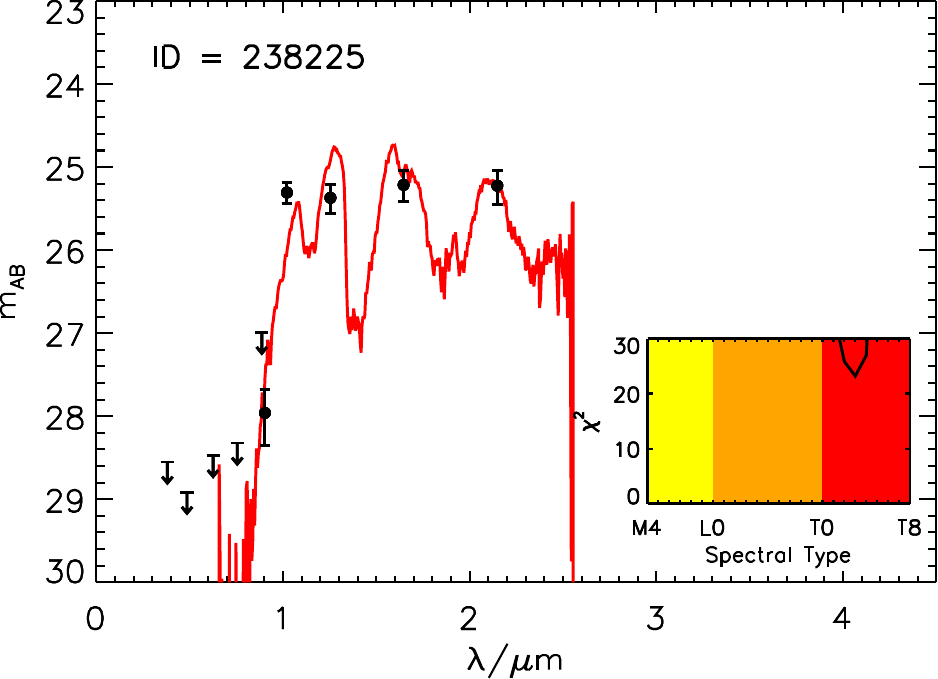} \hspace{0.3cm}\\ 

\contcaption{}
\end{figure*}

\begin{figure*}

\includegraphics[width = 0.265\textwidth, trim = 0cm -0.5cm 0cm 0.0cm, clip = true]{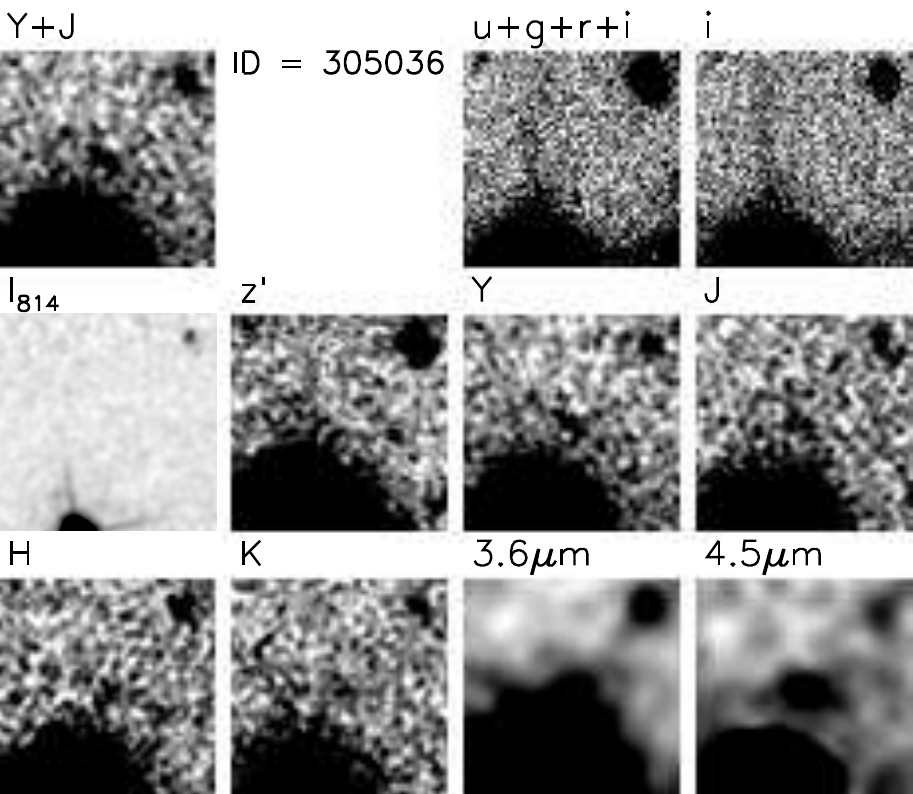} \hspace{0.3cm}
\includegraphics[width = 0.31\textwidth]{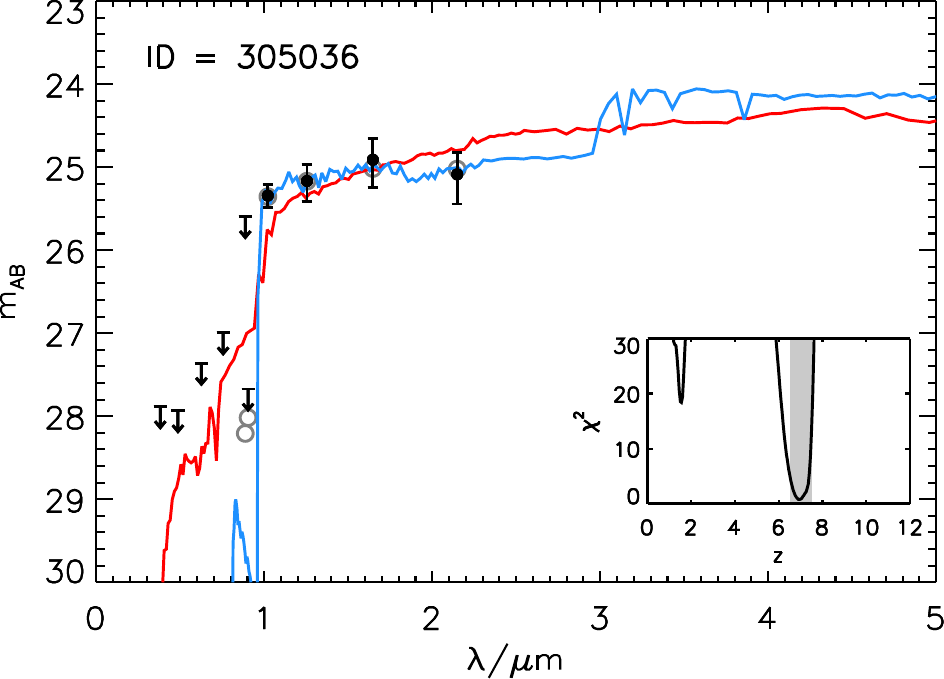} \hspace{0.3cm}
\includegraphics[width = 0.31\textwidth]{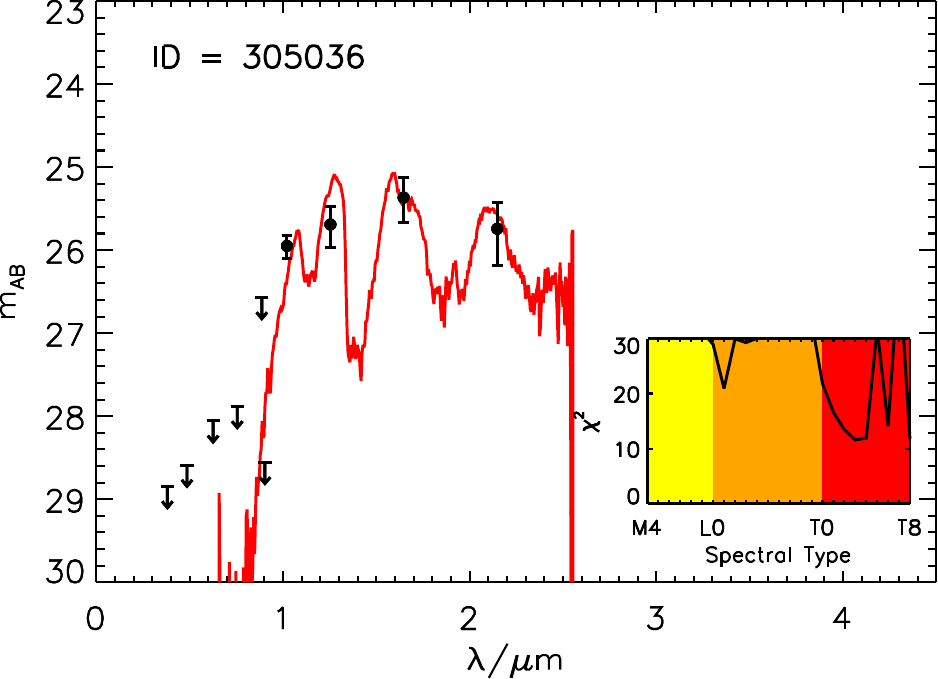} \hspace{0.3cm}\\ 

\includegraphics[width = 0.265\textwidth, trim = 0cm -0.5cm 0cm 0.0cm, clip = true]{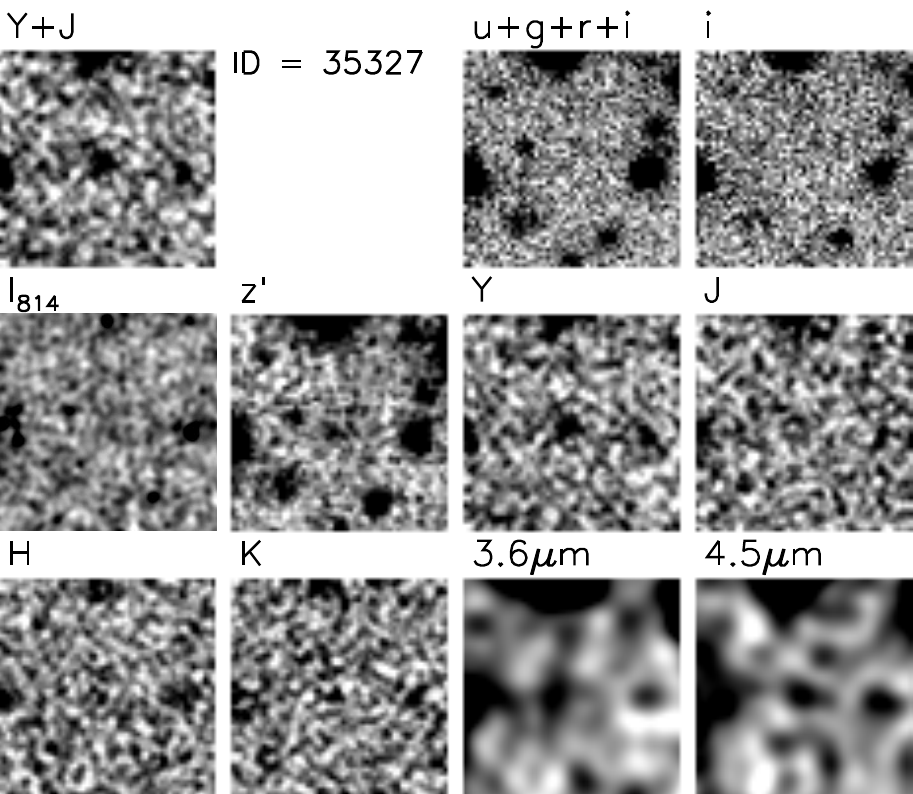} \hspace{0.3cm}
\includegraphics[width = 0.31\textwidth]{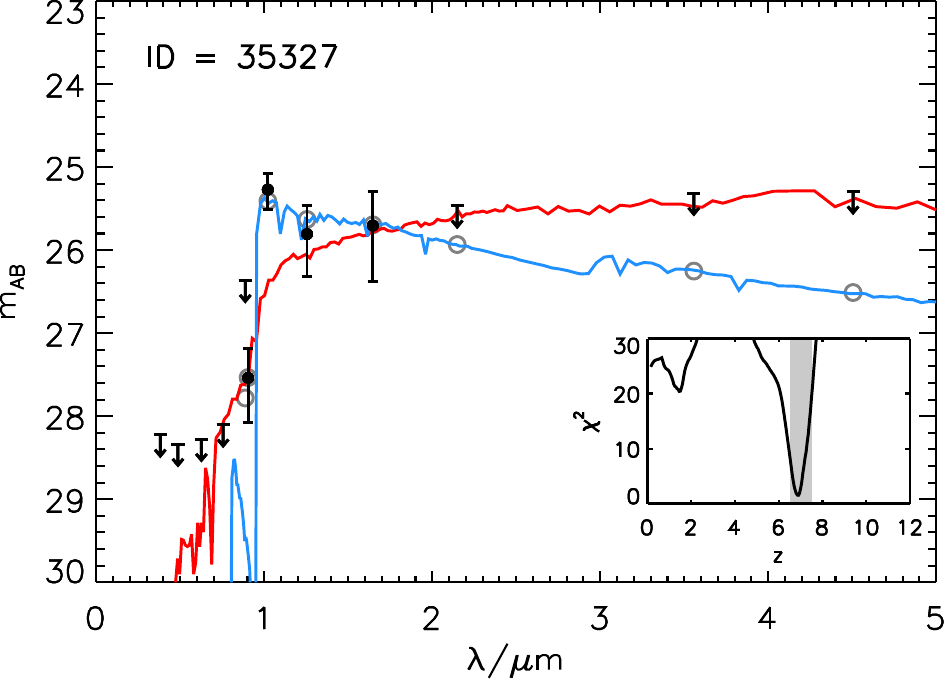} \hspace{0.3cm}
\includegraphics[width = 0.31\textwidth]{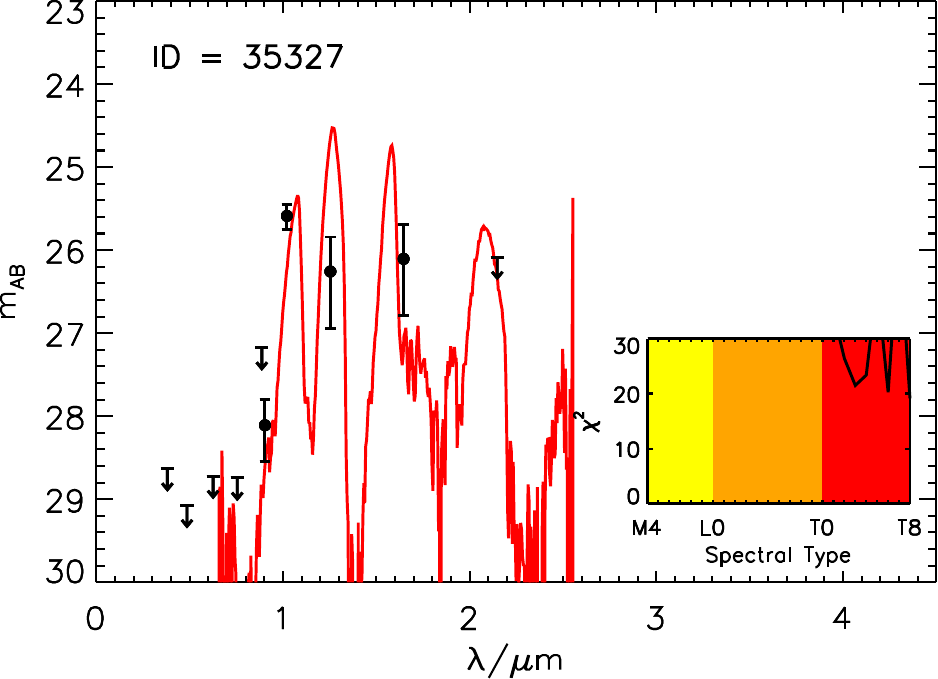} \hspace{0.3cm}\\ 
\vspace{0.2mm}
\includegraphics[width = 0.265\textwidth, trim = 0cm -0.5cm 0cm 0.0cm, clip = true]{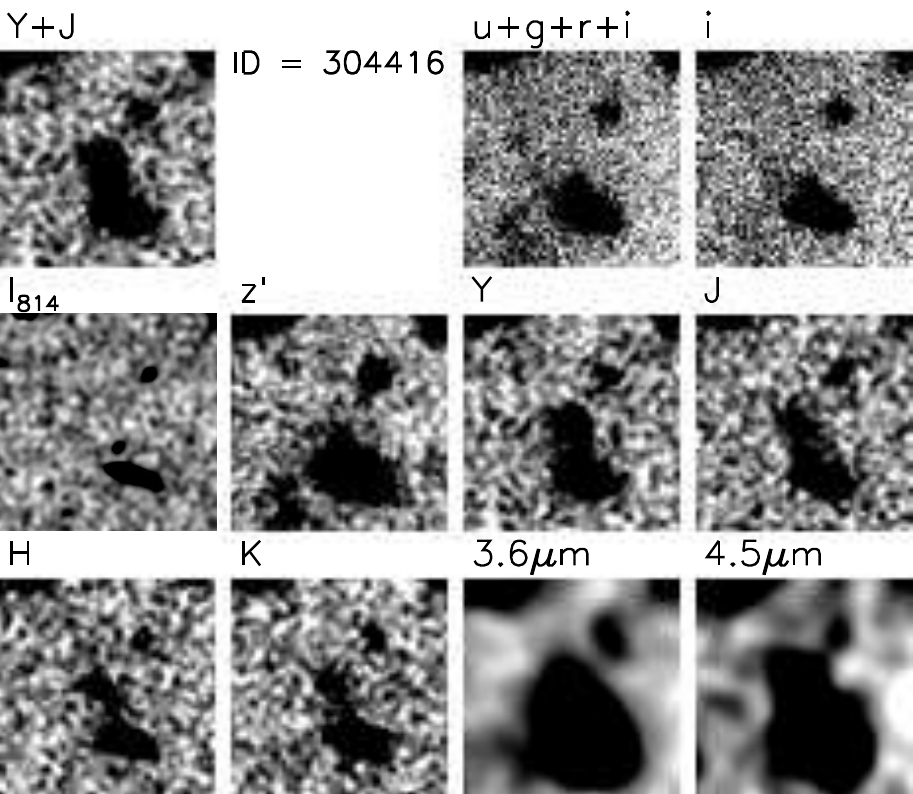} \hspace{0.3cm}
\includegraphics[width = 0.31\textwidth]{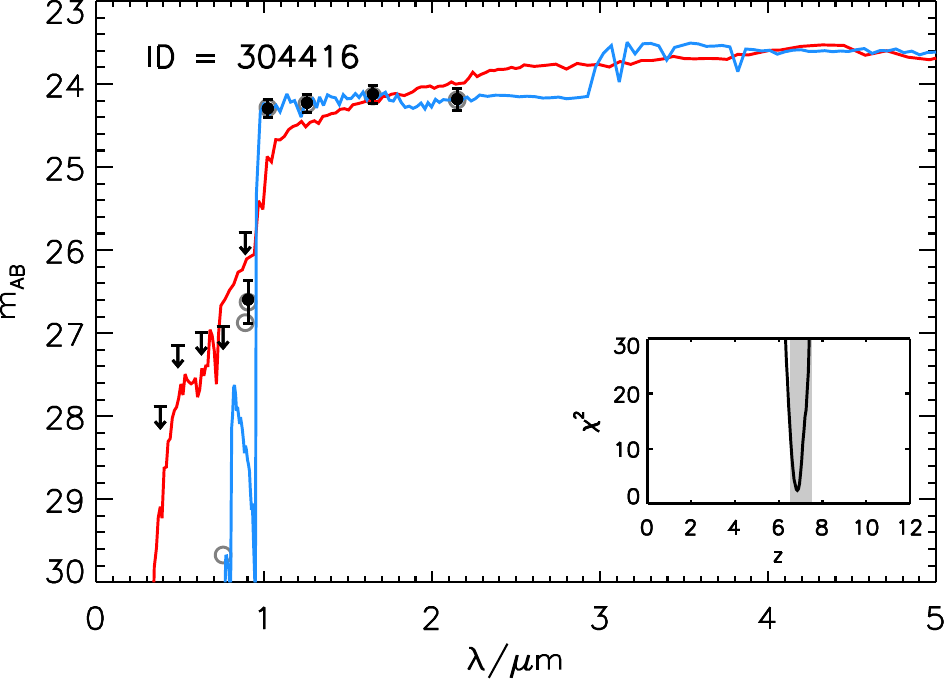} \hspace{0.3cm}
\includegraphics[width = 0.31\textwidth]{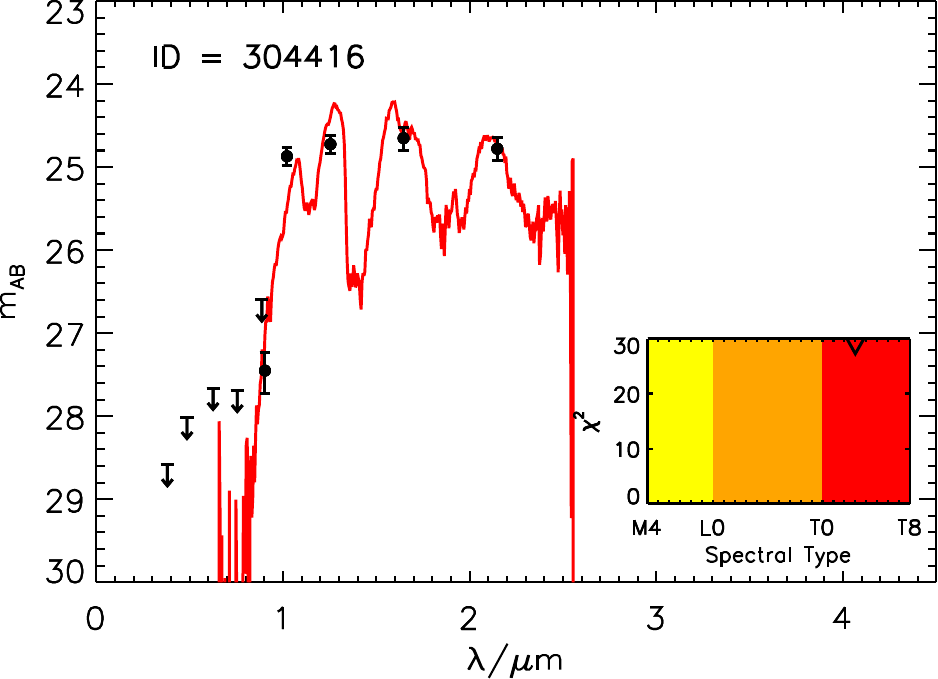} \hspace{0.3cm}\\ 
\vspace{0.2mm}
\includegraphics[width = 0.265\textwidth, trim = 0cm -0.5cm 0cm 0.0cm, clip = true]{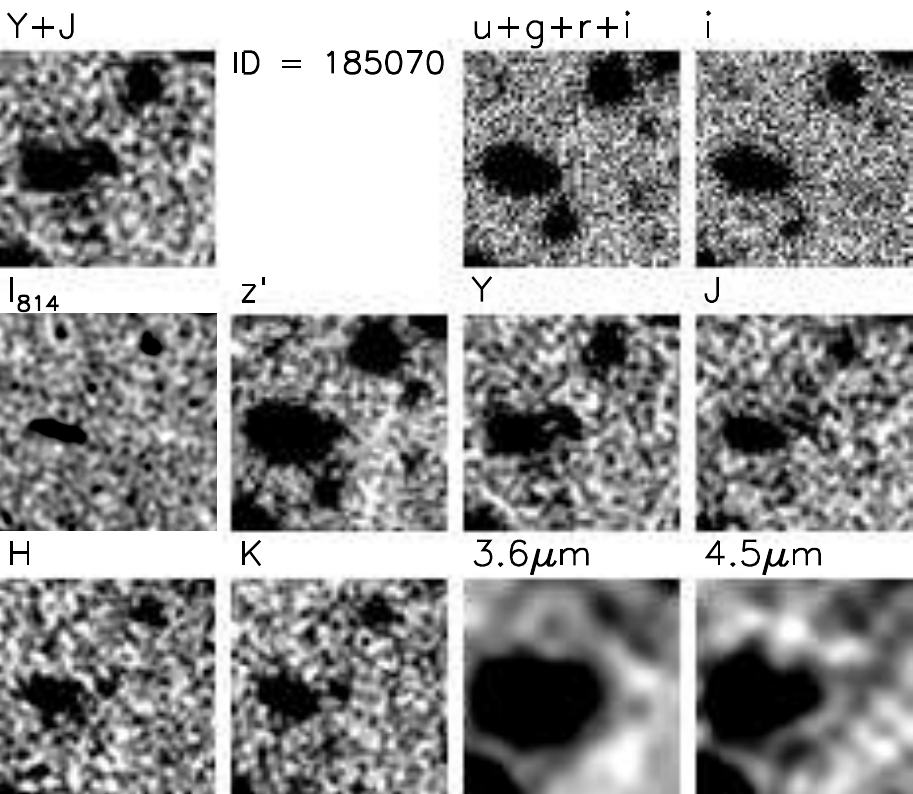} \hspace{0.3cm}
\includegraphics[width = 0.31\textwidth]{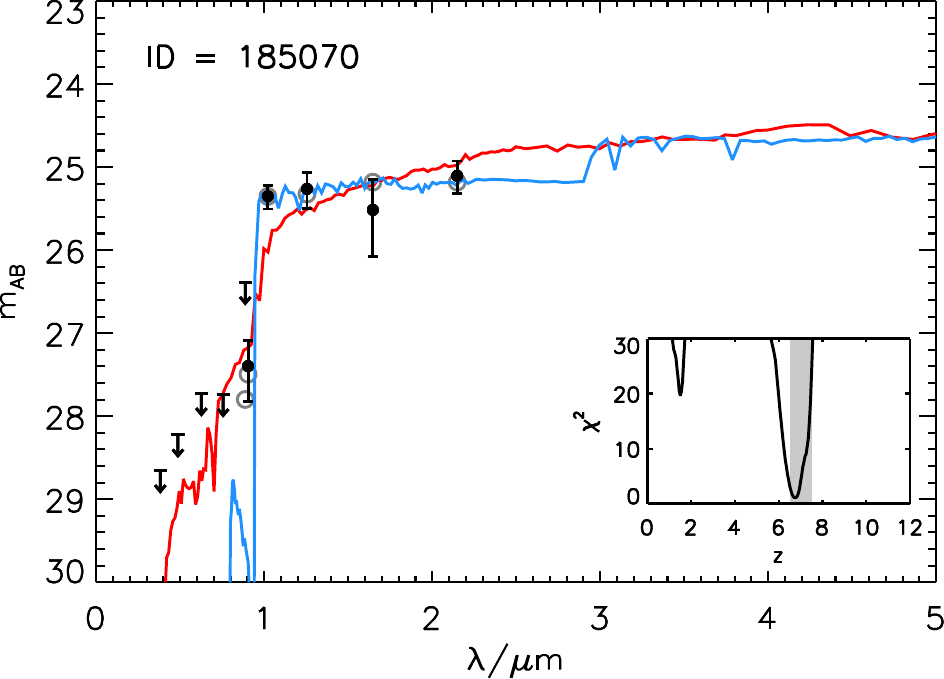} \hspace{0.3cm}
\includegraphics[width = 0.31\textwidth]{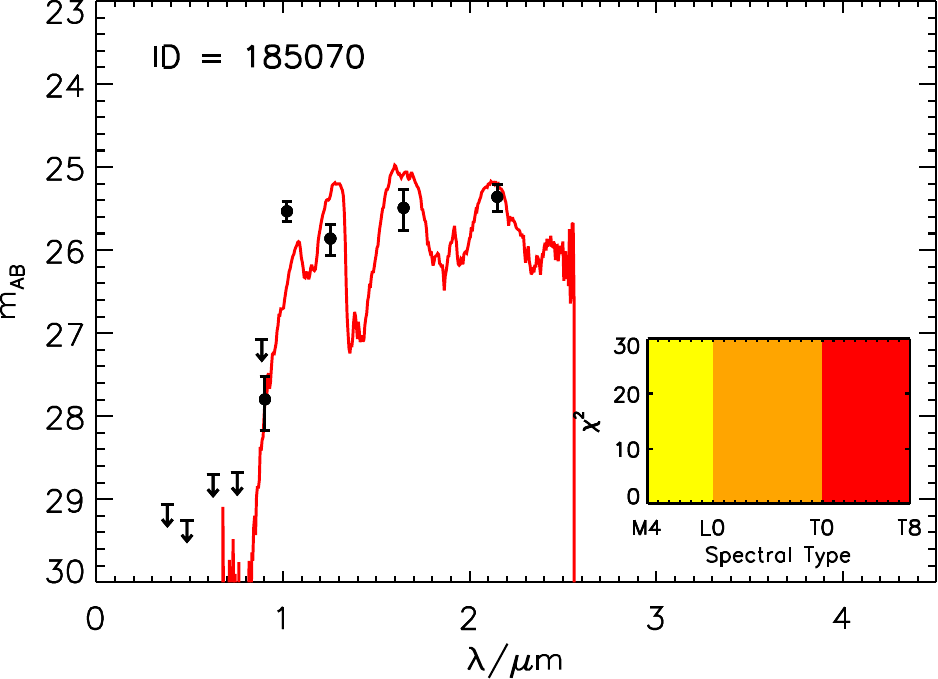} \hspace{0.3cm}\\ 
\vspace{0.2mm}
\includegraphics[width = 0.265\textwidth, trim = 0cm -0.5cm 0cm 0.0cm, clip = true]{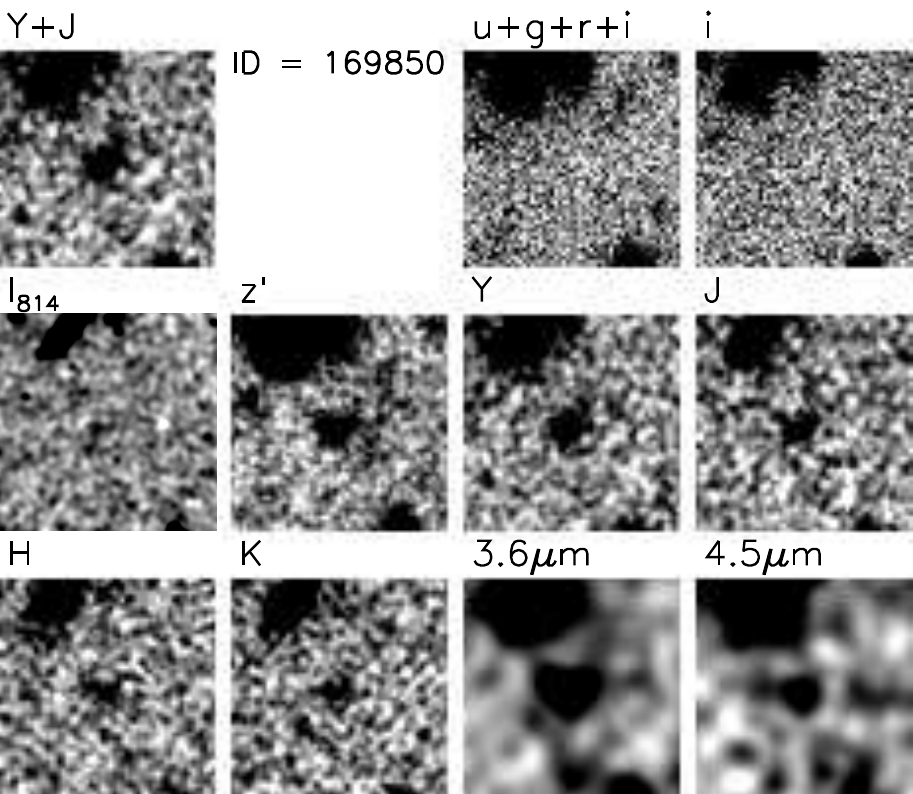} \hspace{0.3cm}
\includegraphics[width = 0.31\textwidth]{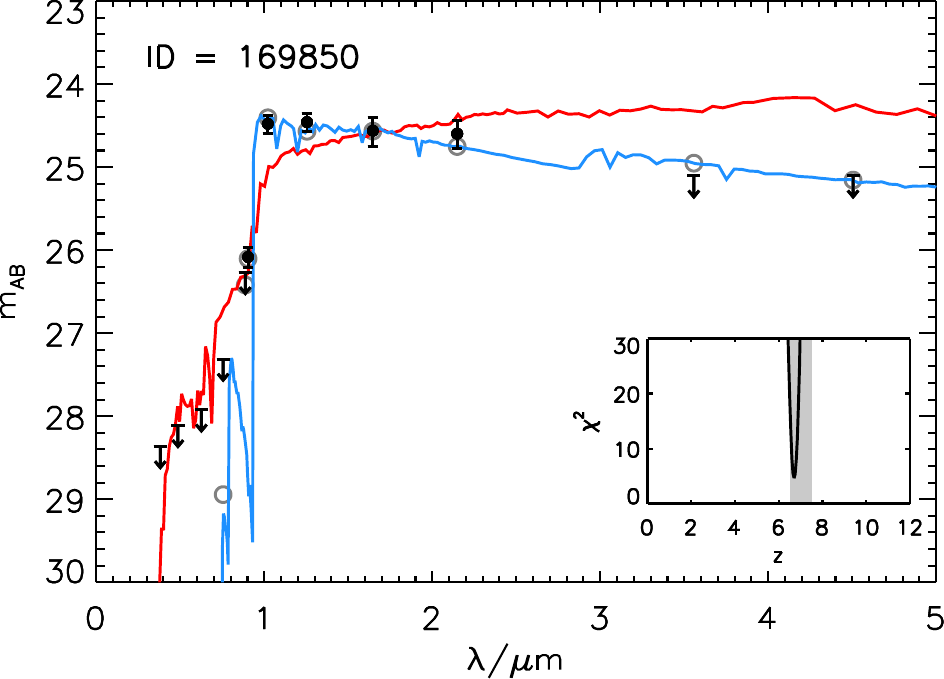} \hspace{0.3cm}
\includegraphics[width = 0.31\textwidth]{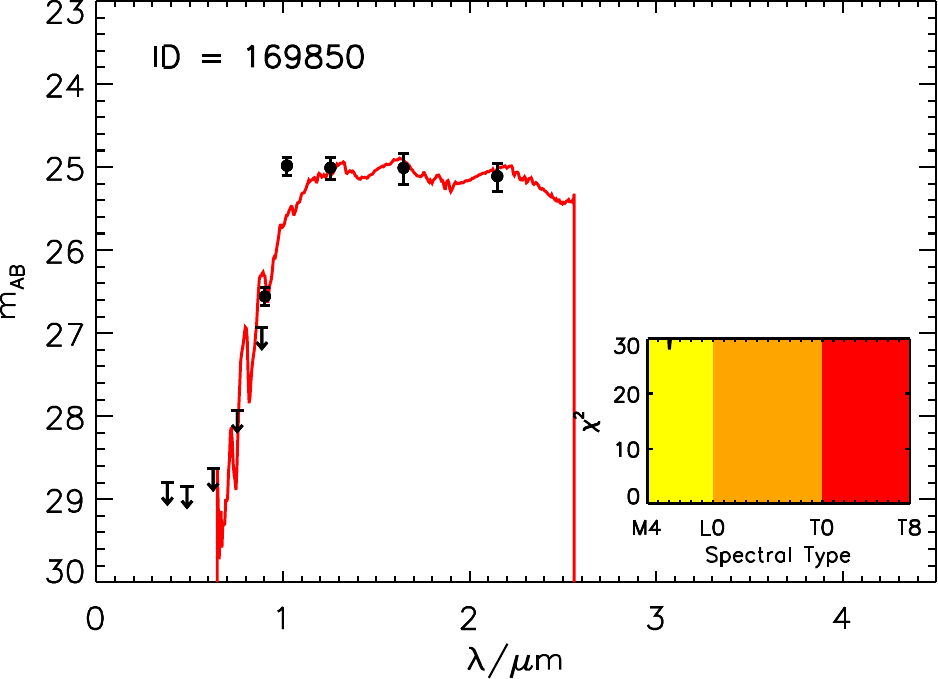} \hspace{0.3cm}\\ 
\vspace{0.2mm}

\contcaption{}
\end{figure*}

\begin{figure*}

\includegraphics[width = 0.265\textwidth, trim = 0cm -0.5cm 0cm 0.0cm, clip = true]{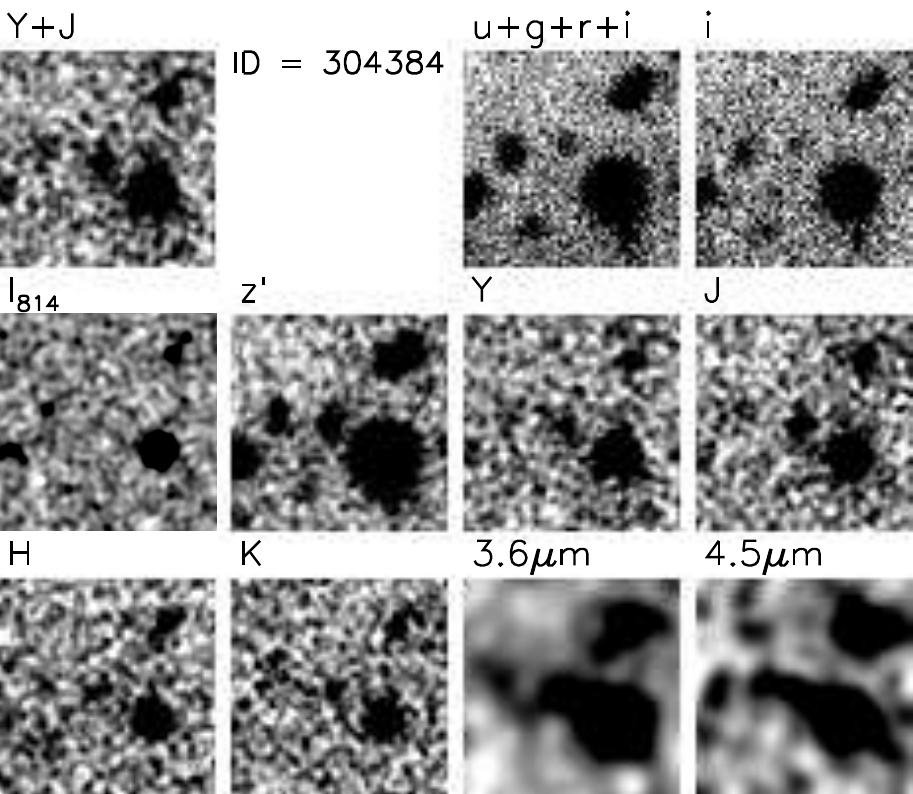} \hspace{0.3cm}
\includegraphics[width = 0.31\textwidth]{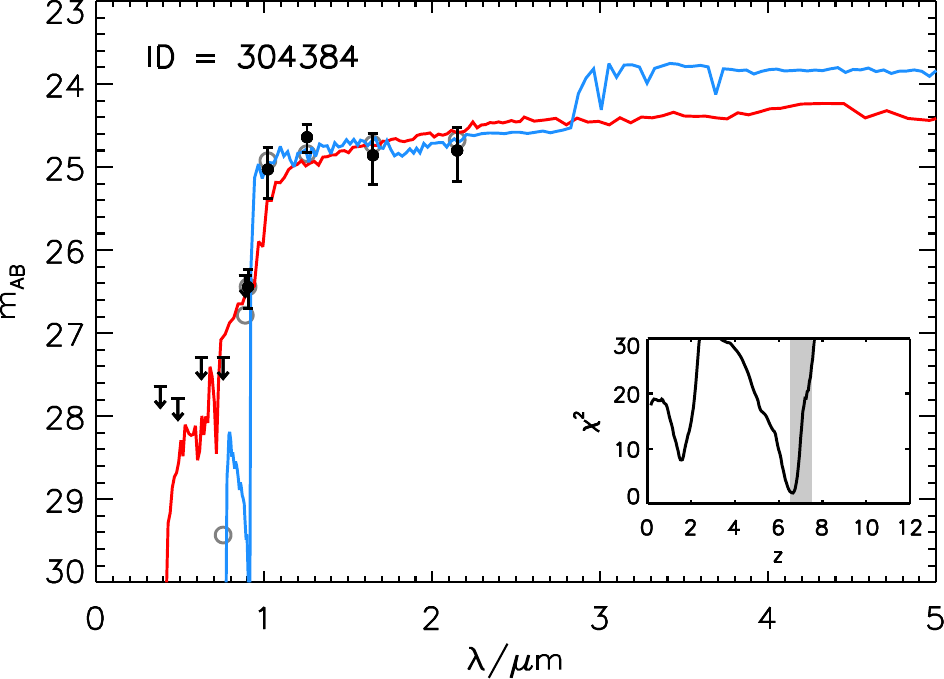} \hspace{0.3cm}
\includegraphics[width = 0.31\textwidth]{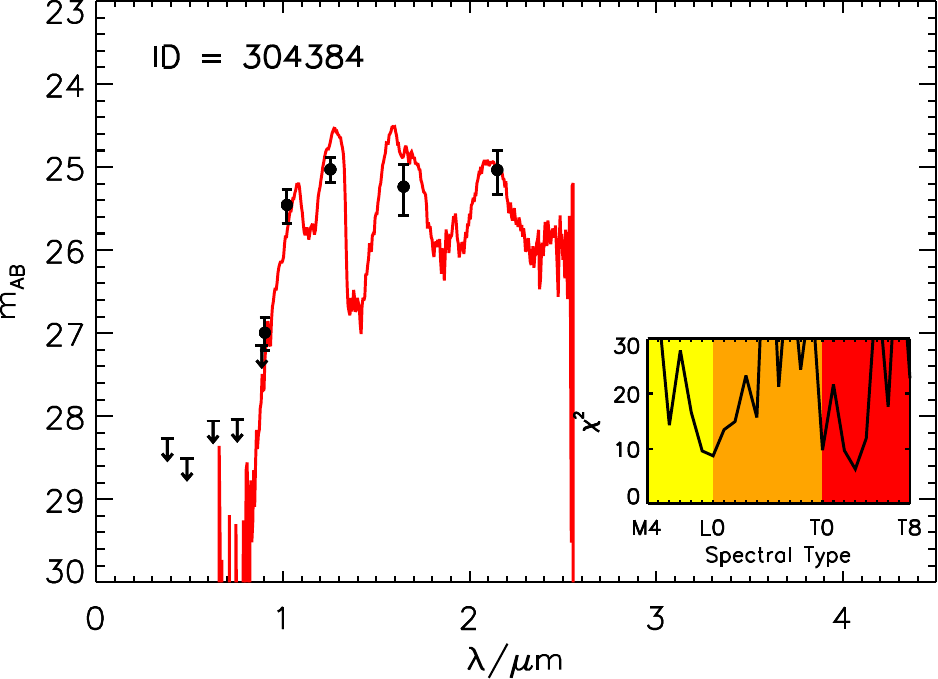} \hspace{0.3cm}\\ 
\vspace{0.2mm}
\includegraphics[width = 0.265\textwidth, trim = 0cm -0.5cm 0cm 0.0cm, clip = true]{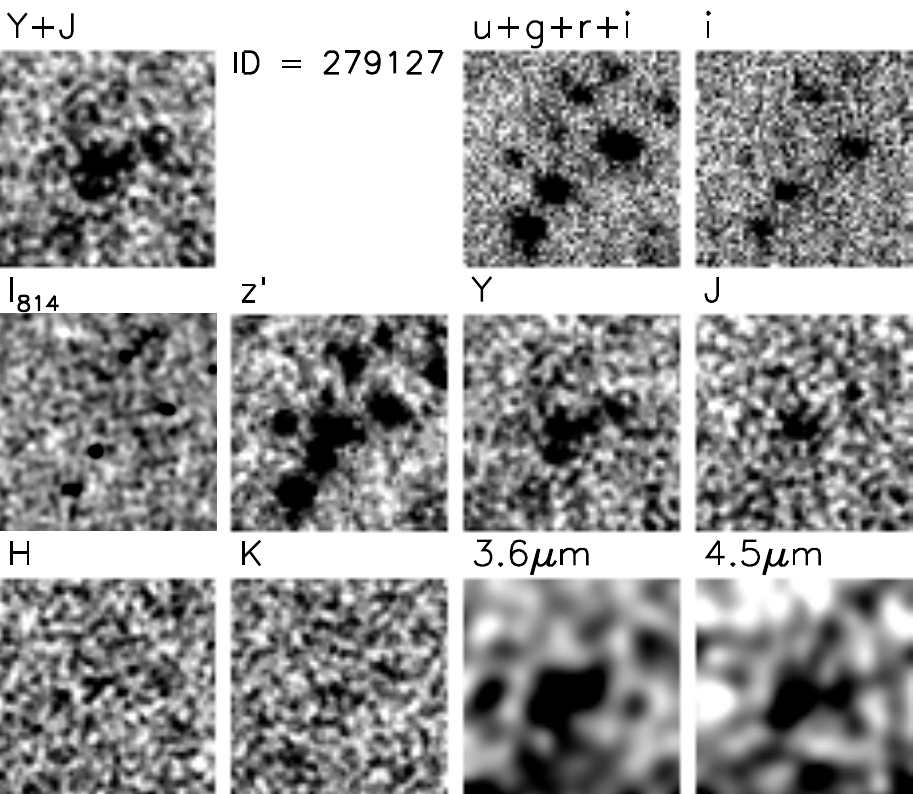} \hspace{0.3cm}
\includegraphics[width = 0.31\textwidth]{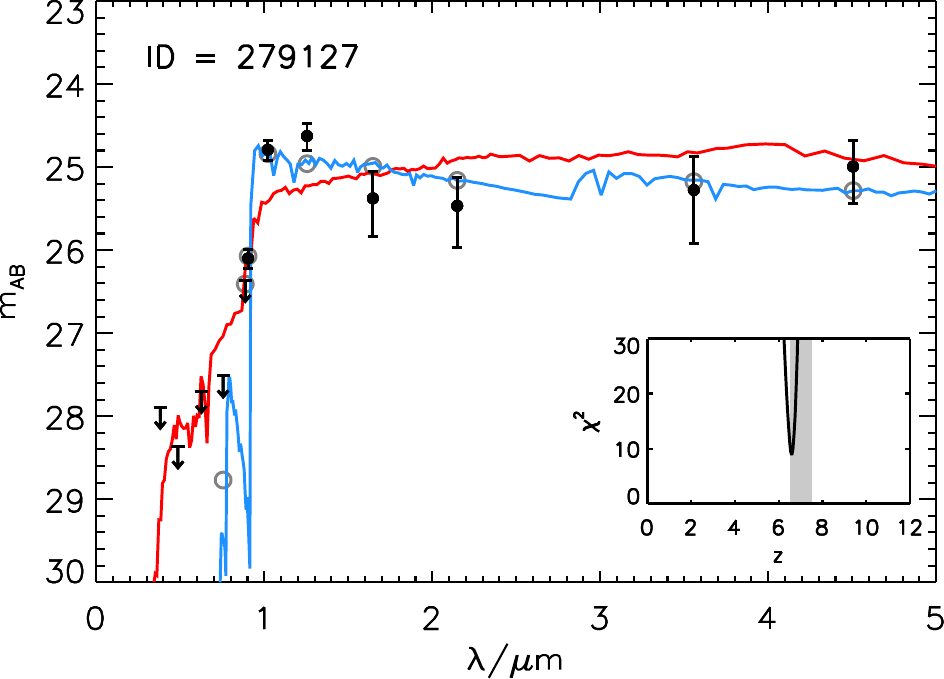} \hspace{0.3cm}
\includegraphics[width = 0.31\textwidth]{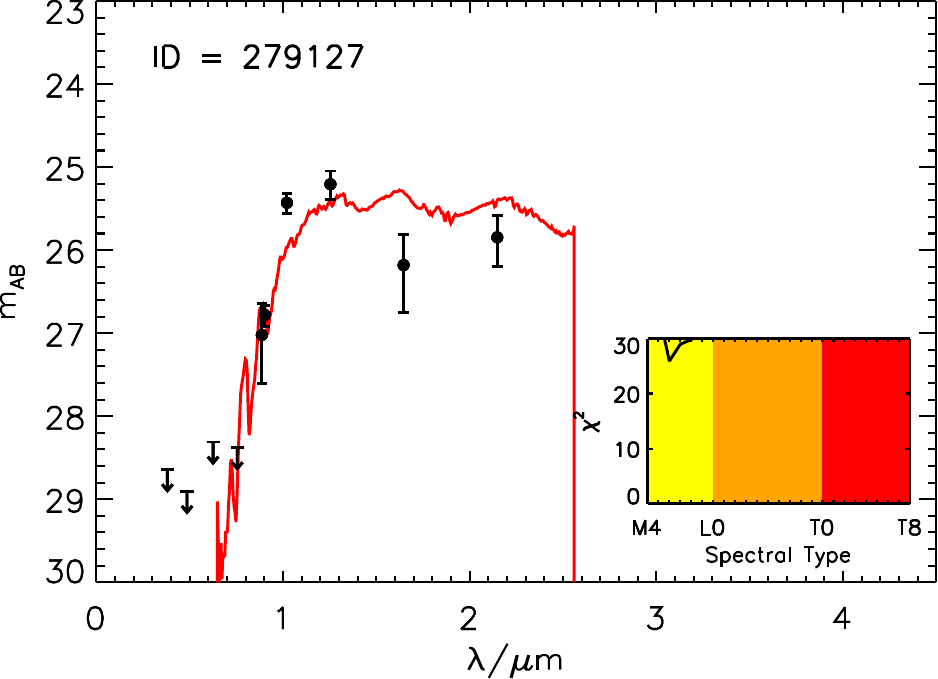} \hspace{0.3cm}\\ 
\vspace{0.2mm}
\includegraphics[width = 0.265\textwidth, trim = 0cm -0.5cm 0cm 0.0cm, clip = true]{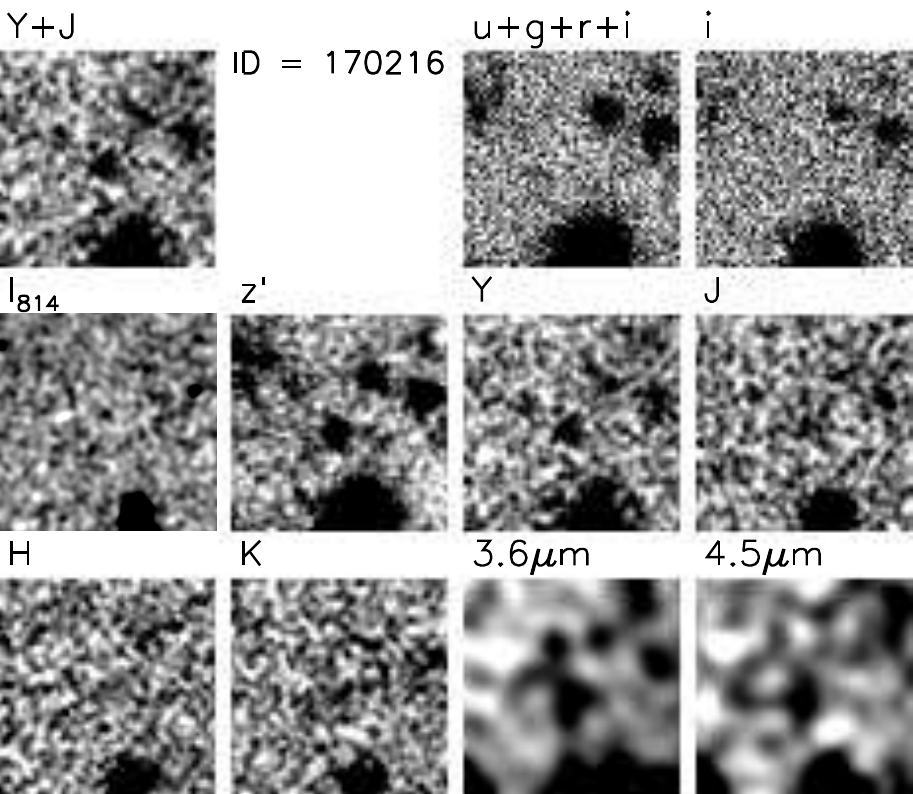} \hspace{0.3cm}
\includegraphics[width = 0.31\textwidth]{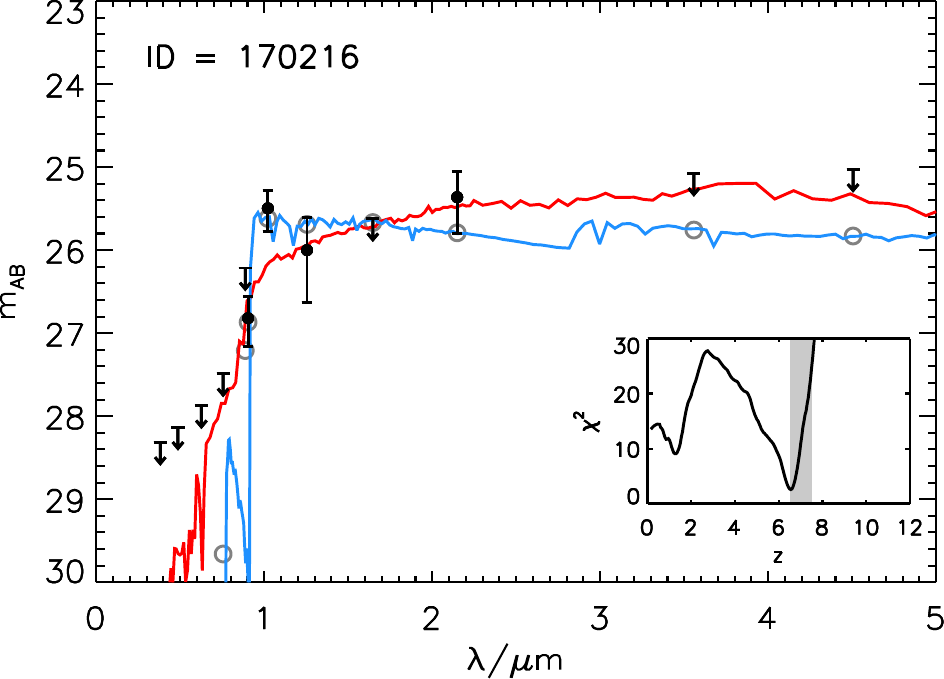} \hspace{0.3cm}
\includegraphics[width = 0.31\textwidth]{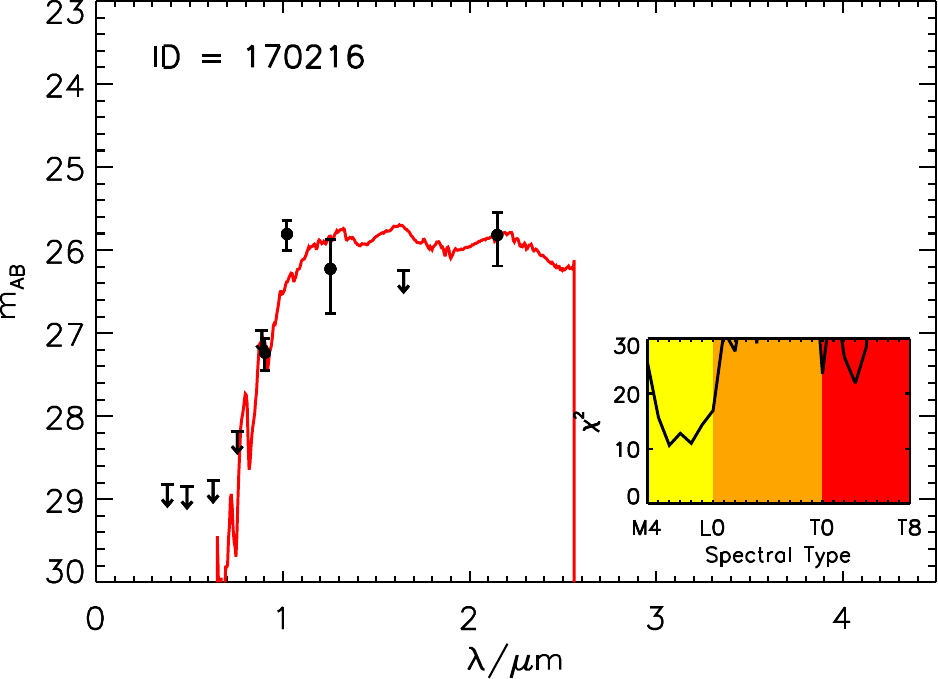} \hspace{0.3cm}\\ 
\vspace{0.2mm}
\includegraphics[width = 0.265\textwidth, trim = 0cm -0.5cm 0cm 0.0cm, clip = true]{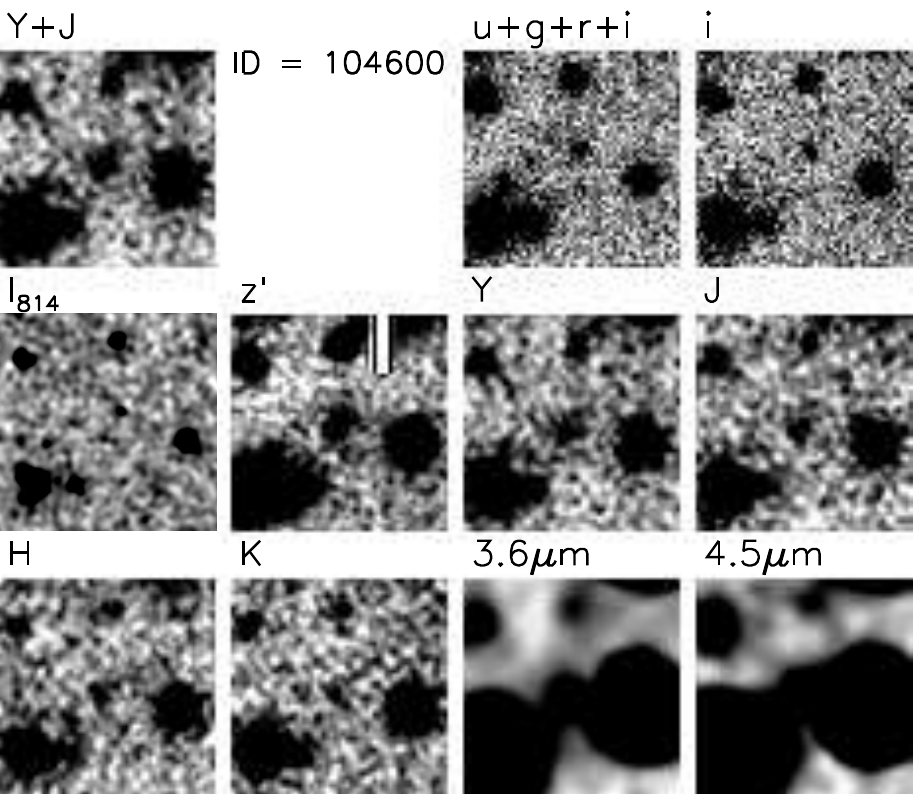} \hspace{0.3cm}
\includegraphics[width = 0.31\textwidth]{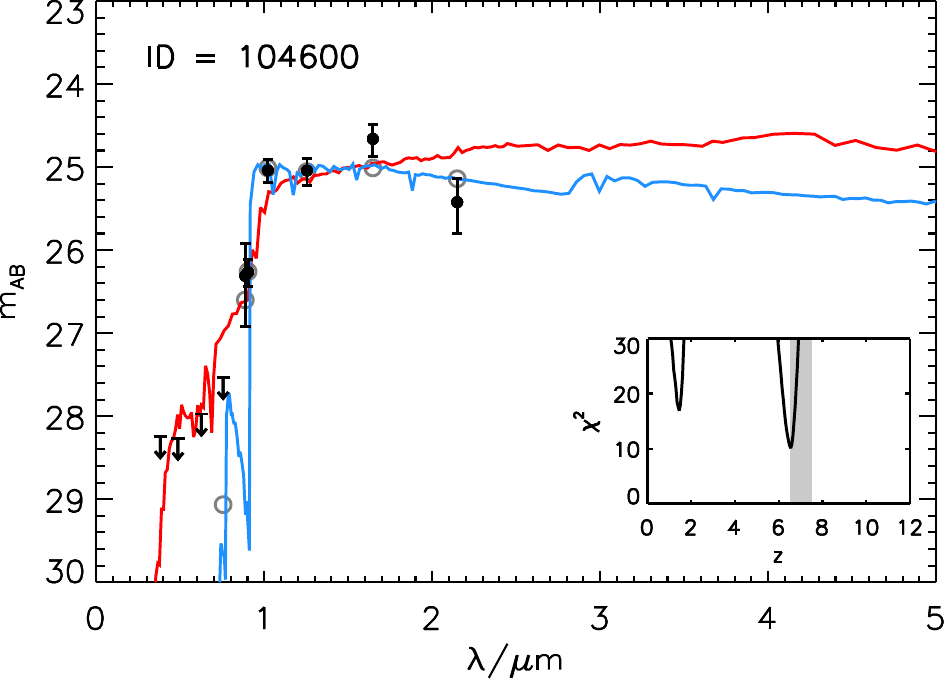} \hspace{0.3cm}
\includegraphics[width = 0.31\textwidth]{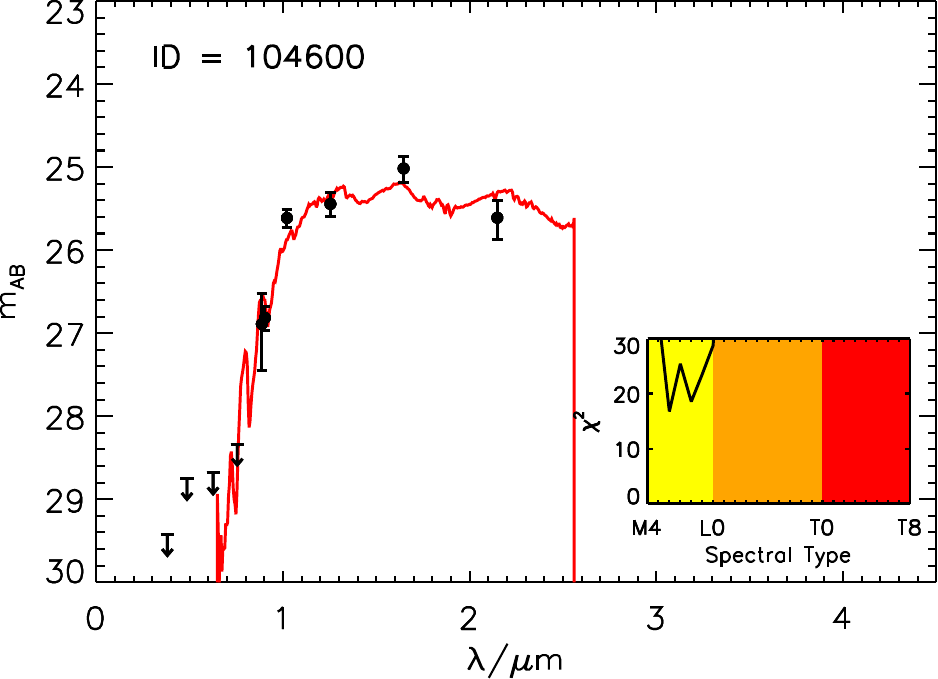} \hspace{0.3cm}\\ 
\vspace{0.2mm}
\includegraphics[width = 0.265\textwidth, trim = 0cm -0.5cm 0cm 0.0cm, clip = true]{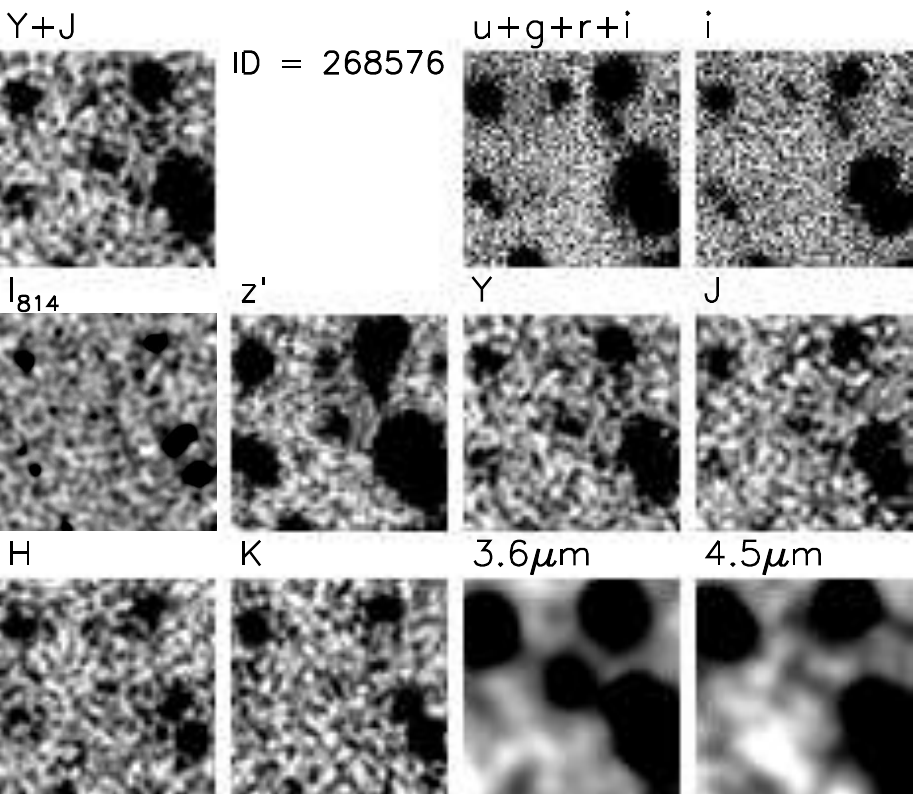} \hspace{0.3cm}
\includegraphics[width = 0.31\textwidth]{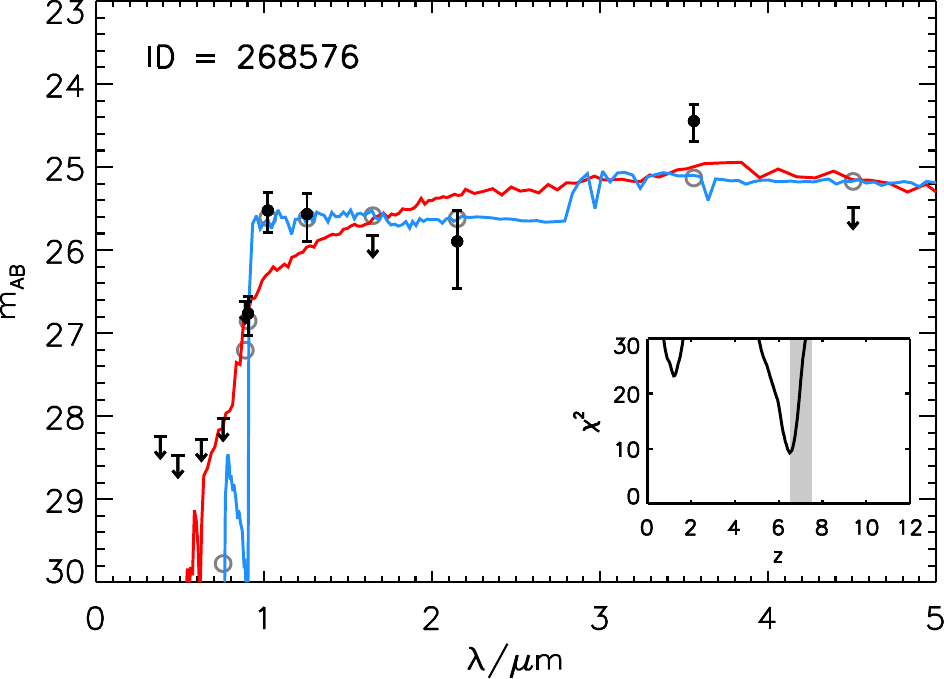} \hspace{0.3cm}
\includegraphics[width = 0.31\textwidth]{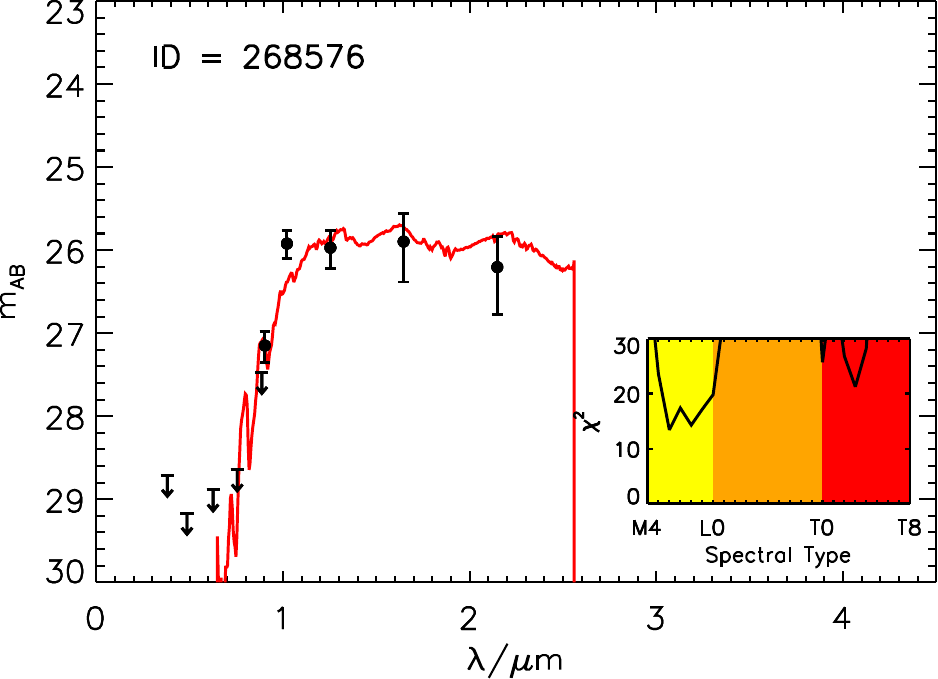} \hspace{0.3cm}\\ 

\contcaption{}
\end{figure*}

\begin{figure*}

\includegraphics[width = 0.265\textwidth, trim = 0cm -0.5cm 0cm 0.0cm, clip = true]{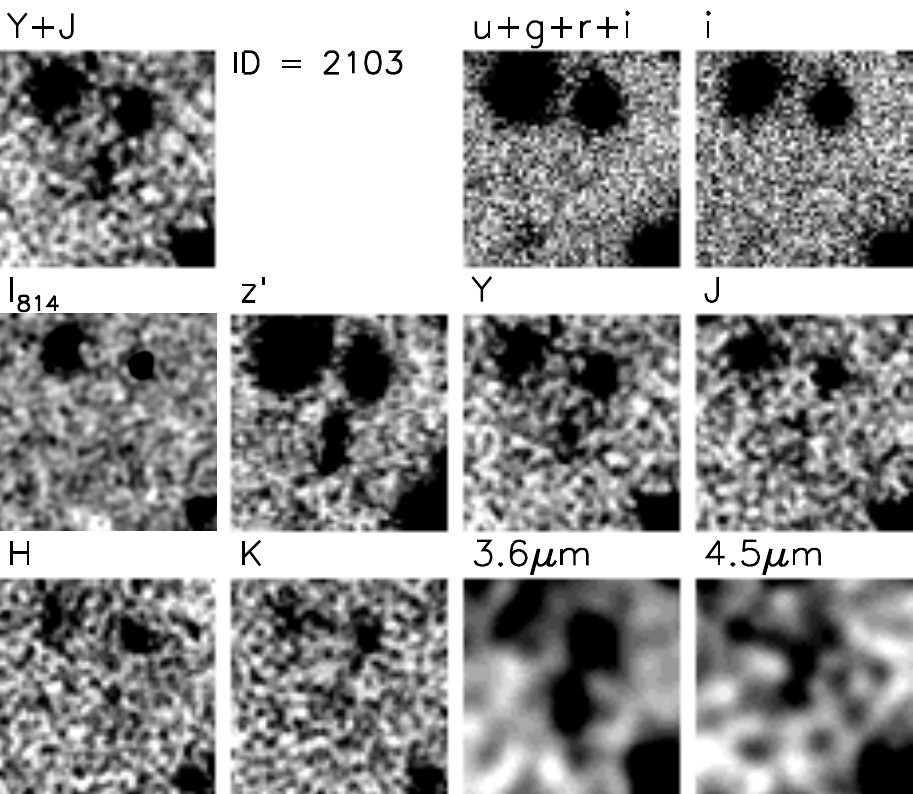} \hspace{0.3cm}
\includegraphics[width = 0.31\textwidth]{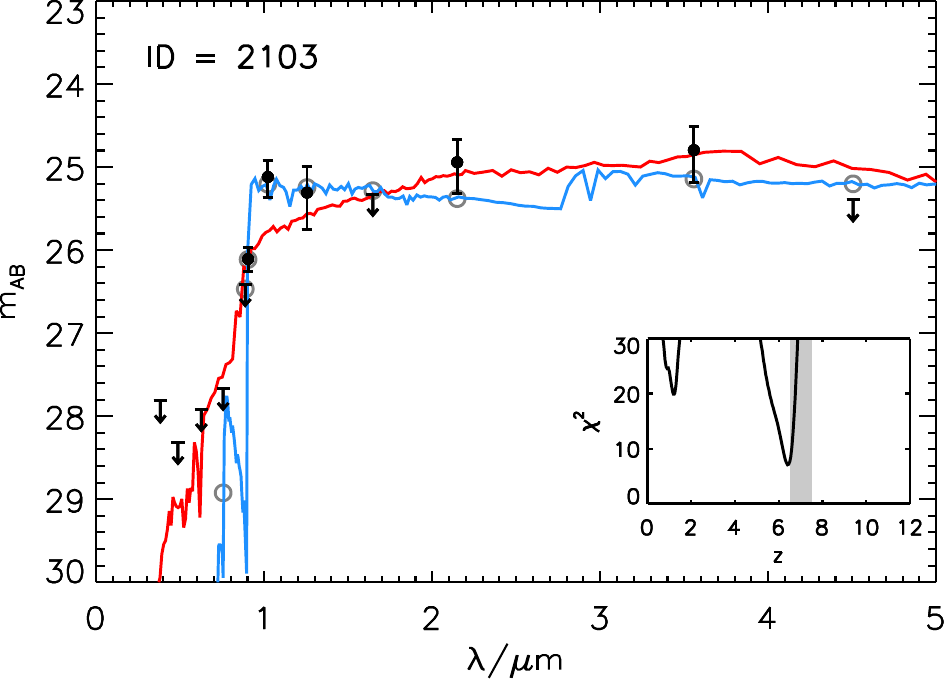} \hspace{0.3cm}
\includegraphics[width = 0.31\textwidth]{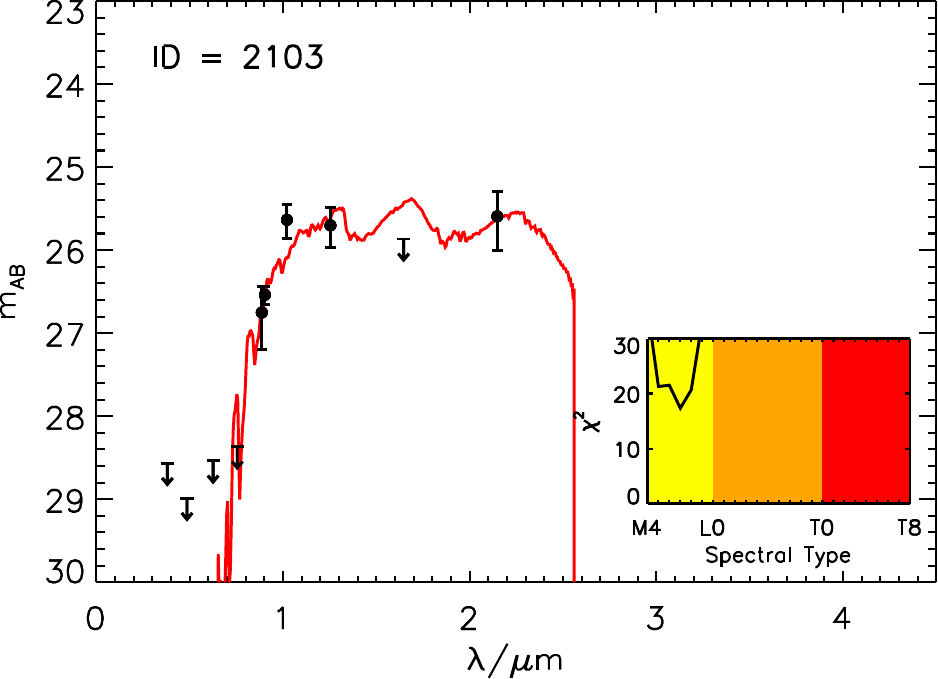} \hspace{0.3cm}\\ 
\vspace{0.2mm}
\includegraphics[width = 0.265\textwidth, trim = 0cm -0.5cm 0cm 0.0cm, clip = true]{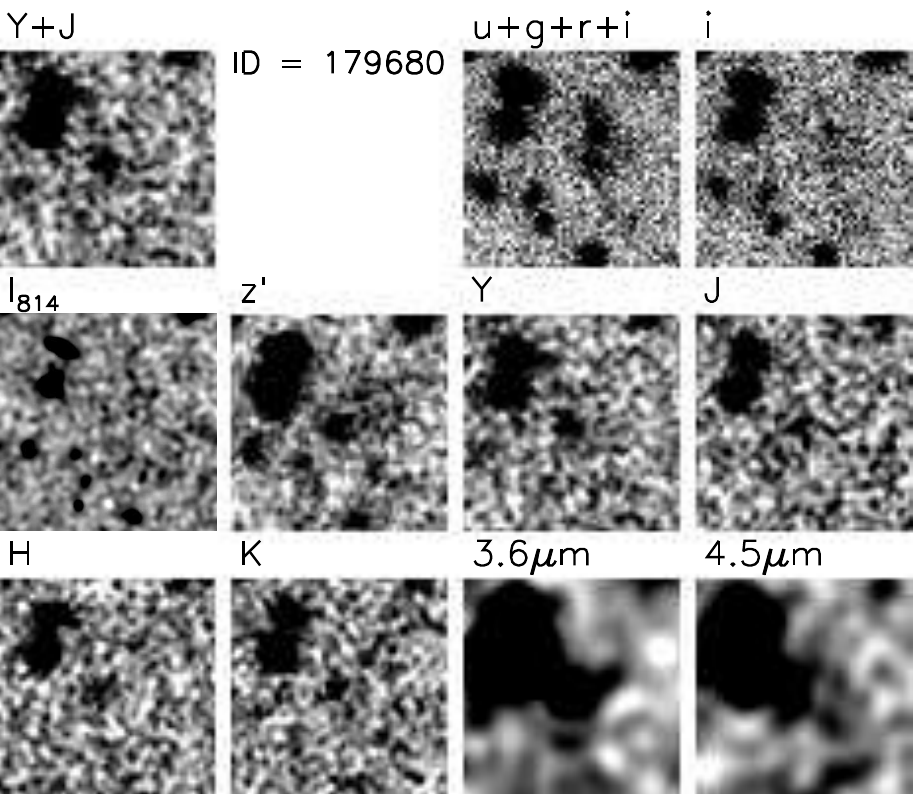} \hspace{0.3cm}
\includegraphics[width = 0.31\textwidth]{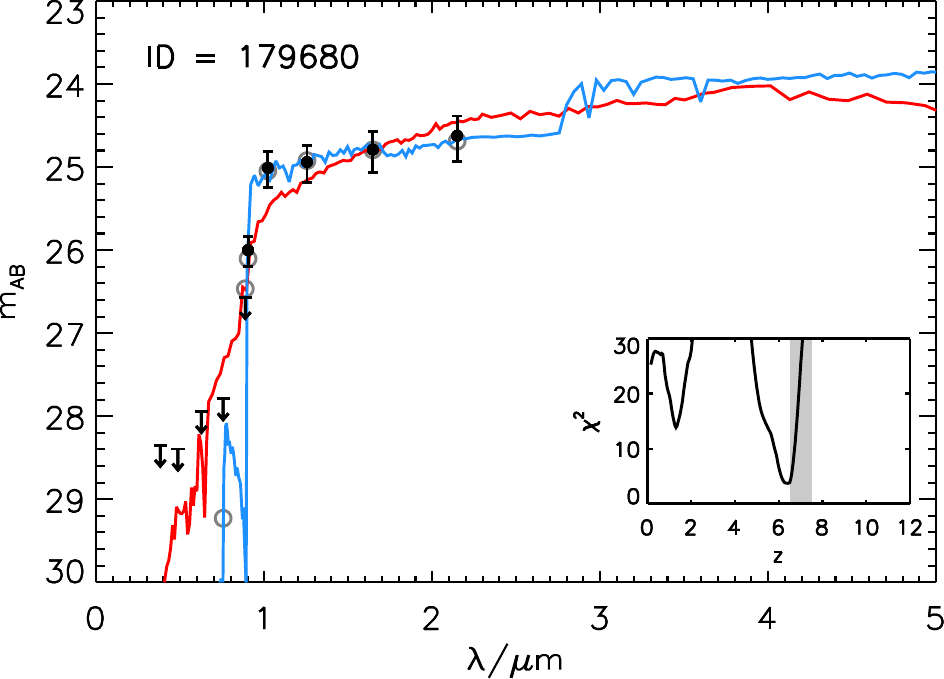} \hspace{0.3cm}
\includegraphics[width = 0.31\textwidth]{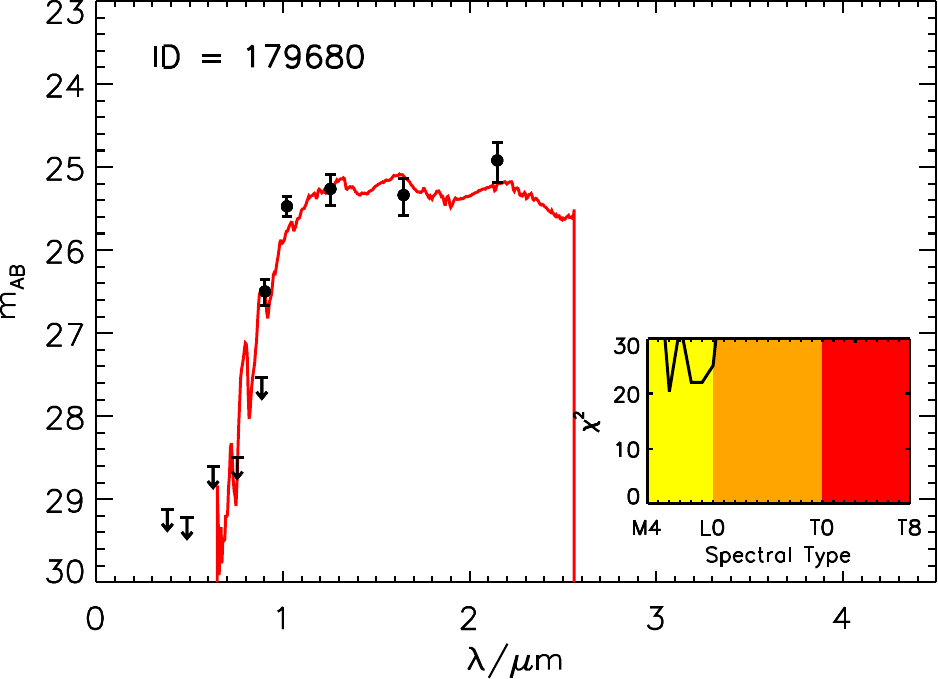} \hspace{0.3cm}\\ 
\vspace{0.2mm}
\includegraphics[width = 0.265\textwidth, trim = 0cm -0.5cm 0cm 0.0cm, clip = true]{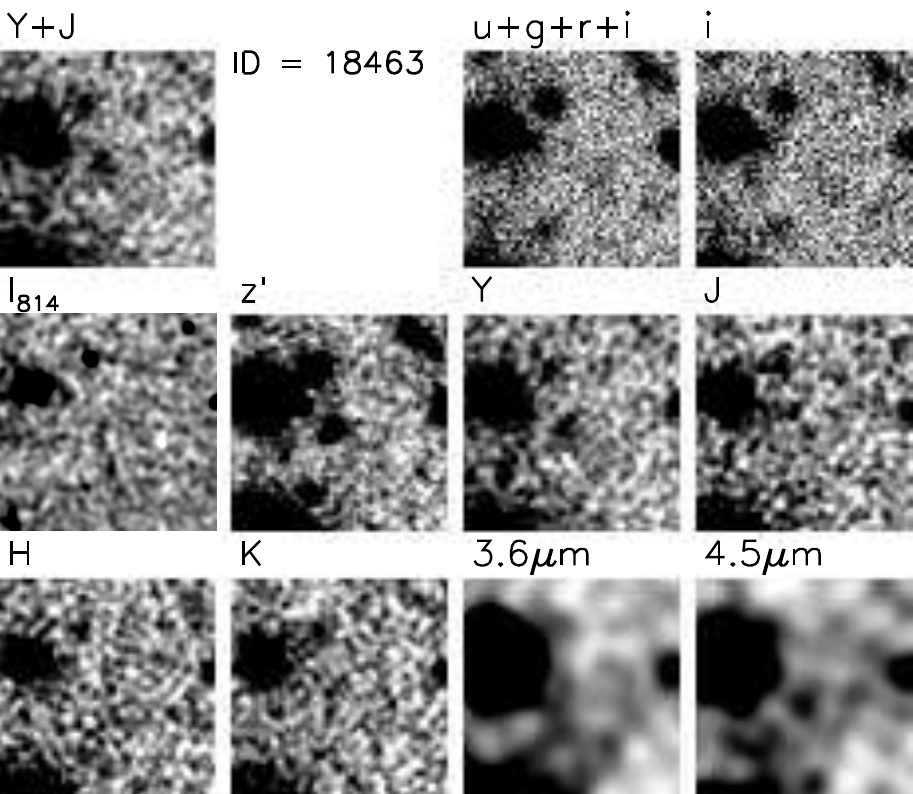} \hspace{0.3cm}
\includegraphics[width = 0.31\textwidth]{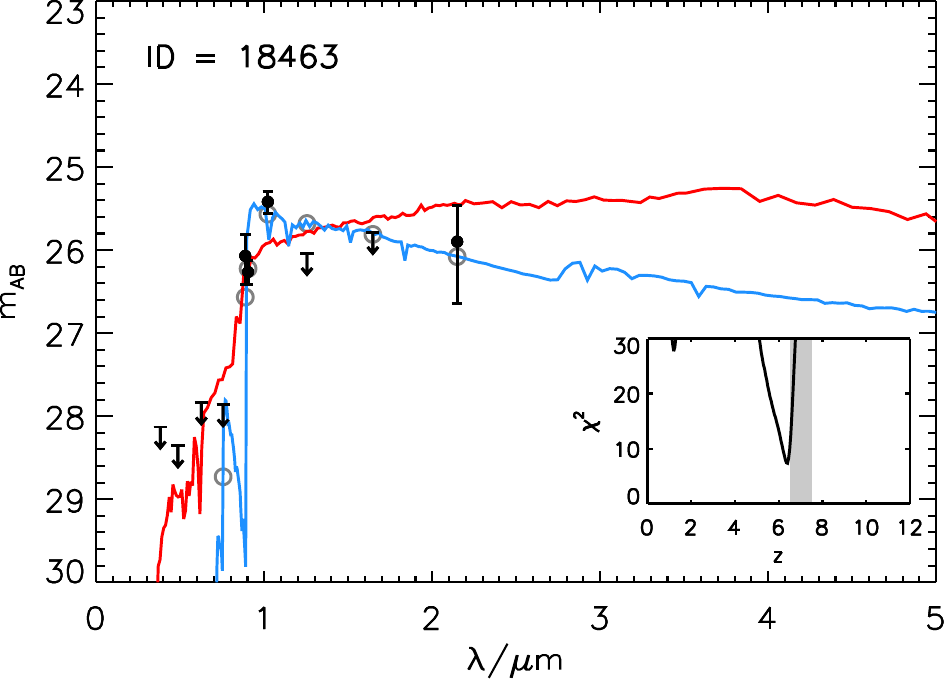} \hspace{0.3cm}
\includegraphics[width = 0.31\textwidth]{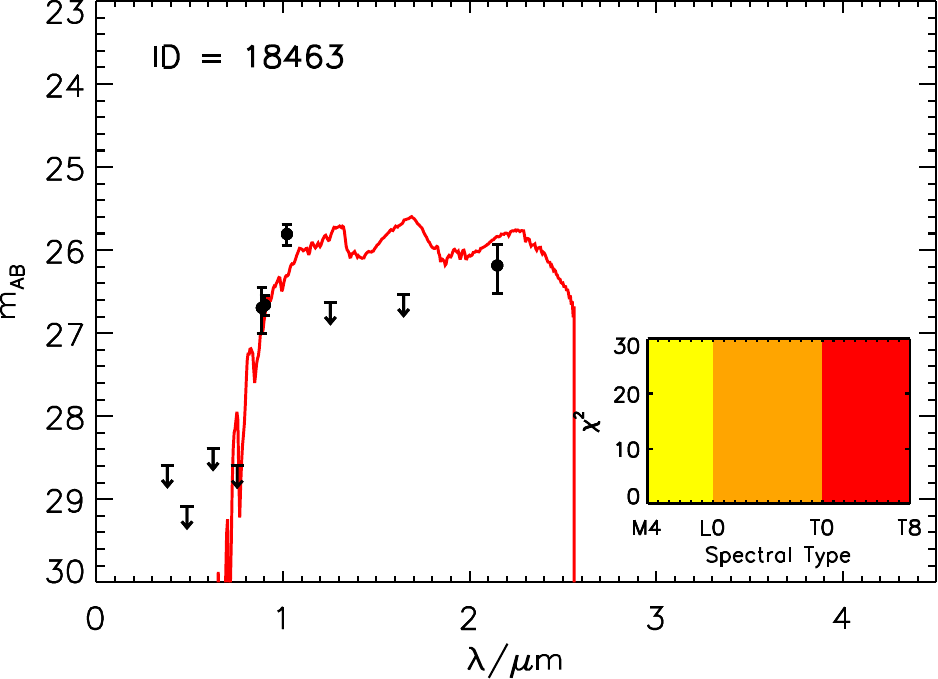} \hspace{0.3cm}\\ 
\vspace{0.2mm}
\includegraphics[width = 0.265\textwidth, trim = 0cm -0.5cm 0cm 0.0cm, clip = true]{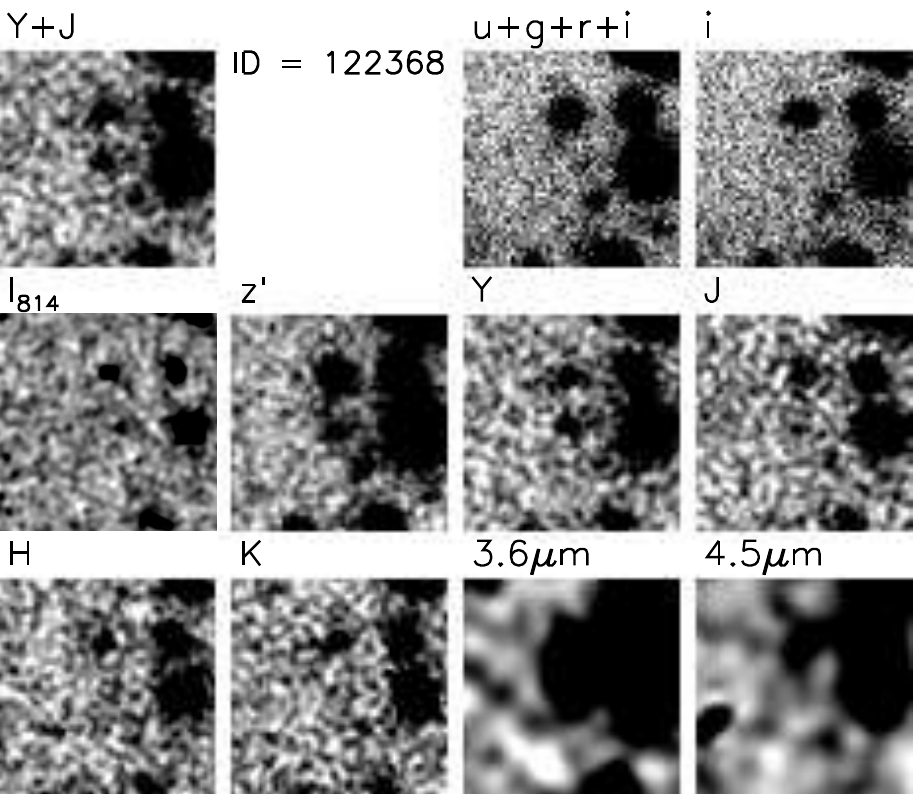} \hspace{0.3cm}
\includegraphics[width = 0.31\textwidth]{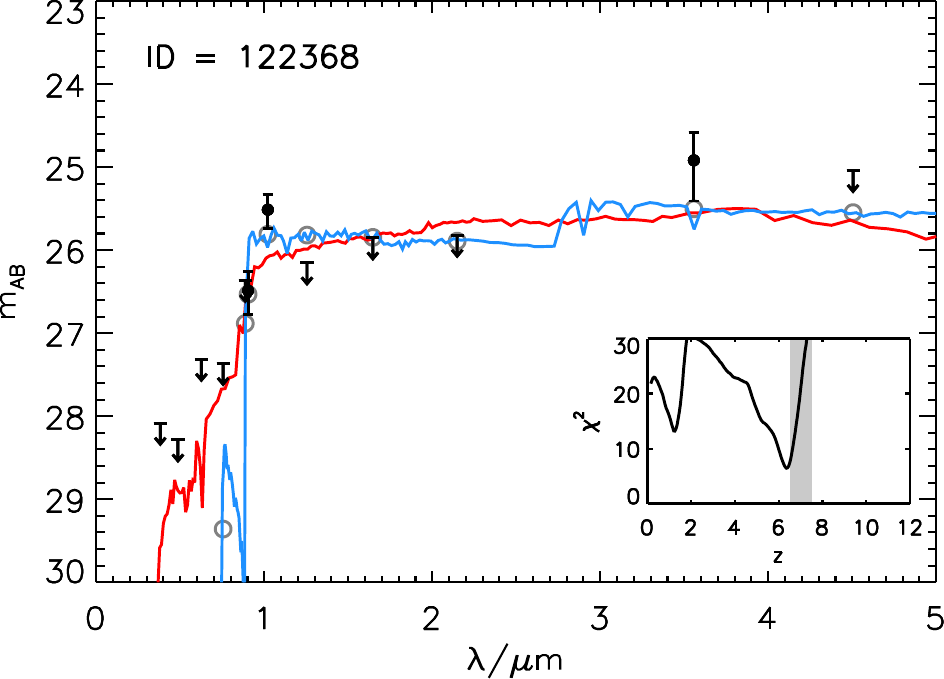} \hspace{0.3cm}
\includegraphics[width = 0.31\textwidth]{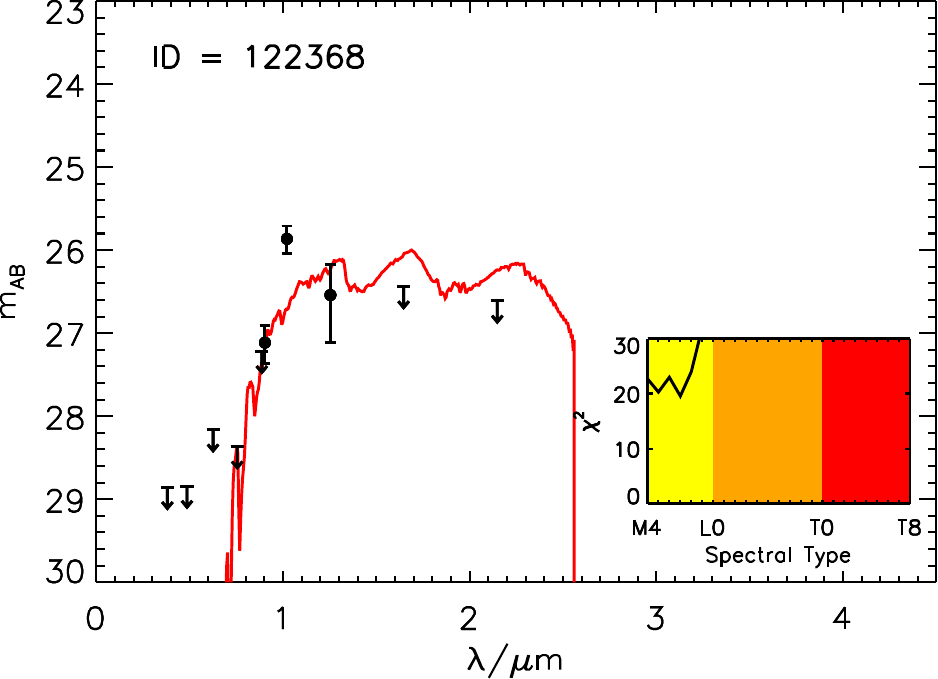} \hspace{0.3cm}\\ 
\vspace{0.2mm}
\includegraphics[width = 0.265\textwidth, trim = 0cm -0.5cm 0cm 0.0cm, clip = true]{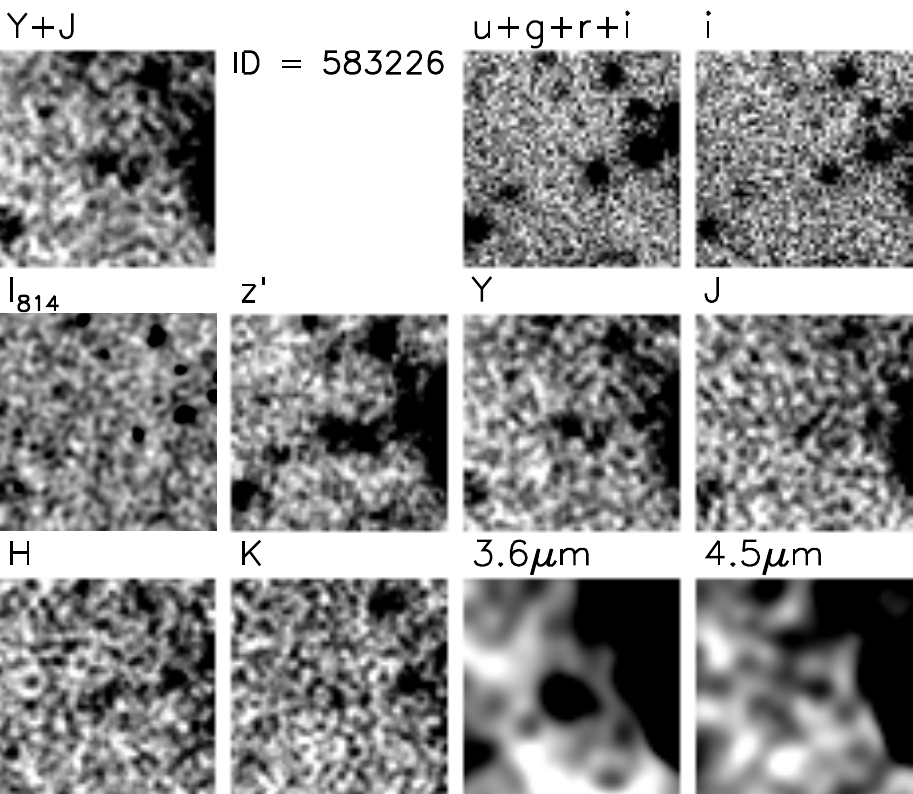} \hspace{0.3cm}
\includegraphics[width = 0.31\textwidth]{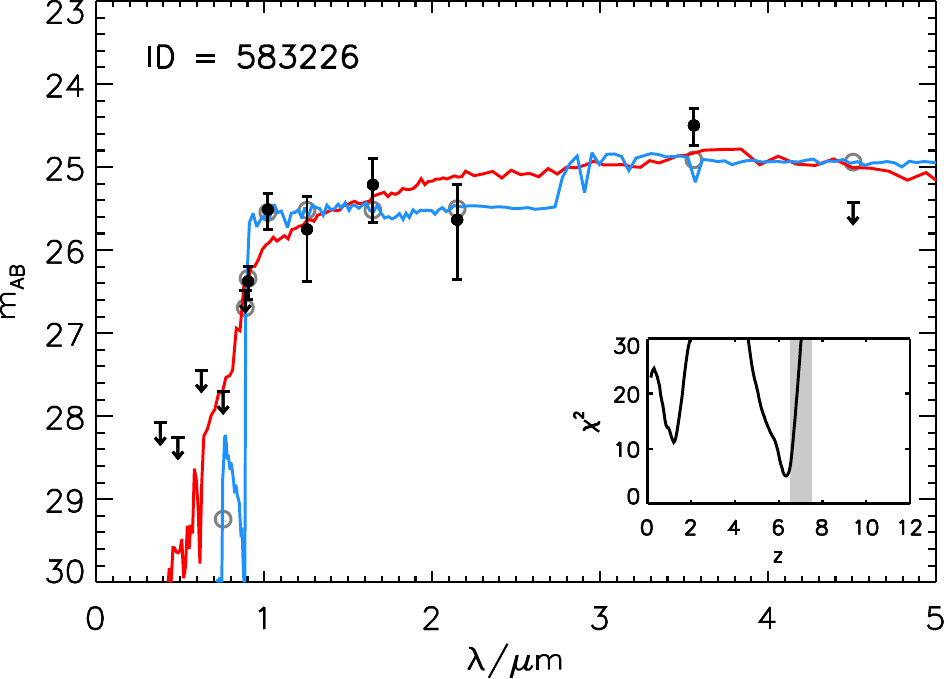} \hspace{0.3cm}
\includegraphics[width = 0.31\textwidth]{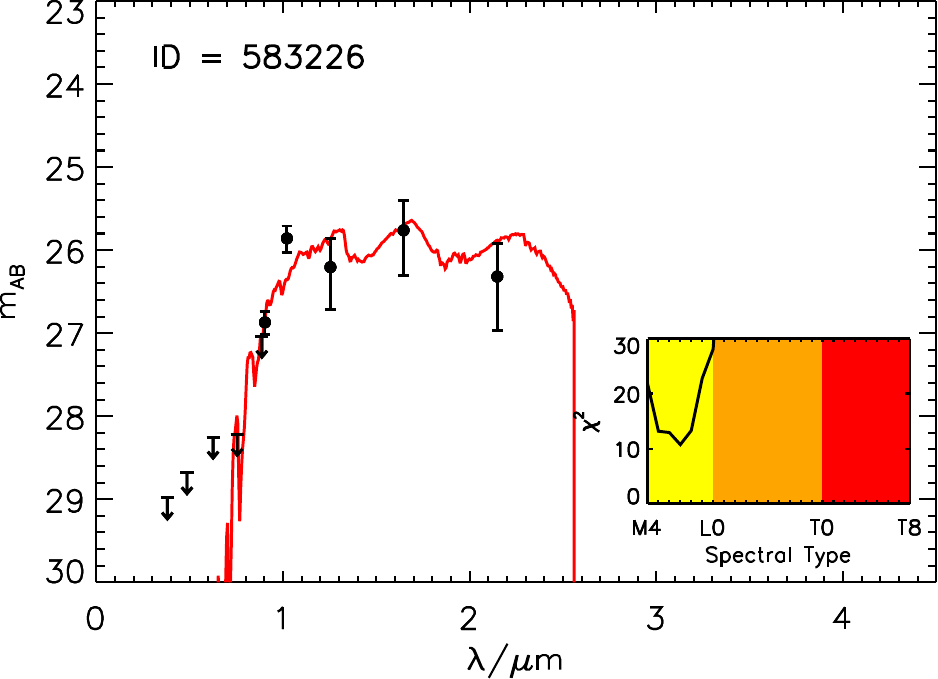} \hspace{0.3cm}\\ 
\contcaption{}
\end{figure*}

\begin{figure*}

\includegraphics[width = 0.265\textwidth, trim = 0cm -0.5cm 0cm 0.0cm, clip = true]{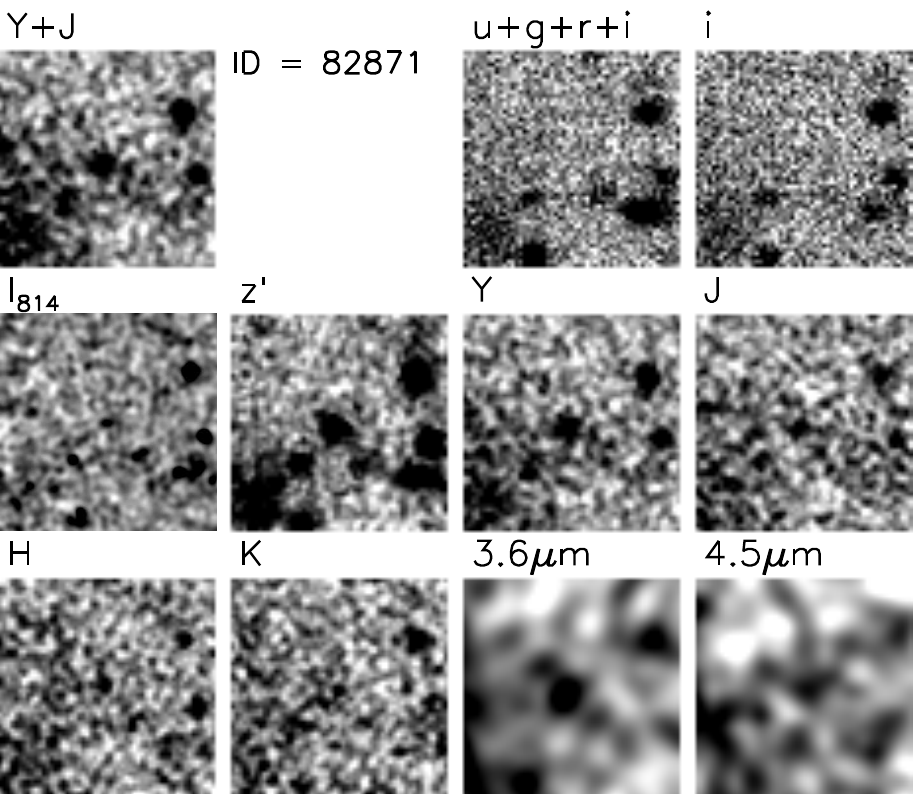} \hspace{0.3cm}
\includegraphics[width = 0.31\textwidth]{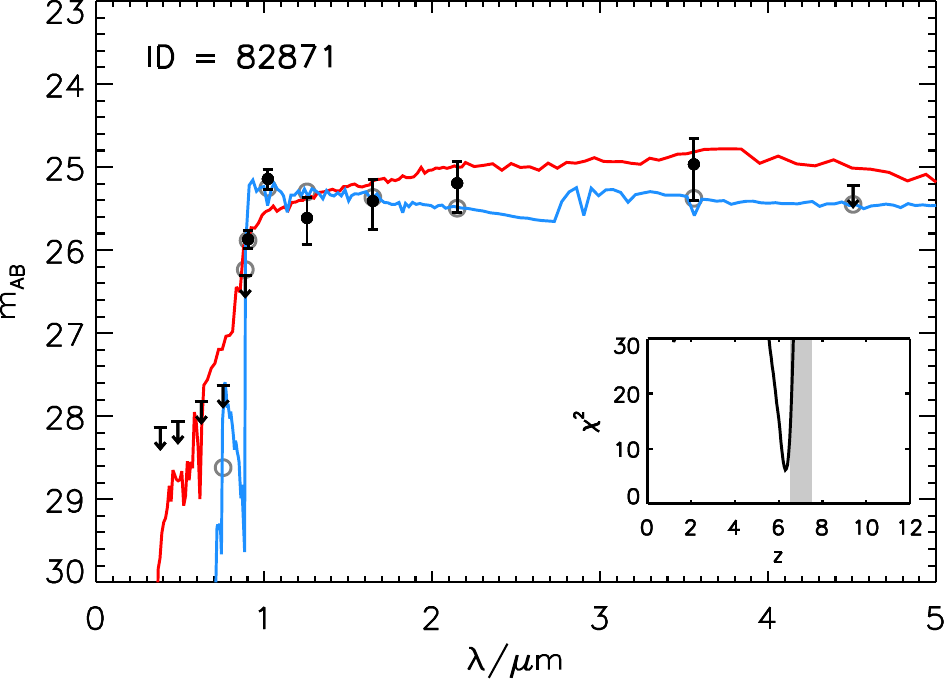} \hspace{0.3cm}
\includegraphics[width = 0.31\textwidth]{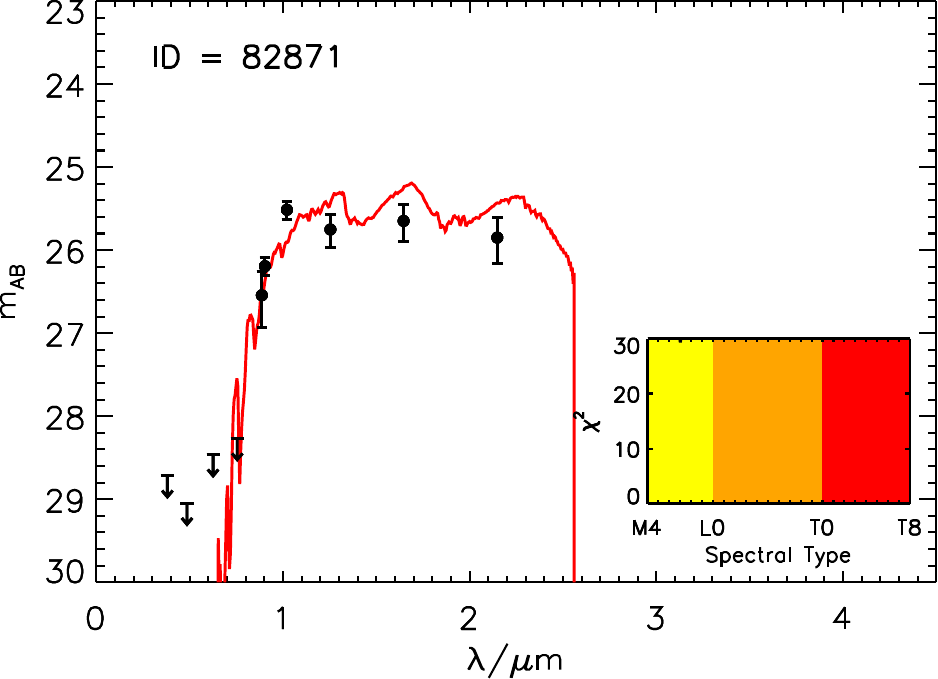} \hspace{0.3cm}\\ 
\vspace{0.2mm}
\includegraphics[width = 0.265\textwidth, trim = 0cm -0.5cm 0cm 0.0cm, clip = true]{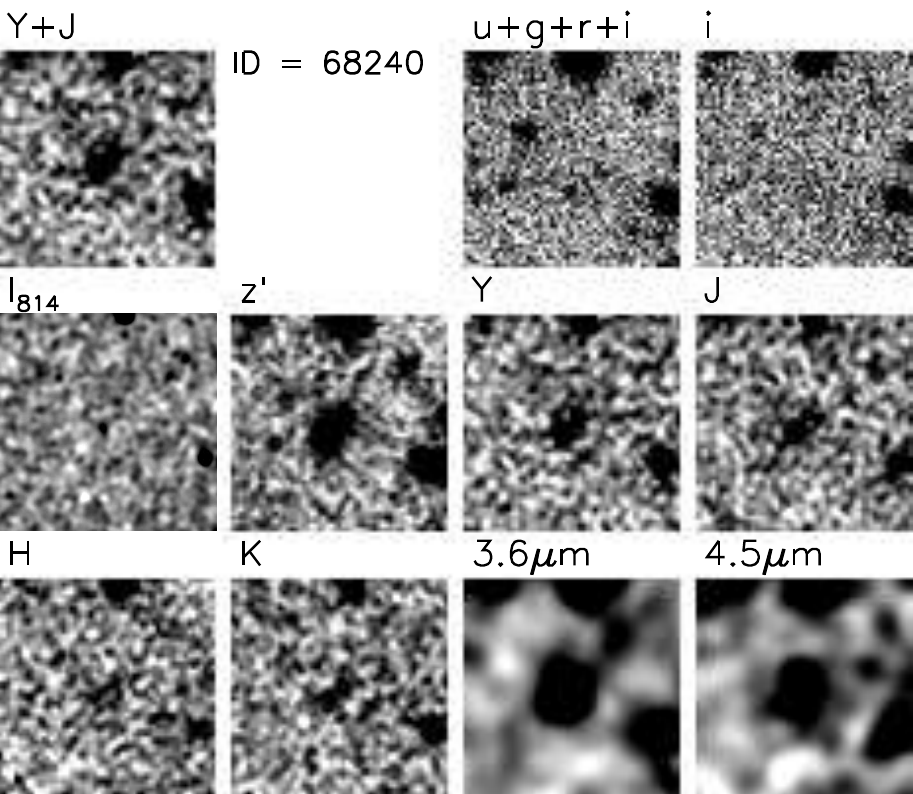} \hspace{0.3cm}
\includegraphics[width = 0.31\textwidth]{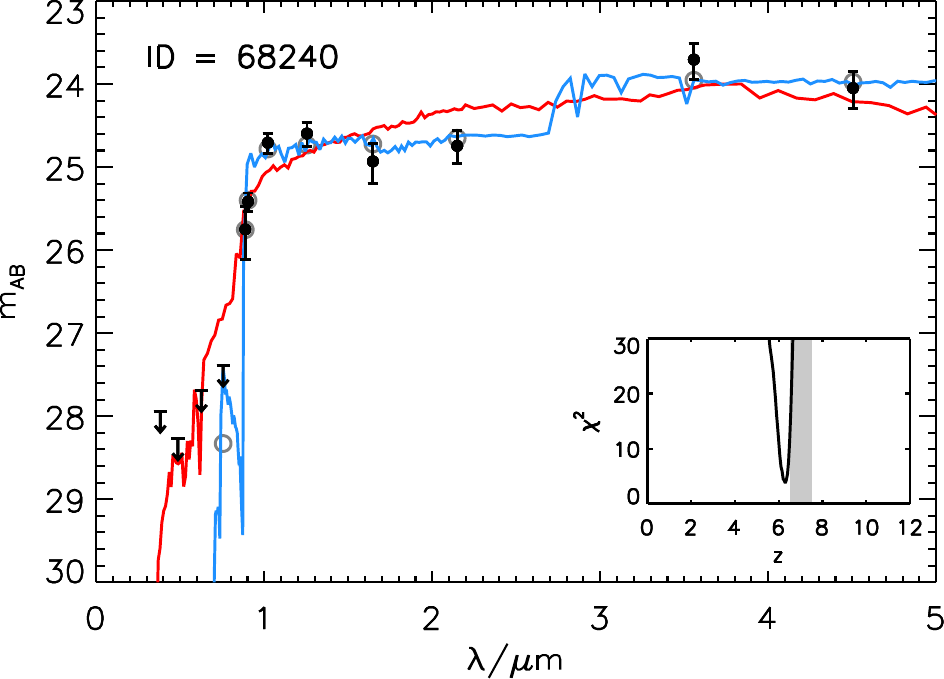} \hspace{0.3cm}
\includegraphics[width = 0.31\textwidth]{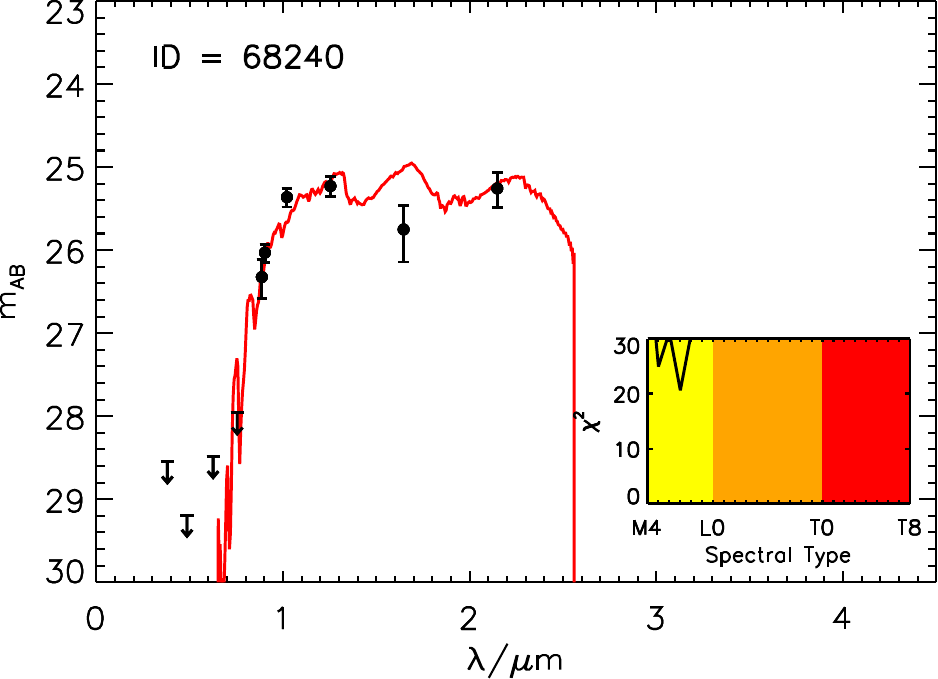} \hspace{0.3cm}\\ 
\vspace{0.2mm}
\includegraphics[width = 0.265\textwidth, trim = 0cm -0.5cm 0cm 0.0cm, clip = true]{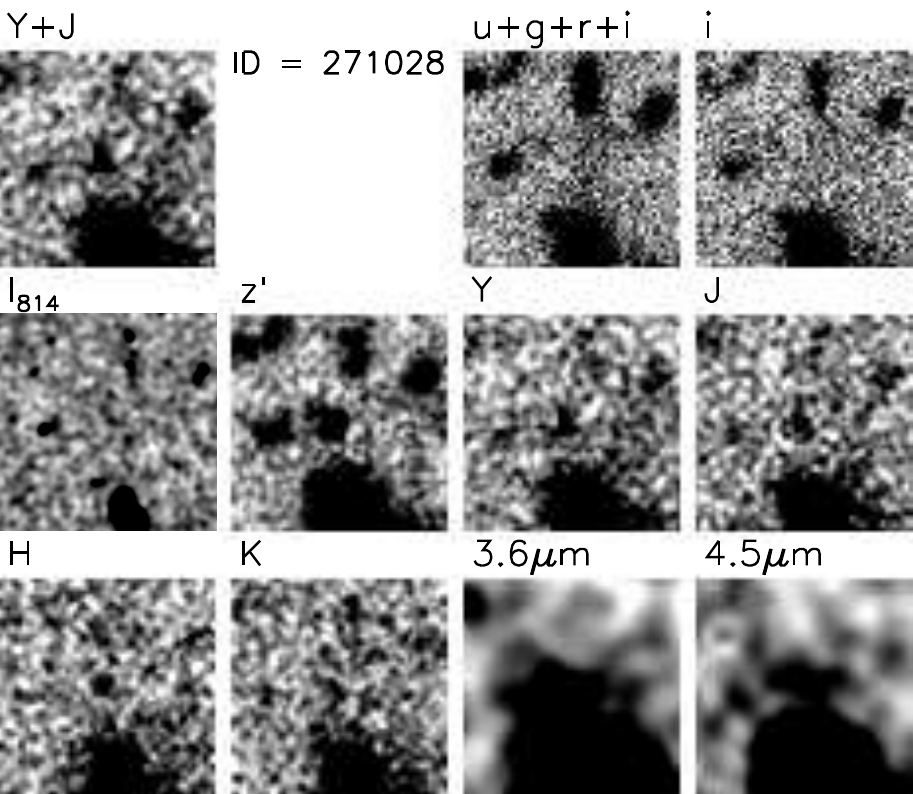} \hspace{0.3cm}
\includegraphics[width = 0.31\textwidth]{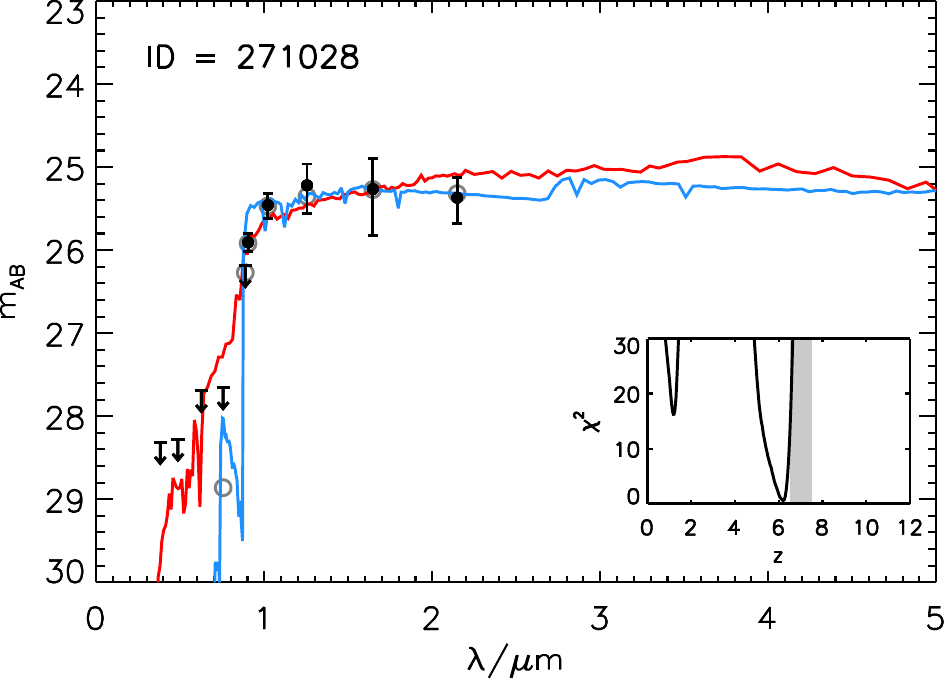} \hspace{0.3cm}
\includegraphics[width = 0.31\textwidth]{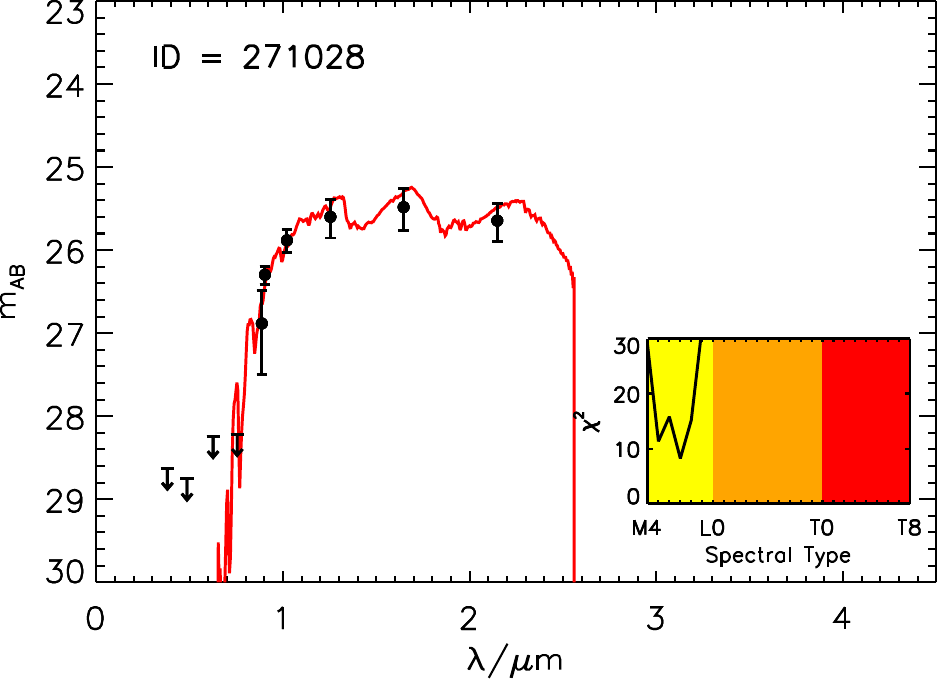} \hspace{0.3cm}\\ 
\vspace{0.2mm}
\includegraphics[width = 0.265\textwidth, trim = 0cm -0.5cm 0cm 0.0cm, clip = true]{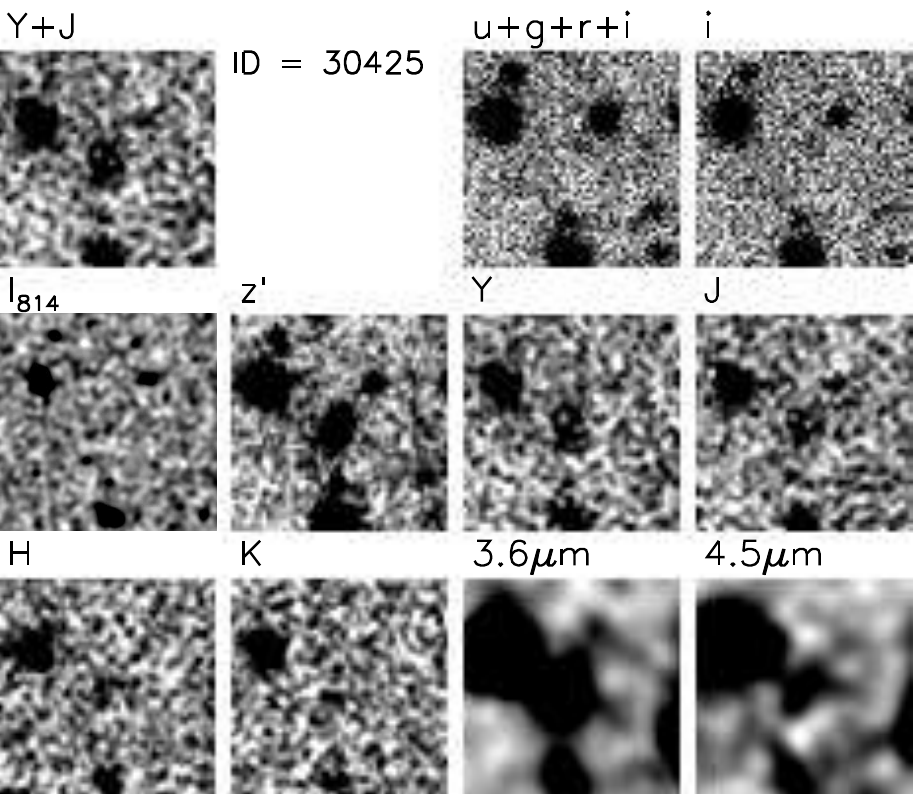} \hspace{0.3cm}
\includegraphics[width = 0.31\textwidth]{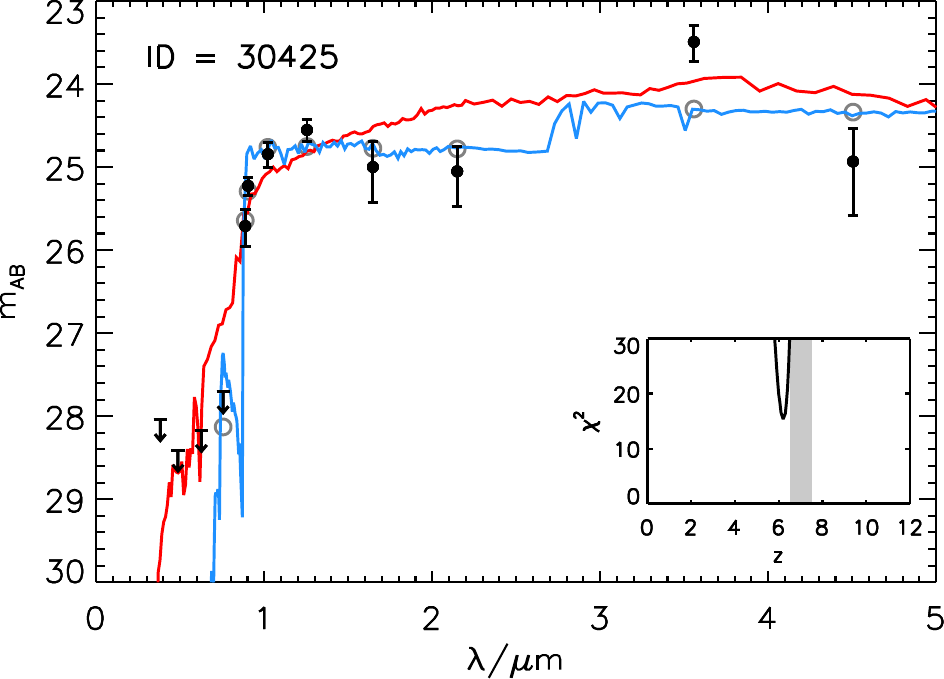} \hspace{0.3cm}
\includegraphics[width = 0.31\textwidth]{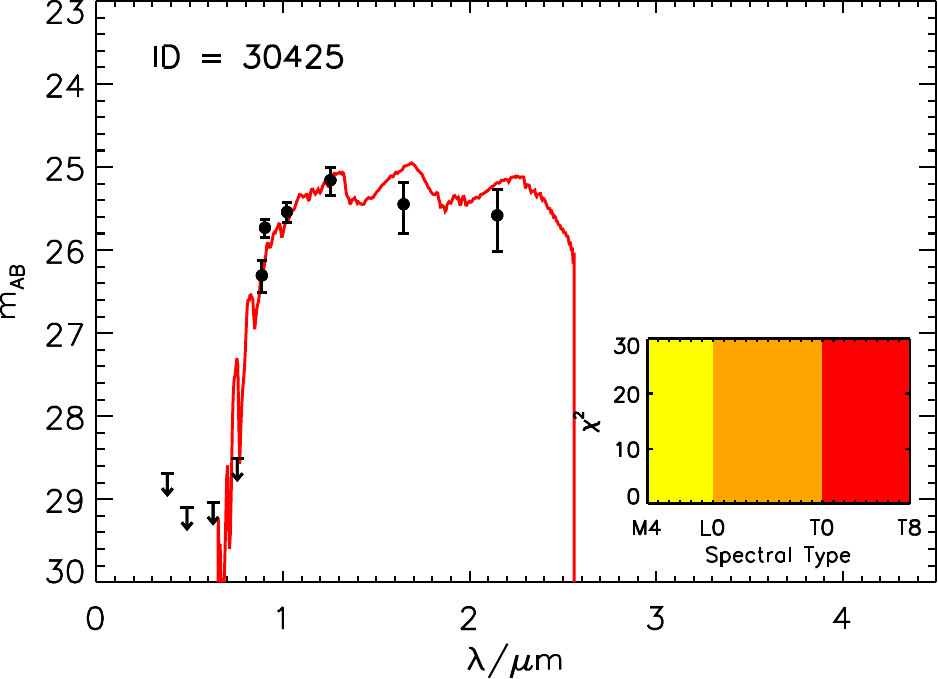} \hspace{0.3cm}\\ 
\vspace{0.2mm}
\includegraphics[width = 0.265\textwidth, trim = 0cm -0.5cm 0cm 0.0cm, clip = true]{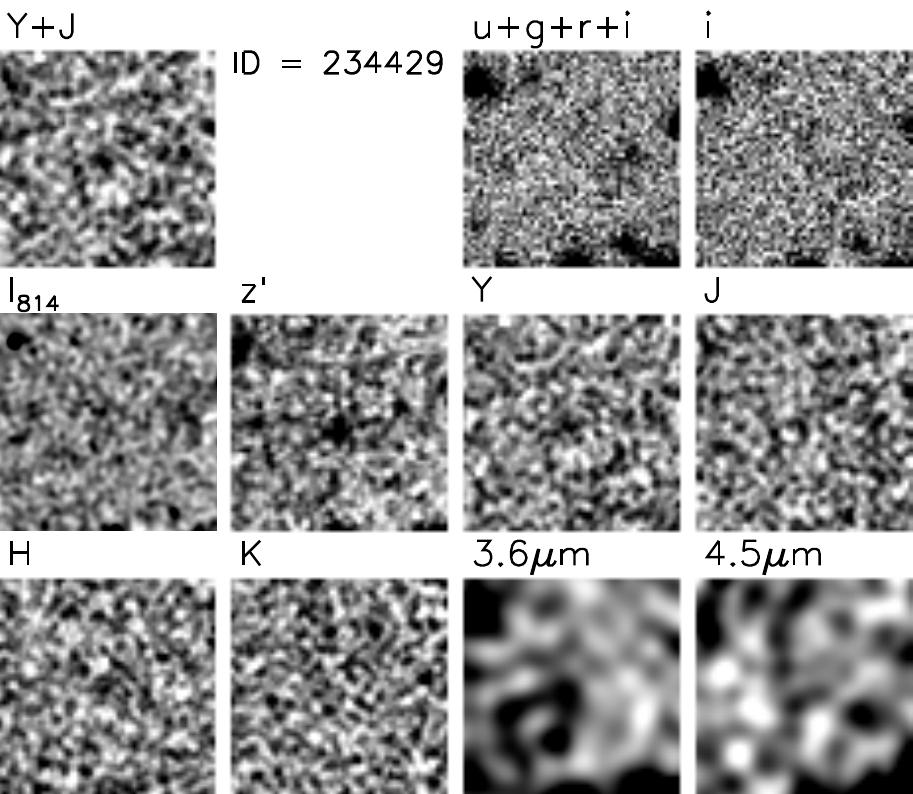} \hspace{0.3cm}
\includegraphics[width = 0.31\textwidth]{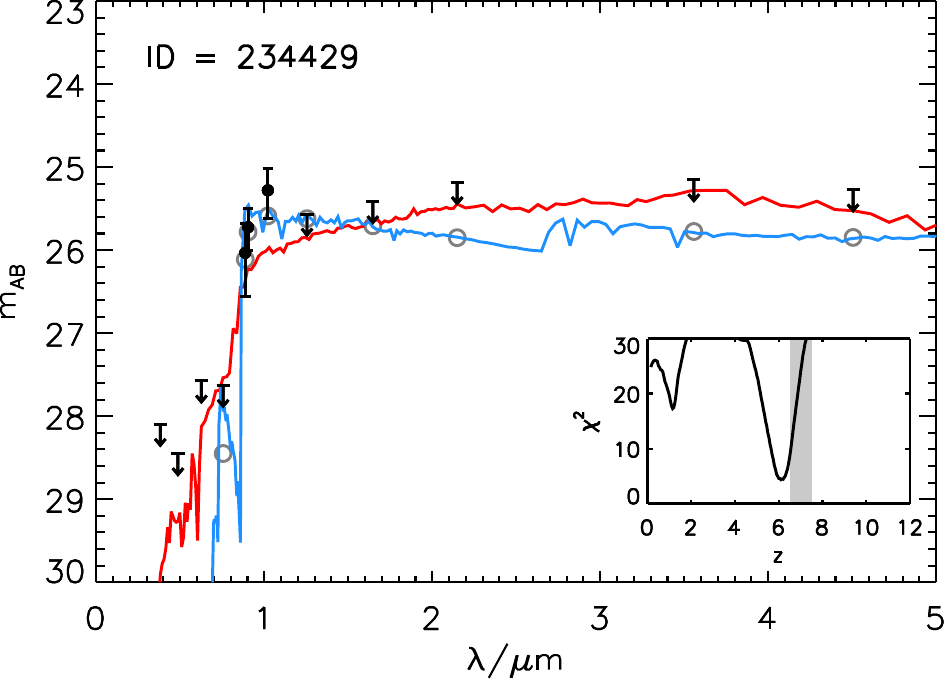} \hspace{0.3cm}
\includegraphics[width = 0.31\textwidth]{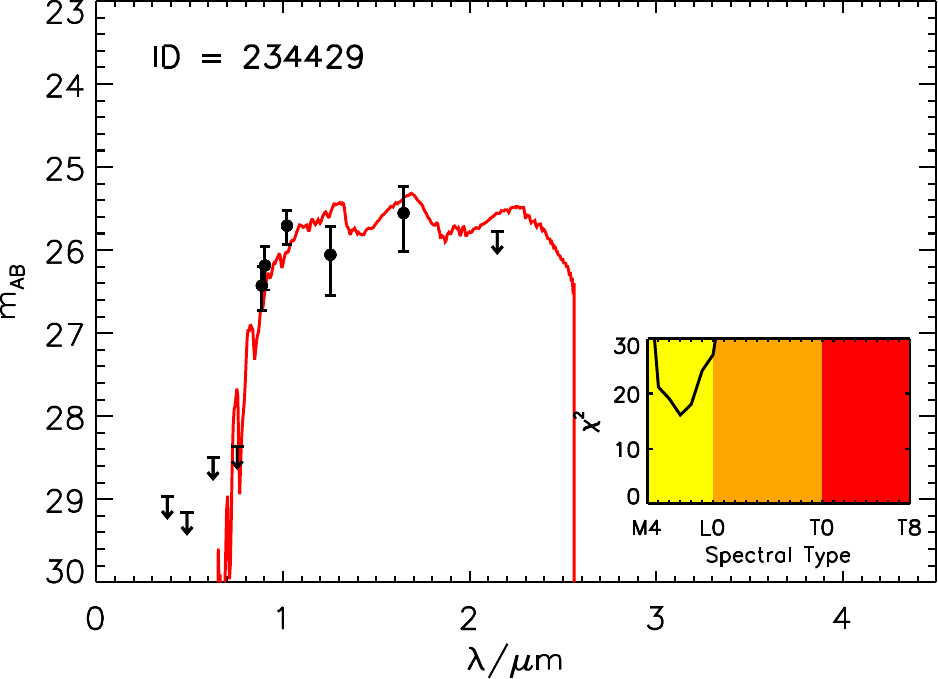} \hspace{0.3cm}\\ 
\vspace{0.2mm}

\contcaption{}
\end{figure*}

\begin{figure*}

\includegraphics[width = 0.265\textwidth, trim = 0cm -0.5cm 0cm 0.0cm, clip = true]{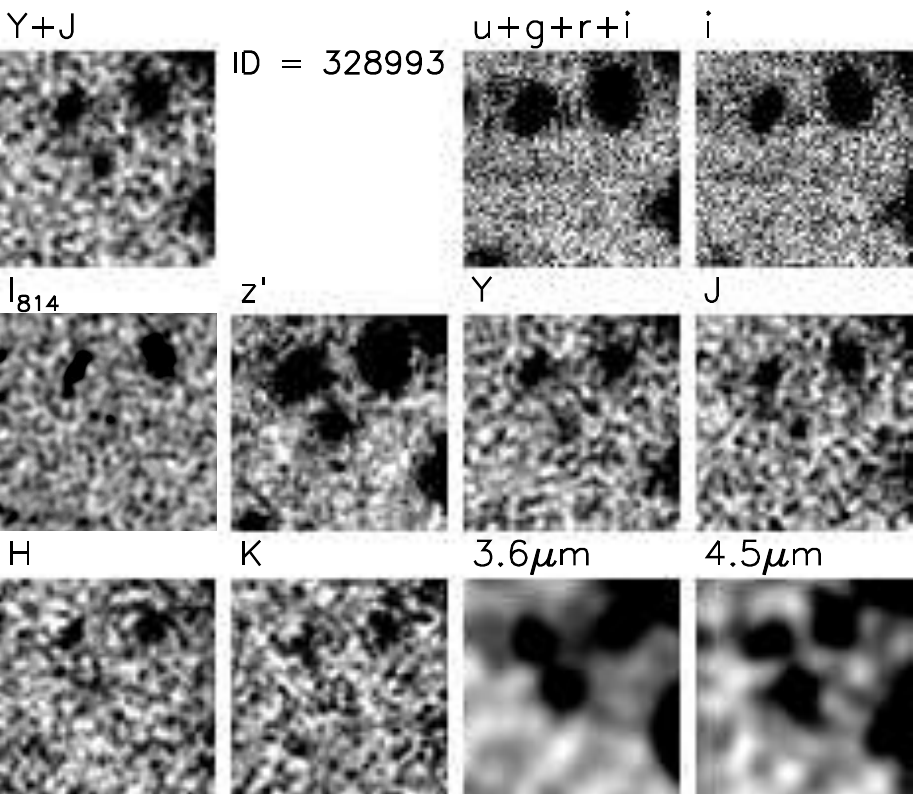} \hspace{0.3cm}
\includegraphics[width = 0.31\textwidth]{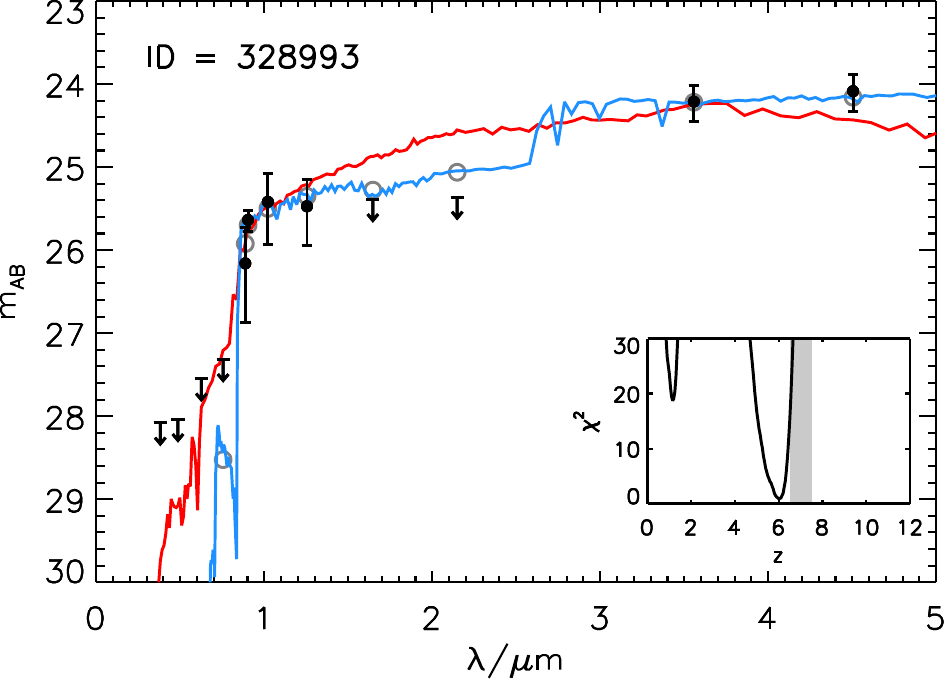} \hspace{0.3cm}
\includegraphics[width = 0.31\textwidth]{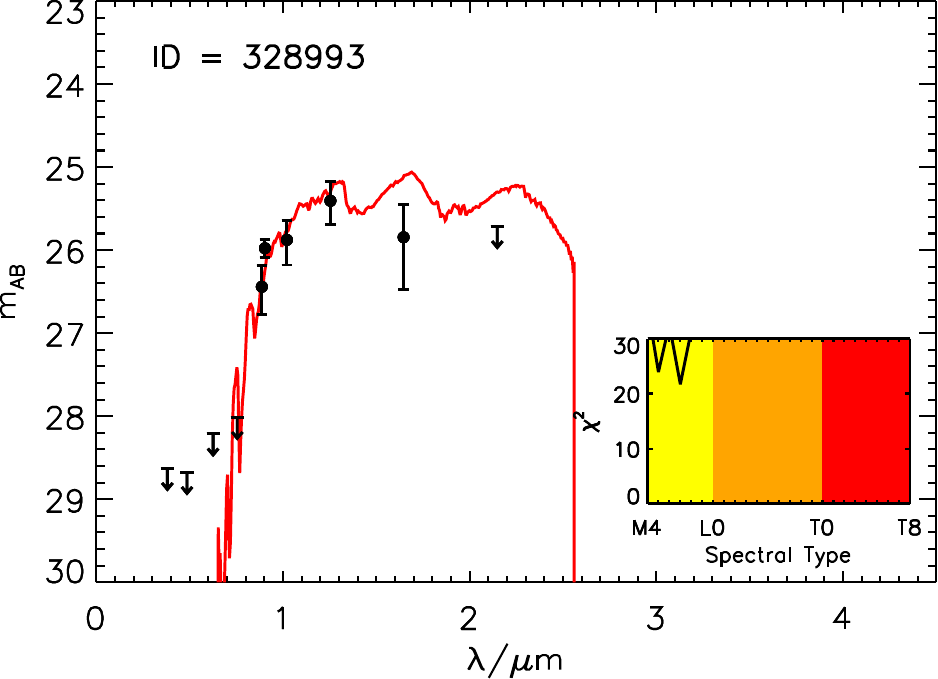} \\

\contcaption{}
\end{figure*}

\nopagebreak

%% For the UDS
\begin{figure*}

\includegraphics[width = 0.265\textwidth, trim = 0cm -0.5cm 0cm 0.0cm, clip = true]{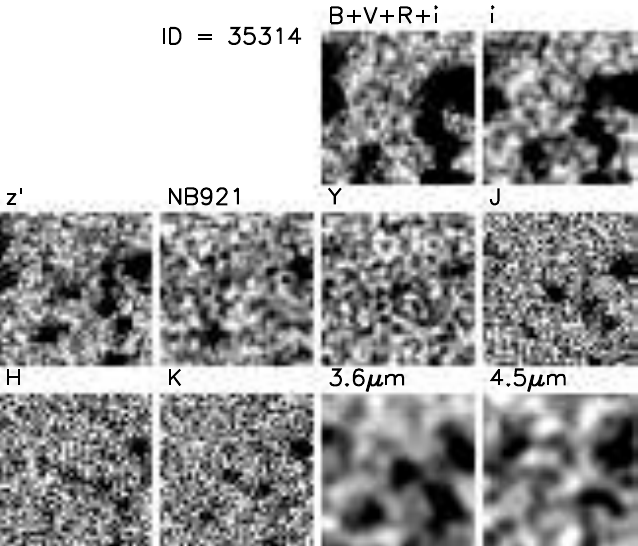}\hspace{0.3cm}
\subfloat{\includegraphics[width = 0.3\textwidth]{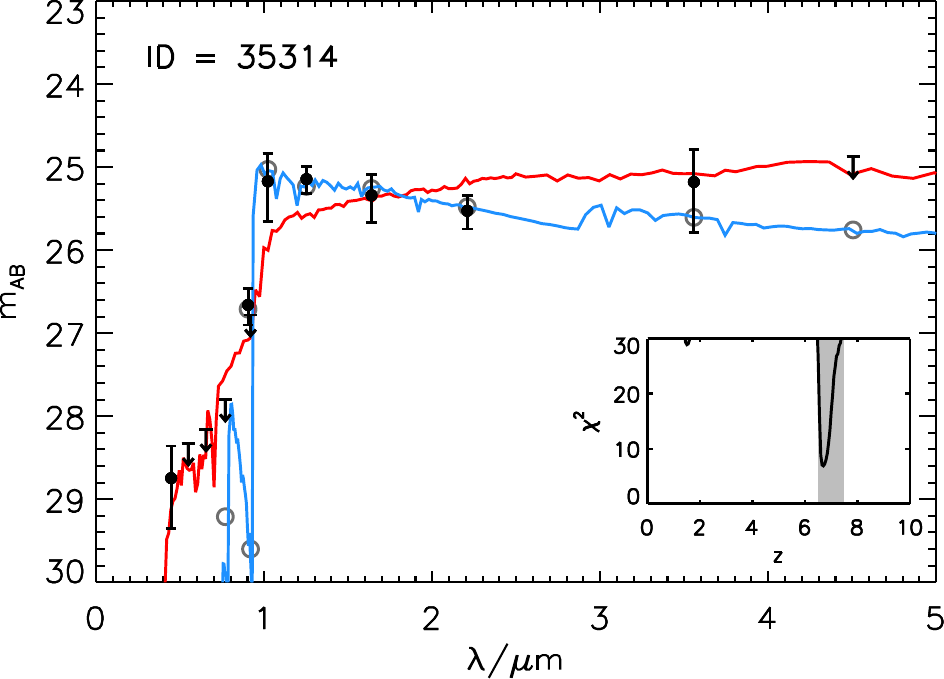}} \hspace{0.3cm}
\subfloat{\includegraphics[width = 0.3\textwidth]{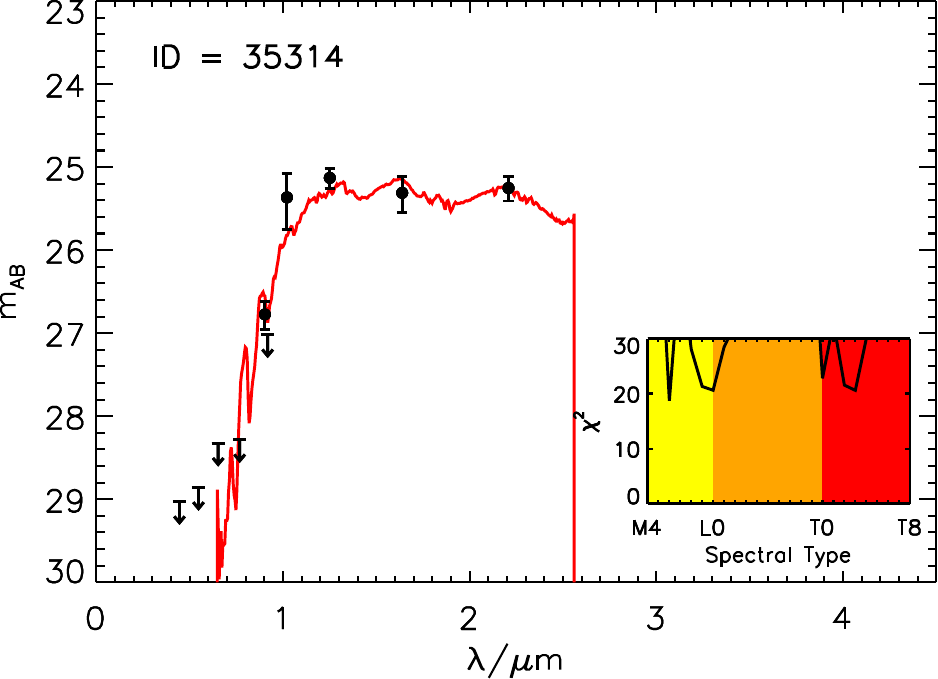}} \\ 
\vspace{0.2cm}
\includegraphics[width = 0.265\textwidth, trim = 0cm -0.5cm 0cm 0.0cm, clip = true]{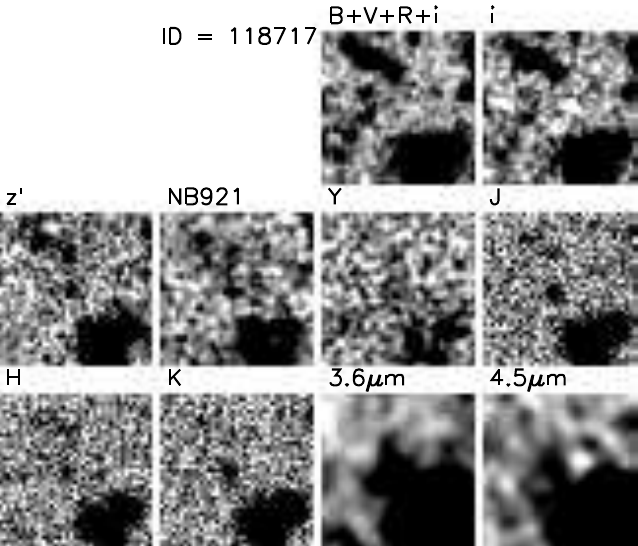}\hspace{0.3cm}
\subfloat{\includegraphics[width = 0.3\textwidth]{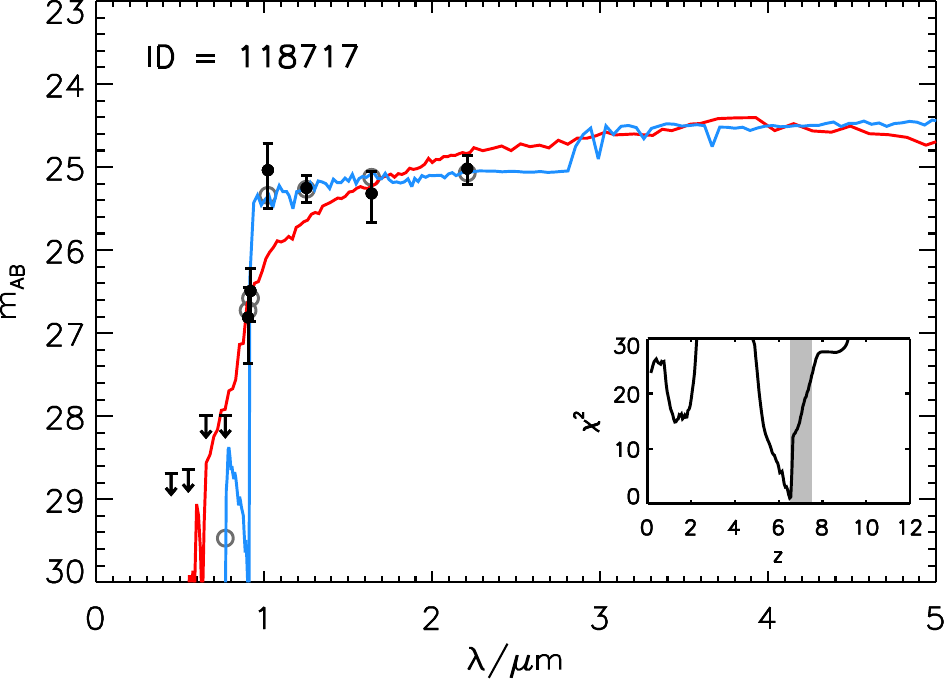}} \hspace{0.3cm}
\subfloat{\includegraphics[width = 0.3\textwidth]{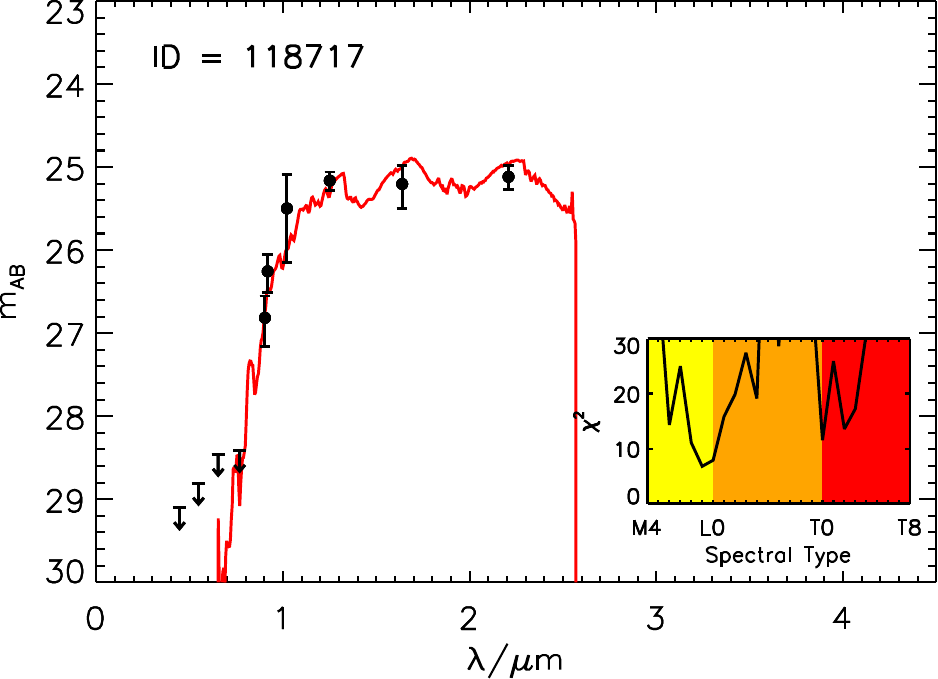}} \\ 
\vspace{0.2cm}
\includegraphics[width = 0.265\textwidth, trim = 0cm -0.5cm 0cm 0.0cm, clip = true]{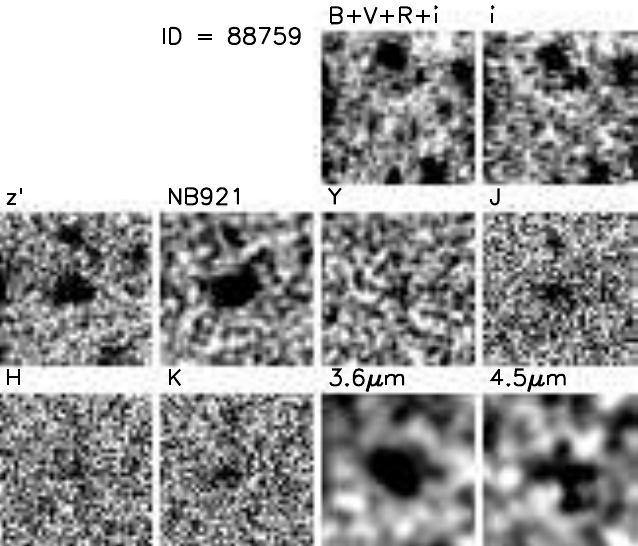}\hspace{0.3cm}
\subfloat{\includegraphics[width = 0.3\textwidth]{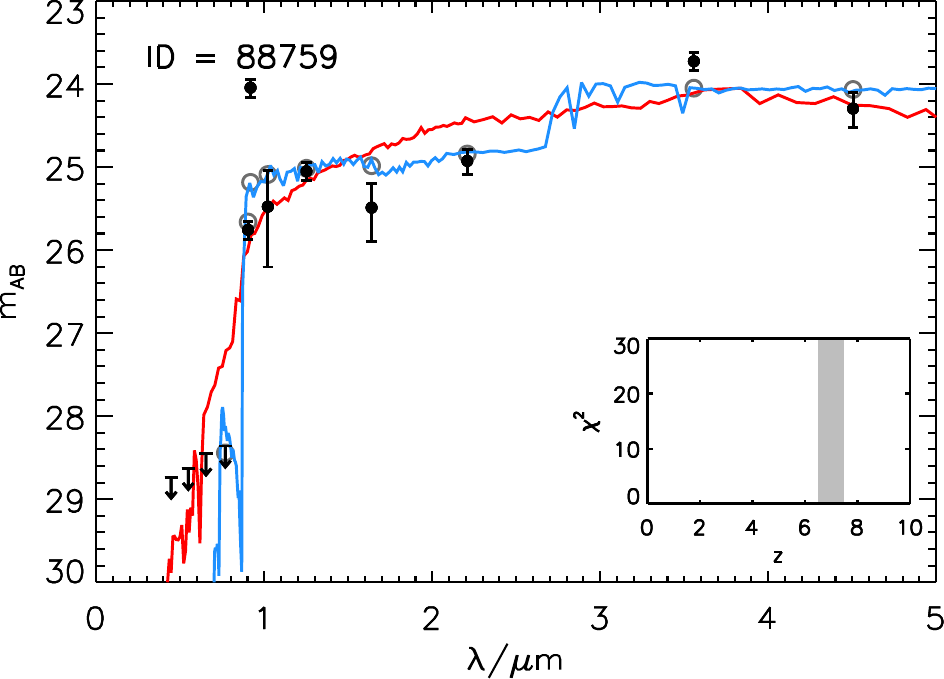}} \hspace{0.3cm}
\subfloat{\includegraphics[width = 0.3\textwidth]{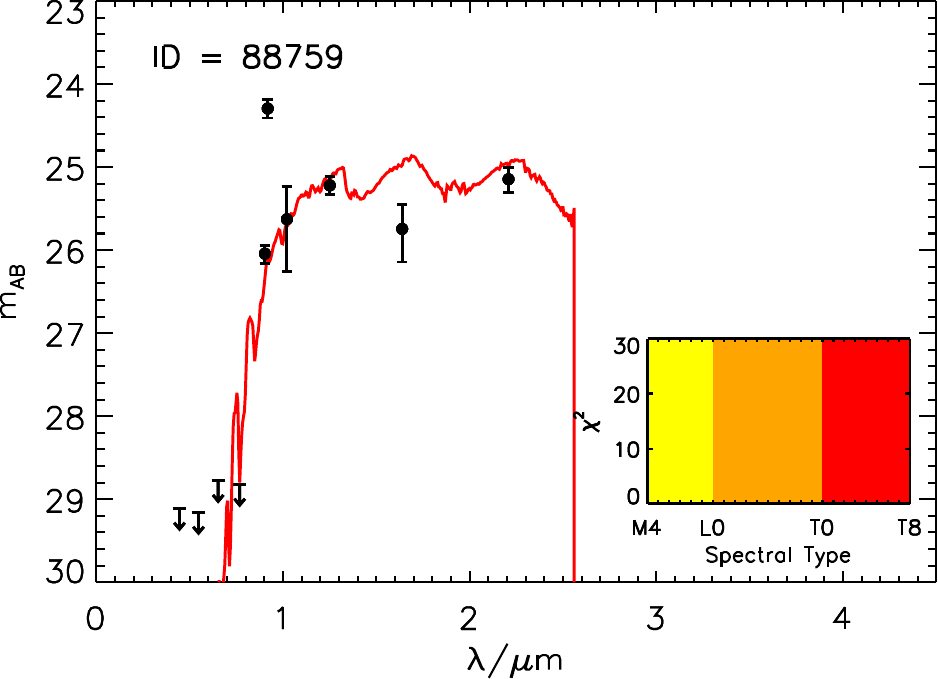}} \\ 
\vspace{0.2cm}

\includegraphics[width = 0.265\textwidth, trim = 0cm -0.5cm 0cm 0.0cm, clip = true]{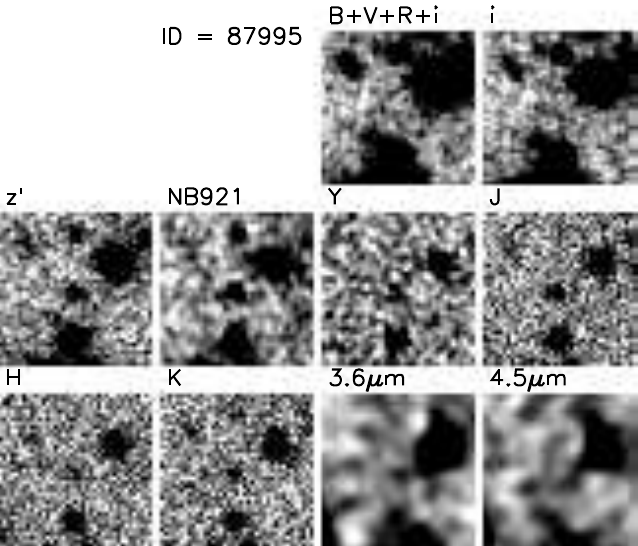}\hspace{0.3cm}
\subfloat{\includegraphics[width = 0.3\textwidth]{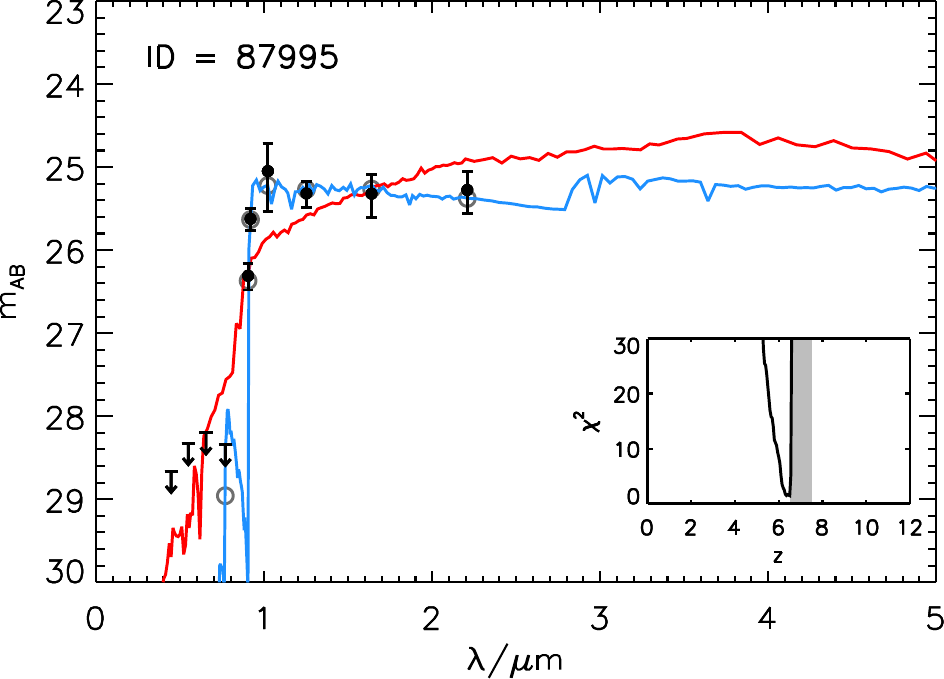}} \hspace{0.3cm}
\subfloat{\includegraphics[width = 0.3\textwidth]{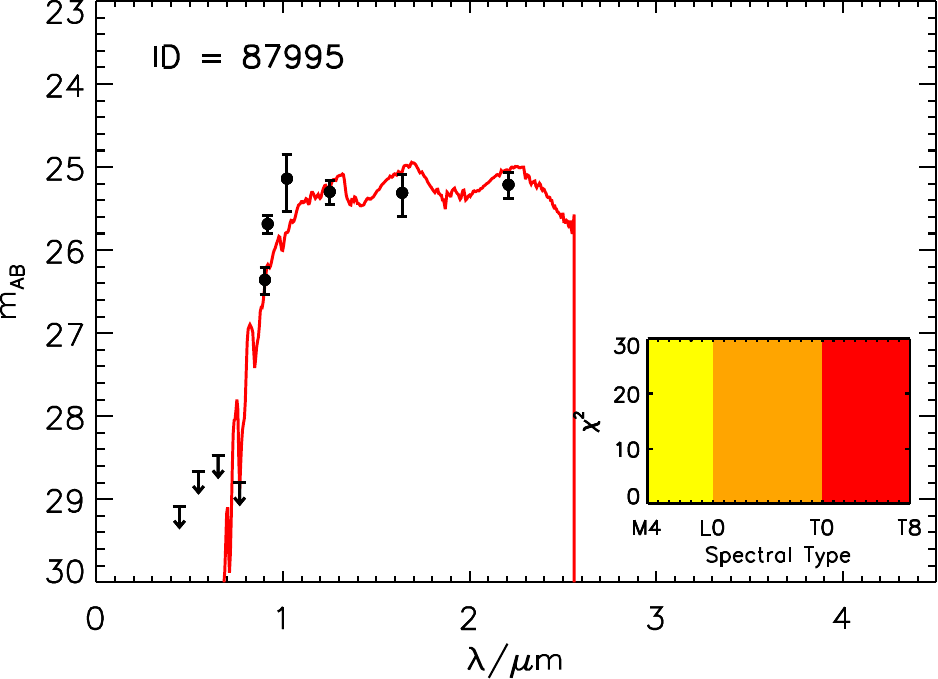}} \\ 
\vspace{-0.2cm}

\caption{Postage-stamp images and galaxy and star SED fits to the four galaxies from the UDS field.
See the caption of Fig.~\ref{figure:seds} for details.
}
\label{figure:udsseds}
\end{figure*} 

%\clearpage

\section{Bowler et al. 2012 improved photometry and SED plots}

\begin{figure*}
\includegraphics[width = 0.265\textwidth, trim = 0cm -0.5cm 0cm 0.0cm, clip = true]{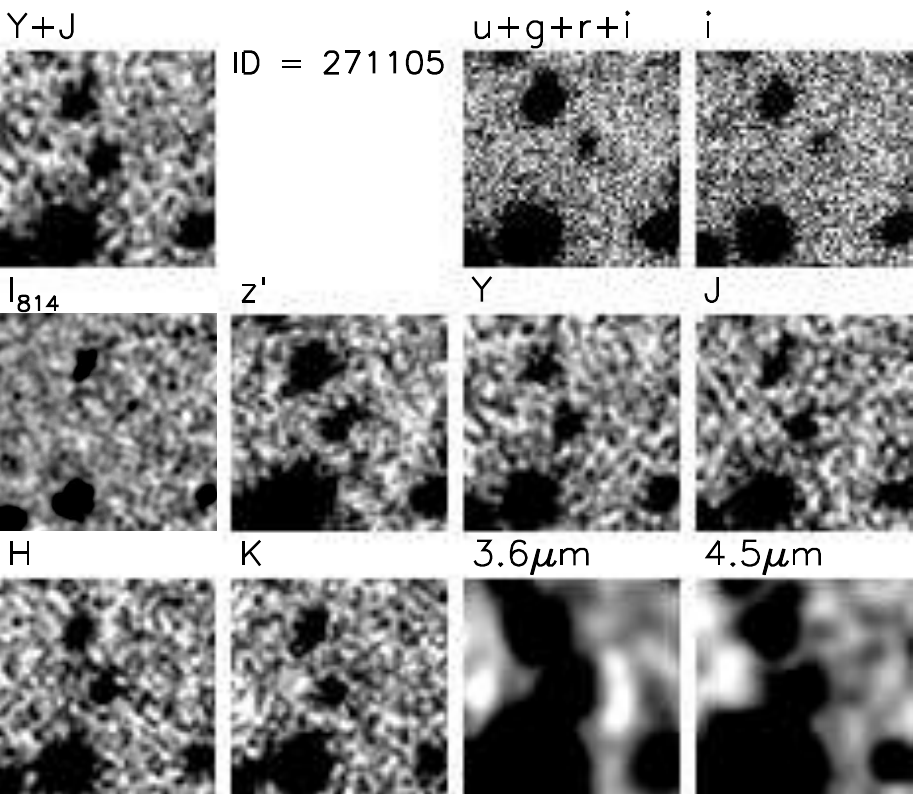} \hspace{0.3cm}
\subfloat{\includegraphics[width=0.31\textwidth]{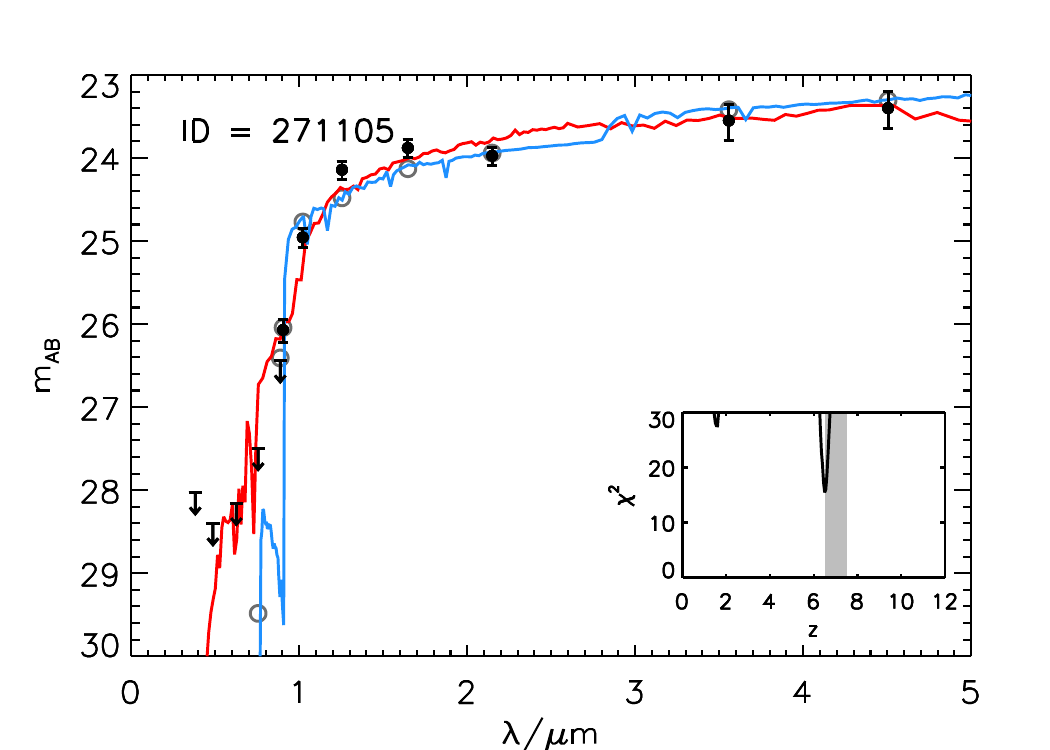}}\hspace{0.4cm}
\subfloat{\includegraphics[width=0.31\textwidth]{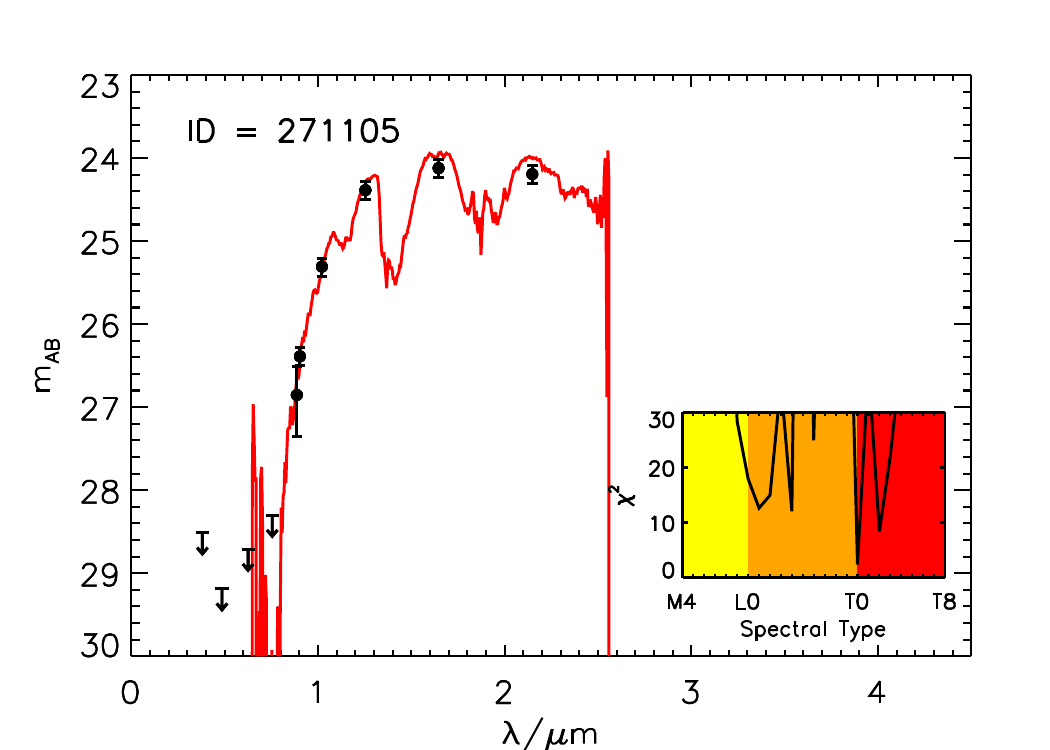}} \\

\includegraphics[width = 0.265\textwidth, trim = 0cm -0.5cm 0cm 0.0cm, clip = true]{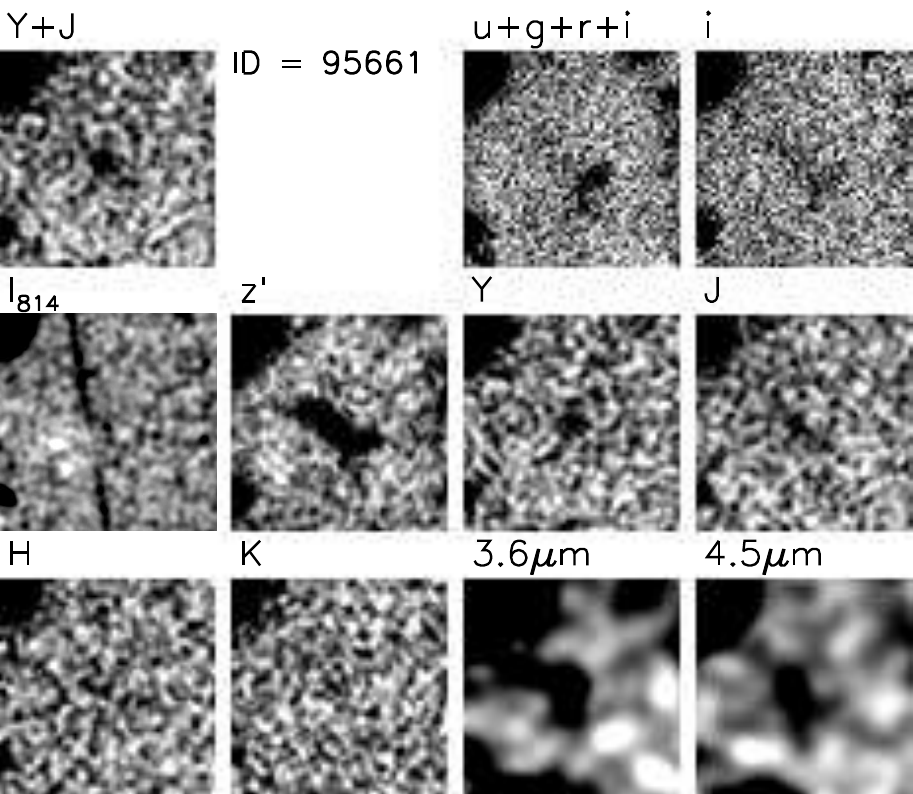} \hspace{0.3cm}
\subfloat{\includegraphics[width=0.31\textwidth]{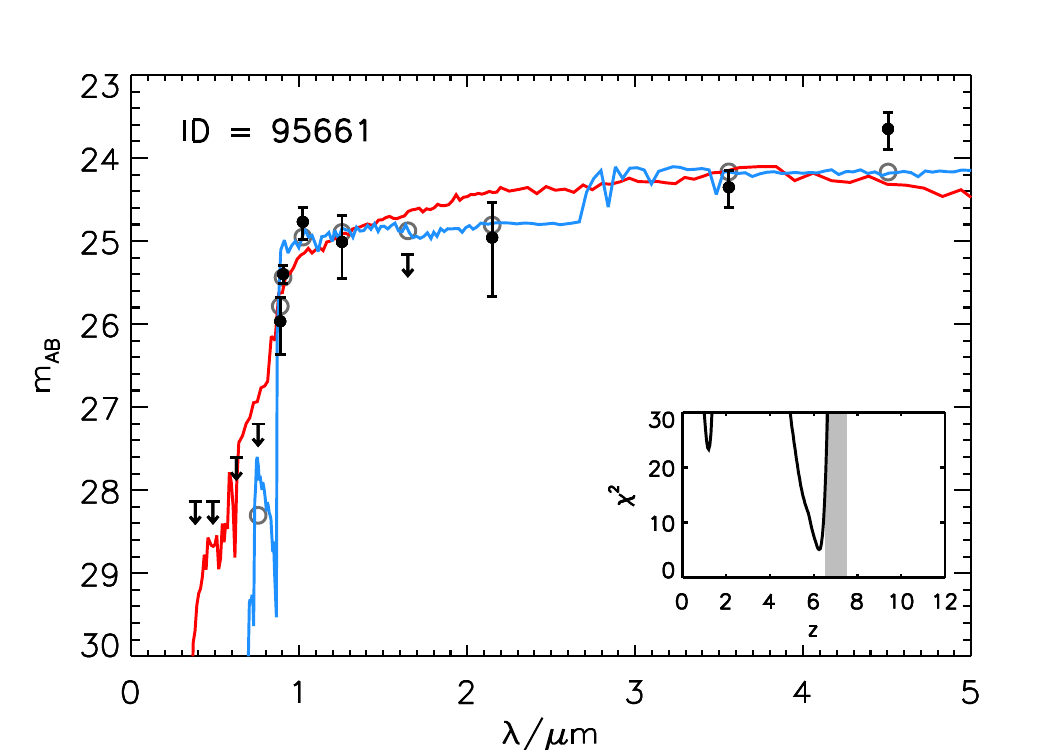}}\hspace{0.4cm}
\subfloat{\includegraphics[width=0.31\textwidth]{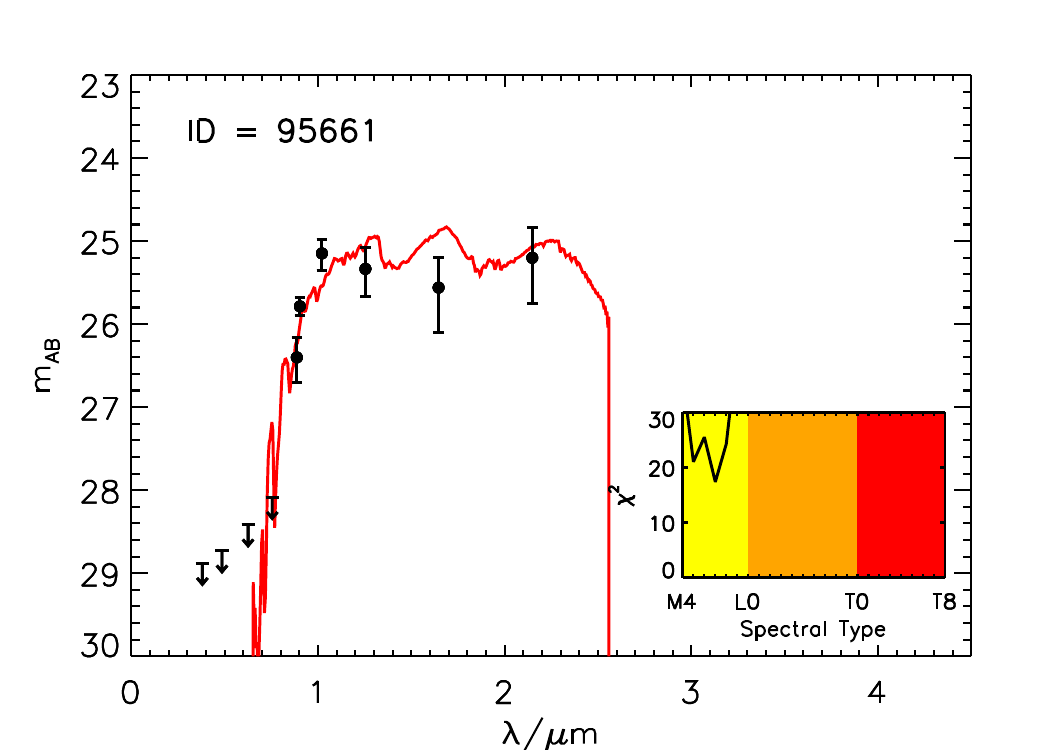}} \\

\includegraphics[width = 0.265\textwidth, trim = 0cm -0.5cm 0cm 0.0cm, clip = true]{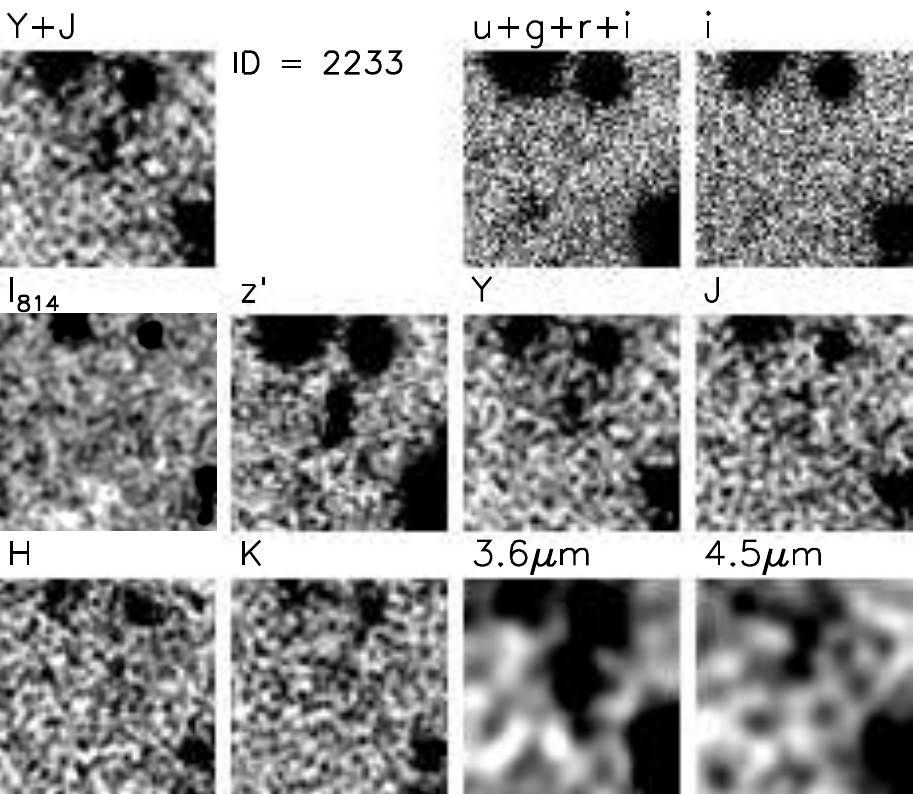} \hspace{0.3cm}
\subfloat{\includegraphics[width=0.31\textwidth]{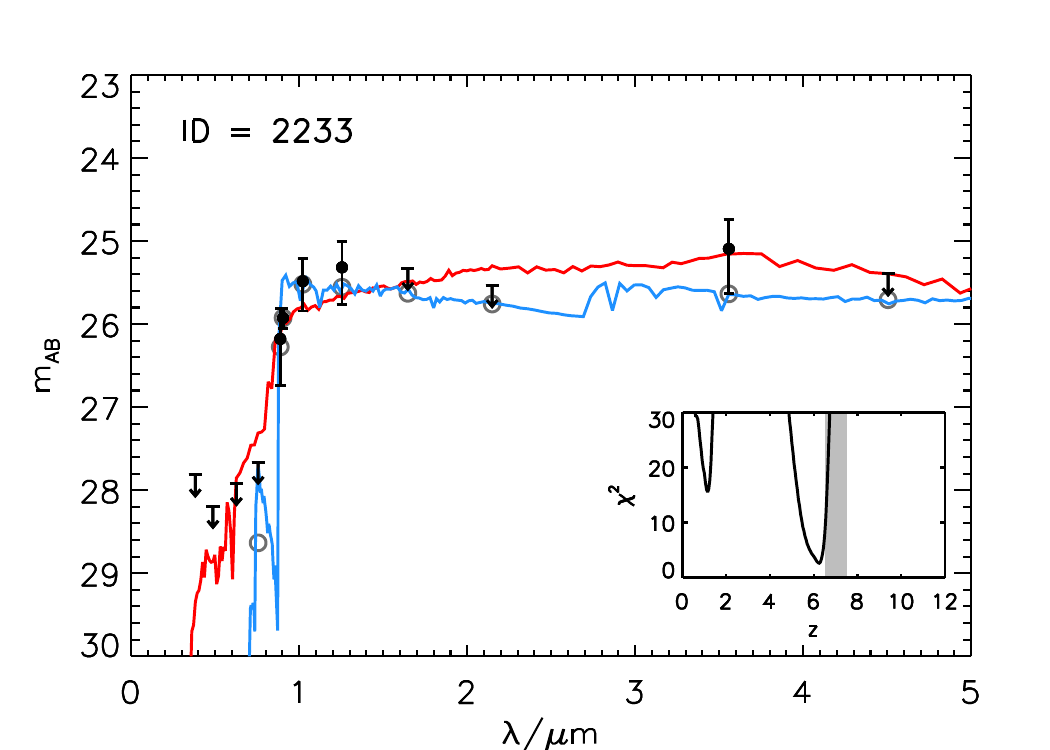}}\hspace{0.4cm}
\subfloat{\includegraphics[width=0.31\textwidth]{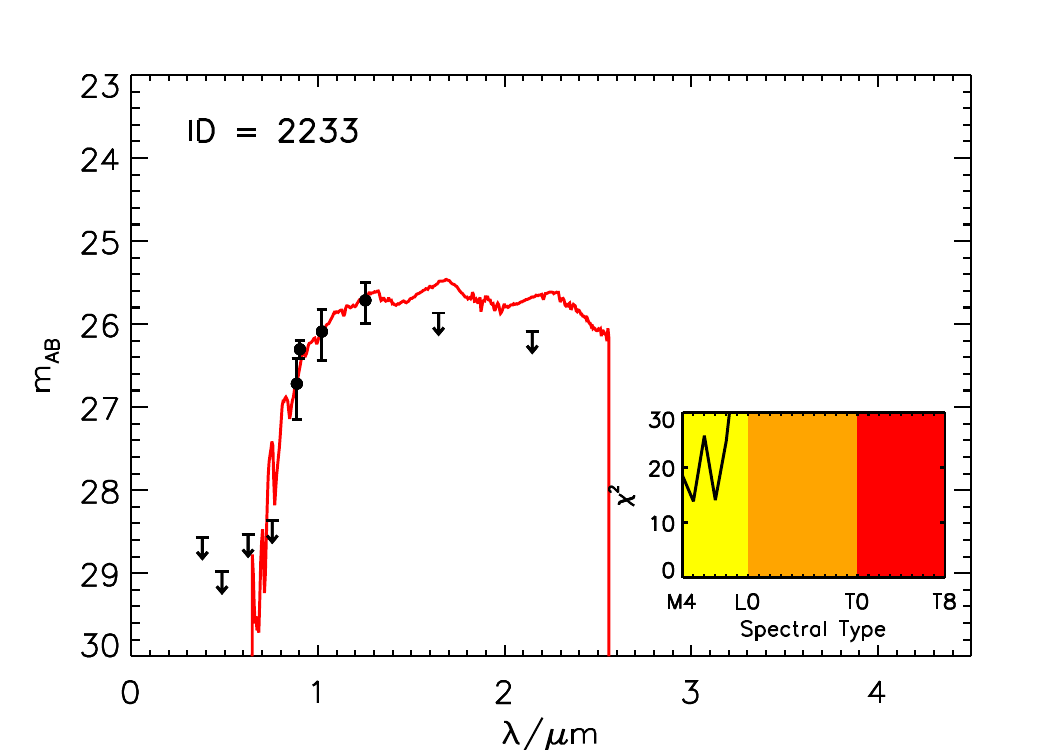}} 

\caption{Postage-stamp images, and galaxy and star SED fits for the three candidates from the~\citet{Bowler2012} sample that are not present in our final sample.
The details of the images and plots are described in the caption for Fig.~\ref{figure:seds}.
With the improved photometry, candidate 271105 is now best-fitted as a type-T0 dwarf star, showing the characteristic hook-like spectrum.
The other candidates now have best-fitting photometric redshifts in the range $6.0 < z < 6.5$.
}
\label{fig:B12sedfits}
\end{figure*}

Here we present improved photometry and SED fitting results for the 10 high-redshift galaxy candidates from~\citet{Bowler2012}, and~\ref{fig:B12sedfits} shows postage-stamp images and SED fits to the three~\citet{Bowler2012} candidates that are not included in our new sample.

\nopagebreak[4]

\begin{table*}
\caption{ The DR2 UltraVISTA magnitudes for the ten galaxy candidates presented in~\citet{Bowler2012}.
The magnitudes were measured in a 1.8-arcsec diameter circular aperture in all cases except the IRAC magnitudes which were measured in a 2.8-arcsec diameter circular aperture.
All magnitudes have been corrected to the 84\% enclosed flux level using appropriate point-source corrections, and the errors presented are determined from the local error method detailed in Section~\ref{imagedepths}.
The signal-to-noise of the detection is presented in brackets after each magnitude.
Where an object was detected at less than $2\sigma$ significance, the magnitude is replaced with the $2\sigma$ local depth as an upper limit.
}
\begin{tabular}{l  r r r r r r r}
\hline
ID  &  \multicolumn{1}{c}{$z'$}&   \multicolumn{1}{c}{$Y$} & \multicolumn{1}{c}{$J$} & \multicolumn{1}{c}{$H$} & \multicolumn{1}{c}{$Ks$} & \multicolumn{1}{c}{$3.6\umu$m} & \multicolumn{1}{c}{$4.5\umu$m} \\
\hline

277912  & $  26.6_{-  0.2}^{+  0.3}\phantom{0}$  (4) & $  24.3_{-  0.1}^{+  0.1}$ (16) & $  24.2_{-  0.1}^{+  0.1}$ (10) & $  24.1_{-  0.1}^{+  0.1}$ (10) & $  24.2_{-  0.1}^{+  0.1}\phantom{0}$ (8) & $  23.4_{-  0.2}^{+  0.2}$ (5) & $  23.4_{-  0.2}^{+  0.2}$ (4) \\
155880  & $  26.1_{-  0.1}^{+  0.1}\phantom{0}$  (9) & $  24.5_{-  0.1}^{+  0.1}$ (11) & $  24.5_{-  0.1}^{+  0.1}$ (10) & $  24.6_{-  0.2}^{+  0.2}\phantom{0}$ (6) & $  24.6_{-  0.2}^{+  0.2}\phantom{0}$ (6) & $ >  25.1\phantom{0}$  (1) & $ >  25.1\phantom{0}$ (0) \\
218467  & $ >  27.6\phantom{00}$ (1) & $  25.0_{-  0.2}^{+  0.2}\phantom{0}$ (5) & $  25.0_{-  0.2}^{+  0.2}\phantom{0}$ (5) & $  25.0_{-  0.2}^{+  0.3}\phantom{0}$ (4) & $  24.9_{-  0.2}^{+  0.3}\phantom{0}$ (4) & $  24.7_{-  0.2}^{+  0.3}$ (4) & $ >  25.2\phantom{0}$ (1) \\
61432 & $ >  27.6\phantom{00}$  (1) & $  24.9_{-  0.1}^{+  0.2}\phantom{0}$ (7) & $  24.7_{-  0.2}^{+  0.2}\phantom{0}$ (5) & $  24.6_{-  0.2}^{+  0.3}\phantom{0}$ (4) & $  24.8_{-  0.2}^{+  0.3}\phantom{0}$ (4) & $ >  25.1\phantom{0}$  (0) & $  24.6_{-  0.2}^{+  0.3}$ (4) \\
\hline
277880 & $  26.4_{-  0.2}^{+  0.3}\phantom{0}$  (4) & $  25.0_{-  0.3}^{+  0.4}\phantom{0}$ (3) & $  24.6_{-  0.2}^{+  0.2}\phantom{0}$ (6) & $  24.9_{-  0.3}^{+  0.4}\phantom{0}$ (3) & $  24.8_{-  0.3}^{+  0.4}\phantom{0}$  (3) & $  24.6_{-  0.2}^{+  0.3}$ (4) & $  24.7_{-  0.3}^{+  0.3}$ (3) \\
268511  & $ >  27.6\phantom{00}$  (0) & $  25.0_{-  0.2}^{+  0.3}\phantom{0}$ (3) & $  25.0_{-  0.3}^{+  0.5}\phantom{0}$ (2) & $  25.4_{-  0.4}^{+  0.6}\phantom{0}$ (2) & $ >  25.0\phantom{00}$ (0) & $ >  25.6\phantom{0}$ (1) & $ >  25.2$ \phantom{0}(1) \\
271105  & $  26.1_{-  0.1}^{+  0.1}\phantom{0}$   (7) & $  25.0_{-  0.1}^{+  0.1}\phantom{0}$ (9) & $  24.1_{-  0.1}^{+  0.1}$ (12) & $  23.9_{-  0.1}^{+  0.1}$ (11) & $  24.0_{-  0.1}^{+  0.1}$ (13) & $  23.5_{-  0.2}^{+  0.2}$ (5) & $  23.4_{-  0.2}^{+  0.2}$ (5) \\
\hline
95661  & $  25.4_{-  0.1}^{+  0.1}$  (15) & $  24.8_{-  0.2}^{+  0.2}\phantom{0}$ (5) & $  25.0_{-  0.3}^{+  0.4}\phantom{0}$ (2) & $ >  25.2\phantom{00}$ (1) & $  25.0_{-  0.4}^{+  0.7}\phantom{0}$ (2) & $  24.4_{-  0.2}^{+  0.2}$ (5) & $  23.6_{-  0.2}^{+  0.2}$ (5) \\
28400 & $  25.2_{-  0.1}^{+  0.1}$  (16) & $  24.8_{-  0.1}^{+  0.2}\phantom{0}$ (7) & $  24.6_{-  0.1}^{+  0.1}\phantom{0}$ (8) & $  25.0_{-  0.3}^{+  0.4} \phantom{0}$ (3) & $  25.1_{-  0.3}^{+  0.4}\phantom{0}$ (3) & $  23.5_{-  0.2}^{+  0.2}$ (5) & $  24.9_{-  0.4}^{+  0.6}$ (2) \\
2233  & $  25.9_{-  0.1}^{+  0.1}\phantom{0}$  (9) & $  25.5_{-  0.3}^{+  0.4}\phantom{0}$ (3) & $  25.3_{-  0.3}^{+  0.4}\phantom{0}$ (2) & $ >  25.3\phantom{00}$ (0) & $ >  25.5\phantom{00}$ (1) & $  25.1_{-  0.4}^{+  0.5}$ (2) & $ >  25.4\phantom{0}$ (0) \\

\hline
\end{tabular}
\label{table:photometry}
\end{table*}

\begin{table*}
\caption{The best-fitting photometric redshift parameters and galaxies sizes derived from the improved UltraVISTA DR2 imaging of the ten galaxy candidates presented in~\citet{Bowler2012}.
The photometric redshift is calculated by fitting to all available photometric bands including the IRAC $3.6\umu$m and $4.5 \umu$m filters.
The object 28400 has an unusually blue [$3.6 - 4.5$] colour and hence cannot be fitted well with our templates (that do not include potential nebular emission), and so this object has a large $\chi^2$-value.
Best-fitting redshifts with \Lya emission included are shown in the centre of the table; note that here we do not include the IRAC photometry.
The FWHM values presented on the right-hand-side were calculated using {\sc SExtractor}; missing values indicate that the object was not significantly detected in that band.}
 \begin{tabular}{l c r c c c r c c c c r c c c }
\hline

& \multicolumn{4}{l}{No \Lya} & \multicolumn{5}{l}{With \Lya}  &  \multicolumn{2}{l}{Star} & \multicolumn{3}{c}{FWHM} \\
\hline
    
    ID & $z $ & $\chi^2$ & $A_V$ & ${\rm Z} $& $z$ & $\chi^2$ & ${\rm EW}_0$ & $A_V$& ${\rm Z} $& Stellar & $\chi^2$ & $z'$ & $ Y$ & $J$ \\ 
 & & & /mag & /${\rm Z}_{\sun}$ & & & /\AA & /${\rm Z}_{\sun}$  & & Type & &  \multicolumn{3}{c}{/arcsec} \\
 \hline

 277912 & $ 6.85_{-0.08}^{+0.08}$ &  2.4 &  0.0 & 1.0 & $ 6.84 $ &  2.4 &   0 & 0.0 &   1.0 & T3 &  27.2 &  -  &   1.7 &   1.9  \\
      155880 & $ 6.70_{-0.06}^{+0.05}$ &  4.6 &  0.2 & 1.0 & $ 6.86 $ &  4.4 &  50 & 0.2 &   1.0 & M6 &  28.0 &   1.4 &   1.7 &   2.2  \\
      218467 & $ 6.98_{-0.12}^{+0.12}$ &  3.1 &  0.5 & 1.0 & $ 7.01 $ &  3.0 &  10 & 0.5 &   1.0 & T3 &  23.2 &   1.0 &   1.0 &   1.6  \\
       61432 & $ 7.04_{-0.11}^{+0.16}$ &  5.5 &  0.4 & 1.0 & $ 7.04 $ &  5.5 &   0 & 0.4 &   1.0 & T4 &  25.0 &   1.7 &   1.1 &   1.6  \\
       \hline
      277880 & $ 6.67_{-0.12}^{+0.11}$ &  2.2 &  0.7 & 0.2 & $ 6.66 $ &  2.2 &   0 & 0.7 &   0.2 & T3 &   6.3 &   1.3 &   1.1 &   1.7  \\      
      268511 & $ 7.12_{-0.11}^{+0.14}$ &  2.6 &  0.0 & 0.2 & $ 7.25 $ &  2.2 &  80 & 0.0 &   0.2 & T8 &   9.2 &   1.7 &   1.4 &   0.5  \\
      271105 & $ 6.51_{-0.04}^{+0.05}$ & 15.5 &  1.4 & 1.0 & $ 6.55 $ & 22.5 &   0 & 1.0 &   1.0 & T0 &   2.4 &   1.6 &   1.1 &   0.9  \\
      \hline
       95661 & $ 6.25_{-0.13}^{+0.10}$ &  5.1 &  0.1 & 0.2 & $ 6.31 $ &  4.9 &  20 & 0.0 &   0.2 & M7 &  17.4 &   2.7 &   0.8 &   1.8  \\
       28400 & $ 6.20_{-0.08}^{+0.10}$ & 15.3 &  0.0 & 1.0 & $ 6.59 $ & 13.6 & 150 & 0.0 &   0.2 & M7 &  31.9 &   1.3 &   2.1 &   1.3  \\
        2233 & $ 6.24_{-0.20}^{+0.12}$ &  2.6 &  0.0 & 0.2 & $ 6.30 $ &  2.6 &  20 & 0.0 &   0.2 & M5 &  13.9 &   1.4 &   2.8 &  -   \\
 \hline
\end{tabular}
\end{table*}

\end{document}